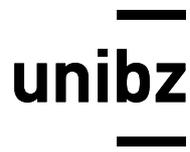

**Freie Universität Bozen**
**Libera Università di Bolzano**
**Università Liedia de Bulsan**

# New Techniques for Graph Edit Distance Computation

*by*

DAVID B. BLUMENTHAL

This dissertation is submitted for the degree of *Doctor of Philosophy* at the *Faculty of Computer Science* of the *Free University of Bozen-Bolzano*.

# Contents

# Acknowledgements

First of all, I would like to thank my Ph. D. supervisor Johann Gamper for his contributions of time, effort, and ideas during the last three years. In particular, I would like to thank him for his invaluable suggestions on how to fine-tune our papers to meet the expectations of the research community and for his trust in my ability to independently pursue my ideas. Moreover, I would like to thank Sébastien Bougleux, Luc Brun, Nicolas Boria, and Évariste Daller for collaborating with me during my stays at the GREYC Research Lab in Digital Sciences in Caen, France. Working with Sébastien, Luc, Nicolas, and Évariste has been truly inspiring and resulted in many ideas which I would not have been able to come up with alone. I would also like to thank the external reviewers Walter G. Kropatsch and Jean-Yves Ramel for their comments on a preliminary version of this thesis, my colleagues from our research group in Bolzano for innumerable helpful discussions and our table football matches after lunch, and my friends and academic advisors back in Berlin, with whom I discovered and deepened my desire to become a professional researcher (although back then I thought that I was going to do a Ph. D. in Philosophy rather than in Computer Science). Last but not least, I would like to thank my family: My sister Mirjam for being sister and best friend in one. My father Andreas for raising me with a love of critical thought and for always encouraging me to follow my passions and interests. My daughters Livia and Romina for countless little moments of joy and happiness. And, most importantly, my wife Clizia for supporting me ever since we met during our Erasmus exchanges in Edinburgh almost nine years ago. Thank you!

<div style="text-align:right">David B. Blumenthal, August 2019</div>

# ABSTRACT

Due to their capacity to encode rich structural information, labeled graphs are often used for modeling various kinds of objects such as images, molecules, and chemical compounds. If pattern recognition problems such as clustering and classification are to be solved on these domains, a (dis-)similarity measure for labeled graphs has to be defined. A widely used measure is the graph edit distance (GED), which, intuitively, is defined as the minimum amount of distortion that has to be applied to a source graph in order to transform it into a target graph. The main advantage of GED is its flexibility and sensitivity to small differences between the input graphs. Its main drawback is that it is hard to compute.

In this thesis, new results and techniques for several aspects of computing GED are presented. Firstly, theoretical aspects are discussed: competing definitions of GED are harmonized, the problem of computing GED is characterized in terms of complexity, and several reductions from GED to the quadratic assignment problem (QAP) are presented. Secondly, solvers for the linear sum assignment problem with error-correction (LSAPE) are discussed. LSAPE is a generalization of the well-known linear sum assignment problem (LSAP), and has to be solved as a subproblem by many GED algorithms. In particular, a new solver is presented that efficiently reduces LSAPE to LSAP. Thirdly, exact algorithms for computing GED are presented in a systematic way, and improvements of existing algorithms as well as a new mixed integer programming (MIP) based approach are introduced. Fourthly, a detailed overview of heuristic algorithms that approximate GED via upper and lower bounds is provided, and eight new heuristics are described. Finally, a new easily extensible C++ library for exactly or approximately computing GED is presented.

— 1 —

# Introduction

## 1.1 Background

Labeled graphs can be used for modeling various kinds of objects, such as chemical compounds, images, molecular structures, and many more. Because of this, labeled graphs have received increasing attention over the past years. One task researchers have focused on is the following: Given a database $\mathcal{G}$ that contains labeled graphs from a domain $\mathbb{G}$, find all graphs $G \in \mathcal{G}$ that are sufficiently similar to a query graph $H$ or find the $k$ graphs from $\mathcal{G}$ that are most similar to $H$ [35, 47, 107]. Being able to quickly and precisely answer graph similarity queries of this kind is crucial for the development of performant pattern recognition techniques in various application domains [104], such as keyword spotting in handwritten documents [103] and cancer detection [81].

The task of answering graph similarity queries can be addressed in several ways [47]. The straightforward approach is to define a graph (dis-)similarity measure $f : \mathbb{G} \times \mathbb{G} \to \mathbb{R}$. Subsequently, $f$ can be used for answering the graph similarity queries in the graph space, e. g., via techniques such as the $k$-nearest neighbors algorithm. If $f$ is defined as a graph kernel, i. e., if $f$ is symmetric and positive semi-definite, more advanced pattern recognition techniques such as support vector machines and principal component analysis can be used, too [51, 78, 79]. An alternative approach consists in defining a graph embedding $g : \mathbb{G} \to \mathbb{R}^d$ that maps graphs to multidimensional real-valued vectors and then answering the graph similarity queries via vector matching techniques [30, 72, 73, 84–86, 109, 110].

A very flexible and therefore widely used graph dissimilarity measure is the graph edit distance (GED) [29, 96]. GED is defined as the minimum cost of an edit path between two graphs. An edit path between graphs $G$ and $H$ is a sequence of edit operations that transforms $G$ into $H$. There are six edit operations, namely, node insertion, deletion, and substitution as well as edge insertion, deletion, and substitution. Each edit operation comes with an associated non-negative edit cost, and the cost of an edit path is defined as the sum of the costs of its edit operations. The disadvantage of GED is that it is $\mathcal{NP}$-hard to compute [111]; its advantage is that it is very sensitive to small differences between the input graphs. GED is therefore mainly used for rather small graphs, where information that is disregarded by rougher dissimilarity measures is crucial [104]. It can be employed either as a stand-alone graph dissimilarity measure, or as a building block for graph kernels [51, 78, 79] or graph embeddings [30, 84–86].

Another problem that can be addressed by computing GED is the error-correcting or inexact graph matching problem [28, 29, 47]. This problem problem asks to align the nodes and edges of two input graphs $G$ and $H$, while allowing that some nodes and edges may be inserted or deleted. It can be solved by exactly or approximately computing GED, because edit paths correspond to error-correcting graph matchings (cf. Section 2.1).

Research on GED has been conducted both in the database and in the pattern recognition community, although the two communities use slightly different problem definitions. While some exact algorithms are available [3, 4, 42, 52, 60, 68, 69, 89], in practice, these algorithms do not scale well and can hence only be used to compute GED on very small graphs. Research has therefore mainly focused on the task of designing heuristics that approximate GED via lower or upper bounds. These heuristics use various techniques such as transformations to the linear sum assignment problem with error-correction (LSAPE) [32, 40, 50, 60, 83, 88, 111, 113, 114], local search [22, 25, 44, 92, 111], linear programming [42, 60, 68, 69], and reductions to the quadratic assignment problem (QAP) [22, 25].

## 1.2　Contributions and Organization

In this thesis, new techniques for exactly and heuristically computing GED are presented. More specifically, it contains the following core contributions:

– We demonstrate that computing and approximating GED is $\mathcal{NP}$-hard

  even on very sparse graphs, harmonize the GED definitions used in the database and in the pattern recognition communities, and show how to reduce quasimetric GED to compact instances of QAP.
- We suggest `FLWC`, an efficient, generally applicable, and easily implementable LSAPE solver.
- We speed-up and generalize the existing exact algorithms `A*`, `DFS-GED`, and `CSI-GED` and suggest `COMPACT-MIP`, a new compact mixed integer linear (MIP) formulation of the problem of computing GED which is smaller than all existing MIP formulations.
- We present the new LSAPE based heuristics `BRANCH`, `BRANCH-FAST`, `RING`, and `RING-ML`, as well as the post-processing technique `MULTI-SOL` for tightening the upper bounds computed by all LSAPE based heuristics. While `BRANCH` and `BRANCH-FAST` yield excellent tradeoffs between runtime and accuracy of the produced lower bounds, `RING` outperforms all existing LSAPE based heuristics in terms of tightness of the produced upper bounds.
- We propose `BRANCH-TIGHT`, an anytime algorithm which computes the tightest available lower bounds in settings where editing edges is more expensive than editing nodes.
- We present the local search algorithm `K-REFINE` and the framework `RANDPOST` for stochastically generating promising initial solution to be used by local search algorithms. On small graphs, `K-REFINE` is as accurate as but much faster than the most accurate existing local search algorithms; on larger graphs, it yields an excellent tradeoff between runtime and tightness of the produced upper bounds. `RANDPOST` is particularly effective on larger graphs, where it significantly tightens the upper bounds of all local search algorithms.
- We present GEDLIB, an easily extensible C++ library for exactly or heuristically computing GED.

  The thesis is divided into one preliminary chapter, four main chapters, one concluding chapter, and one appendix. In the following paragraphs, we provide concise summaries of all chapters and of the appendix.

**Chapter 2 – Preliminaries.** In this chapter, concepts, notations, and datasets are introduced that are used throughout the thesis.

**Chapter 3 – Theoretical Aspects.** In this chapter, it is shown that computing and approximating GED is $\mathcal{NP}$-hard not only on general, but also on very sparse graphs, namely, unlabeled graphs with maximum degree two. Furthermore, the equivalence of two competing definitions of GED is established, which are used, respectively, in the database and in the pattern recognition community. Finally, a new compact reduction from GED to the quadratic assignment problem (QAP) for quasimetric edit costs is presented. Unlike existing reductions to QAP, the newly proposed reduction yields small instances of the standard version of QAP, and hence makes a wide range of methods that were originally designed for QAP available for efficiently computing GED.

**Chapter 4 – The Linear Sum Assignment Problem with Error-Correction.** LSAPE is a polynomially solvable combinatorial optimization problem, which occurs as a subproblem in many exact and heuristic algorithms for GED. In this chapter, the new LSAPE solver `FLWC` (fast solver for LSAPE without cost constraints) is presented and evaluated empirically. `FLWC` is the first available algorithm that works for general instances of LSAPE and is both efficient and easily implementable.

**Chapter 5 – Exact Algorithms.** In this chapter, a systematic overview of algorithms for exactly computing GED is provided. Existing exact algorithms are either modeled as node based tree search algorithms, as edge based tree search algorithms designed for constant, triangular node edit costs, or as algorithms based on mixed integer linear programming (MIP). Furthermore, we show how to speed-up the existing node based tree search algorithms `A*` and `DFS-GED` for constant, triangular edit costs, and generalize the existing edge based tree search algorithm `CSI-GED` to arbitrary edit costs. We also present a new MIP formulation `COMPACT-MIP` of GED, which is smaller than all existing MIP formulations. Finally, the newly proposed techniques are evaluated empirically. Note that, as the problem of computing GED is $\mathcal{NP}$-hard, in practice, all exact algorithms only work on very small graphs.

**Chapter 6 – Heuristic Algorithms.** In this chapter, we provide a systematic overview of algorithms that heuristically approximate GED via lower or upper bounds. Whenever possible, the presented heuristics are modeled as transformations to LSAPE, variants of local search, or linear programming (LP) based algorithms. Moreover, we present two new LSAPE based algorithms `BRANCH` and `BRANCH-FAST` for lower bounding GED which assume different positions in the range between fast-but-rather-loose and tight-but-rather-slow, as well as the anytime algorithm algorithm `BRANCH-TIGHT` which further improves the lower bound computed by `BRANCH`. We also present the LSAPE based heuristics `RING` and `RING-ML` for upper bounding GED, the post-processing technique `MULTI-SOL` for tightening LSAPE based upper bounds, the new local search based heuristic `K-REFINE`, and the framework `RANDPOST` for generating good initial solutions for local search based heuristics. Extensive experiments are carried out to test the newly proposed techniques and to compare them with a wide range of competitors. The experiments show that all newly proposed techniques significantly improve the state of the art.

**Chapter 7 – Conclusions and Future Work.** In this chapter, we summarize the thesis' main contributions and point out to future work.

**Appendix A – GEDLIB: A C++ Library for Graph Edit Distance Computation.** In this appendix, we present GEDLIB, an easily extensible C++ library for exact and approximate GED computation. GEDLIB facilitates the implementation of competing GED algorithm within the same environment and hence fosters fair empirical comparisons. To the best of our knowledge, no currently available software is designed for this purpose. GEDLIB is freely available on GitHub.

## 1.3 Publications

The thesis extends and consolidates the following articles:

**Journal Publications**

– D. B. Blumenthal, N. Boria, J. Gamper, S. Bougleux, and L. Brun, "Comparing heuristics for graph edit distance computation", *VLDB J.*, 2019, in press. DOI: `10.1007/s00778-019-00544-1`
– D. B. Blumenthal and J. Gamper, "On the exact computation of the graph edit distance", *Pattern Recognit. Lett.*, 2018, in press. DOI: `10.1016/j.patrec.2018.05.002`

**International Conference and Workshop Publications**

**Submitted Manuscripts**

# — 2 —
# Preliminaries

In this chapter, we introduce the most important concepts and notations, and provide an overview of the datasets we used to empirically evaluate the techniques that are proposed in this thesis. In Section 2.1, we provide two definitions of GED that are used, respectively, in the database and in the pattern recognition community. In Section 2.2, we introduce miscellaneous concepts and notations that are used throughout the thesis. In Section 2.3, we introduce the linear sum assignment problem with and without error-correction (LSAPE and LSAP). LSAPE is a polynomially solvable combinatorial optimization problem that has to be solved as a subproblem by many exact and heuristics algorithms for GED. In Section 2.4, we present the test datasets.

## 2.1 Two Definitions of GED

Since the graphs for which GED based methods are applied are mostly undirected [2, 87, 104], most heuristics for GED are presented for undirected labeled graphs, although they can usually be easily generalized to directed graphs. For the ease of presentation, we restrict to undirected graphs also in this thesis.

**Definition 2.1 (Graph).** An *undirected labeled graph* $G$ is a 4-tuple $G = (V^G, E^G, \ell_V^G, \ell_E^G)$, where $V^G$ and $E^G$ are sets of nodes and edges, $\Sigma_V$ and $\Sigma_E$ are label alphabets, and $\ell_V^G : V^G \to \Sigma_V$, $\ell_E^G : E^G \to \Sigma_E$ are labeling functions. The *dummy symbol* $\epsilon$ denotes dummy nodes and edges as well as their labels.

Throughout the thesis, it is tacitly assumed that the label alphabets $\Sigma_V$ and $\Sigma_E$ are fixed. With this assumption, we introduce $\mathbb{G} := \{G \mid \ell_V^G[V^G] \subseteq$ $\Sigma_V \wedge \ell_E^G[E^G] \subseteq \Sigma_E\}$ as the domain of graphs with node labels from $\Sigma_V$ and edge labels from $\Sigma_E$. All graphs considered in this thesis are elements of $\mathbb{G}$.

Table 2.1. Edit operations and edit costs.

| edit operations | edit costs |
| --- | --- |
| substitute $\alpha$-labeled node by $\alpha'$-labeled node | $c_V(\alpha, \alpha')$ |
| delete isolated $\alpha$-labeled node | $c_V(\alpha, \epsilon)$ |
| insert isolated $\alpha$-labeled node | $c_V(\epsilon, \alpha)$ |
| substitute $\beta$-labeled edge by $\beta'$-labeled edge | $c_E(\beta, \beta')$ |
| delete $\beta$-labeled edge | $c_E(\beta, \epsilon)$ |
| insert $\beta$-labeled edge between existing nodes | $c_E(\epsilon, \beta)$ |

**Definition 2.2 (Edit Cost Function).** A function $c_V : \Sigma_V \cup \{\epsilon\} \times \Sigma_V \cup \{\epsilon\} \to \mathbb{R}_{\geq 0}$ is called *node edit cost function for* $\mathbb{G}$, if and only if $c_V(\alpha, \alpha) = 0$ holds for all $\alpha \in \Sigma_V \cup \{\epsilon\}$. A function $c_E : \Sigma_E \cup \{\epsilon\} \times \Sigma_E \cup \{\epsilon\} \to \mathbb{R}_{\geq 0}$ is called *edge edit cost function for* $\mathbb{G}$, if and only if $c_E(\beta, \beta) = 0$ holds for all $\beta \in \Sigma_E \cup \{\epsilon\}$. A node edit cost function $c_V$ is called *constant* just in case there are constants $c_V^{\text{sub}}, c_V^{\text{del}}, c_V^{\text{ins}} \in \mathbb{R}$ such that $c_V(\alpha, \alpha') = c_V^{\text{sub}}$, $c_V(\alpha, \epsilon) = c_V^{\text{del}}$, and $c_V(\epsilon, \alpha') = c_V^{\text{ins}}$ holds for all $(\alpha, \alpha') \in \Sigma_V \times \Sigma_V$ with $\alpha \neq \alpha'$; and *triangular* or *quasimetric* just in case $c_V(\alpha, \alpha') \leq c(\alpha, \epsilon) + c(\epsilon, \alpha')$ holds for all $(\alpha, \alpha') \in \Sigma_V \times \Sigma_V$. Constant and triangular edge edit cost functions are define analogously. Edit cost functions $c_V$ and $c_E$ are called *uniform* if and only if both of them are constant and, additionally, there is a constant $c \in \mathbb{R}$ such that $c = c_V^{\text{sub}} = c_V^{\text{del}} = c_V^{\text{ins}} = c_E^{\text{sub}} = c_E^{\text{del}} = c_E^{\text{ins}}$.

Given two graphs $G, H \in \mathbb{G}$ and edit cost functions $c_V$ and $c_E$, $\text{GED}(G, H)$ is defined as the minimum cost of an *edit path* between $G$ and $H$. An edit path $P$ between two graphs $G, H \in \mathbb{G}$ is a sequence $P = (o_i)_{i=1}^r$ of *edit operations* that satisfies $(o_r \circ \ldots \circ o_1)(G) \simeq H$, i.e., transforms $G$ into a graph $H'$ which is isomorphic to $H$. There are six different edit operations (deleting, inserting, and substituting nodes or edges), whose associated *edit costs* are defined in terms of $c_V$ and $c_E$, as detailed in Table 2.1. The cost $c(P)$ of an edit path $P = (o_i)_{i=1}^r$ is defined as the sum $c(P) := \sum_{i=1}^r c(o_i)$ of its contained edit operations. We can now give a first, intuitive definition of GED.

**Definition 2.3 (GED – First Definition).** The *graph edit distance* (GED) between two graphs $G, H \in \mathbb{G}$ is defined as $\text{GED}(G, H) := \min\{c(P) \mid P \in \Psi(G, H)\}$, where $\Psi(G, H)$ is the set of all edit paths between $G$ and $H$.

This definition of GED was proposed in [60] and is now used in the database community. Note that it slightly differs from the original definition of GED due to Bunke and Allermann [29], which is employed in the pattern recognition community. This is because Bunke and Allermann require edit paths between graphs $G$ and $H$ to transform $G$ into a graph that is identical to $H$, and only allow edit operations that only involve nodes and edges contained in $G$ or $H$.

Definition 2.3 is very intuitive but unsuited for algorithmic purposes. This is because, firstly, there are infinitely many edit paths between two graphs $G$ and $H$. Secondly, there is no known polynomial algorithm for deciding whether two graphs are isomorphic [6], which implies that edit paths cannot be recognized as such in polynomial time. Therefore, all existing exact or approximate algorithms for GED restrict their attention to those edit paths that are induced by a node map between $G$ and $H$. Intuitively, a node map $\pi$ between $G$ and $H$ is a relation $\pi \subseteq (V^G \cup \{\epsilon\}) \times (V^H \cup \{\epsilon\})$ that covers all nodes $u \in V^G$ and $v \in V^H$ exactly once. Clearly, there are only finitely many node maps between $G$ and $H$. This implies that GED is effectively computable, if it can be shown that it suffices to consider edit paths induced by node maps.

**Definition 2.4 (Node Map).** Let $G, H \in \mathbb{G}$ be two graphs. A relation $\pi \subseteq (V^G \cup \{\epsilon\}) \times (V^H \cup \{\epsilon\})$ is called *node map* between $G$ and $H$ if and only if $|\{v \mid v \in (V^H \cup \{\epsilon\}) \wedge (u,v) \in \pi\}| = 1$ holds for all $u \in V^G$ and $|\{u \mid u \in (V^G \cup \{\epsilon\}) \wedge (u,v) \in \pi\}| = 1$ holds for all $v \in V^H$. We write $\pi(u) = v$ just in case $(u,v) \in \pi$ and $u \neq \epsilon$, and $\pi^{-1}(v) = u$ just in case $(u,v) \in \pi$ and $v \neq \epsilon$. $\Pi(G, H)$ denotes the set of all node maps between $G$ and $H$. For edges $e = (u_1, u_2) \in E^G$ and $f = (v_1, v_2) \in E^H$, we introduce the short-hand notations $\pi(e) := (\pi(u_1), \pi(u_2))$ and $\pi^{-1}(f) := (\pi^{-1}(v_1), \pi^{-1}(v_2))$.

A node map $\pi \in \Pi(G, H)$ specifies for all nodes and edges of $G$ and $H$ whether they are substituted, deleted, or inserted. Table 2.2 details these edit operations and introduces short-hand notations for their edit costs.

**Definition 2.5 (Induced Edit Path).** Let $G, H \in \mathbb{G}$ be graphs, $\pi \in \Pi(G, H)$ be a node map between them, and $O$ be the set of $\pi$'s induced edit operations as specified in Table 2.2. Then an ordering $P_\pi := (o_i)_{i=0}^{|O|-1}$ of $O$ is called *induced edit path* of the node map $\pi$ if and only if edge deletions come first and edge insertions come last, i.e., if there are indices $i_1$ and $i_2$ such that $o_i$ is an edge deletion just in case $0 \leq i \leq i_1$ and $o_i$ is an edge insertion just in case $i_2 \leq i \leq |O|$.

**Table 2.2.** Edit operations and edit costs induced by node map $\pi \in \Pi(G, H)$; $u \in V^G$ and $v \in V^H$ are nodes, $e \in E^G$ and $f \in E^H$ are edges.

| case | edit operations | short-hand notation for edit costs |
|---|---|---|
| *node edit operations* | | |
| $\pi(u) = v$ | substitute node $u$ by node $v$ | $c_V(u, v) := c_V(\ell_V^G(u), \ell_V^H(v))$ |
| $\pi(u) = \epsilon$ | delete node $u$ | $c_V(u, \epsilon) := c_V(\ell_V^G(u), \epsilon)$ |
| $\pi^{-1}(v) = \epsilon$ | insert node $v$ | $c_V(\epsilon, v) := c_V(\epsilon, \ell_V^H(v))$ |
| *edge edit operations* | | |
| $\pi(e) = f$ | substitute edge $e$ by edge $f$ | $c_E(e, f) := c_E(\ell_E^G(e), \ell_E^H(f))$ |
| $\pi(e) \notin E^H$ | delete edge $e$ | $c_E(e, \epsilon) := c_E(\ell_E^G(e), \epsilon)$ |
| $\pi^{-1}(f) \notin E^G$ | insert edge $f$ | $c_E(\epsilon, f) := c_E(\epsilon, \ell_E^H(f))$ |

It has been shown that induced edit paths are indeed edit paths, i.e., that $P_\pi \in \Psi(G, H)$ holds for all $\pi \in \Pi(G, H)$ [22]: starting the edit path with the edge deletions ensures that nodes are isolated when they are deleted; ending it with the edge insertions ensures that edges are inserted only between existing nodes. The cost $c(P_\pi)$ of an edit path $P_\pi$ induced by a node map $\pi \in \Pi(G, H)$ is given as follows:

$$c(P_\pi) = \underbrace{\sum_{\substack{u \in V^G \\ \pi(u) \neq \epsilon}} c_V(u, \pi(u))}_{\text{node substitutions}} + \underbrace{\sum_{\substack{u \in V^G \\ \pi(u) = \epsilon}} c_V(u, \epsilon)}_{\text{node deletions}} + \underbrace{\sum_{\substack{v \in V^H \\ \pi^{-1}(v) = \epsilon}} c_V(\epsilon, v)}_{\text{node insertions}} \quad (2.1)$$

$$+ \underbrace{\sum_{\substack{e \in E^G \\ \pi(e) \neq \epsilon}} c_E(e, \pi(e))}_{\text{edge substitutions}} + \underbrace{\sum_{\substack{e \in E^G \\ \pi(e) = \epsilon}} c_E(e, \epsilon)}_{\text{edge deletions}} + \underbrace{\sum_{\substack{f \in E^H \\ \pi^{-1}(f) = \epsilon}} c_E(\epsilon, f)}_{\text{edge insertions}}$$

Note that, by Definition 2.5, a node map $\pi$ generally induces more than one edit path. However, all of these edit paths are equivalent, as they differ only w.r.t. the ordering of the contained edit operations. Throughout the thesis, we will therefore identify all edit paths induced by $\pi$. We can now give an alternative definition of GED.

**Definition 2.6 (GED – Second Definition).** The *graph edit distance* (GED) between two graphs $G, H \in \mathbb{G}$ is defined as $\text{GED}(G, H) := \min\{c(P_\pi) \mid \pi \in \Pi(G, H)\}$, where $P_\pi \in \Psi(G, H)$ is the edit path induced by the node map $\pi$.

In [21, 22], it is shown that Definition 2.6 is equivalent to the original GED definition due to Bunke and Allermann [29], which is used in the pattern recognition community. Furthermore, it has been demonstrated in [60] that, if the underlying edit cost functions $c_V$ and $c_E$ are metrics, Definition 2.6 is equivalent to Definition 2.3, which is used in the database community. In Section 3.4 below, we show that Definition 2.3 and Definition 2.6 are equivalent also for general edit costs, and hence harmonize the GED definitions used in the different research communities.

Also note that Definition 2.6 provides a direct link between GED and the error-correcting graph matching problem mentioned in Section 1.1. Since each node map specifies for all nodes and edges whether they are substituted, deleted, or inserted (cf. Table 2.2), node maps can be interpreted as feasible solutions to the error-correcting graph matching problem. An optimal node map $\pi \in \Pi(G, H)$ with $c(P_\pi) = \text{GED}(G, H)$ is hence an optimal solution to the error-correcting graph matching problem w. r. t. the edit costs $c_V$ and $c_E$. Or put less formally: $\pi$ aligns the input graphs' nodes and edges as conservatively as possible.

**Example 2.1 (Illustration of Basic Definitions).** Consider the graphs $G$ and $H$ shown in Figure 2.1a. $G$ and $H$ are taken from the LETTER (H) dataset and represent distorted letter drawings [87]. Their nodes are labeled with two-dimensional, non-negative Euclidean coordinates. Edges are unlabeled. Hence, we have $\Sigma_V = \mathbb{R}_{\geq 0} \times \mathbb{R}_{\geq 0}$ and $\Sigma_E = \{1\}$. In [84], it is suggested that the edit cost functions $c_V$ and $c_E$ for LETTER (H) should be defined as follows: $c_E(1, \epsilon) := c_E(\epsilon, 1) := 0.425$, $c_V(\alpha, \alpha') := 0.75 \|\alpha - \alpha'\|_2$, and $c_V(\alpha, \epsilon) := c_V(\epsilon, \alpha) := 0.675$ for all node labels $\alpha, \alpha' \in \Sigma_V$. Now consider the node map $\pi \in \Pi(G, H)$ also shown in Figure 2.1a. Its induced edit operations and edit costs are detailed in Figure 2.1b. By summing the induced edit costs, we obtain that $\pi$'s induced edit path $P_\pi \in \Psi(G, H)$ has cost $c(P_\pi) = 2.623179$, which implies $\text{GED}(G, H) \leq 2.623179$.

## 2.2 Miscellaneous Definitions

The following Definitions 2.7 to 2.24 introduce concepts and notations in alphabetic order that are used throughout the thesis.

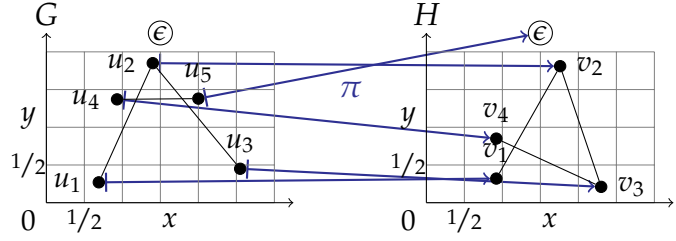

**(a)** Graphs $G$ and $H$ and the node map $\pi$ used in Example 2.1.

| edit operations | edit costs |
|---|---|
| *node edit operations* | |
| substitute node $u_1$ by node $v_1$ | $c_V(u_1, v_1) = 0.75 \left\| \begin{bmatrix} 0.69 \\ 0.27 \end{bmatrix} - \begin{bmatrix} 0.92 \\ 0.32 \end{bmatrix} \right\|_2$ |
| substitute node $u_2$ by node $v_2$ | $c_V(u_2, v_2) = 0.75 \left\| \begin{bmatrix} 1.40 \\ 1.85 \end{bmatrix} - \begin{bmatrix} 1.76 \\ 1.81 \end{bmatrix} \right\|_2$ |
| substitute node $u_3$ by node $v_3$ | $c_V(u_3, v_3) = 0.75 \left\| \begin{bmatrix} 2.55 \\ 0.45 \end{bmatrix} - \begin{bmatrix} 2.30 \\ 0.21 \end{bmatrix} \right\|_2$ |
| substitute node $u_4$ by node $v_4$ | $c_V(u_4, v_4) = 0.75 \left\| \begin{bmatrix} 0.93 \\ 1.37 \end{bmatrix} - \begin{bmatrix} 0.92 \\ 0.85 \end{bmatrix} \right\|_2$ |
| delete node $u_5$ | $c_V(u_5, \epsilon) = 0.675$ |
| *edge edit operations* | |
| substitute edge $(u_1, u_2)$ by edge $(v_1, v_2)$ | $c_E((u_1, u_2), (v_1, v_2)) = 0$ |
| substitute edge $(u_2, u_3)$ by edge $(v_2, v_3)$ | $c_E((u_2, u_3), (v_2, v_3)) = 0$ |
| delete edge $(u_4, u_5)$ | $c_E((u_4, u_5), \epsilon) = 0.425$ |
| insert edge $(v_3, v_4)$ | $c_E(\epsilon, (v_3, v_4)) = 0.425$ |

**(b)** Edit operations and edit costs induced by node map $\pi$ shown in Figure 2.1a, given the edit cost functions defined in Example 2.1.

**Figure 2.1.** Illustration of basic definitions.

**Definition 2.7 (Constants $\Delta_{\min}^{G,H}$ and $\Delta_{\max}^{G,H}$).** Let $G, H \in \mathbb{G}$ be graphs. The constants $\Delta_{\min}^{G,H}$ and $\Delta_{\max}^{G,H}$ are defined as $\Delta_{\min}^{G,H} := \min\{\max\deg(G), \max\deg(H)\}$ and $\Delta_{\max}^{G,H} := \max\{\max\deg(G), \max\deg(H)\}$.

**Definition 2.8 (Error-Correcting Matching).** For all $n, m \in \mathbb{N}$, the set of all *error-correcting matchings* between $[n]$ and $[m]$ is defined as

$$\Pi_{n,m,\epsilon} := \{\mathbf{X} \in \{0,1\}^{(n+1)\times(m+1)} \mid \mathbf{X}\mathbf{1}_{m+1} = (\mathbf{1}_n^\mathsf{T}, \star)^\mathsf{T} \wedge \mathbf{1}_{n+1}^\mathsf{T}\mathbf{X} = (\mathbf{1}_m^\mathsf{T}, \star)\},$$

where $\star$ is a placeholder that may be substituted by any natural number. The relational representation $\pi \subseteq [n+1] \times [m+1]$ of an error-correcting matching $\mathbf{X} \in \Pi_{n,m,\epsilon}$ is defined as $\pi := \{(i, k) \in [n+1] \times [m+1] \mid x_{i,k} = 1\}$. Depending on the context, we identify $\Pi_{n,m,\epsilon}$ with the set of all error-correcting matchings in relational representation.

**Definition 2.9 (Graph Diameter).** The *diameter of a graph* $G \in \mathbb{G}$ is defined as $\mathrm{diam}(G) := \max_{u \in V^G} \mathrm{e}^G(u)$.

**Definition 2.10 (Index Set $[n]$).** For all $n \in \mathbb{N}$, the index set $[n] \subset \mathbb{N}$ is defined as $[n] := \{i \in \mathbb{N} \mid 1 \leq i \leq n\}$.

**Definition 2.11 (Indicator Function $\delta$).** The indicator function $\delta : \{\texttt{false}, \texttt{true}\} \to \{0, 1\}$ is defined as $\delta_{\texttt{true}} := 1$ and $\delta_{\texttt{false}} := 0$.

**Definition 2.12 (Induced Subgraph).** Let $G \in \mathbb{G}$ be a graph and $V' \subseteq V^G$ be a subset of $G$'s nodes. Then $G[V'] := (V', E^G \cap (V' \times V'), \ell_V^G, \ell_E^G)$ denotes $G$'s induced subgraph on $V'$.

**Definition 2.13 (Matrices $\mathbf{1}_{n,m}$ and $\mathbf{0}_{n,m}$, Vectors $\mathbf{1}_n$ and $\mathbf{0}_n$).** For all $n, m \in \mathbb{N}$, $\mathbf{1}_{n,m}$ denotes a matrix of ones of size $n \times m$, and $\mathbf{1}_n$ denotes a column vector of ones of size $n$. The matrix $\mathbf{0}_{n,m}$ and the column vector $\mathbf{0}_n$ are defined analogously.

**Definition 2.14 (Maximum Degree).** The *maximum degree of a graph $G \in \mathbb{G}$* is defined as $\max \deg(G) := \max_{u \in V^G} \deg^G(u)$.

**Definition 2.15 (Maximum Matching).** For all $n, m \in \mathbb{N}$, the set of all *maximum matchings* between $[n]$ and $[m]$ is defined as follows:

$$\Pi_{n,m} := \{\mathbf{X} \in \{0,1\}^{n \times m} \mid \mathbf{X}\mathbf{1}_m \leq \mathbf{1}_n \wedge \mathbf{1}_n^\mathsf{T} \mathbf{X} \leq \mathbf{1}_m^\mathsf{T} \wedge \sum_{i=1}^n \sum_{k=1}^m x_{i,k} = \min\{n,m\}\}$$

The relational representation $\pi \subseteq [n] \times [m]$ of a maximum matching $\mathbf{X} \in \Pi_{n,m}$ is defined as $\pi := \{(i,k) \in [n] \times [m] \mid x_{i,k} = 1\}$. Depending on the context, we identify $\Pi_{n,m}$ with the set of all maximum matchings in relational representation.

**Definition 2.16 (Multiset Intersection Operator $\Gamma$).** Let $A$ and $B$ be multisets and $c^{\text{sub}}, c^{\text{del}}, c^{\text{ins}} \in \mathbb{R}$ be constants. Then the operator $\Gamma$ is defined as $\Gamma(A, B, c^{\text{sub}}, c^{\text{del}}, c^{\text{ins}}) := c^{\text{sub}}(\min\{|A|, |B|\} - |A \cap B|) + c^{\text{del}} \max\{|A| - |B|, 0\} + c^{\text{ins}} \max\{|B| - |A|, 0\}$.

**Definition 2.17 (Node Degree).** Let $G \in \mathbb{G}$ be a graph. The *degree of a node* $u \in V^G$ in $G$ is defined as $\deg^G(u) := |N^G(u)|$.

**Definition 2.18 (Node Distance).** Let $G \in \mathbb{G}$ be a graph. The *distance between two nodes $u, u' \in V^G$ in $G$*, is defined as $\mathrm{d}^G(u, u') := 0$, if $u = u'$, as $\mathrm{d}^G(u, u') := \min\{|P| \mid P \text{ is path between } u \text{ and } u'\}$, if $u \neq u'$ and $u$ and $u'$ are in the same connected component of $G$, as $\mathrm{d}^G(u, u') := \infty$, if $u$ and $u'$ are in different connected components of $G$.

**Definition 2.19 (Node Eccentricity).** Let $G \in \mathbb{G}$ be a graph. The *eccentricity of a node* $u \in V^G$ in $G$ is defined as $e^G(u) := \max_{u' \in V^G} d^G(u, u')$.

**Definition 2.20 (Node Incidences).** Let $G \in \mathbb{G}$ be a graph. The *set of edges that are incident with a node* $u \in V^G$ in $G$ is denoted as $E^G(u) := \{(u, u') \in E^G \mid u' \in N^G(u)\}$.

**Definition 2.21 (Node Neighborhood).** Let $G \in \mathbb{G}$ be a graph. The $k^{th}$ *neighborhood of a node* $u \in V^G$ in $G$ is defined as $N_k^G(u) := \{u' \in V^G \mid d^G(u, u') = k\}$. The 1$^{st}$ neighborhood is called *neighborhood of node u* and abbreviated as $N^G(u) := N_1^G(u)$.

**Definition 2.22 (Path).** Let $G \in \mathbb{G}$ be a graph. A walk between two nodes $u, u' \in V^G$ is called *path between u and u'*, if and only if no node is encountered more than once.

**Definition 2.23 (Set Image and Multiset Image).** Let $f : X \to Y$ be a function and $A \subseteq X$ be a subset of its domain. The *image of A under f* is denoted by $f[A]$, and the *multiset image of A under f* is denoted by $f[\![A]\!]$. For $f[X]$, we also write $\text{img}(f)$.

**Definition 2.24 (Walk).** Let $G \in \mathbb{G}$ be a graph. An edge sequence $((u_{i_1}, u_{i_2}))_{i=1}^{k}$, $(u_{i_1}, u_{i_2}) \in E^G$ for all $i \in [k]$, is called *walk of length k* between the nodes $u_{i_1}, u_{k_2} \in V^G$, if and only if $u_{i_2} = u_{i+1_1}$ holds for all $i \in [k-1]$.

## 2.3 Definitions of LSAP and LSAPE

The *linear sum assignment problem* (LSAP), also known as minimum cost maximum matching problem in bipartite graphs, is a classical combinatorial optimization problem [62, 77]. LSAP is defined on complete bipartite graphs $K_{n,m} = (U, V, U \times V)$, where $U := \{u_i \mid i \in [n]\}$ and $V := \{v_k \mid k \in [m]\}$ are sets of nodes. An edge set $\pi \subseteq U \times V$ is called a *maximum matching* for $K_{n,m}$ if and only if $|\pi| = \min\{n, m\}$ and all nodes in $U$ in $V$ are incident with at most one edge in $\pi$. A maximum matching $\pi$ can be encoded with a binary matrix $\overline{\mathbf{X}} \in \Pi_{n,m}$ by setting $\overline{x}_{i,j} := 1$ just in case $(u_i, v_k) \in \pi$ (cf. Definition 2.15 above).

**Definition 2.25 (LSAP).** Given a cost matrix $\overline{\mathbf{C}} \in \mathbb{R}^{n \times m}$, LSAP consists in the task of finding a maximum matching $\overline{\mathbf{X}}^\star \in \arg\min_{\overline{\mathbf{X}} \in \Pi_{n,m}} \overline{\mathbf{C}}(\overline{\mathbf{X}})$, where $\overline{\mathbf{C}}(\overline{\mathbf{X}}) := \sum_{i=1}^{n} \sum_{k=1}^{m} \overline{c}_{i,k} \overline{x}_{i,k}$.

LSAP can be solved in polynomial time and space with several algorithms. For instance, if the cost matrix $\overline{\mathbf{C}}$ is balanced, i.e., if $n = m$, it can be solved in $O(n^3)$ time and $O(n^2)$ space with the Hungarian Algorithm [62, 77]. For the unbalanced case, it can be solved in $O(n^2 m)$ time [26]. See [31, 63] for more details.

Given a maximum matching $\overline{\mathbf{X}}$ for $K_{n,m} = (U, V, U \times V)$ and a cost matrix $\overline{\mathbf{C}} \in \mathbb{R}^{n \times m}$, a cell $\overline{x}_{i,j}$ of $\overline{\mathbf{X}}$ with $\overline{x}_{i,j} = 1$ can be interpreted as a substitution of the node $u_i \in U$ by the node $v_j \in V$ with associated substitution cost $\overline{c}_{i,j}$. LSAP can hence be viewed as the task to substitute all the nodes of $U$ by pairwise distinct nodes of $V$, such that the substitution cost is minimized. Note that, under this interpretation, LSAP does not require all nodes in $U$ and $V$ to be taken care of, as, if $n \neq m$, there are always nodes that are not substituted.

There are scenarios such as approximation of GED, where one is faced with a slightly different matching problem: Firstly, in addition to node substitutions, one also wants to allow node insertions and node deletions. Secondly, one wants to enforce that all the nodes in $U$ and $V$ are taken care of. The *linear sum assignment problem with error-correction* (LSAPE) models these settings. To this purpose, the sets $U$ and $V$ are extended to $U_\epsilon := U \cup \{\epsilon\}$ and $V_\epsilon := V \cup \{\epsilon\}$, where $\epsilon$ is a dummy node. An edge set $\pi = S \cup R \cup D \subseteq U_\epsilon \times V_\epsilon$ is called *error-correcting matching* for $K_{n,m,\epsilon} := (U_\epsilon, V_\epsilon, U_\epsilon \times V_\epsilon)$ if and only if all nodes in $U$ and $V$ are incident with exactly one edge in $\pi$. The set $S \subseteq U \times V$ contains all node substitutions, $R \subseteq U \times \{\epsilon\}$ contains all removals, and $D \subseteq \{\epsilon\} \times V$ contains all deletions. A error-correcting matching $\pi$ can be encoded with a binary matrix $\mathbf{X} \in \Pi_{n,m,\epsilon}$ by setting $x_{i,k} := 1$ just in case $(u_i, v_k) \in \pi$, $x_{i,m+1} := 1$ just in case $(u_i, \epsilon) \in \pi$, and $x_{n+1,k} := 1$ just in case $(\epsilon, v_k) \in \pi$ (cf. Definition 2.8 above).

**Definition 2.26 (LSAPE).** Given a cost matrix $\mathbf{C} \in \mathbb{R}_{\geq 0}^{(n+1) \times (m+1)}$, LSAPE consists in the task of finding an error-correcting matching $\mathbf{X}^\star \in \arg\min_{\mathbf{X} \in \Pi_{n,m,\epsilon}} \mathbf{C}(\mathbf{X})$, where $\mathbf{C}(\mathbf{X}) := \sum_{i=1}^{n+1} \sum_{k=1}^{m+1} c_{i,k} x_{i,k}$.

The following Proposition 2.1 states that node maps $\pi \in \Pi(G, H)$ can be identified with error-correcting matchings $\pi \in \Pi_{|V^G|, |V^H|, \epsilon}$. By Definition 2.6, it hence implies that each feasible solution $\pi$ for an LSAPE instance $\mathbf{C} \in \mathbb{R}_{\geq 0}^{(|V^G|+1) \times (|V^H|+1)}$ yields an upper bound $c(P_\pi)$ for $\text{GED}(G, H)$. This fact is used my many exact and approximate algorithms for GED.

**Proposition 2.1 (Equivalence of Node Maps and Error-Correcting Matchings).** *For all graphs G and H, the function $f^{G,H} : \Pi(G,H) \to \Pi_{|V^G|,|V^H|,\epsilon}$, defined as $\{(u,v) \mid (u,v) \in \pi\} \mapsto \{(f^G(u), f^H(v)) \mid (u,v) \in \pi\}$ is bijective. The function $f^G : V^G \cup \{\epsilon\} \to [|V^G|+1]$ is defined as $f^G(\epsilon) := |V^G|+1$ and $f^G(u_i) := i$, for all $u_i \in V^G$. The function $f^H : V^H \cup \{\epsilon\} \to [|V^H|+1]$ is defined analogously.*

*Proof.* The proposition follows from Definition 2.4 and Definition 2.8.    □

## 2.4  Test Datasets and Edit Costs

To evaluate the techniques proposed in this thesis, we tested on the datasets AIDS, MUTA, PROTEIN, LETTER (H), GREC, and FINGERPRINT from the IAM Graph Database Repository [84, 87], and on the datasets ALKANE, ACYCLIC, MAO, and PAH from GREYC's Chemistry Dataset.[1] All of these datasets are widely used in the research community [22, 25, 32, 41, 49, 83, 91, 100, 102, 112].

The graphs contained in the datasets AIDS, MUTA, MAO, PAH, ALKANE, and ACYCLIC represent molecular compounds, which, in the case of AIDS, MUTA, MAO, and PAH, are divided into two classes: molecules that do or do not exhibit activity against HIV (AIDS), molecules that do or do not cause genetic mutation (MUTA), molecules that do or do not inhibit the monoamine oxidase (MAO), and molecules that are cancerous or are not cancerous (PAH). The dataset PROTEIN contains graphs which represent proteins and are annotated with their EC classes from the BRENDA enzyme database [98]. The graphs contained in the dataset LETTER (H) represent highly distorted drawings of the capital letters A, E, F, H, I, K, L, M, N, T, V, W, X, Y, and Z. The graphs contained in the dataset GREC represent 22 different symbols from electronic and architectural drawings, which consist of geometric primitives such as lines, arcs, and cycles. And the graphs contained in the dataset FINGERPRINT represent fingerprint images, annotated with their classes from the Galton-Henry classification system [56].

Table 2.3 and Table 2.4 summarize the most important properties of the test datasets. The dataset MUTA-<100 contains all graphs from MUTA with less than 100 nodes. We included this dataset in the tables, because we noticed that in all graphs contained in MUTA all nodes whose ID is greater or equal to 100 are isolated. We then contacted the main author of the IAM Graph

---

[1] Available at `https://iapr-tc15.greyc.fr/links.html`.

**Table 2.3.** Properties of test datasets.

| dataset | node/edge labels | avg.(max.) $|V^G|$ | avg.(max.) $|E^G|$ | classes |
|---|---|---|---|---|
| *IAM Graph Database Repository* | | | | |
| AIDS | yes/yes | 15.7(95) | 16.2(103) | 2 |
| MUTA | yes/yes | 30.3(417) | 30.8(112) | 2 |
| MUTA-<100 | yes/yes | 29.2(98) | 30.1(105) | 2 |
| PROTEIN | yes/yes | 32.6(126) | 62.1(149) | 6 |
| LETTER (H) | yes/no | 4.7(9) | 4.5(9) | 15 |
| GREC | yes/yes | 11.5(26) | 12.2(30) | 22 |
| FINGERPRINT | no/yes | 5.4(26) | 4.4(25) | 4 |
| *GREYC's Chemistry Dataset* | | | | |
| MAO | yes/yes | 18.4(27) | 19.6(29) | 2 |
| PAH | yes/yes | 20.7(28) | 24.4(34) | 2 |
| ALKANE | yes/yes | 8.9(10) | 7.9(9) | 1 |
| ACYCLIC | yes/yes | 8.2(11) | 7.2(10) | 1 |

**Table 2.4.** Further properties of test datasets.

| dataset | avg.(max.) components | acyclic graphs | planar graphs |
|---|---|---|---|
| *IAM Graph Database Repository* | | | |
| AIDS | 1.12(7) | 21 % | 100 % |
| MUTA | 1.61(319) | 17 % | 100 % |
| MUTA-<100 | 1.08(5) | 17 % | 100 % |
| PROTEIN | 1.24(84) | 1 % | 41 % |
| LETTER (H) | 1.08(3) | 35 % | 100 % |
| GREC | 1.52(3) | 22 % | 100 % |
| FINGERPRINT | 1.01(2) | 99 % | 100 % |
| *GREYC's Chemistry Dataset* | | | |
| MAO | 1.00(1) | 0 % | 100 % |
| PAH | 1.00(1) | 0 % | 100 % |
| ALKANE | 1.00(1) | 100 % | 100 % |
| ACYCLIC | 1.00(1) | 100 % | 100 % |

Database Repository [84, 87] and he confirmed our suspicion that this is probably an error in the data. However, since more than 99 % of the graphs contained in MUTA have less than 100 nodes, this error should not have a very big impact on the experiments conducted on MUTA.

Table 2.3 shows that all datasets contain sparse, small to medium-sized graphs. More details about the topology of the graphs are given in Table 2.4. We see that all datasets except for PROTEIN contain only planar graphs. Moreover, the datasets from the IAM Graph Database Repository contain graphs that are usually — but not always — connected and cyclic. The datasets from GREYC's Chemistry Dataset can be categorized more clearly: All graphs in all datasets are connected, all graphs contained in ALKANE and ACYCLIC are acyclic (i.e., trees), and all graphs contained in MAO and PAH are cyclic.

In [2, 84], edit cost functions are defined for all test datasets. In [1], slightly different edit costs are defined for the datasets AIDS, MUTA, MAO, PAH, ALKANE, and ACYCLIC that contain graphs which model molecular compounds. Both edit costs are used for experiments in this thesis. Because of this, we briefly present them in the following Sections 2.4.1 to 2.4.5.

### 2.4.1 Edit Costs for AIDS, MUTA, MAO, PAH, ALKANE, and ACYCLIC

The nodes of the graphs contained in AIDS, MUTA, MAO, PAH, ALKANE, and ACYCLIC are labeled with chemical symbols. Edges are labeled with a valence (either 1 or 2).

In [2, 84], node edit costs are defined as

$$c_V(\alpha, \alpha') := 5.5 \cdot \delta_{\alpha \neq \alpha'}$$
$$c_V(\alpha, \epsilon) := 2.75$$
$$c_V(\epsilon, \alpha') := 2.75,$$

for all $(\alpha, \alpha') \in \Sigma_V \times \Sigma_V$. Similarly, edge edit costs are defined as

$$c_E(\beta, \beta') := 1.65 \cdot \delta_{\beta \neq \beta'}$$
$$c_E(\beta, \epsilon) := 0.825$$
$$c_E(\epsilon, \beta') := 0.825,$$

for all $(\beta, \beta') \in \Sigma_E \times \Sigma_E$.

In [1], slightly different edit costs are suggested. Given constants $(c_V^s, c_V^d, c_V^i, c_E^s, c_E^d, c_E^i) \in \mathbb{R}^6_{>0}$, node edit costs are defined as

$$c_V(\alpha, \alpha') := c_V^s \cdot \delta_{\alpha \neq \alpha'}$$
$$c_V(\alpha, \epsilon) := c_V^d$$
$$c_V(\epsilon, \alpha') := c_V^i,$$

for all $(\alpha, \alpha') \in \Sigma_V \times \Sigma_V$. Edge edit costs are defined as

$$c_E(\beta, \beta') := c_E^s \cdot \delta_{\beta \neq \beta'}$$
$$c_E(\beta, \epsilon) := c_E^d$$

$$c_E(\epsilon, \beta') := c_E^i,$$

for all $(\beta, \beta') \in \Sigma_E \times \Sigma_E$. The constants are suggested to be chosen as $(c_V^s, c_V^d, c_V^i, c_E^s, c_E^d, c_E^i) := (2, 4, 4, 1, 1, 1)$, $(c_V^s, c_V^d, c_V^i, c_E^s, c_E^d, c_E^i) := (2, 4, 4, 1, 2, 2)$, or $(c_V^s, c_V^d, c_V^i, c_E^s, c_E^d, c_E^i) := (6, 2, 2, 3, 1, 1)$.

### 2.4.2 Edit Costs for PROTEIN

The nodes of the graphs contain in PROTEIN are labeled with tuples $(t, s)$, where $t$ is the node's type (helix, sheet, or loop) and $s$ is its amino acid sequence. Nodes are connected via structural or sequential edges or both, i.e., edges $(u_i, u_j)$ are labeled with tuples $(t_1, t_2)$, where $t_1$ is the type of the first edge connecting $u_i$ and $u_j$ and $t_2$ is the type of the second edge connecting $u_i$ and $u_j$ (possibly null).

In [2, 84], node edit costs are defined as

$$c_V(\alpha, \alpha') := 16.5 \cdot \delta_{\alpha.t \neq \alpha'.t} + 0.75 \cdot \delta_{\alpha.t = \alpha'.t} \cdot \mathrm{LD}(\alpha.s, \alpha'.s))$$
$$c_V(\alpha, \epsilon) := 8.25$$
$$c_V(\epsilon, \alpha') := 8.25,$$

for all $(\alpha, \alpha') \in \Sigma_V \times \Sigma_V$, where $\mathrm{LD}(\cdot, \cdot)$ is Levenshtein's string edit distance. Edge edit costs are defined as

$$c_E(\beta, \beta') := 0.25 \cdot \mathrm{LSAPE}(\mathbf{C}^{\beta, \beta'})$$
$$c_E(\beta, \epsilon) := 0.25 \cdot f(\beta)$$
$$c_E(\epsilon, \beta') := 0.25 \cdot f(\beta'),$$

for all $(\beta, \beta') \in \Sigma_E \times \Sigma_E$, where $f(\beta) := 1 + \delta_{\beta.t_2 \neq \text{null}}$ and $\mathbf{C}^{\beta, \beta'} \in \mathbb{R}^{(f(\beta)+1) \times (f(\beta')+1)}$ is constructed as $c_{r,s}^{\beta,\beta'} := 2 \cdot \delta_{\beta.t_r \neq \beta'.t_s}$ and $c_{r,f(\beta')+1}^{\beta,\beta'} := c_{f(\beta)+1,s}^{\beta,\beta'} := 1$, for all $(r, s) \in [f(\beta)] \times [f(\beta')]$.

### 2.4.3 Edit Costs for LETTER (H)

The nodes of the graphs contain in LETTER (H) are labeled with two-dimensional Euclidean coordinates. Edges are unlabeled.

In [2, 84], node edit costs are defined as

$$c_V(\alpha, \alpha') := 0.75 \cdot \|\alpha - \alpha'\|$$
$$c_V(\alpha, \epsilon) := 0.675$$

$$c_V(\epsilon, \alpha') := 0.675,$$

for all $(\alpha, \alpha') \in \Sigma_V \times \Sigma_V$, where $\|\cdot\|$ is the Euclidean norm. The edge edit costs $c_E$ are defined as $c_E(1, \epsilon) := c_E(\epsilon, 1) := 0.425$.

### 2.4.4 Edit Costs for GREC

The nodes of the graphs contain in GREC are labeled with tuples $(t, x, y)$, where $t$ equals one of four node types and $(x, y)$ is a two-dimensional Euclidean coordinate. Nodes are connected via line or arc edges or both, i.e., edges $(u_i, u_j)$ are labeled with tuples $(t_1, t_2)$, where $t_1$ is the type of the first edge connecting $u_i$ and $u_j$ and $t_2$ is the type of the second edge connecting $u_i$ and $u_j$ (possibly null).

In [2, 84], node edit costs are defined as

$$c_V(\alpha, \alpha') := 0.5 \cdot \|\alpha.(x,y) - \alpha'.(x,y)\| \cdot \delta_{\alpha.t=\alpha'.t} + 90 \cdot \delta_{\alpha.t \neq \alpha'.t}$$
$$c_V(\alpha, \epsilon) := 45$$
$$c_V(\epsilon, \alpha') := 45,$$

for all $(\alpha, \alpha') \in \Sigma_V \times \Sigma_V$. Edge edit costs are defined as

$$c_E(\beta, \beta') := 0.5 \cdot \mathrm{LSAPE}(\mathbf{C}^{\beta, \beta'})$$
$$c_E(\beta, \epsilon) := 0.5 \cdot f(\beta)$$
$$c_E(\epsilon, \beta') := 0.5 \cdot f(\beta'),$$

for all $(\beta, \beta') \in \Sigma_E \times \Sigma_E$, where $f(\beta) := 1 + \delta_{\beta.t_2 \neq \mathtt{null}}$ and $\mathbf{C}^{\beta, \beta'} \in \mathbb{R}^{(f(\beta)+1) \times (f(\beta')+1)}$ is constructed as $c^{\beta, \beta'}_{r,s} := 30 \cdot \delta_{\beta.t_r \neq \beta'.t_s}$ and $c^{\beta, \beta'}_{r, f(\beta')+1} := c^{\beta, \beta'}_{f(\beta)+1, s} := 15$ for all $(r, s) \in [f(\beta)] \times [f(\beta')]$.

### 2.4.5 Edit Costs for FINGERPRINT

The nodes of the graphs contain in FINGERPRINT are unlabeled and edges are labeled with an orientation $\beta \in \mathbb{R}$ with $-\pi/2 < \beta \leq \pi/2$.

In [2, 84], node edit costs are defined as $c_V(1, \epsilon) := c_V(\epsilon, 1) := 0.525$. Edge edit costs are defined as

$$c_E(\beta, \beta') := 0.5 \cdot \min\{|\beta - \beta'|, \pi - |\beta - \beta'|\}$$
$$c_E(\beta, \epsilon) := 0.375$$
$$c_E(\epsilon, \beta') := 0.375,$$

for all $(\beta, \beta') \in \Sigma_E \times \Sigma_E$.

# — 3 —
# Theoretical Aspects

In this chapter, we summarize known results that characterize the problem of computing GED in terms of complexity and extend these results to settings where the input graphs are very simple. Furthermore, we present a new compact reduction from GED to the quadratic assignment problem (QAP) for quasimetric edit costs. While existing reductions from GED to QAP either use very large QAP instances [21, 22, 82] or reduce GED to non-standard versions of QAP [25], the newly proposed reduction yields small instances of the standard version QAP. The practical consequence of this is that, with the new reduction, a wide range of methods that were originally defined for the standard version of QAP can efficiently be used for computing GED.

The third technical contribution presented in this chapter consists in the harmonization of the two alternative Definitions 2.3 (used in the database community) and 2.6 (used in the pattern recognition community) of GED. In [60], it has already been shown that Definition 2.3 and Definition 2.6 coincide if the underlying edit cost functions are metric. We here extend this result to general edit costs.

The results presented in this chapter have previously been presented in the following articles:

– D. B. Blumenthal and J. Gamper, "On the exact computation of the graph edit distance", *Pattern Recognit. Lett.*, 2018, in press. DOI: `10.1016/j.patrec.2018.05.002`
– D. B. Blumenthal, É. Daller, S. Bougleux, L. Brun, and J. Gamper, "Quasimetric graph edit distance as a compact quadratic assignment problem", in *ICPR 2018*, IEEE Computer Society, 2018, pp. 934–939. DOI: `10.1109/`

The remainder of this chapter is organized as follows: In Section 3.1, the problem of computing GED is characterized in terms of complexity and existing reductions from GED to QAP are summarized. In Section 3.2, it is shown that GED is hard to compute and approximate also on very simple graphs — namely, forests of undirected paths. In Section 3.3 the new compact reduction is presented. In Section 3.4, the two definitions of GED are harmonized. In Section 3.5, the effect of the newly proposed reduction on QAP based heuristics is empirically evaluated. Section 3.6 concludes the chapter.

## 3.1 State of the Art

In this section, we provide an overview of known results related to theoretical aspects of computing GED. In Section 3.1.1, we summarize results that characterize the problem of computing GED in terms of complexity. In Section 3.1.2, existing reductions from GED to QAP are presented.

### 3.1.1 Complexity of GED Computation

We begin with characterizing the problem of computing GED in terms of complexity. More specifically, four proposition are provided, which state that GED is not only hard to compute, but also hard to approximate. All of these propositions are known results from the literature or follow immediately from simple observations. For the sake of completeness, we nonetheless include short sketches of their proofs.

**Proposition 3.1 (Hardness of Uniform GED Computation).** *Exactly computing GED is $\mathcal{NP}$-hard even for uniform edit costs.*

*Proof.* This result was established by Zeng, Tung, Wang, Feng, and Zhou [111] via a polynomial reduction from the subgraph isomorphism problem (SGI), which is known to be $\mathcal{NP}$-complete. SGI is defined as follows: Given two unlabeled, undirected graphs $G$ and $H$, determine whether $G$ is subgraph isomorphic to $H$, i.e., whether there is an injection $f : V^G \to V^H$ such that $G \simeq H[f[V^G]]$. To this purpose, $G$ and $H$ are transformed into labeled graphs $G'$ and $H'$ by assigning the label 1 to each node and edge, and the edit cost functions are defined to be uniform, i.e., as $c_V(\alpha', \alpha'') := \delta_{\alpha' \neq \alpha''}$ and $c_E(\beta, \beta') := \delta_{\beta \neq \beta'}$. It can easily be verified that $G$ is subgraph isomorphic to $H$, if and only if $|V^G| \leq |V^H|$, $|E^G| \leq |E^H|$, and $\text{GED}(G', H') = (|V^H| - |V^G|) + (|E^H| - |E^G|)$. This proves the proposition. □

The next question which naturally arises is whether there are $\alpha$-approximation algorithms for GED for some approximation guarantees $\alpha$. Recall that an $\alpha$-approximation algorithm for GED is an algorithm that runs in polynomial time and returns an edit path $P \in \Psi(G, H)$ with cost $c(P) \leq \alpha \, \text{GED}(G, H)$.

**Proposition 3.2 (Non-Approximability of Non-Metric GED).** *If arbitrary, non-metric edit cost functions are allowed, there is no $\alpha$-approximation algorithm for GED for any $\alpha$, unless $\mathcal{P} = \mathcal{NP}$.*

*Proof.* For proving this proposition, it suffices to slightly modify the proof of Proposition 3.1. Given an instance $(G, H)$ of SGI, like before, we define labeled graphs $G'$ and $H'$ by assigning the label 1 to each node and edge. However, we now define the edit cost functions as $c_V(\alpha', \alpha'') := \delta_{(\alpha', \alpha'') = (\epsilon, 1)}$ and $c_E(\beta, \beta') := \delta_{(\beta, \beta') = (\epsilon, 1)}$, such that only node and edge insertions incur a non-zero edit cost. It is easy to see that, with this construction, $G$ is subgraph isomorphic to $H$ just in case $\text{GED}(G', H') = 0$. But then, any $\alpha$-approximation algorithm for GED can decide SGI, which implies the desired result. □

**Proposition 3.3 (Non-Approximability of Metric GED).** *Even if non-metric edit cost functions are forbidden, there is no polynomial time approximation scheme for GED, i.e., there is a constant $c \in \mathbb{R}_{\geq 0}$ such that there is no $(1+\varepsilon)$-approximation algorithm for GED for any $0 < \varepsilon < c$, unless $\mathcal{P} = \mathcal{NP}$.*

*Proof.* This result follows from a work by Lin [70], who proved the same result for the graph transformation problem (GT). In GT, we are given two unlabeled, undirected graphs $G$ and $H$ with $|V^G| = |V^H|$ and $|E^G| = |E^H| =: m$, and are asked to determine $\text{GT}(G, H) \in \mathbb{N}$, which is defined as the smallest $k \in \mathbb{N}$ such that $G$ can be made isomorphic to $H$ by reallocating $k$ edges of $G$, i.e., by deleting $k$ existing edges from and adding $k$ new edges to $G$. Note that, $\text{GT}(G, H) \leq m$ holds by definition of GT.

Let $c \in \mathbb{R} \geq 0$ be a constant such that, unless $\mathcal{P} = \mathcal{NP}$, there is no $(1+\varepsilon)$-approximation algorithm for GT for any $\varepsilon \in (0, c)$. In [70], it has been shown that such a constant exists. Like above, let $G'$ and $H'$ denote the labeled graphs obtained from $G$ and $H$ by assigning the label 1 to all nodes and edges. Moreover, let the edit cost functions be defined as $c_V(\alpha, \alpha') := (1+c)m\delta_{\alpha \neq \alpha'}$ and $c_E(\beta, \beta') := \delta_{\beta \neq \beta'}/2$. Then each $k$-sized set of edge allocations that transforms $G$ into $H$ corresponds to an edit path $P \in \Psi(G', H')$ without node insertions and deletions and cost $c(P) = k$, and vice versa. This implies $\text{GED}(G', H') \leq \text{GT}(G, H)$. Moreover, we know that an optimal edit path $P \in \Psi(G', H')$ does not contain any node insertions or deletions, as this would imply $\text{GED}(G', H') = c(P) \geq (1+c)m > \text{GT}(G, H) \geq \text{GED}(G', H')$. We therefore have $\text{GED}(G', H') = \text{GT}(G, H)$.

Now assume that there is an $\varepsilon \in (0, c)$ such that there is a $(1+\varepsilon)$-approximation algorithm `ALG` for GED with metric edit costs. Let $P \in \Psi(G', H')$ be the edit path between $G'$ and $H'$ computed by `ALG`. We have $c(P) \leq (1+\epsilon)\text{GED}(G', H') \leq (1+\epsilon)m < (1+c)m$, from which we can conclude that $P$ does not contain node insertions or deletions. But then $P$ corresponds to a $c(P)$-sized set of edge allocations that transforms $G$ into $H$, which implies that `ALG` is a $(1+\varepsilon)$-approximation algorithm for GT. This contradicts the result established in [70]. □

**Proposition 3.4 (Non-Approximability of Uniform GED).** *Even if only uniform edit cost functions are allowed, there is no $\alpha$-approximation algorithm for GED for any $\alpha$, unless the graph isomorphism problem (GI) is in $\mathcal{P}$.*

*Proof.* Recall that GI is defined as the task to determine whether two unlabeled, undirected graphs $G$ and $H$ are isomorphic to each other. Like in the proof of Proposition 3.1, we transform $G$ and $H$ into labeled graphs $G'$ and $H'$ by assigning the label 1 to each node and edge, and consider uniform edit cost functions $c_V$ and $c_E$, i.e., $c_V(\alpha', \alpha'') := \delta_{\alpha' \neq \alpha''}$ and $c_E(\beta, \beta') := \delta_{\beta \neq \beta'}$. With this construction, we have $G \simeq H$ just in case $\text{GED}(G', H') = 0$, which implies that any $\alpha$-approximation algorithm for (uniform) GED can decide GI and hence proves the statement of the proposition. □

The question whether $\text{GI} \in \mathcal{P}$ has been open for decades; the fastest currently available algorithm runs in quasi-polynomial time [6]. In fact, GI is a prime candidate for an $\mathcal{NP}$-intermediate problem that is neither in $\mathcal{P}$ nor $\mathcal{NP}$-complete. In view of this and because of Proposition 3.4, it is unrealistic to hope for an $\alpha$-approximation algorithm for any variant of GED. Rather, when it comes to approximating GED, all one can do is to design heuristics that compute lower or upper bounds and empirically test the tightness of the produced bounds.

### 3.1.2 Reductions to QAP and QAPE

We present two reductions of GED to, respectively, the quadratic assignment problem (QAP) and the quadratic assignment problem with error-correction (QAPE). QAP and QAPE are defined as follows:

**Definition 3.1 (QAP).** Given natural numbers $n, m \in \mathbb{N}$ and a real-valued matrix $\mathbf{C} \in \mathbb{R}^{(n \cdot m) \times (n \cdot m)}$, the *quadratic assignment problem* (QAP) consists in the task to compute a maximum matching $\mathbf{X}^\star := \arg\min\{Q(\mathbf{X}, \mathbf{C}) \mid \mathbf{X} \in \Pi_{n,m}\}$ between $[n]$ and $[m]$ that minimizes the quadratic cost $Q(\mathbf{X}, \mathbf{C}) := \text{vec}(\mathbf{X})^\mathsf{T} \mathbf{C} \, \text{vec}(\mathbf{X})$, where vec is a matrix-vectorization operator. The cost $Q(\mathbf{X}^\star, \mathbf{C})$ of an optimal solution $\mathbf{X}^\star \in \Pi_{n,m}$ for $\mathbf{C}$ is denoted by QAP($\mathbf{C}$).

**Definition 3.2 (QAPE).** Given natural numbers $n, m \in \mathbb{N}$ and a real-valued matrix $\mathbf{C} \in \mathbb{R}^{((n+1) \cdot (m+1)) \times ((n+1) \cdot (m+1))}$, the *quadratic assignment problem with error-correction* (QAPE) consists in the task to compute an error-correcting matching $\mathbf{X}^\star := \arg\min\{Q(\mathbf{X}, \mathbf{C}) \mid \mathbf{X} \in \Pi_{n,m,\epsilon}\}$ between $[n]$ and $[m]$ that minimizes the quadratic cost $Q(\mathbf{X}, \mathbf{C}) := \text{vec}(\mathbf{X})^\mathsf{T} \mathbf{C} \, \text{vec}(\mathbf{X})$, where vec is a matrix-vectorization operator. The cost $Q(\mathbf{X}^\star, \mathbf{C})$ of an optimal solution $\mathbf{X}^\star \in \Pi_{n,m,\epsilon}$ for $\mathbf{C}$ is denoted by QAPE($\mathbf{C}$).

We first present the reduction from GED to QAP, which has been proposed independently in [82] and [21, 22] (Section 3.1.2.1). Subsequently, the reduction to QAPE suggested in [25] is presented (Section 3.1.2.2). The main practical use of the reductions is that they render well-performing heuristics that have been designed for QAP applicable to GED, in particular, the integer fixed point method suggested in [67]. Adaptions of this heuristic to GED are presented below in Section 3.5 and Section 6.1.3.

For the remainder of this section, we fix graphs $G, H \in \mathbb{G}$ as well as edit cost functions $c_V$ and $c_E$, define $n := |V^G|$ and $m := |V^H|$, and assume w.l.o.g. that $V^G = [n]$ and $V^H = [m]$.

#### 3.1.2.1 Baseline Reduction to QAP

The reduction from GED to QAP proposed in [82] and [21, 22] works on enlarged graphs $(V^{G+\epsilon}, E^G)$ and $(V^{H+\epsilon}, E^H)$, where $V^{G+\epsilon} := V^G \cup \mathcal{E}^G$, $V^{H+\epsilon} := V^H \cup \mathcal{E}^H$, and $\mathcal{E}^G := \{\epsilon_1 := n+1, \ldots, \epsilon_m := n+m\}$ and $\mathcal{E}^H := \{\epsilon_1 := m+1, \ldots, \epsilon_n := m+n\}$ contain $m$ and $n$ isolated dummy nodes, respectively. Let $\mathbf{X} \in \Pi_{(n+m),(m+n)}$ be a maximum matching between $V^{G+\epsilon}$ and $V^{H+\epsilon}$. $\mathbf{X}$ can be written in the following form:

$$\mathbf{X} = \begin{array}{c} \\ V^G \\ \mathcal{E}^G \end{array} \begin{array}{c} V^H \quad \mathcal{E}^H \\ \left[ \begin{array}{cc} \mathbf{X}^s & \mathbf{X}^d \\ \mathbf{X}^i & \mathbf{X}^0 \end{array} \right] \end{array}$$

The north-west quadrant $\mathbf{X}^s$ encodes node substitutions, the south-west quadrant $\mathbf{X}^i$ encodes node insertions, the north-east quadrant $\mathbf{X}^d$ encodes node deletions, and the south-east quadrant $\mathbf{X}^0$ encodes assignments of dummy nodes to dummy nodes. Since nodes can be inserted and removed only once, assignments $x_{i,\epsilon_j} = 1$ and $x_{\epsilon_l,k} = 1$ with $i \neq j$ and $k \neq l$ are forbidden. The notation $i \nrightarrow k$ is used to denote that assigning $i \in V^{G+\epsilon}$ to $k \in V^{H+\epsilon}$ is forbidden.

Next, auxiliary matrices $\mathbf{C}^V, \mathbf{C}^E \in \mathbb{R}^{(n+m)^2 \times (n+m)^2}$ are constructed. For the ease of notation, we use two-dimensional indices $(i,k) \in V^{G+\epsilon} \times V^{H+\epsilon}$ for referring to their elements, and a special vectorization operator vec that first concatenates the columns of $V^G \times V^H$, then the columns of $\mathcal{E}^G \times V^H$, followed by the columns of $V^G \times \mathcal{E}^H$, and the columns of $\mathcal{E}^G \times \mathcal{E}^H$.

$\mathbf{C}^V$ contains the costs of the node edit operations. It is defined as

$$c^V_{(i,k),(j,l)} := \begin{cases} 0 & \text{if } i \neq j \vee k \neq l \vee i \nrightarrow k \\ c'_V(i,k) & \text{otherwise,} \end{cases} \quad (3.1)$$

where $c'_V$ is defined as follows:

$$c'_V(i,k) := \delta_{i \in V^G} \delta_{k \in V^H} c_V(i,k) \quad (3.2)$$
$$+ (1 - \delta_{i \in V^G}) \delta_{k \in V^H} c_V(\epsilon, k) + \delta_{i \in V^G} (1 - \delta_{k \in V^H}) c_V(i, \epsilon)$$

$\mathbf{C}^E$ contains the costs of the edge edit operations. It is defined as

$$c^E_{(i,k),(j,l)} := \begin{cases} \omega & \text{if } i \nrightarrow k \vee j \nrightarrow l \\ c'_E(i,k,j,l) & \text{otherwise,} \end{cases} \quad (3.3)$$

where $\omega$ can be set to any upper bound for $\text{GED}(G,H)$ and $c'_E$ is defined as follows:

$$c'_E(i,k,j,l) := \delta_{(i,j) \in E^G} \delta_{(k,l) \in E^H} c_E((i,j),(k,l)) \quad (3.4)$$

$$+ (1 - \delta_{(i,j) \in E^G}) \delta_{(k,l) \in E^H} c_E(\epsilon, (k,l))$$
$$+ \delta_{(i,j) \in E^G} (1 - \delta_{(k,l) \in E^H}) c_E((i,j), \epsilon)$$

The following theorem provides a reduction from GED to QAP:

**Theorem 3.1 (Baseline Reduction to QAP).** *Let $\mathbf{C} \in \mathbb{R}^{(n+m)^2 \times (n+m)^2}$ be defined as $\mathbf{C} := \frac{1}{2}\mathbf{C}^E + \mathbf{C}^V$, where $\mathbf{C}^V$ is defined as specified in equation (3.1) and $\mathbf{C}^E$ is defined as specified in equation (3.3). Then it holds that $\mathrm{GED}(G, H) = \mathrm{QAP}(\mathbf{C})$.*

#### 3.1.2.2 Reduction to QAPE

The reduction from GED to QAPE suggested in [25] constructs auxiliary matrices $\mathbf{C}^V, \mathbf{C}^E \in \mathbb{R}^{((n+1) \cdot (m+1)) \times ((n+1) \cdot (m+1))}$. Like above, $\mathbf{C}^V$ is defined as specified in equation (3.1). The definition of $\mathbf{C}^E$ is slightly different, since there are no forbidden node assignments. So we have

$$c^E_{(i,k),(j,l)} := c'_E(i, k, j, l), \tag{3.5}$$

where $c'_E$ is defined as specified in equation (3.4). The following theorem provides a reduction from GED to QAPE:

**Theorem 3.2 (Reduction to QAPE).** *Let $\mathbf{C}' \in \mathbb{R}^{((n+1) \cdot (m+1)) \times ((n+1) \cdot (m+1))}$ be defined as $\mathbf{C}' := \frac{1}{2}\mathbf{C}^E + \mathbf{C}^V$, where $\mathbf{C}^V$ is defined as specified in equation (3.1) and $\mathbf{C}^E$ is defined as specified in equation (3.5). Then it holds that $\mathrm{GED}(G, H) = \mathrm{QAPE}(\mathbf{C}')$.*

The advantage of this reduction to QAPE w. r. t. the baseline reduction to QAP presented in the previous section is that the constructed QAPE instance is significantly smaller than the constructed QAP instance. The drawback is that QAPE is a non-standard variant of QAP. Hence, many methods that can be used off-the-shelf with the reduction to QAP provided by Theorem 3.1 have to be adapted manually if the reduction to QAPE provided by Theorem 3.2 is employed. This requires a thorough knowledge of matching theory on the side of the implementer.

## 3.2 Hardness on Very Sparse Graphs

In Section 3.1.1 above, we summarized results which show that GED is hard to compute and approximate even if we restrict to special classes of edit cost functions. In this section, we discuss a related question: What if, instead of restricting to special classes of edit cost functions, we restrict to special classes of graphs? Does this reduce the complexity of computing or approximating GED?

On a first glance, there seem to be good reasons to assume that this is the case. For instance, on planar graphs, GI is known to be polynomially solvable [38] and solving SGI (the problem of determining whether $G$ is subgraph isomorphic to $H$) is linear in the size $|V^H|$ of the target graph [43] and even polynomial in the size $|V^G|$ of the pattern for special classes of planar graphs [57]. This is especially relevant, because the proofs of Propositions 3.1 to 3.2 given in Section 3.1.1 above employ reductions from SGI.

Another reason for conjecturing that computing GED might be tractable on simple graphs is that the tree edit distance (TED) can be computed in polynomial time on ordered trees [7]. TED is similar to GED in that it is a distance measure defined in terms of edit operations and edit costs, namely, insertions, deletions, and substitutions of nodes. However, edges are not edited directly. Rather, they are inserted and deleted simultaneously when editing the nodes in order to maintain the tree structure at each step of the edit sequence. In spite of the similarities, TED and GED on trees are hence different distance measures, i.e., we in general have $\text{TED}(G, H) \neq \text{GED}(G, H)$, even if both $G$ and $H$ are trees.

The following Propositions 3.5 to 3.6 show that, despite these prima facie arguments to support the contrary, computing and approximating GED is $\mathcal{NP}$-hard even on very sparse graphs. More precisely, Proposition 3.5 shows that computing GED is $\mathcal{NP}$-hard even if we restrict to unlabeled graphs with maximum degree two and uniform edit cost functions. Proposition 3.6 demonstrates that, if we allow arbitrary, non-metric edit cost functions, then there are no $\alpha$-approximation algorithms for GED on unlabeled graphs with maximum degree two, unless $\mathcal{P} = \mathcal{NP}$. In other words, the two propositions show that the Propositions 3.1 to 3.2 are valid also on forests of unlabeled paths.

**Proposition 3.5 (Hardness of Uniform GED Computation on Very Sparse Graphs).** *Exactly computing GED on unlabeled graphs with maximum degree two is $\mathcal{NP}$-hard even for uniform edit costs.*

*Proof.* We prove the proposition via a polynomial reduction from the 3-

partition problem (3-PARTITION), which is known to be strongly $\mathcal{NP}$-hard [108]. Given a natural number $B \in \mathbb{N}$ and natural numbers $(a_i)_{i=1}^{3n}$ with $\sum_{i=1}^{3n} a_i = n \cdot B$ and $B/4 < a_i < B/2$ for all $i \in [3n]$, 3-PARTITION asks to decide whether the index set $[3n]$ can be partitioned into blocks $(\{i_{k_1}, i_{k_2}, i_{k_3}\})_{k=1}^{n}$ such that $\sum_{l=1}^{3} a_{i_{k_l}} = B$ holds for all $k \in [n]$.

Given an instance $(B, (a_i)_{i=1}^{3n})$ of 3-PARTITION, we construct two graphs $G$ and $H$ as forests of paths without node or edge labels. For each $i \in [3n]$, $G$ contains a path $p_i^G$ of length $a_i$. Similarly, for each $k \in [n]$, $H$ contains a path $p_k^H$ of length $B$. The edit cost functions $c_V$ and $c_E$ are assumed to be uniform. Note that this construction is polynomial in the size of the 3-PARTITION instance, because 3-PARTITION is not only $\mathcal{NP}$-hard but strongly $\mathcal{NP}$-hard.

By construction, both $G$ and $H$ have $n \cdot B$ nodes, $G$ has $n \cdot B - 3n$ edges, and $H$ has $n \cdot B - n$ edges. Hence, at least $2n$ edges have to be inserted for transforming $G$ into $H$, which implies $\text{GED}(G, H) \geq 2n$. We claim that $\text{GED}(G, H) = 2n$ if and only if $(B, (a_i)_{i=1}^{3n})$ is a yes-instance of 3-PARTITION. This claim proves the theorem.

First assume that $(B, (a_i)_{i=1}^{3n})$ is a yes-instance of 3-PARTITION. Let $(\{i_{k_1}, i_{k_2}, i_{k_3}\})_{k=1}^{n}$ be a partition of the index set $[3n]$ such that $\sum_{l=1}^{3} a_{i_{k_l}} = B$ holds for all $k \in [n]$. For all $k \in [n]$, we insert two edges into $G$: one between the last node of $p_{i_{k_1}}^G$ and the first node of $p_{i_{k_2}}^G$, the other one between the last node of $p_{i_{k_2}}^G$ and the first node of $p_{i_{k_3}}^G$. After these edit operations, $G$ consists of $n$ paths of length $B$ and is hence isomorphic to $H$. We have thus found an edit path $P$ between $G$ and $H$ with $c(P) = 2n$, which, together with $\text{GED}(G, H) \geq 2n$, proves the first direction of the claim.

For proving the other direction, assume that there is an edit path $P$ between $G$ and $H$ with $c(P) = 2n$. Then $P$ can contain only edge insertions between start or terminal nodes of some of the paths $p_i^G$. More precisely, since $B/4 < a_i < B/2$, we know that there are exactly $n$ paths $(p_{i_{k_2}}^G)_{k=1}^{n}$ such that $P$ joins the first node of $p_{i_{k_2}}^G$ to the last node of another path $p_{i_{k_1}}^G$ and the last node of $p_{i_{k_2}}^G$ to the first node of yet another path $p_{i_{k_3}}^G$. Moreover, each concatenated path $(p_{i_{k_1}}^G, p_{i_{k_2}}^G, p_{i_{k_3}}^G)$ has length $B$. Hence, $(\{i_{k_1}, i_{k_2}, i_{k_3}\})_{k=1}^{n}$ is a partition of the index set $[3n]$ that satisfies $\sum_{l=1}^{3} a_{i_{k_l}} = B$ for all $k \in [n]$. □

**Proposition 3.6 (Non-Approximability of Non-Metric GED on Very Sparse Graphs).** *If arbitrary, non-metric edit cost functions are allowed, there is no α-approximation algorithm for GED on unlabeled graphs with maximum degree two* *for any $\alpha$, unless $\mathcal{P} = \mathcal{NP}$.*

*Proof.* The proof is very similar to the proof of Proposition 3.5; we again use a polynomial reduction from 3-PARTITION. Given an instance $(B, (a_i)_{i=1}^{3n})$ of 3-PARTITION, $G$ and $H$ are constructed as before. The only difference is that we now define non-uniform edit costs that allow to insert edges for free, i. e., we set $c_E(1, \epsilon) := 0$ and $c_E(\epsilon, 1) := 1$. By using analogous arguments to the ones provided in the proof of Proposition 3.5, we can show that $(B, (a_i)_{i=1}^{3n})$ is a yes-instance of 3-PARTITION if and only if $\text{GED}(G, H) = 0$. This proves the proposition. □

In view of the Propositions 3.5 to 3.6, we conclude that, even if one restricts to very sparse graphs, it is unreasonable to look for $\alpha$-approximation algorithms or even computationally tractable exact algorithms for GED. However, given the existence of polynomially computable distance measures such as TED, it might be a good idea to use these distance measures instead of GED if the input graphs fall within their scope. Yet, discussing when exactly it is beneficial to move from GED to more easily computable distance measures is outside the scope of this thesis, as the thesis' objective is not to justify GED but to provide new techniques for its exact and heuristic computation.

## 3.3 Compact Reduction to QAP for Quasimetric Edit Costs

In this section, we introduce a new, compact reduction from GED to QAP. The new reduction combines the advantages of the existing reductions presented above. Like the baseline reduction presented in Section 3.1.2.1, the new reduction uses the standard quadratic assignment problem QAP rather than the non-standard variant QAPE and hence renders off-the-shelf QAP methods applicable to GED. And like the reduction to QAPE presented in Section 3.1.2.2, the new reduction avoids to blow up the input graphs and produces a compact instance of QAP.

The only restriction of the new reduction is that it is only applicable in settings where the triangle inequalities

$$c_V(\alpha, \alpha') \leq c_V(\alpha, \epsilon) + c_V(\epsilon, \alpha') \tag{3.6}$$

$$c_E(\beta, \beta') \leq c_E(\beta, \epsilon) + c_E(\epsilon, \beta') \tag{3.7}$$

are satisfied for all $(\alpha, \alpha') \in \Sigma_V \times \Sigma_V$ and all $(\beta, \beta') \in \Sigma_E \times \Sigma_E$, i.e., for settings where the edge and node edit costs $c_V$ and $c_E$ are quasimetric. Note that equation (3.6) can be assumed to hold w.l.o.g.: Since edge substitutions can always be replaced by a removal and an insertion, we can enforce equation (3.6) by setting $c_E(\beta, \beta') := \min\{c_E(\beta, \beta'), c_E(\beta, \epsilon) + c_E(\epsilon, \beta')\}$ (cf. Section 3.4 below for more details). This is not true of equation (3.6), which effectively constrains the node edit costs $c_V$. However, equation (3.6) is met in many application scenarios [81, 104].

Like in Section 3.1.2, throughout this section, we fix graphs $G, H \in \mathbf{G}$ as well as edit cost functions $c_V$ and $c_E$, define $n := |V^G|$ and $m := |V^H|$, and assume w.l.o.g. that $V^G = [n]$ and $V^H = [m]$. Furthermore, we assume that $c_V$ and $c_E$ are quasimetric. The following Theorem 3.3 establishes our compact reduction from GED to QAP. Note that Theorem 3.3 implies that, if the edit costs are quasimetric and $n \leq m$, then there is an optimal edit path between $G$ and $H$ that contains no node removals and hence no edge removals induced by node removals. Analogously, $n \geq m$ entails that there is an optimal edit path between $G$ and $H$ that contains no node insertions and hence no edge insertions induced by node insertions.

**Theorem 3.3 (Compact Reduction to QAP).** *Let $\mathbf{C} \in \mathbb{R}^{(n+m)^2 \times (n+m)^2}$ be the QAP instance defined by the baseline reduction in Theorem 3.1, and let $\widehat{\mathbf{C}} \in \mathbb{R}^{(n \cdot m) \times (n \cdot m)}$ be defined as follows:*

$$\widehat{c}_{(i,k),(j,l)} := c_{(i,k),(j,l)} - \delta_{i=j}\delta_{k=l}\left[\delta_{n<m}c_V(\epsilon, k) + \delta_{n>m}c_V(i, \epsilon)\right] \tag{3.8}$$
$$- \frac{3}{2}\left[\delta_{n<m}\delta_{(k,l)\in E^H}c_E(\epsilon, (k,l)) + \delta_{n>m}\delta_{(i,j)\in E^G}c_E((i,j), \epsilon)\right]$$

*Then it holds that*

$$\begin{aligned}\text{QAP}(\mathbf{C}) = \text{QAP}(\widehat{\mathbf{C}}) & \tag{3.9} \\ + \delta_{n<m}\Bigg[\sum_{(k,l) \in E^H} c_E(\epsilon, (k,l)) &+ \sum_{k \in V^H} c_V(\epsilon, k)\Bigg] \\ + \delta_{n>m}\Bigg[\sum_{(i,j) \in E^G} c_E((i,j), \epsilon) &+ \sum_{i \in V^G} c_V(i, \epsilon)\Bigg],\end{aligned}$$

*which, by Theorem 3.1, implies that $\widehat{\mathbf{C}}$ is a QAP formulation of GED.*

In Section 3.3.1, we prove that the reduction stated in Theorem 3.3 is correct. In Section 3.3.2, we explain how to turn it into a paradigm for (approximately) computing GED.

### 3.3.1 Proving the Correctness of the Reduction

Throughout this section, we assume w. l. o. g. that $n \leq m$. The case $n \geq m$ is analogous. The first observation, of which we will make continuous use, is that, by applying the vectorization operator vec defined in Section 3.1.2.1, the original QAP formulation $\mathbf{C}$ defined in Theorem 3.1 can be rewritten in the following way:

$$\mathbf{C} = \begin{array}{c} \\ V^G \times V^H \\ \mathcal{E}^G \times V^H \\ V^G \times \mathcal{E}^H \\ \mathcal{E}^G \times \mathcal{E}^H \end{array} \begin{bmatrix} \overset{V^G \times V^H}{\mathbf{C}^{ss}} & \overset{\mathcal{E}^G \times V^H}{\mathbf{C}^{si}} & \overset{V^G \times \mathcal{E}^H}{\mathbf{C}^{sr}} & \overset{\mathcal{E}^G \times \mathcal{E}^H}{\mathbf{C}^{s0}} \\ \mathbf{C}^{is} & \mathbf{C}^{ii} & \mathbf{C}^{ir} & \mathbf{C}^{i0} \\ \mathbf{C}^{rs} & \mathbf{C}^{ri} & \mathbf{C}^{rr} & \mathbf{C}^{r0} \\ \mathbf{C}^{0s} & \mathbf{C}^{0i} & \mathbf{C}^{0r} & \mathbf{C}^{00} \end{bmatrix}$$

Recall that the cell $c_{(i,k),(j,l)}$ of $\mathbf{C}$ contains the cost induced by simultaneously assigning the node $i \in V^G \cup \mathcal{E}^G$ to the node $k \in V^H \cup \mathcal{E}^H$ and the node $j \in V^G \cup \mathcal{E}^G$ to the node $l \in V^H \cup \mathcal{E}^H$. Moreover, assignments $(i,k) \in V^G \times V^H$ are node substitutions, assignments $(i,k) \in V^G \times \mathcal{E}^H$ are node removals, assignments $(i,k) \in \mathcal{E}^G \times V^H$ are node insertions, and assignments $(i,k) \in \mathcal{E}^G \times \mathcal{E}^H$ are assignments of dummy nodes to dummy nodes. Hence, the submatrix $\mathbf{C}^{ss}$ contains the costs induced by two node substitutions, $\mathbf{C}^{is}$ contains the costs induced by a node insertion and a node substitution, $\mathbf{C}^{rs}$ contains the costs induced by a node removal and a node substitution, and $\mathbf{C}^{os}$ contains the costs induced by an assignment of a dummy node to another dummy node and a node substitution. For the other submatrices of $\mathbf{C}$, analogous statements hold.

The first Lemma 3.1 tells us that, w. l. o. g., we can focus on matchings without forbidden assignments.

**Lemma 3.1.** *Let* $\mathbf{X}^\star \in \Pi_{(n+m),(m+n)}$ *be optimal for* $\mathbf{C}$. *Then* $x^\star_{i,k} = 0$ *for all* $i \not\to k$, *i.e.,* $\mathbf{X}^\star$ *does not contain any forbidden assignments.*

*Proof.* Assume that $\mathbf{X}^\star$ contains a forbidden assignment $i \not\to k$. Then we have, $Q(\mathbf{X}^\star, \mathbf{C}) \geq c_{(i,k),(i,k)} x^\star_{i,k} x^\star_{i,k} = \omega$. This contradicts the choice of $\omega$ as an upper bound for $\text{GED}(G, H)$ and the fact that, by Theorem 3.1, it holds that $Q(\mathbf{X}^\star, \mathbf{C}) = \text{QAP}(\mathbf{C}) = \text{GED}(G, H)$. □

Next, we observe that some parts of $\mathbf{C}$ can be ignored when computing the quadratic cost of matchings without forbidden assignments.

**Lemma 3.2.** *Let $\mathbf{X} \in \Pi_{(n+m),(m+n)}$ be a maximum matching without forbidden assignments. Then its quadratic cost $Q(\mathbf{X}, \mathbf{C})$ can be rewritten as follows:*

$$\begin{aligned} Q(\mathbf{X}, \mathbf{C}) = &\ \text{vec}(\mathbf{X}^s)^T \mathbf{C}^{ss} \text{vec}(\mathbf{X}^s) \\ &+ \text{vec}(\mathbf{X}^s)^T \mathbf{C}^{si} \text{vec}(\mathbf{X}^i) + \text{vec}(\mathbf{X}^s)^T \mathbf{C}^{sr} \text{vec}(\mathbf{X}^r) \\ &+ \text{vec}(\mathbf{X}^i)^T \mathbf{C}^{is} \text{vec}(\mathbf{X}^s) + \text{vec}(\mathbf{X}^i)^T \mathbf{C}^{ii} \text{vec}(\mathbf{X}^i) \\ &+ \text{vec}(\mathbf{X}^r)^T \mathbf{C}^{rs} \text{vec}(\mathbf{X}^s) + \text{vec}(\mathbf{X}^r)^T \mathbf{C}^{rr} \text{vec}(\mathbf{X}^r) \end{aligned}$$

*Proof.* The lemma immediately follows from the construction of $\mathbf{C}$. □

We now construct a function $f$ that maps maximum matching $\widehat{\mathbf{X}} \in \Pi_{n,m}$ for $\widehat{\mathbf{C}}$ to a maximum matching for $\mathbf{C}$ without node removals. For dummy nodes $(\epsilon_k, \epsilon_i) \in \mathcal{E}^G \times \mathcal{E}^H$, we introduce the notation $\epsilon_k \xrightarrow{\widehat{\mathbf{X}}} \epsilon_i$ to denote the condition that $k$ is the $i^{\text{th}}$ node in $V^H$ to which $\widehat{\mathbf{X}}$ assigns a node from $V^G$. The mapping $f$ is defined as follows:

$$f(\widehat{\mathbf{X}})_{i,k} := \begin{cases} \widehat{x}_{i,k} & \text{if } (i,k) \in V^G \times V^H \\ 1 - \sum_{j \in V^G} \widehat{x}_{j,k} & \text{if } i = \epsilon_k \\ 1 & \text{if } (i,k) \in \mathcal{E}^G \times \mathcal{E}^H \wedge i \xrightarrow{\widehat{\mathbf{X}}} k \\ 0 & \text{otherwise} \end{cases} \quad (3.10)$$

**Lemma 3.3.** *For each $\widehat{\mathbf{X}} \in \Pi_{n,m}$, it holds that $f(\widehat{\mathbf{X}}) \in \Pi_{n+m,m+n}$, that $f(\widehat{\mathbf{X}})^r = \mathbf{0}_{n,n}$, and that $f(\widehat{\mathbf{X}})_{i,k} = 0$ for all $i \not\to k$.*

*Proof.* For proving the first part of the lemma, note that, since $\widehat{\mathbf{X}} \in \Pi_{n,m}$ and $n \leq m$, the first two lines in the definition of $f$ ensure that $f(\widehat{\mathbf{X}})$ covers all rows in $V^G$, all columns in $V^H$, and leaves exactly $n$ rows in $\mathcal{E}^G$ uncovered. These rows as well as all columns in $\mathcal{E}^H$ are covered by the third line of the definition of $f$. The second and the third part of the lemma immediately follow from the definition of $f$. □

The next lemma shows that restricting to matchings without node removals and restricting to matchings that are contained in $\text{img}(f)$ is equivalent.

**Lemma 3.4.** *Let $\mathbf{X} \in \Pi_{(n+m),(m+n)}$ be a maximum matching without forbidden assignments that satisfies $\mathbf{X}^r = \mathbf{0}_{n,n}$. Then there is a maximum matching $\mathbf{X}' \in \mathrm{img}(f)$ such that $Q(\mathbf{X}, \mathbf{C}) = Q(\mathbf{X}', \mathbf{C})$.*

*Proof.* Since $\mathbf{X}^r = \mathbf{0}_{n,n}$ and $\mathbf{X} \in \Pi_{(n+m),(m+n)}$, we know that $\mathbf{X}^s \in \Pi_{n,m}$. This allows us to define $\mathbf{X}' := f(\mathbf{X}^s)$. Since $\mathbf{X}$ and $\mathbf{X}'$ do not contain forbidden assignments, we know from Lemma 3.2 that $\mathbf{X}^0$ and $\mathbf{X}'^0$ do not contribute to $Q(\mathbf{X}, \mathbf{C})$ and $Q(\mathbf{X}', \mathbf{C})$, respectively. Furthermore, we have $\mathbf{X}^s = \mathbf{X}'^s$ and $\mathbf{X}^r = \mathbf{0}_{n,n} = \mathbf{X}'^r$ by construction. The lemma thus follows if we can show that $\mathbf{X}^i = \mathbf{X}'^i$. So let $(i,k) \in \mathcal{E}^G \times V^H$. If $i \neq \epsilon_k$, we have $i \not\to k$ and hence $x'_{i,k} = x_{i,k} = 0$. Otherwise, it holds that $x'_{i,k} = 1 - \sum_{j \in V^G} x_{i,j} = x_{i,k}$, where the last equality follows from $\sum_{j \in V^{G+\epsilon}} x_{j,k} = 1$ and $x_{j,k} = 0$ for all $j \in \mathcal{E}^G$ with $j \neq \epsilon_k$. □

Next, we show that, for quasimetric edit costs, it suffices to optimize over the matchings contained in $\mathrm{img}(f)$.

**Lemma 3.5.** *If the edit costs $c_V$ and $c_E$ are quasimetric, then it holds that $\mathrm{QAP}(\mathbf{C}) = \min_{\mathbf{X} \in \mathrm{img}(f)} Q(\mathbf{X}, \mathbf{C})$.*

*Proof.* Because of Lemma 3.1, Lemma 3.3, and Lemma 3.4, it suffices to show that, for each matching $\mathbf{X} \in \Pi_{(n+m),(m+n)}$ without forbidden assignments and $r > 0$ node removals, i.e., $|\mathrm{supp}(\mathbf{X}^r)| = r$, there is a matching $\mathbf{X}' \in \Pi_{(n+m),(m+n)}$ without forbidden assignments and $r-1$ node removals such that $Q(\mathbf{X}', \mathbf{C}) \leq Q(\mathbf{X}, \mathbf{C})$. So assume that $\mathbf{X} \in \Pi_{(n+m),(m+n)}$ contains no forbidden assignments and the node removal $x_{i,\epsilon_i} = 1$ for some node $i \in V^G$. Since $\mathbf{X}$ is a matching without forbidden assignment and $n \leq m$, we then know that $\mathbf{X}$ also contains a node insertion $x_{\epsilon_k,k} = 1$ for some node $k \in V^H$.

We now define the matrix $\mathbf{X}' \in \Pi_{(n+m),(m+n)}$ as

$$x'_{j,l} := \begin{cases} 1 & \text{if } (j,l) = (i,k) \vee (j,l) = (\epsilon_k, \epsilon_i) \\ 0 & \text{if } (j,l) = (i, \epsilon_i) \vee (j,l) = (\epsilon_k, k) \\ x_{j,l} & \text{otherwise,} \end{cases}$$

and introduce $\Delta := Q(\mathbf{X}, \mathbf{C}) - Q(\mathbf{X}', \mathbf{C})$. Since $\mathbf{X}' \in \Pi_{(n+m),(m+n)}$ is immediately implied by $\mathbf{X} \in \Pi_{(n+m),(m+n)}$, the lemma follows if we can show that $\Delta \geq 0$.

Let $I := \{(i,k), (i, \epsilon_i), (\epsilon_k, k), (\epsilon_k, \epsilon_i)\}$ be the set of all indices $(j,l) \in V^{G+\epsilon} \times V^{H+\epsilon}$ with $x_{j,l} \neq x'_{j,l}$. It is easy to see that we have

$$c_{(j,l),(j',l')} x_{j,l} x_{j',l'} \quad = \quad c_{(j,l),(j',l')} x'_{j,l} x'_{j',l'} \quad = \quad 0 \quad \text{for all}$$

$((j,l),(j',l')) \in I \times I \setminus \{((i,k),(i,k)),((i,\epsilon_i),(i,\epsilon_i)),((\epsilon_k,k),(\epsilon_k,k))\}$. For this reason, since $\mathbf{C}$ is symmetric, and $\mathbf{X}$ does not contain forbidden assignments, we can write $\Delta$ as follows:

$$\Delta = \Delta_V + \sum_{(j,l) \in (V^{G+\epsilon} \times V^{H+\epsilon}) \setminus (I \cup F)} \Delta_{j,l} x_{j,l},$$

where $F := \{(j,l) \in V^{G+\epsilon} \times V^{H+\epsilon} \mid j \not\to l\}$ is the set of all forbidden assignments and $\Delta_V$ and $\Delta_{j,l}$ are defined as follows:

$$\Delta_V := c_{(i,\epsilon_i),(i,\epsilon_i)} + c_{(\epsilon_k,k),(\epsilon_k,k)} - c_{(i,k),(i,k)}$$

$$\Delta_{j,l} := 2(c_{(i,\epsilon_i),(j,l)} + c_{(\epsilon_k,k),(j,l)} - c_{(i,k),(j,l)} - c_{(\epsilon_k,\epsilon_i),(j,l)})$$

By definition of $\mathbf{C}$ and $c'_V$ given in equation (3.2), we have $\Delta_V = c_V(i,\epsilon) + c_V(\epsilon,k) - c_V(i,k) \geq 0$, where the inequality follows from the fact that $c_V$ is quasimetric. Similarly, the definition of $\mathbf{C}$ and $c'_E$ given in equation (3.4) and the fact that $(j,l) \notin F$ gives us $\Delta_{j,l} = \delta_{(i,j) \in E^G} \delta_{(k,l) \in E^H} [c_E((i,j),\epsilon) + c_E(\epsilon,(k,l)) - c_E((i,j),(k,l))] \geq 0$ for all $(j,l) \in (V^{G+\epsilon} \times V^{H+\epsilon}) \setminus (I \cup F)$, where the inequality follows from $c_E$ being quasimetric. □

The next lemma simplifies the quadratic cost $Q(\mathbf{X},\mathbf{C})$ for maximum matching $\mathbf{X} \in \text{img}(f)$.

**Lemma 3.6.** *The quadratic cost $Q(\mathbf{X},\mathbf{C})$ of a maximum matchings $\mathbf{X} \in \text{img}(f)$ can be written as follows:*

$$Q(\mathbf{X},\mathbf{C}) = \text{vec}(\mathbf{X}^s)^T \mathbf{C}^{ss} \text{vec}(\mathbf{X}^s)$$

$$+ \sum_{i \in V^G} \sum_{k \in V^H} \sum_{l \in V^H} \frac{\delta_{(k,l) \in E^H}}{2} c_E(\epsilon,(k,l)) x_{i,k} x_{\epsilon_i,l} \quad (3.11)$$

$$+ \sum_{k \in V^H} \sum_{j \in V^G} \sum_{l \in V^H} \frac{\delta_{(k,l) \in E^H}}{2} c_E(\epsilon,(k,l)) x_{\epsilon_k,k} x_{j,l} \quad (3.12)$$

$$+ \sum_{k \in V^H} \sum_{l \in V^H} \frac{\delta_{(k,l) \in E^H}}{2} c_E(\epsilon,(k,l)) x_{\epsilon_k,k} x_{\epsilon_l,l} \quad (3.13)$$

$$+ \sum_{k \in V^H} c_V(\epsilon,k) x_{\epsilon_k,k} \quad (3.14)$$

*Proof.* By Lemma 3.3, $\mathbf{X}$ does not contain forbidden assignments, which implies $\text{vec}(\mathbf{X}^s)^T \mathbf{C}^{si} \text{vec}(\mathbf{X}^i) = (3.11)$, $\text{vec}(\mathbf{X}^i)^T \mathbf{C}^{is} \text{vec}(\mathbf{X}^s) = (3.12)$, and $\text{vec}(\mathbf{X}^i)^T \mathbf{C}^{ii} \text{vec}(\mathbf{X}^i) = (3.13) + (3.14)$. Furthermore, we have $\mathbf{X}^r = \mathbf{0}_{n,n}$ from Lemma 3.3. Therefore, the lemma follows from Lemma 3.2. □

The next step is to relate the cost of a maximum matching $\widehat{\mathbf{X}} \in \Pi_{n,m}$ to the cost of its image under $f$.

**Lemma 3.7.** *The equation* $Q(f(\widehat{\mathbf{X}}), \mathbf{C}) = Q(\widehat{\mathbf{X}}, \widehat{\mathbf{C}}) + \delta_{n<m}[\sum_{(k,l) \in E^H} c_E(\epsilon, (k,l)) + \sum_{k \in V^H} c_V(\epsilon, k)]$ *holds for each* $\widehat{\mathbf{X}} \in \Pi_{n,m}$.

*Proof.* Let $\mathbf{X} = f(\widehat{\mathbf{X}})$ and $c_{k,l} = \frac{\delta_{(k,l) \in E^H}}{2} c_E(\epsilon, (k,l))$. If $n = m$, we obtain $\mathbf{X}^i = \mathbf{0}_{m,m}$ from Lemma 3.3 and the definition of $\Pi_{n+m,m+n}$. Since $\widehat{\mathbf{C}} = \mathbf{C}^{ss}$, this implies the statement of the lemma. So we can focus on the case $n < m$.

By definition of $\widehat{\mathbf{C}}$, $Q(\widehat{\mathbf{X}}, \widehat{\mathbf{C}})$ can be written as follows:

$$Q(\widehat{\mathbf{X}}, \widehat{\mathbf{C}}) = \sum_{i \in V^G} \sum_{k \in V^H} \sum_{j \in V^G} \sum_{l \in V^H} c_{(i,k),(j,l)} \widehat{x}_{i,k} \widehat{x}_{j,l} \tag{3.15}$$

$$- \sum_{i \in V^G} \sum_{k \in V^H} \sum_{l \in V^H} c_{k,l} \widehat{x}_{i,k} \sum_{j \in V^G} \widehat{x}_{j,l} \tag{3.16}$$

$$- \sum_{k \in V^H} \sum_{j \in V^G} \sum_{l \in V^H} c_{k,l} \widehat{x}_{j,l} \sum_{i \in V^G} \widehat{x}_{i,k} \tag{3.17}$$

$$- \sum_{k \in V^H} \sum_{l \in V^H} c_{k,l} \sum_{i \in V^G} \widehat{x}_{i,k} \sum_{j \in V^G} \widehat{x}_{j,l} \tag{3.18}$$

$$- \sum_{k \in V^H} c_V(\epsilon, k) \sum_{i \in V^G} \widehat{x}_{i,k} \tag{3.19}$$

Since $\widehat{x}_{i,k} = x_{i,k}$ for all $(i,k) \in V^G \times V^H$, it holds that:

$$(3.15) = \text{vec}(\mathbf{X}^s)^T \mathbf{C}^{ss} \text{vec}(\mathbf{X}^s) \tag{3.20}$$

Furthermore, by substituting $-\sum_{j \in V^G} \widehat{x}_{j,l}$ with $1 - \sum_{j \in V^G} \widehat{x}_{j,l} + 1 = x_{\epsilon_l, l} - 1$ in (3.16) and (3.18) and substituting $-\sum_{i \in V^G} \widehat{x}_{i,k}$ with $1 - \sum_{i \in V^G} \widehat{x}_{i,k} + 1 = x_{\epsilon_k, k} - 1$ in (3.17), (3.18), and (3.19), we obtain the following equalities:

$$(3.16) = (3.11) - \sum_{i \in V^G} \sum_{k \in V^H} \sum_{l \in V^H} c_{k,l} \widehat{x}_{i,k} \tag{3.21}$$

$$(3.17) = (3.12) - \sum_{k \in V^H} \sum_{j \in V^G} \sum_{l \in V^H} c_{k,l} \widehat{x}_{j,l} \tag{3.22}$$

$$(3.18) = (3.13) + \sum_{i \in V^G} \sum_{k \in V^H} \sum_{l \in V^H} c_{k,l} \widehat{x}_{i,k} \tag{3.23}$$

$$+ \sum_{k \in V^H} \sum_{j \in V^G} \sum_{l \in V^H} c_{k,l} \widehat{x}_{j,l} - \sum_{k \in V^H} \sum_{l \in V^H} c_{k,l}$$

$$(3.19) = (3.14) - \sum_{k \in V^H} c_V(\epsilon, k) \tag{3.24}$$

Since $\sum_{k \in V^H} \sum_{l \in V^H} c_{k,l} = \sum_{(k,l) \in E^H} c_E(\epsilon, (k,l))$, the lemma follows from summing the equations (3.20) to (3.24) and applying Lemma 3.6. $\square$

We are now in the position to prove the main theorem.

*Proof of Theorem 3.3.* Let $\widehat{\mathbf{X}}^\star \in \Pi_{n,m}$ be optimal for $\widehat{\mathbf{C}}$, i.e., $Q(\widehat{\mathbf{X}}^\star, \widehat{\mathbf{C}}) = \text{QAP}(\widehat{\mathbf{C}})$. Then we know from Lemma 3.7 that $Q(f(\widehat{\mathbf{X}}^\star), \mathbf{C}) = \text{QAP}(\widehat{\mathbf{C}}) + \sum_{(k,l) \in E^H} c_E(\epsilon, (k,l)) + \sum_{k \in V^H} c_V(\epsilon, k)$, which implies $\text{QAP}(\mathbf{C}) \leq \text{QAP}(\widehat{\mathbf{C}}) + \sum_{(k,l) \in E^H} c_E(\epsilon, (k,l)) + \sum_{k \in V^H} c_V(\epsilon, k)$. On the other hand, Lemma 3.5 implies that there is a matching $\mathbf{X}^\star \in \text{img}(f)$ with $Q(\mathbf{X}^\star, \mathbf{C}) = \text{QAP}(\mathbf{C})$. Let $\widehat{\mathbf{X}} \in \Pi_{n,m}$ such that $f(\widehat{\mathbf{X}}) = \mathbf{X}^\star$. Then $\text{QAP}(\mathbf{C}) = Q(\widehat{\mathbf{X}}, \widehat{\mathbf{C}}) + \delta_{n<m}[\sum_{(k,l) \in E^H} c_E(\epsilon, (k,l)) + \sum_{k \in V^H} c_V(\epsilon, k)]$ follows from Lemma 3.7. Since $n \leq m$, this proves the theorem. □

### 3.3.2 Turning the Reduction into a GED Paradigm

Given the definition of $\widehat{\mathbf{C}}$ in Theorem 3.3, it might well happen that $\widehat{c}_{(i,k),(j,l)} < 0$ for some $(i,j) \in V^G \times V^G$, $(k,l) \in V^H \times V^H$. However, some QAP methods only work on non-negative cost matrices. In order to render these methods applicable to $\widehat{\mathbf{C}}$, $\widehat{\mathbf{C}}$ can be transformed into a non-negative cost matrix $\widehat{\mathbf{C}}^+ \in \mathbb{R}^{(n \cdot m) \times (n \cdot m)}$ defined as

$$\widehat{c}^+_{(i,k),(j,l)} := \widehat{c}_{(i,k),(j,l)} + c, \tag{3.25}$$

where $c = \min\{\widehat{c}_{(i,k),(j,l)} \mid (i,j) \in V^G \times V^G \wedge (k,l) \in V^H \times V^H\}$. Clearly a maximum matching $\widehat{\mathbf{X}} \in \Pi_{n,m}$ is optimal for $\widehat{\mathbf{C}}$ just in case it is optimal for $\widehat{\mathbf{C}}^+$, as

$$Q(\widehat{\mathbf{C}}, \widehat{\mathbf{X}}) = Q(\widehat{\mathbf{C}}^+, \widehat{\mathbf{X}}) - \min\{n, m\}^2 c$$

holds for all $\widehat{\mathbf{X}} \in \Pi_{n,m}$.

Figure 3.1 shows how to use Theorem 3.3 as a paradigm for (suboptimally) computing GED: Given two graphs $G$ and $H$ and a QAP method M that yields a maximum matching $\widehat{\mathbf{X}}$ for a given cost matrices $\widehat{\mathbf{C}} \in \mathbb{R}^{(n \cdot m) \times (n \cdot m)}$, in a first step, a compact QAP instance $\widehat{\mathbf{C}}$ as specified in Theorem 3.3 (line 1). Subsequently, $\widehat{\mathbf{C}}$ is rendered non-negative by applying equation (3.25), if this is required by M (lines 2 to 3). Next, the paradigm runs M on $\widehat{\mathbf{C}}$ to obtain a maximum matching $\widehat{\mathbf{X}} \in \Pi_{n,m}$ (line 4), transforms $\widehat{\mathbf{X}}$ into a feasible solution $\mathbf{X} \in \Pi_{n+m,m+n}$ for the baseline reduction $\mathbf{C}$ from GED to QAP defined in Theorem 3.1 (line 5), and returns $Q(\mathbf{X}, \mathbf{C})$ (line 6). We have $Q(\mathbf{X}, \mathbf{C}) = \text{GED}(G, H)$, if the QAP method M is guaranteed to return optimal solutions. Otherwise, $Q(\mathbf{X}, \mathbf{C})$ is an upper bound for $\text{GED}(G, H)$.

**Input**: Two graphs $G$ and $H$, quasimetric edit costs, and a QAP method M that yields maximum matchings $\widehat{\mathbf{X}} \in \Pi_{n,m}$ for given cost matrices $\widehat{\mathbf{C}} \in \mathbb{R}^{(n \cdot m) \times (n \cdot m)}$.
**Output**: The exact edit distance GED$(G, H)$, if M yields optimal solutions; an upper bound for GED$(G, H)$, otherwise.

1 construct compact QAP instance $\widehat{\mathbf{C}}$ as specified in Theorem 3.3;
2 **if** M *requires non-negative cost matrices* **then**
3   apply equation (3.25) to render $\widehat{\mathbf{C}}$ non-negative;
4 $\widehat{\mathbf{X}} \leftarrow \text{M}(\widehat{\mathbf{C}})$;
5 $\mathbf{X} \leftarrow f(\widehat{\mathbf{X}})$, where $f$ is defined according to equation (3.10);
6 **return** $Q(\mathbf{X}, \mathbf{C})$, where $\mathbf{C}$ is constructed as specified in Theorem 3.1;

**Figure 3.1.** QAP based computation of GED.

## 3.4 Harmonization of GED Definitions

Recall that in Section 2.1, we have given two definitions of GED. Definition 2.3 defines GED as a minimization problem over the set of all edit paths. It was introduced by [60] and is nowadays used mainly in the database community. Definition 2.6 defines GED as a minimization problem over the set of all edit paths that are induced by node maps. In [21, 22], this definition is shown to be equivalent to the original GED definition by Bunke and Allermann [29]. Today, it is used in the pattern recognition community. In [60], it is shown that the two definitions coincide if the edit cost functions $c_V$ and $c_E$ are metric. Here, we generalize this result to arbitrary edit costs. More precisely, we show that the following theorem holds:

**Theorem 3.4.** *Let* $G, H \in \mathbb{G}$ *be graphs,* $c_V : \Sigma_V \cup \{\epsilon\} \times \Sigma_V \cup \{\epsilon\} \to \mathbb{R}_{\geq 0}$ *be a node edit cost function, and* $c_E : \Sigma_E \cup \{\epsilon\} \times \Sigma_E \cup \{\epsilon\} \to \mathbb{R}_{\geq 0}$ *be an edge edit cost function. Furthermore, let* $\text{GED}(G, H) := \min\{c(P) \mid P \in \Psi(G, H)\}$ *be defined as in Definition 2.3. Then there is a preprocessing routine for transforming the edit costs* $c_V$ *and* $c_E$ *that runs in* $\Theta(\min\{|V^G| + |V^H|, |\Sigma_V|\}^3 + \min\{|E^G| + |E^H|, |\Sigma_E|\}^3)$ *time, leaves* $\text{GED}(G, H)$ *invariant, and ensures that, after the preprocessing, it holds that* $\text{GED}(G, H) = \min\{c(P_\pi) \mid \pi \in \Pi(G, H)\}$.

The remainder of this section is dedicated to the proof of Theorem 3.4. Our proof crucially builds upon Bougleux et al.'s result that the set of edit paths induced by node maps is exactly the set of restricted edit paths [22].

**Definition 3.3 (Restricted Edit Path).** A *restricted edit path* between the graphs $G$ and $H$ is an edit path that edits each node and edge at most once and does not delete and reinsert edges between nodes that have been substituted.

For proving Theorem 3.4, it hence suffices to show that GED can equivalently be defined as the minimum cost of a restricted edit path. To this purpose, we define the notions of irreducible and strongly irreducible edit operations. Note that all edit operations are strongly irreducible, if and only if the edit cost functions $c_V$ and $c_E$ respect the triangle inequality.

**Definition 3.4 (Irreducible Edit Operations).** An edit operation $o$ with associated edit cost $c_S(\alpha, \beta)$, $(\alpha, \beta) \in \Sigma_S \times \Sigma_S$, $S \in \{V, E\}$, is called *irreducible*, if and only if $c_S(\alpha, \beta) \leq c_S(\alpha, \gamma) + c_S(\gamma, \beta)$ holds for all $\gamma \in \Sigma_S \setminus \{\epsilon\}$. It is called *strongly irreducible*, if and only if it is irreducible and $c_S(\alpha, \beta) \leq c_S(\alpha, \epsilon) + c_S(\epsilon, \beta)$ holds, too.

The following Lemma 3.8 and Lemma 3.9 provide links between irreducible edit operations and restricted and optimal edit paths.

**Lemma 3.8.** *Let $G, H \in \mathbb{G}$ be graphs and assume that all node substitutions are irreducible and that all node deletions and insertions as well as all edge edit operations are strongly irreducible. Then $\mathrm{GED}(G, H)$ equals the minimum cost of a restricted edit path.*

*Proof.* Let $P$ be an optimal edit path between $G$ and $H$ which is minimally non-restricted in the sense that, among all optimal edit paths, it contains a minimal number of edit operations which are forbidden for restricted edit paths. The lemma follows if $P$ is restricted. Assume that this is not the case. Then $P$ deletes and reinserts an edge between nodes that have been substituted, or there is a node or an edge that is edited twice by $P$. Assume that we are in the first case and let $e = (u_1, u_2) \in E^G$ and $f = (v_1, v_2) \in E^H$ be edges such that $P$ substitutes $u_1$ by $v_1$ and $u_2$ by $v_2$, but deletes $e$ and reinserts $f$. Let $P'$ be the edit path between $G$ and $H$ which, instead of deleting $e$ and reinserting $f$, substitutes $e$ by $f$. Since edge deletions and insertions can always be replaced by edge substitutions, $P'$ is indeed an edit path. Moreover, $P'$ contains fewer forbidden edit operations than $P$. Since edge substitutions are strongly irreducible, we have $c_E(e, f) \leq c_E(e, \epsilon) + c_E(\epsilon, f)$ and hence $c(P') \leq c(P)$. So $P'$ is optimal, which contradicts $P$ being minimally non-restricted. The other cases follow similarly. □

**Lemma 3.9.** *Optimal edit paths between graphs $G, H \in \mathbb{G}$ contain only strongly irreducible edge edit operations, strongly irreducible node deletions and insertions, and irreducible node substitutions.*

*Proof.* Let $P$ be an optimal edit path between $G$ and $H$. Assume that $P$ contains an edge edit operation $o$ with associated edit cost $c_E(\alpha, \beta)$ that is not strongly irreducible. Then there is an edge label $\gamma \in \Sigma_E \cup \{\epsilon\}$, which might be the dummy label $\epsilon$, with $c_E(\alpha, \beta) > c_E(\alpha, \gamma) + c_E(\gamma, \beta)$. Let $P'$ be the edit path that, instead of $o$, contains the edit operations associated with $c_E(\alpha, \gamma)$ and $c_E(\gamma, \beta)$. Since edge substitutions can always be replaced by edge deletions and insertions and vice versa, $P'$ is indeed an edit path between $G$ and $H$. Furthermore, it holds that $c(P') < c(P)$, which contradicts the optimality of $P$. The proofs for showing that an optimal edit path $P$ cannot contain a node insertion or deletion that is not strongly irreducible or a node substitutions that is not irreducible are analogous. □

Note that Lemma 3.9 cannot be generalized to also show that optimal edit paths only contain strongly irreducible node substitutions. The reason is that, as we can delete and insert only isolated nodes, it is in general illegal to replace a node substitution by a deletion and an insertion. With the help of Lemma 3.8 and Lemma 3.9, we can now prove Theorem 3.4.

*Proof of Theorem 3.4.* We describe a preprocessing routine for the edit costs $c_V$ and $c_E$ that leaves $\mathrm{GED}(G, H)$ invariant and ensures that, after the preprocessing, all edit operations meet the constraints of Lemma 3.8. Since it has been shown that an edit path is restricted just in case it is induced by a node map [22], this proves the theorem.

Let $\Sigma_E^{G,H} := \mathrm{img}(\ell_E^G) \cup \mathrm{img}(\ell_E^H)$. For each edge edit operation $o$ with associated edit cost $c_E(\alpha, \beta)$, the preprocessing routine computes a shortest path $p$ between $\alpha$ and $\beta$ in the complete directed graph $(\Sigma_E^{G,H} \cup \{\epsilon\}) \times (\Sigma_E^{G,H} \cup \{\epsilon\})$ with edge costs $c_E$, and then updates $c_E(\alpha, \beta)$ to $c_E(p)$. After this update, $o$ is strongly irreducible, since otherwise $p$ would not be a shortest path. Furthermore, it holds that $p = (\alpha, \beta)$, if and only if $o$ is strongly irreducible before the update. Lemma 3.9 and the fact that the costs of edge operations that were strongly irreducible before the update have not been changed imply that $\mathrm{GED}(G, H)$ is left invariant. The same technique is used to enforce the strong irreducibility of node insertions and deletions and the irreducibility of node substitutions. If one uses the Floyd-Warshall algorithm for computing the shortest paths, the complexity of this preprocessing routine is cubic in $|\operatorname{img}(\ell_V^G) \cup \operatorname{img}(\ell_V^H) \cup \{\epsilon\}| = O(\min\{|V^G| + |V^H|, |\Sigma_V|\})$ and $|\operatorname{img}(\ell_E^G) \cup \operatorname{img}(\ell_E^H) \cup \{\epsilon\}| = O(\min\{|E^G| + |E^H|, |\Sigma_E|\})$. □

## 3.5 Empirical Evaluation

In this section, we empirically evaluate the compact QAP formulation $\widehat{\mathbf{C}}$ suggested in Section 3.3. In Section 3.5.1, we describe the setup of the experiments; in Section 3.5.2, we present the results.

### 3.5.1 Setup and Datasets

**Datasets and Edit Costs.** We tested on the datasets ALKANE, ACYCLIC, MAO, and PAH. For all datasets, we used the quasimetric edit costs suggested in [1], i. e., set all node substitutions costs to 2, all node deletion and insertion costs to 4, and all edge edit costs to 4 (cf. Section 2.4). This choice guarantees that the results of our experiments are comparable to the experimental results reported in [1].

**Compared Methods.** We evaluated how employing our QAP formulation $\widehat{\mathbf{C}}$ instead of the baseline formulation $\mathbf{C}$ and the QAPE formulation $\mathbf{C}'$ affects the performance of the QAP based method IPFP with MULTI-START [41] (cf. Section 6.1.3.4 and Section 6.1.3.5 below). IPFP with MULTI-START is the best performing currently available QAP based method for upper-bounding GED [1]. In the following, IPFP-C-QAP, IPFP-B-QAP, and IPFP-QAPE denote IPFP set up with our compact QAP formulation $\widehat{\mathbf{C}}$, with the baseline QAP formulation $\mathbf{C}$, and with the QAPE formulation $\mathbf{C}'$, respectively.

**Protocol and Test Metrics.** We ran all compared methods on all pairs of graphs from the test datasets. We recorded the mean upper bound for GED ($d$), the mean runtime in seconds ($t$), and the mean error w. r. t. the exact GED ($e$), which we computed with the standard exact algorithm A* [89].

**Implementation and Hardware Specifications.** All methods were implemented in C++ and experiments were executed on a Linux Ubuntu machine with 512 GB of main memory and four AMD Opteron processors with 2.6

**Table 3.1.** Effect of different QAP and QAPE formulations on `IPFP` with `MULTI-START`. Test metrics: mean upper bound ($d$), mean error w.r.t. exact GED ($e$), mean runtime in seconds ($t$).

| algorithm | $d$ | $e$ | $t$ | $d$ | $e$ | $t$ |
|---|---|---|---|---|---|---|
| | ALKANE | | | ACYCLIC | | |
| IPFP-B-QAP [21, 22, 49] | 15.37 | 0.023 | 0.41 | 16.77 | 0.035 | 0.24 |
| IPFP-QAPE [25] | 15.34 | 0.009 | 0.22 | 16.73 | 0.008 | 0.13 |
| IPFP-C-QAP [Section 3.3] | 15.39 | 0.062 | 0.15 | 16.81 | 0.079 | 0.06 |
| | MAO | | | PAH | | |
| IPFP-B-QAP [21, 22, 49] | 33.4 | — | 2.9 | 36.7 | — | 3.14 |
| IPFP-QAPE [25] | 33.3 | — | 0.8 | 36.6 | — | 1.17 |
| IPFP-C-QAP [Section 3.3] | 39.7 | — | 1.5 | 36.7 | — | 0.89 |

GHz and 64 cores, four of which where used to run all compared methods in parallel.

### 3.5.2 Results of the Experiments

Table 3.1 shows the results of our experiments. Note that the graphs contained in MAO and PAH are too large (more than 16 nodes on average, cf. Section 5.6) to allow for an exact computation of GED, and so, for these datasets, no mean errors are reported. We see that, on all datasets expect MAO, `IPFP-C-QAP` is significantly faster than both `IPFP-B-QAP` and `IPFP-QAPE`. On MAO, `IPFP-C-QAP` is still faster than `IPFP-B-QAP` but slower than `IPFP-QAPE`.

In terms of accuracy, on the datasets ALKANE, ACYCLIC, and PAH, `IPFP-C-QAP` performs only marginally worse than `IPFP-B-QAP` and `IPFP-QAPE`. Furthermore, we see from the results for ALKANE and ACYCLIC that all three algorithms return an upper bound which is very close to the exact GED. The only exception is again MAO, where the relative deviation of the upper bound returned by `IPFP-C-QAP` from the ones returned by `IPFP-B-QAP` and `IPFP-QAPE` is around 20 %. The slight accuracy loss of `IPFP-C-QAP` w.r.t. `IPFP-B-QAP` and `IPFP-QAPE` can be explained by the fact that, since $\widehat{\mathbf{C}}$ is denser that $\mathbf{C}$ and $\mathbf{C}'$, `IPFP-C-QAP` reached the maximum number of iterations $I$ before reaching the convergence threshold $\beta$ more often than the other two algorithms.

## 3.6 Conclusions and Future Work

In this chapter, we showed that the two definitions of GED, which are used, respectively, in the database and in the pattern recognition community, are equivalent. Furthermore, we proposed a compact QAP formulation $\widehat{\mathbf{C}}$ of GED with quasimetric edit costs. Experiments show that running the state of the art algorithm IPFP with $\widehat{\mathbf{C}}$ instead of the baseline formulation $\mathbf{C}$ suggested in [21, 22, 49] leads to a speed-up by a factor between 2 and 4, while the accuracy loss is negligible on most datasets. In comparison to the QAPE formulation $\mathbf{C}'$ [25], the speed-up obtained by using $\widehat{\mathbf{C}}$ is smaller. However, implementing IPFP with $\widehat{\mathbf{C}}$ is much easier than implementing it with $\mathbf{C}'$: For implementing IPFP with $\widehat{\mathbf{C}}$, one can use an off-the-shelf solver for the linear sum assignment problem; for implementing it with $\mathbf{C}'$, one has to implement a solver for the linear sum assignment problem with error-correction.

# — 4 —

# The Linear Sum Assignment Problem with Error-Correction

In this chapter, we present `FLWC` (<u>f</u>ast solver for <u>LSAPE</u> <u>w</u>ithout <u>c</u>onstraints), a new efficient solver for the linear sum assignment problem with error-correction (LSAPE). Recall that LSAPE is a polynomially solvable combinatorial optimization problem that has to be solved as a subproblem by many exact and heuristic algorithms for GED. Consequently, all of these algorithms directly benefit if LSAPE can be solved quickly.

Exact solvers for LSAPE can be divided into two classes: Solvers of the first kind reduce instances of LSAPE to instances of the linear sum assignment problem (LSAP) and then solve the resulting LSAP instance with classical methods such as the Hungarian Algorithm [62, 77]. Solvers of the second kind adapt these classical methods to directly solve the LSAPE instances. The main disadvantage of these solvers w. r. t. solvers of the first kind is that they are much more difficult to implement: While implementations of solvers for LSAP such as the Hungarian Algorithm exist for all major programming languages, adapting these algorithms to LSAPE requires a thorough theoretical knowledge on the side of the implementer.

The newly proposed LSAPE solver `FLWC` is of the first kind. However, unlike existing reductions to LSAP, `FLWC` neither blows up the LSAPE instances nor assumes that the instances respect certain constraints. `FLWC` hence constitutes the first easily implementable, efficient, and generally applicable solver for LSAPE.

The results presented in this chapter have previously been presented in the following article:

- S. Bougleux, B. Gaüzère, D. B. Blumenthal, and L. Brun, "Fast linear sum assignment with error-correction and no cost constraints", *Pattern Recognit. Lett.*, 2018, in press. DOI: 10.1016/j.patrec.2018.03.032

Throughout this chapter, we make constant use of the fact that, by Definition 2.8 and Definition 2.26, LSAPE instances $\mathbf{C} \in \mathbb{R}^{(n+1)\times(m+1)}$ and error-correcting matchings $\mathbf{X} \in \Pi_{n,m,\epsilon}$ are of the form given in equation (4.1). $\mathbf{C}_{\text{sub}} = (c_{i,k}) \in \mathbb{R}^{n\times m}$ contains the substitution costs, $\mathbf{c}_{\text{del}} = (c_{i,\epsilon}) \in \mathbb{R}^{n\times 1}$ contains the deletion costs, and $\mathbf{c}_{\text{ins}} = (c_{\epsilon,k}) \in \mathbb{R}^{1\times m}$ contains the insertion costs. Analogously, $\mathbf{X}_{\text{sub}} = (x_{i,k}) \in \{0,1\}^{n\times m}$ encodes the substitutions, $\mathbf{x}_{\text{del}} = (x_{i,\epsilon}) \in \{0,1\}^{n\times 1}$ encodes the deletions, and $\mathbf{x}_{\text{ins}} = (x_{\epsilon,k}) \in \{0,1\}^{1\times m}$ encodes the insertions. Moreover, we assume w.l.o.g. that we are given LSAPE instances $\mathbf{C} \in \mathbb{R}^{(n+1)\times(m+1)}$ with $n \leq m$. This can easily be enforced by transposing $\mathbf{C}$.

$$\mathbf{C} = \begin{bmatrix} \mathbf{C}_{\text{sub}} & \mathbf{c}_{\text{del}} \\ \mathbf{c}_{\text{ins}} & 0 \end{bmatrix} \quad \mathbf{X} = \begin{bmatrix} \mathbf{X}_{\text{sub}} & \mathbf{x}_{\text{del}} \\ \mathbf{x}_{\text{ins}} & 0 \end{bmatrix} \tag{4.1}$$

The remainder of this chapter is organized as follows: In Section 4.1, state of the art solvers for LSAPE are reviewed. In Section 4.2, FLWC is presented. In Section 4.3, FLWC is evaluated empirically. Section 4.4 concludes the chapter.

## 4.1 State of the Art

In this section, we provide an overview of state of the art solvers for LSAPE. In Section 4.1.1 we present a reduction from LSAPE to LSAP that works for general LSAPE instances. In Section 4.1.2, we present three smaller reductions that are designed for LSAPE instances which respect the triangle inequality. In Section 4.1.3, we review solvers that use adaptions of algorithms originally designed for LSAP.

### 4.1.1 Unconstrained Reduction to LSAP

The standard algorithm *extended bipartite matching (EBP)* [82, 83] solves LSAPE by reducing it to LSAP. Given an instance $\mathbf{C}$ of LSAPE, EBP constructs an instance $\overline{\mathbf{C}} \in \mathbb{R}^{(n+m)\times(m+n)}$ of LSAP as

$$\overline{\mathbf{C}} := \begin{bmatrix} \mathbf{C}_{\text{sub}} & \omega(\mathbf{1}_{n\times n} - \mathbf{I}_n) + \text{diag}(\mathbf{c}_{\text{del}}) \\ \omega(\mathbf{1}_{m\times m} - \mathbf{I}_m) + \text{diag}(\mathbf{c}_{\text{ins}}) & \mathbf{0}_{m\times n} \end{bmatrix},$$

where $\omega$ denotes a very large value and the operator diag maps a vector $(v_i)_{i=1}^r$ to the diagonal matrix $(d_{i,k})_{i,k=1}^r$ with $d_{i,i} = v_i$ and $d_{i,k} = 0$ for all $i \neq k$. EBP then calls an LSAP solver to compute an optimal maximum matching $\overline{\mathbf{X}}^\star \in \Pi_{n+m,m+n}$ for LSAP. By construction, $\overline{\mathbf{X}}^\star$ is of the form

$$\overline{\mathbf{X}}^\star = \begin{bmatrix} \overline{\mathbf{X}}^\star_{\text{sub}} & \overline{\mathbf{X}}^\star_{\text{del}} \\ \overline{\mathbf{X}}^\star_{\text{ins}} & \overline{\mathbf{X}}^\star_\epsilon \end{bmatrix}, \qquad (4.2)$$

where $\overline{\mathbf{X}}^\star_{\text{sub}} \in \{0,1\}^{n \times m}$, $\overline{\mathbf{X}}^\star_{\text{del}} \in \{0,1\}^{n \times n}$, and $\overline{\mathbf{X}}^\star_{\text{ins}} \in \{0,1\}^{m \times m}$. The southeast quadrant $\overline{\mathbf{X}}^\star_\epsilon \in \{0,1\}^{m \times n}$ contains assignments from dummy nodes to dummy nodes which are not needed for encoding node removals or insertions but are anyway computed by LSAP algorithms. $\overline{\mathbf{X}}^\star$ is then transformed into an optimal error-correcting matching $\mathbf{X}^\star \in \Pi_{n,m,\epsilon}$ for LSAPE with $\mathbf{C}(\mathbf{X}^\star) = \overline{\mathbf{C}}(\overline{\mathbf{X}}^\star)$ by setting $\mathbf{X}^\star_{\text{sub}} := \overline{\mathbf{X}}^\star_{\text{sub}}$, $\mathbf{x}^\star_{\text{del}} := \overline{\mathbf{X}}^\star_{\text{del}} \mathbf{1}_n$, and $\mathbf{x}^\star_{\text{ins}} := \mathbf{1}_m^\top \overline{\mathbf{X}}^\star_{\text{ins}}$. The time complexity of EBP is dominated by the complexity of solving the LSAP instance $\overline{\mathbf{C}}$. Therefore, EBP runs in $O((n+m)^3)$ time.

### 4.1.2 Cost-Constrained Reductions to LSAP

The algorithms *fast bipartite matching (FBP, FBP-0)* [100] and *square fast bipartite matching (SFBP)* [99, 101] build upon more compact reductions of LSAPE to LSAP than EBP. However, FBP, FBP-0, and SFBP are applicable only to those instances $\mathbf{C}$ of LSAPE which respect the following triangle inequalities:

$$c_{i,k} \leq c_{i,\epsilon} + c_{\epsilon,k} \quad \forall (i,k) \in [n] \times [m] \qquad (4.3)$$

In other terms, FBP, FBP-0, and SFBP can be used for instances of LSAPE where substituting a node $u_i \in U$ by a node $v_k \in V$ is never more expensive than deleting $u_i$ and inserting $v_k$. The following proposition is the key-ingredient of FBP, FBP-0, and SFBP (for its proof, cf. [99–101]).

**Proposition 4.1 (Correctness of Cost-Constrained Reductions to LSAP).** *Let $\mathbf{X}^\star \in \Pi_{n,m,\epsilon}$ be an optimal error-correcting matching for an instance $\mathbf{C} \in \mathbb{R}^{(n+1) \times (m+1)}$ of LSAPE with $n \leq m$ which satisfies equation (4.3). Then $\mathbf{X}$ contains no node deletion, i.e., satisfies $\mathbf{x}_{\text{del}} = \mathbf{0}_n$. Note that this implies that $\mathbf{X}^\star$ contains exactly $m - n$ node insertions and hence that $\mathbf{x}_{\text{ins}} = \mathbf{0}_m$, if $n = m$.*

Given an instance $\mathbf{C}$ of LSAPE that satisfies equation (4.3), FBP constructs an instance $\overline{\mathbf{C}} \in \mathbb{R}^{n \times m}$ of LSAP by setting $\overline{\mathbf{C}} := \mathbf{C}_{\text{sub}} - \mathbf{c}_{\text{del}} \mathbf{1}_n^\top - \mathbf{1}_m \mathbf{c}_{\text{ins}}$. Subsequently, FBP computes an optimal maximum matching $\overline{\mathbf{X}}^\star \in \Pi_{n,m}$ for $\overline{\mathbf{C}}$.

Because of Proposition 4.1, it then holds that the matrix $\mathbf{X}^\star \in \Pi_{n,m,\epsilon}$ defined by $\mathbf{X}^\star_{\text{sub}} := \overline{\mathbf{X}}^\star$, $\mathbf{x}^\star_{\text{del}} := \mathbf{0}_n$, and $\mathbf{x}^\star_{\text{ins}} := \mathbf{1}_m^\top - \mathbf{1}_n^\top \overline{\mathbf{X}}^\star$ is an optimal error-correcting matching for $\mathbf{C}$.

The variant FBP-0 of FBP transforms $\mathbf{C}$ into a balanced instance $\overline{\mathbf{C}}_0 \in \mathbb{R}^{m \times m}$ of LSAP, by adding the matrix $\mathbf{0}_{m-n,n}$ below $\overline{\mathbf{C}}$ defined for FBP. After solving this instance, FBP-0 transforms the resulting optimal maximum matching

$$\overline{\mathbf{X}}^\star = \begin{bmatrix} \overline{\mathbf{X}}^\star_{\text{sub}} \\ \overline{\mathbf{X}}^\star_0 \end{bmatrix}$$

for $\overline{\mathbf{C}}_0$ into an optimal error-correcting matching $\mathbf{X}^\star \in \Pi_{n,m,\epsilon}$ for $\mathbf{C}$ by setting $\mathbf{X}^\star_{\text{sub}} := \overline{\mathbf{X}}^\star_{\text{sub}}$, $\mathbf{x}^\star_{\text{del}} := \mathbf{0}_n$, and $\mathbf{x}^\star_{\text{ins}} := \mathbf{1}_m^\top - \mathbf{1}_n^\top \overline{\mathbf{X}}^\star_{\text{sub}}$.

Like FBP-0, SFBP transforms an instance $\mathbf{C}$ into a balanced instance $\overline{\mathbf{C}} \in \mathbb{R}^{m \times m}$ of LSAP. However, $\overline{\mathbf{C}}$ is now defined as follows:

$$\overline{\mathbf{C}} := \begin{bmatrix} \mathbf{C}_{\text{sub}} \\ \mathbf{1}_{m-n} \mathbf{c}_{\text{ins}} \end{bmatrix}$$

In the next step, SFBP computes an optimal maximum matching

$$\overline{\mathbf{X}}^\star = \begin{bmatrix} \overline{\mathbf{X}}^\star_{\text{sub}} \\ \overline{\mathbf{X}}^\star_{\text{ins}} \end{bmatrix}$$

for $\overline{\mathbf{C}}$. SFBP then constructs the matrix $\mathbf{X}^\star \in \Pi_{n,m,\epsilon}$ by setting $\mathbf{X}^\star_{\text{sub}} := \overline{\mathbf{X}}^\star_{\text{sub}}$, $\mathbf{x}^\star_{\text{del}} := \mathbf{0}_n$, and $\mathbf{x}^\star_{\text{ins}} := \mathbf{1}_m^\top \overline{\mathbf{X}}^\star_{\text{ins}}$. Again, Proposition 4.1 ensures that $\mathbf{X}^\star$ is indeed an optimal error-correcting matching for LSAPE.

The time complexities of FBP, FBP-0, and SFBP are dominated by the complexities of solving the LSAP instances $\overline{\mathbf{C}}$. Therefore, FBP runs in $O(n^2 m)$ time, while FBP-0 and SFBP run in $O(m^3)$ time. These are significant improvements over EBP. However, recall that FBP, FBP-0, and SFBP can be used only if the cost matrix $\mathbf{C}$ respects the triangle inequalities equation (4.3).

### 4.1.3 Adaptions of Classical Algorithms

LSAPE can also be solved directly by adapting algorithms originally designed for LSAP. An adaptation of the Jonker-Volgenant Algorithm is proposed in [59]. An adaption of the Hungarian Algorithm, denoted HNG-E in this thesis, has been suggested in [24]. Both modifications lead to an overall time complexity of $O(n^2 m)$.

## 4.2 A Fast Solver Without Cost Constraints

In this section, it is shown that LSAPE without cost constraints can be reduced to an instance of LSAP of size $n \times m$. The reduction translates into the algorithm FLWC, which, like FBP, runs in $O(n^2 m)$ time, but, unlike FBP, FBP-0, and SFBP, does not assume the costs to respect the triangle inequalities (4.3). The following Theorem 4.1 states the reduction principle. It relies on a cost-dependent factorization of substitutions, deletions, and insertions. In Section 4.2.1, we discuss special cases and present FLWC. In Section 4.2.2, we present the proof of Theorem 4.1.

**Theorem 4.1 (Reduction Principle of FLWC).** *Let $\mathbf{C} \in \mathbb{R}^{(n+1)\times(m+1)}$ be an instance of LSAPE and let $\overline{\mathbf{X}}^\star \in \Pi_{n,m}$ be an optimal maximum matching for the instance $\overline{\mathbf{C}}$ of LSAP defined by setting*

$$\bar{c}_{i,k} := \delta^{\mathbf{C}}_{i,k,\epsilon} c_{i,k} + (1 - \delta^{\mathbf{C}}_{i,k,\epsilon})(c_{i,\epsilon} + c_{\epsilon,k}) - \delta_{n<m} c_{\epsilon,k}, \tag{4.4}$$

*for all $(i,k) \in [n] \times [m]$, where $\delta^{\mathbf{C}}_{i,k,\epsilon}$ is set to 1 if the triangle inequality $c_{i,k} \leq c_{i,\epsilon} + c_{\epsilon,k}$ holds, and to 0 otherwise. Furthermore, let the function $f_{\mathbf{C}} : \Pi_{n,m} \to \Pi_{n,m,\epsilon}$ be defined as*

$$\left(f_{\mathbf{C}}(\overline{\mathbf{X}})\right)_{i,k} := \delta^{\mathbf{C}}_{i,k,\epsilon} \bar{x}_{i,k}, \tag{4.5}$$

$$\left(f_{\mathbf{C}}(\overline{\mathbf{X}})\right)_{i,\epsilon} := 1 - \sum_{k=1}^{m} \delta^{\mathbf{C}}_{i,k,\epsilon} \bar{x}_{i,k}, \tag{4.6}$$

$$\left(f_{\mathbf{C}}(\overline{\mathbf{X}})\right)_{\epsilon,k} := 1 - \sum_{i=1}^{n} \delta^{\mathbf{C}}_{i,k,\epsilon} \bar{x}_{i,k}, \tag{4.7}$$

*for all $(i,k) \in [n] \times [m]$. Then $\mathbf{X}^\star := f_{\mathbf{C}}(\overline{\mathbf{X}}^\star)$ is an optimal error-correcting matching for $\mathbf{C}$ with cost $\mathbf{C}(\mathbf{X}^\star) = \overline{\mathbf{C}}(\overline{\mathbf{X}}^\star) + \delta_{n<m} \sum_{k=1}^{m} c_{\epsilon,k}$.*

**Example 4.1 (Illustration of Reduction Principle of FLWC).** Assume that $n = 2$, $m = 3$, and consider the instance $\mathbf{C}$ of LSAPE and the induced instance $\overline{\mathbf{C}}$ of LSAP:

$$\mathbf{C} = \begin{array}{c} i\backslash k \\ 1 \\ 2 \\ \epsilon \end{array} \begin{array}{cccc} 1 & 2 & 3 & \epsilon \\ \left[\begin{array}{cccc} 3 & 5 & 1 & 4 \\ 8 & 9 & 4 & 4 \\ 2 & 4 & 0 & 0 \end{array}\right] \end{array} \xrightarrow{\text{apply (4.4)}} \overline{\mathbf{C}} = \begin{array}{c} i\backslash k \\ 1 \\ 2 \end{array} \begin{array}{ccc} 1 & 2 & 3 \\ \left[\begin{array}{ccc} 1 & 1 & 1 \\ 4 & 4 & 4 \end{array}\right] \end{array}$$

For instance, we have $\bar{c}_{1,1} = \delta^{\mathbf{C}}_{1,1,\epsilon} 3 + (1 - \delta^{\mathbf{C}}_{1,1,\epsilon})(2+4) - \delta_{2<3} 2 = 3 - 2 = 1$ and $\bar{c}_{2,2} = \delta^{\mathbf{C}}_{2,2,\epsilon} 9 + (1 - \delta^{\mathbf{C}}_{2,2,\epsilon})(4+4) - \delta_{2<3} 4 = 8 - 4 = 4$. Figure 4.1a shows an optimal maximum matching $\overline{\mathbf{X}}^\star \in \Pi_{n,m}$ for $\overline{C}$, and Figure 4.1b shows an optimal error-correcting matching $\mathbf{X}^\star = f_{\mathbf{C}}(\overline{X}^\star) \in \Pi_{n,m,\epsilon}$ for $\mathbf{C}$. Note that $f_{\mathbf{C}}$ factorizes the substitution $(u_2, v_2)$ into the removal $(u_2, \epsilon)$ and the insertion $(\epsilon, v_2)$, since $c_{2,2} > c_{2,\epsilon} + c_{\epsilon,2}$ (Figure 4.1c). On the other hand, the substitution $(u_1, v_1)$ is not factorized, since $c_{1,1} \leq c_{1,\epsilon} + c_{\epsilon,1}$ (Figure 4.1d). Furthermore, we have $\mathbf{C}(\mathbf{X}^\star) = 11$, $\overline{C}(\overline{\mathbf{X}}^\star) = 5$, and $\delta_{n<m} \sum_{k=1}^{m} c_{\epsilon,k} = 6$. Therefore, it holds that $\mathbf{C}(\mathbf{X}^\star) = \overline{C}(\overline{\mathbf{X}}^\star) + \delta_{n<m} \sum_{k=1}^{m} c_{\epsilon,k}$, as stated by our reduction principle.

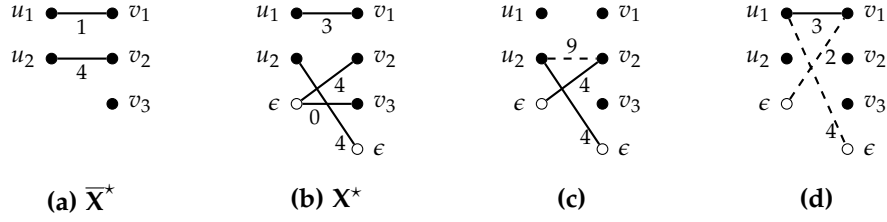

**Figure 4.1.** Illustration of the reduction principle of FLWC.

### 4.2.1 Discussion of Special Cases and Presentation of FLWC

Theorem 4.1 states that a general instance $\mathbf{C} \in \mathbb{R}^{(n+1)\times(m+1)}$ of LSAPE, which is not required to respect the triangle inequalities (4.3), can be reduced to a $(n \times m)$-sized instance of LSAP. However, there is still room for improvement if $\mathbf{C}$ does respect the triangle inequalities.

**Proposition 4.2 (LSAPE with Triangular Costs).** *Let $\mathbf{C} \in \mathbb{R}^{(n+1)\times(m+1)}$ be an instance of LSAPE with $n \leq m$ that respects the triangle inequalities (4.3). Then the reduction principle specified in Theorem 4.1 can be carried out without knowledge of the deletion costs $\mathbf{c}_{\text{del}}$. If $n = m$ holds, too, then knowledge of the insertion costs $\mathbf{c}_{\text{ins}}$ is not necessary, either.*

*Proof.* The proposition immediately follows from the facts that $\delta^{\mathbf{C}}_{i,k,\epsilon} = 1$ holds for all $(i,k) \in [n] \times [m]$ if $\mathbf{C}$ respects (4.3), and that $\delta_{n<m} = 0$ if $n = m$. □

Proposition 4.2 is useful, because in many application scenarios for LSAPE, $\mathbf{C}$ is not given but has to be computed. Furthermore, $\mathbf{C}$ is often known a priori to respect the triangle inequalities, for instance, because its entries contain the distances between elements in a metric space. In such settings, the overall performance of an algorithm that calls a LSAPE solver as a subroutine can improve significantly if only parts of the cost matrix $\mathbf{C}$ have

**Table 4.1.** Required parts of **C** for reduction from LSAPE to LSAP employed by `FLWC`.

|       | triangle inequalities hold | triangle inequalities do not hold |
|-------|---------------------------|-----------------------------------|
| $n = m$ | $\mathbf{C}_{\text{sub}}$ | $\mathbf{C}_{\text{sub}}, \mathbf{c}_{\text{ins}}, \mathbf{c}_{\text{del}}$ |
| $n < m$ | $\mathbf{C}_{\text{sub}}, \mathbf{c}_{\text{ins}}$ | $\mathbf{C}_{\text{sub}}, \mathbf{c}_{\text{ins}}, \mathbf{c}_{\text{del}}$ |
| $n > m$ | $\mathbf{C}_{\text{sub}}, \mathbf{c}_{\text{del}}$ | $\mathbf{C}_{\text{sub}}, \mathbf{c}_{\text{ins}}, \mathbf{c}_{\text{del}}$ |

**Input**: An instance $\mathbf{C} \in \mathbb{R}^{(n+1)\times(m+1)}$ of LSAPE.
**Output**: An optimal error-correcting matching $\mathbf{X}^\star \in \Pi_{n,m,\epsilon}$ for $\mathbf{C}$.

1 **if** $n > m$ **then** $\mathbf{C} \leftarrow \mathbf{C}^\mathsf{T}$; $(n,m) \leftarrow (m,n)$;
2 initialize $\overline{\mathbf{C}} \in \mathbb{R}^{n \times m}$;
3 **for** $i \in [n]$ **do**
4     **for** $k \in [m]$ **do**
5         **if** $\mathbf{C}$ *respects triangle inequalities* **then**
6             **if** $n = m$ **then** $\overline{c}_{i,k} \leftarrow c_{i,k}$ **else** $\overline{c}_{i,k} \leftarrow c_{i,k} - c_{\epsilon,k}$;
7         **else**
8             $\overline{c}_{i,j} \leftarrow \delta^{\mathbf{C}}_{i,k,\epsilon} c_{i,k} + (1 - \delta^{\mathbf{C}}_{i,k,\epsilon})(c_{i,\epsilon} + c_{\epsilon,k}) - \delta_{n<m} c_{\epsilon,k}$;

9 call LSAP solver to compute optimal maximum matching $\overline{\mathbf{X}}^\star \in \Pi_{n,m}$ for $\overline{\mathbf{C}}$;
10 $\mathbf{X}^\star \leftarrow f_{\mathbf{C}}(\overline{\mathbf{X}}^\star)$;
11 **if** *cost matrix* $\mathbf{C}$ *was transposed in line 1* **then** $\mathbf{X}^\star \leftarrow \mathbf{X}^{\star\mathsf{T}}$;
12 **return** $\mathbf{X}^\star$;

**Figure 4.2.** The LSAPE solver `FLWC`.

to be computed. Table 4.1 summarizes which parts of **C** have to be known for our reduction from LSAPE to LSAP. The case $n > m$ can straightforwardly be obtained from the case $n < m$ by transposing **C**. Recall that both `FBP` and `FBP-0` always require the entire cost matrix **C** to be known. `SFBP` requires the same parts of **C** as our approach, but reduces LSAPE to a larger instance of LSAP ($\max\{n,m\} \times \max\{n,m\}$ vs. $n \times m$).

Figure 4.2 shows the algorithm `FLWC`, which turns our reduction from LSAPE to LSAP into a method for computing an optimal error-correcting matching. `FLWC` only uses those parts of **C** which are really required by the reduction (cf. Table 4.1). If the adaption of the Hungarian Algorithm to unbalanced instances of LSAP is used in line 9, `FLWC` runs in $O(\min\{n,m\}^2 \max\{n,m\})$ time and $O(nm)$ space.

Table 4.2 compares `FLWC`'s time and space complexities to the complexities of existing competitors. Note that our reduction principle can also be used for the fast computation of a suboptimal solution for LSAPE. To this end, it suffices to replace the optimal LSAP solver in line 9 of Algorithm 4.2 by a suboptimal one such as one of the greedy heuristics suggested in [91].

**Table 4.2.** Time and space complexities of existing algorithms for LSAPE under the assumptions that reductions to LSAP use the Hungarian Algorithm for solving LSAP.

| method | time | space |
|---|---|---|
| *cost-constrained methods* | | |
| FBP [100] | $O(\min\{n,m\}^2 \max\{n,m\})$ | $O(nm)$ |
| FBP-0 [100] | $O(\max\{n,m\}^3)$ | $O(\max\{n,m\}^2)$ |
| SFBP [99, 101] | $O(\max\{n,m\}^3)$ | $O(\max\{n,m\}^2)$ |
| *general methods* | | |
| EBP [82, 83] | $O((n+m)^3)$ | $O((n+m)^2)$ |
| HNG-E [24] | $O(\min\{n,m\}^2 \max\{n,m\})$ | $O(nm)$ |
| FLWC [our approach] | $O(\min\{n,m\}^2 \max\{n,m\})$ | $O(nm)$ |

### 4.2.2 Correctness of the Reduction Principle

The first step towards the proof is the following Lemma 4.1, which constitutes a relation between error-correcting matchings and maximum matchings.

**Lemma 4.1.** *Let* $\mathbf{X} \in \Pi_{n,m,\epsilon}$ *be an error-correcting matching. Furthermore, let* $Z_{\mathbf{X}} := (U, V, \{(u_i, v_k) \in U \times V \mid x_{i,\epsilon} x_{\epsilon,k} = 1\})$ *be the bipartite graph between $U$ and $V$ whose edges encode all combinations of node removals and insertions, let $\mathcal{Z}^\star$ be the set of maximum matchings for $Z_{\mathbf{X}}$, and let the set $\mathcal{Y}_{\mathbf{X}}$ be defined as follows:*

$$\mathcal{Y}_{\mathbf{X}} := \{\mathbf{X}_{\text{sub}} + \mathbf{Z}^\star \mid \mathbf{Z}^\star \in \mathcal{Z}^\star\} \tag{4.8}$$

*Then* $\mathbf{Y} \in \Pi_{n,m}$ *holds for each* $\mathbf{Y} \in \mathcal{Y}_{\mathbf{X}}$.

*Proof.* Let $I_{\text{sub}} := \{i \in [n] \mid \sum_{k=1}^{m} x_{i,k} = 1\}$ and $K_{\text{sub}} := \{k \in [m] \mid \sum_{i=1}^{n} x_{i,k} = 1\}$ be the set of indices of those nodes of $U$ and $V$ that are substituted by $\mathbf{X}$. Furthermore, let $s = |I_{\text{sub}}|$ be the number of substitutions encoded by $\mathbf{X}$, and let $\mathbf{Z}^\star$ be a maximum matching for $Z_{\mathbf{X}}$. We observe that $Z_{\mathbf{X}}$ can be viewed as the complete bipartite graph between the nodes $U_{\text{del}} := \{u_i \mid i \in [n] \setminus I_{\text{sub}}\}$ and $V_{\text{ins}} := \{v_k \mid k \in [m] \setminus K_{\text{sub}}\}$ that are deleted and inserted by $\mathbf{X}$, and that we have $|U_{\text{del}}| = n - s \leq m - s = |V_{\text{ins}}|$. These observations imply that we have $\sum_{k=1}^{m} z_{i,k}^\star = 0$ for each $i \in I_{\text{sub}}$ and $\sum_{k=1}^{m} z_{i,k}^\star = 1$ for each $i \in [n] \setminus I_{\text{sub}}$.

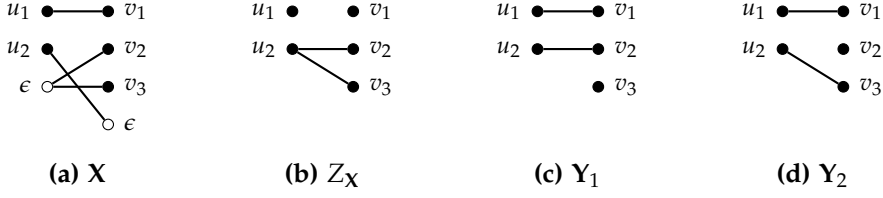

(a) **X**  (b) $Z_\mathbf{X}$  (c) $\mathbf{Y_1}$  (d) $\mathbf{Y_2}$

**Figure 4.3.** Illustration of Lemma 4.1.

Similarly, we have $\sum_{i=1}^{n} z_{i,k}^\star = 0$ for each $k \in K_{\text{sub}}$ and $\sum_{i=1}^{n} z_{i,k}^\star \leq 1$ for each $k \in [m] \setminus K_{\text{sub}}$. This gives us $\sum_{k=1}^{m} x_{i,k} + z_{i,k}^\star = 1$ for each $i \in [n]$ and $\sum_{i=1}^{n} x_{i,k} + z_{i,k}^\star \leq 1$ for each $k \in [m]$, which implies $\mathbf{X}_{\text{sub}} + \mathbf{Z}^\star \in \Pi_{n,m}$. □

**Example 4.2 (Illustration of Lemma 4.1).** Consider the error-correcting matching **X** shown in Figure 4.3a. Since **X** removes $u_2$ and inserts $v_2$ and $v_3$, $Z_\mathbf{X}$ contains the edges $(u_2, v_2)$ and $(u_2, v_3)$ (Figure 4.3b). There are exactly two maximum matchings for $Z_\mathbf{X}$: $\mathbf{Z}_1^\star := \{(u_2, v_2)\}$ and $\mathbf{Z}_2^\star := \{(u_2, v_3)\}$. This implies that $\mathbf{Y}_\mathbf{X} = \{\mathbf{Y_1}, \mathbf{Y_2}\}$ with $\mathbf{Y_1} = \{(u_1, v_1), (u_2, v_2)\}$ (Figure 4.3c) and $\mathbf{Y_2} = \{(u_1, v_1), (u_2, v_3)\}$ (Figure 4.3d).

We now introduce the notion of a minimally-sized error-correcting matching. To this purpose, we call two error-correcting matchings $\mathbf{X}, \mathbf{X}' \in \Pi_{n,m,\epsilon}$ *equivalent* w.r.t. an instance $\mathbf{C}$ of LSAPE (in symbols: $\mathbf{X} \sim_\mathbf{C} \mathbf{X}'$) if and only if, for all $(i,k) \in [n] \times [m]$, $x_{i,k} = x'_{i,k}$ or $c_{i,k} = c_{i,\epsilon} + c_{\epsilon,k}$ and $x_{i,k} = x'_{i,\epsilon} x'_{\epsilon,k}$ or $x'_{i,k} = x_{i,\epsilon} x_{\epsilon,k}$. By definition of $\sim_\mathbf{C}$, the cost $\mathbf{C}(\cdot)$ is invariant on the equivalence classes induced by $\sim_\mathbf{C}$.

**Definition 4.1 (Minimally-Sized Error-Correcting Matching).** Let $\mathbf{C} \in \mathbb{R}^{(n+1)\times(m+1)}$ be an instance of LSAPE and $\mathbf{X} \in \Pi_{n,m,\epsilon}$ be an error-correcting matching. Then **X** is called *minimally-sized* if and only if $|\text{supp}(\mathbf{X})| < |\text{supp}(\mathbf{X}')|$ holds for all $\mathbf{X}' \in [\mathbf{X}]_{\sim_\mathbf{C}}$, where $\text{supp}(\mathbf{X})$ is the support of **X**.

In other words, **X** is minimally-sized just in case it always favors substitution over removal plus insertion, if the costs are the same. By construction, each equivalence class $[\mathbf{X}]_{\sim_\mathbf{C}}$ contains exactly one minimally-sized error-correcting matching. In particular, there is always a minimally-sized optimal error-correcting matching.

The next step is to characterize a subset $\Pi_{n,m,\epsilon}(\mathbf{C}) \subseteq \Pi_{n,m,\epsilon}$ which contains all optimal minimally-sized error-correcting matchings for a given instance $\mathbf{C}$ of LSAPE.

**Lemma 4.2.** *Let* $\mathbf{C} \in \mathbb{R}^{(n+1) \times (m+1)}$ *be an instance of LSAPE, and let the set* $\Pi_{n,m,\epsilon}(\mathbf{C}) \subseteq \Pi_{n,m,\epsilon}$ *of cost-dependent error-correcting matchings be defined as follows:*

$$\Pi_{n,m,\epsilon}(\mathbf{C}) := \{\mathbf{X} \in \Pi_{n,m,\epsilon} \mid (1 - \delta_{i,k,\epsilon}^{\mathbf{C}}) x_{i,k} + \delta_{i,k,\epsilon}^{\mathbf{C}} x_{i,\epsilon} x_{\epsilon,k} = 0 \ \forall (i,k) \in [n] \times [m]\}$$

*Then* $\Pi_{n,m,\epsilon}(\mathbf{C})$ *contains all minimally-sized optimal error-correcting matchings for the LSAPE instance* $\mathbf{C}$.

*Proof.* Assume that there is a minimally-sized optimal error-correcting matching $\mathbf{X}^\star \in \Pi_{n,m,\epsilon} \setminus \Pi_{n,m,\epsilon}(\mathbf{C})$. Then there is a pair $(i,k) \in [n] \times [m]$ such that $(1 - \delta_{i,k,\epsilon}^{\mathbf{C}}) x_{i,k}^\star = 1$ or $\delta_{i,k,\epsilon}^{\mathbf{C}} x_{i,\epsilon}^\star x_{\epsilon,k}^\star = 1$. Assume that we are in the first case, i.e., that $\delta_{i,k,\epsilon}^{\mathbf{C}} = 0$ and $x_{i,k}^\star = 1$. This implies $c_{i,k} > c_{i,\epsilon} + c_{\epsilon,k}$. Now consider the error-correcting matching $\mathbf{X}'$, which, instead of substituting $u_i$ by $v_k$, deletes $u_i$ and inserts $v_k$ ($x'_{i,\epsilon} x'_{\epsilon,k} = 1$). Since $\delta_{i,k,\epsilon}^{\mathbf{C}} = 0$, $\mathbf{X}'$ is cheaper than $\mathbf{X}^\star$. This contradicts $\mathbf{X}^\star$'s optimality.

If we are in the second case, we have $x_{i,\epsilon}^\star x_{\epsilon,k}^\star = 1$ and $c_{i,k} \leq c_{i,\epsilon} + c_{\epsilon,k}$. Since $\mathbf{X}^\star$ is minimally-sized, we can strengthen the last inequality to $c_{i,k} < c_{i,\epsilon} + c_{\epsilon,k}$. Consider the error-correcting matching $\mathbf{X}'$, which, instead of deleting $u_i$ and inserting $v_k$, substitutes $u_i$ by $v_k$ ($x'_{i,k} = 1$). Again, $\mathbf{X}'$ is cheaper than $\mathbf{X}^\star$, which is a contradiction to $\mathbf{X}^\star$'s optimality. □

The following Lemma 4.3 shows that the transformation function $f_\mathbf{C}$ defined in Theorem 4.1 indeed maps maximum matchings to error-correcting matchings and that it is surjective on $\Pi_{n,m,\epsilon}$.

**Lemma 4.3.** *Let* $f_\mathbf{C} : \Pi_{n,m} \to \Pi_{n,m,\epsilon}$ *be defined as in Theorem 4.1. Then it holds that* $\mathrm{img}(f_\mathbf{C}) \subseteq \Pi_{n,m,\epsilon}$ *and that* $\mathrm{img}(f_\mathbf{C}) \supseteq \Pi_{n,m,\epsilon}(\mathbf{C})$.

*Proof.* Consider a maximum matching $\overline{\mathbf{X}} \in \Pi_{n,m}$ and let $\mathbf{X} := f_\mathbf{C}(\overline{\mathbf{X}})$. From (4.5) and (4.6), we have $x_{i,\epsilon} + \sum_{k=1}^{m} x_{i,k} = 1$ for each $i \in [n]$. From (4.5) and (4.7), we have $x_{\epsilon,k} + \sum_{i=1}^{n} x_{i,k} = 1$ for each $k \in [m]$. This implies $\mathbf{X} \in \Pi_{n,m,\epsilon}$ and thus $\mathrm{img}(f_\mathbf{C}) \subseteq \Pi_{n,m,\epsilon}$.

For showing $\mathrm{img}(f_\mathbf{C}) \supseteq \Pi_{n,m,\epsilon}(\mathbf{C})$, we fix an error-correcting matching $\mathbf{X} \in \Pi_{n,m,\epsilon}(\mathbf{C})$. From Lemma 4.1, we know that there is a set $\mathcal{Y}_\mathbf{X}$ of maximum matchings representing $\mathbf{X}$. Consider a maximum matching $\overline{\mathbf{X}} \in \mathcal{Y}_{\mathbf{X}}$, i.e., a maximum matching $\overline{\mathbf{X}}$ that can be written as $\overline{\mathbf{X}} = \mathbf{X}_{\text{sub}} + \mathbf{Z}^{\star}$ for some $\mathbf{Z}^{\star} \in \mathcal{Z}_{\mathbf{X}}^{\star}$. We will show that $\mathbf{X} = f_{\mathbf{C}}(\overline{\mathbf{X}})$, which proves the proposition. To this end, we first show that

$$\delta^{\mathbf{C}}_{i,k,\epsilon} z^{\star}_{i,k} = 0 \tag{4.9}$$

holds for all $(i,k) \in [n] \times [m]$. Consider a pair $(i,k) \in [n] \times [m]$ with $z^{\star}_{i,k} = 1$. From $Z_{\mathbf{X}} \in \mathcal{Z}_{\mathbf{X}'}^{\star}$ we know that $x_{i,\epsilon} x_{\epsilon,k} = 1$. As $\mathbf{X} \in \Pi_{n,m,\epsilon}(\mathbf{C})$, this implies $\delta^{\mathbf{C}}_{i,k,\epsilon} = 0$ and hence proves (4.9).

Now consider an arbitrary pair $(i,k) \in [n] \times [m]$. It holds that $f_{\mathbf{C}}(\overline{\mathbf{X}})_{i,k} = \delta^{\mathbf{C}}_{i,k,\epsilon} x_{i,k} + \delta^{\mathbf{C}}_{i,k,\epsilon} z^{\star}_{i,k} = \delta^{\mathbf{C}}_{i,j,\epsilon} x_{i,k} = x_{i,k}$, where the first equality follows from the definitions of $f_{\mathbf{C}}$ and $\overline{\mathbf{X}}$, the second equality follows from (4.9), and the third equality follows from $\mathbf{X} \in \Pi_{n,m,\epsilon}(\mathbf{C})$. We have hence shown that $\mathbf{X}_{\text{sub}} = f_{\mathbf{C}}(\overline{\mathbf{X}})_{\text{sub}}$. Next, we show that $f_{\mathbf{C}}(\overline{\mathbf{X}})_{\text{ins}} = \mathbf{x}_{\text{ins}}$. For all $k \in [m]$, we have $f_{\mathbf{C}}(\overline{\mathbf{X}})_{\epsilon,k} = 1 - \sum_{i=1}^{n} \delta^{\mathbf{C}}_{i,k,\epsilon} x_{i,k} - \sum_{i=1}^{n} \delta^{\mathbf{C}}_{i,k,\epsilon} z^{\star}_{i,k} = 1 - \sum_{i=1}^{n} \delta^{\mathbf{C}}_{i,k,\epsilon} x_{i,k} = 1 - \sum_{i=1}^{n} x_{i,k} = x_{\epsilon,k}$, as required. Again, the first equality follows from the definitions of $f_{\mathbf{C}}$ and $\overline{\mathbf{X}}$, the second equality follows from (4.9), and the third equality follows from $\mathbf{X} \in \Pi_{n,m,\epsilon}(\mathbf{C})$. The last equality follows from the fact that $\mathbf{X}$ is an error-correcting matching. The argument for showing that $f_{\mathbf{C}}(\overline{\mathbf{X}})_{\text{rem}} = \mathbf{x}_{\text{del}}$ is analogous. $\square$

We can now prove the correctness of our reduction principle.

*Proof of Theorem 4.1.* Let $\overline{\mathbf{X}} \in \Pi_{n,m}$ be a maximum matching. We have:

$$\begin{aligned}
\overline{\mathbf{C}}(\overline{\mathbf{X}}) &= \sum_{i=1}^{n} \sum_{k=1}^{m} (\delta^{\mathbf{C}}_{i,k,\epsilon} c_{i,k} + (1 - \delta^{\mathbf{C}}_{i,k,\epsilon})(c_{i,\epsilon} + c_{\epsilon,k}) - \delta_{n<m} c_{\epsilon,k}) \overline{x}_{i,k} \\
&= \sum_{i=1}^{n} \sum_{k=1}^{m} c_{i,k} \delta^{\mathbf{C}}_{i,k,\epsilon} \overline{x}_{i,k} + (c_{i,\epsilon} + c_{\epsilon,k})(1 - \delta^{\mathbf{C}}_{i,k,\epsilon}) \overline{x}_{i,k} \\
&\quad - \delta_{n<m} \left( \sum_{k=1}^{m} c_{\epsilon,k} \sum_{i=1}^{n} \overline{x}_{i,k} + \sum_{k=1}^{m} c_{\epsilon,k} - \sum_{k=1}^{m} c_{\epsilon,k} \right) \\
&= \sum_{i=1}^{n} \sum_{k=1}^{m} c_{i,k} \underbrace{(\delta^{\mathbf{C}}_{i,k,\epsilon} \overline{x}_{i,k})}_{= f_{\mathbf{C}}(\overline{\mathbf{X}})_{i,k}} + \sum_{i=1}^{n} c_{i,\epsilon} \underbrace{\left( \sum_{k=1}^{m} (1 - \delta^{\mathbf{C}}_{i,k,\epsilon}) \overline{x}_{i,k} \right)}_{=: A_i} \\
&\quad + \sum_{k=1}^{m} c_{\epsilon,k} \underbrace{\left( \delta_{n<m} + \sum_{i=1}^{n} (1 - \delta^{\mathbf{C}}_{i,k,\epsilon} - \delta_{n<m}) \overline{x}_{i,k} \right)}_{=: B_k} - \delta_{n<m} \sum_{k=1}^{m} c_{\epsilon,k}
\end{aligned}$$

Since $\overline{\mathbf{X}}$ is a maximum matching for $\Pi_{n,m}$, we know that $A_i = 1 - \sum_{k=1}^{m} \delta_{i,k,\epsilon}^{\mathbf{C}} \overline{x}_{i,k} = f_{\mathbf{C}}(\overline{\mathbf{X}})_{i,\epsilon}$ for each $i \in [n]$. We now distinguish the cases $\delta_{n<m} = 1$ and $\delta_{n<m} = 0$. In the first case, we immediately have $B_k = 1 - \sum_{i=1}^{n} \delta_{i,k,\epsilon}^{\mathbf{C}} \overline{x}_{i,k} = f_{\mathbf{C}}(\overline{\mathbf{X}})_{\epsilon,k}$ for each $k \in [m]$. In the second case, we have $n = m$ and $B_k = f_{\mathbf{C}}(\overline{\mathbf{X}})_{\epsilon,k}$ holds, too, since, for balanced instances, a maximum matching contains an edge $(u_i, v_k)$ for each $k \in [m]$. We have thus shown that the following equality holds for all $\overline{\mathbf{X}} \in \Pi_{n,m}$:

$$\overline{\mathbf{C}}(\overline{\mathbf{X}}) = \mathbf{C}(f_{\mathbf{C}}(\overline{\mathbf{X}})) - \delta_{n<m} \sum_{k=1}^{m} c_{\epsilon,k} \qquad (4.10)$$

Now let $\mathbf{X}^\star$ be a minimally-sized optimal error-correcting matching for $\mathbf{C}$, $\overline{\mathbf{X}}^\star$ be an optimal maximum matching for $\overline{\mathbf{C}}$, and $\mathbf{X}' := f_{\mathbf{C}}(\overline{\mathbf{X}}^\star)$. From (4.10), we know that $\mathbf{C}(\mathbf{X}') = \overline{\mathbf{C}}(\overline{\mathbf{X}}^\star) + \delta_{n<m} \sum_{k=1}^{m} c_{\epsilon,k}$. Therefore, the theorem follows if we can show that $\mathbf{C}(\mathbf{X}') = \mathbf{C}(\mathbf{X}^\star)$. The $\geq$-direction of the desired equality follows from $\mathbf{X}^\star$'s optimality and the fact that, from Lemma 4.3, we know that $\mathbf{X}' \in \Pi_{n,m,\epsilon}$. For showing the $\leq$-direction, we pick a maximum matching $\overline{\mathbf{X}}' \in \Pi_{n,m}$ with $f_{\mathbf{C}}(\overline{\mathbf{X}}') = \mathbf{X}^\star$. Such a matching exists because, from Lemma 4.2, we know that $\overline{\mathbf{X}}^\star \in \Pi_{n,m,\epsilon}(\mathbf{C})$, and, from Lemma 4.3, we have $\text{img}(f_{\mathbf{C}}) \supseteq \Pi_{n,m,\epsilon}(\mathbf{C})$. Assume that $\mathbf{C}(\mathbf{X}') > \mathbf{C}(\mathbf{X}^\star)$. Then (4.10) implies that $\overline{\mathbf{C}}(\overline{\mathbf{X}}^\star) > \overline{\mathbf{C}}(\overline{\mathbf{X}}')$, which contradicts $\overline{\mathbf{X}}^\star$'s optimality. □

## 4.3 Empirical Evaluation

In our experiments, we compared `FLWC` to all existing competitors mentioned in Table 4.2. We evaluated the runtime of the different methods (Section 4.3.1) and their performances when used within approximate approaches for the computation of GED (Section 4.3.2). All procedures use the same C++ implementation of the Hungarian Algorithm as LSAP solver, which is based on the version presented in [31, 63]. `HNG-E` is also based on this version, so that all procedures are comparable.[1]

### 4.3.1 Runtime Comparison

To illustrate the differences between the methods, we recorded their execution time on three types of instances, some of which do and some of which do not respect the triangle inequalities: Machol-Wien instances, flat instances, and random instances. Experiments on further types yielded similar results. For each instance type, the tested methods' execution time is reported for several values of $n$ and $m$. In the first and third columns of the figures, $n$ is fixed and $m$ is varied, while in the second and fourth columns, $m$ is fixed and $n$ is varied. In any case, we have $n \leq m$. The general methods FLWC, EBP, and HNG-E were tested on all instances, the cost-constrained algorithms FBP, FBP-0, and SFBP only on those that respect the triangle inequalities. Since FBP-0 was slower than FBP across all instances, our plots do not contain curves for FBP-0.

---

[1] Procedures are written both in C++ and MATLAB (or GNU Octave). The source code is available at `https://bougleux.users.greyc.fr/lsape/`.

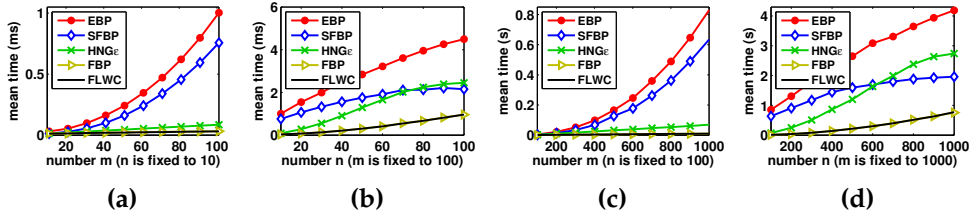

**Figure 4.4.** Time comparisons for Machol-Wien instances.

**Machol-Wien Instances.** Figure 4.4 shows the results for Machol-Wien instances. Machol-Wien instances $\mathbf{C} \in \mathbb{R}^{(n+1)\times(m+1)}$ are defined as $c_{i,k} := (i-1)(k-1)$ for all $(i,k) \in [n+1] \times [m+1]$ and thus satisfy the triangle inequalities [74, 75]. They are known to be hard to optimize for classical LSAP solvers such as the Hungarian Algorithm. This is also the case for the LSAPE solver HNG-E, but not for the other methods. This is explained by the fact that all methods except for HNG-E transform an instance $\mathbf{C}$ of LSAPE into an instance $\overline{\mathbf{C}}$ of LSAP, which reduces the difficulty of the problem. We observe that FLWC and FBP perform very similarly, provide the best results in all cases, and are more stable than the other methods. EBP and SFBP are more time consuming, in particular when $n$ and $m$ increases.

**Flat Instances.** Figure 4.5 shows the results for flat instances. In a first series of experiments, we considered flat instances of the form $\mathbf{C} = \alpha \mathbf{1}_{n,m}$, with $\alpha = 10$. These instances satisfy the triangle inequalities. Moreover, they are easy to solve, which implies that a large part of the execution time is spent on the initialization step of the Hungarian Algorithm and the cost transformations. The results for instances of this kind are displayed Figures 4.5a to 4.5d. As before, FLWC and FBP are more stable and efficient than the other methods. In a second series of experiments, we varied the construction of **C** in order to enforce that $p$ % of the nodes contained in the smaller set be deleted. This can be achieved by randomly selecting $m - n$ nodes in the larger set and setting their insertion cost to $0$. Subsequently, $p$ % of the remaining elements are selected and their insertion cost is set to $(\alpha/2) - 1$. Similarly, the deletion costs of $p$ % of the nodes contained in the smaller set are set to $(\alpha/2) - 1$. The resulting instances do not satisfy the triangle inequalities, and thus FBP, FBP-0, and SFBP are not tested (they do not compute an optimal solution). The results for $p = 25$ and $p = 25$ are shown in Figures 4.5e to 4.5h and Figures 4.5i to 4.5l, respectively. Both FLWC and HNG-E clearly outperform EBP, and FLWC is slightly more stable than HNG-E.

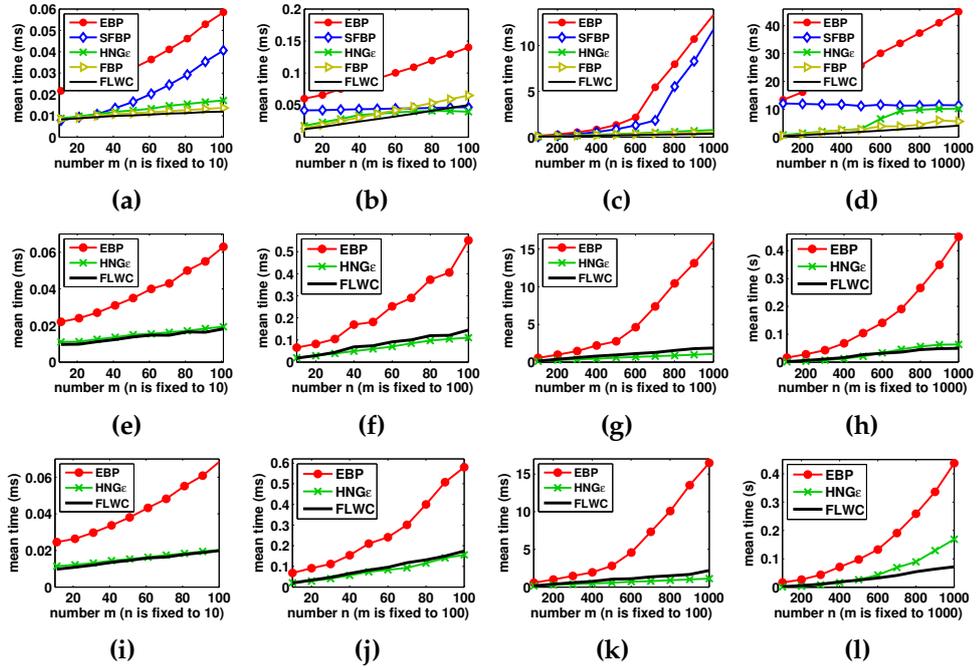

**Figure 4.5.** Time comparisons for flat instances. Figures 4.5a to 4.5d: 0 % deletions. Figures 4.5e to 4.5h: 25 % deletions. Figures 4.5i to 4.5l: 50 % deletions.

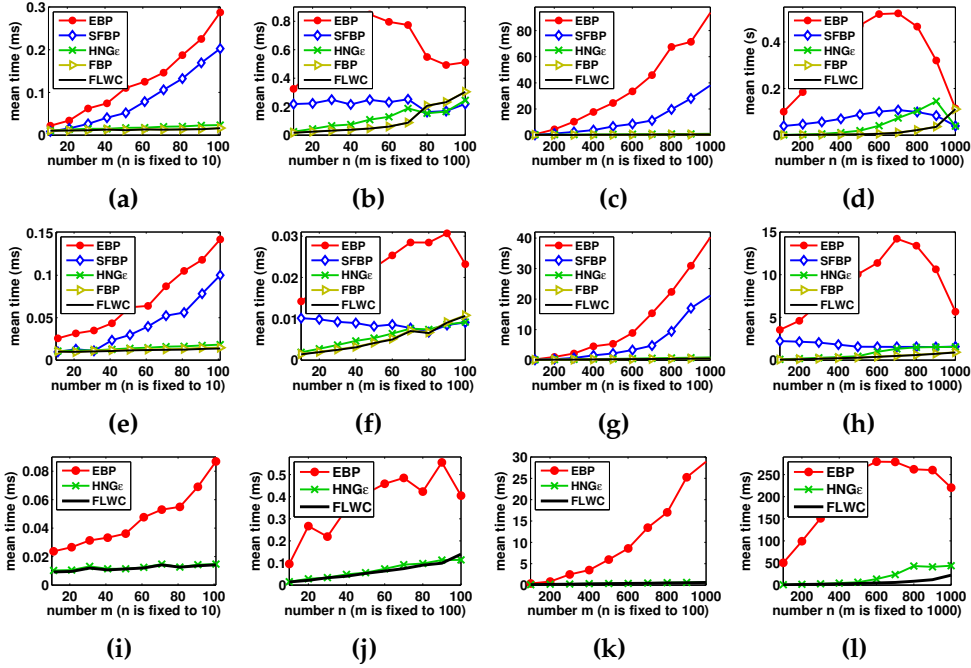

**Figure 4.6.** Time comparisons for random instances. Figures 4.6a to 4.6d: $(c_{\text{sub}}, c_V, c_E) = (40, 20, 20)$. Figures 4.6e to 4.6h: $(c_{\text{sub}}, c_V, c_E) = (20, 40, 0)$. Figures 4.6i to 4.6l: $(c_{\text{sub}}, c_V, c_E) = (40, 2, 2)$.

**Random Instances.** We also carried out experiments on random instances similar to the ones presented in [99, 101]. These instances are very similar to the ones that occur in the context of approximation of GED. The cost $c_{i,k}$ encodes the cost of substituting a node $u_i$ and its set of incident edges in a graph $G$ by a node $v_k$ and its set of incident edges in a graph $H$. Graphs are constructed locally by assigning node degrees randomly from 1 to $(3/10)|V_G|$, where $|V_G|$ is the number of nodes in the graph $G$. Nodes are labeled randomly with an integer value in $\{1, \ldots, \lfloor\sqrt{nm}/10\rfloor\}$. Similarly, edges are labeled with a binary value. Then $c_{i,k}$ is defined as the cost of substituting the labels (0 if they are the same, or the constant $c_{\text{sub}}$ else) plus the cost of an optimal error-correcting matching between the sets of incident edges. The cost of substituting two edges is defined as before, and the cost of deleting or inserting an edge is defined by a constant $c_E \in \mathbb{R}_{\geq 0}$. Deletion and insertion costs are defined as $c_{i,\epsilon} := c_V + \deg^G(u_i)c_E$ and $c_{\epsilon,k} := c_V + \deg^H(v_k)c_E$, respectively, where $c_V \in \mathbb{R}_{\geq 0}$ is a constant.

Figure 4.6 shows the results for random instances. Three parameter settings $(c_{\mathrm{sub}}, c_V, c_E)$ were tested in the experiments: $(40, 20, 20)$, $(20, 40, 0)$, and $(40, 2, 2)$. The first parameter setting (Figures 4.6a to 4.6d) satisfies the triangle inequality. The second (Figures 4.6e to 4.6h) and the third parameter setting (Figures 4.6i to 4.6l) do not satisfy the triangle inequalities. However, for the second setting, none of the computed optimal error-correcting contains removals and insertions, and so all methods were tested. On the contrary, optimal error-correcting matchings for the third setting indeed contain up to 20 % of removals, which is why we did not carry out tests for the cost-constrained methods `FBP`, `FBP-0`, and `SFBP`. We again observe that our method `FLWC` is globally the most stable algorithm and obtains the best results on average.

### 4.3.2 Effect on Approximation of GED

As mentioned in above, many exact or approximate algorithms for GED have to solve LSAPE as a subproblem (cf. Chapters 5 to 6). In particular, given input graphs $G$ and $H$, one prominent class of algorithms constructs LSAPE instances $\mathbf{C} \in {(|V^G|+1) \times (|V^H|+1)}$, which are then employed for computing upper and/or lower bounds for GED [11, 13, 15, 32, 50, 83, 111, 113]. Given an error-correcting matching for $\mathbf{C}$ computed by any LSAPE algorithm, an upper bound for GED is derived by computing the distance associated to the edit path induced by the considered matching. Since this edit path may be suboptimal, the computed distance might be an overestimation of the exact GED, and thus constitutes an upper bound. With this paradigm, each error-correcting matching for $\mathbf{C}$ yields a valid upper bound, although the ones induced by optimal matchings are usually tighter. Note that, if the LSAPE instance $\mathbf{C}$ has only one optimal solution, each optimal LSAPE solver yields the same upper bound for GED. If $\mathbf{C}$ has more than one optimal solution, there might be differences in accuracy, since different LSAPE solvers might pick different optimal solutions depending on the organization of the input data. However, these differences in accuracy are arbitrary and hence disappear once the upper bounds are averaged across enough pairs of input graphs. In other words, when it comes to the approximation of GED, the only relevant property of an exact LSAPE solver is its runtime behavior.

This observation is empirically confirmed by the experiments reported in Table 4.3, which shows how different methods for solving LSAPE affect the performance of the algorithm `BP` suggested in [83], which computes an upper bound for GED. Given two graphs $G$ and $H$, BP constructs an LSAPE instance $\mathbf{C} \in \mathbb{R}^{(|V^G|+1)\times(|V^H|+1)}$. For computing the cell $c_{i,k}$ for $(i,k) \in [|V^G|] \times [|V^H|]$, BP has to solve another LSAPE instance of size $(\deg^G(u_i) + 1) \times (\deg^H(v_k) + 1)$. So altogether, BP has to solve $1 + |V^G||V^H|$ instances of LSAPE. The tests were carried out on the datasets ACYCLIC and MAO from the ICPR GED contest [1]. In order to ensure that also the cost-constrained LSAPE solvers compute optimal solutions, we defined metric edit costs by setting the cost of substituting nodes and edges to 1 and the cost of deleting and inserting nodes and edges to 3 (cf. Section 2.4). We see that, as expected, there are no significant differences between the different solvers w. r. t. the tightness of the produced upper bounds. In terms of runtime, our solver FLWC performs best, followed by FBP and HNG-E.

**Table 4.3.** Effect of LSAPE solvers on algorithm BP suggested in [83].

| method | ACYCLIC | | MAO | |
|---|---|---|---|---|
| | avg. time in μs | avg. UB | avg. time in μs | avg. UB |
| *cost-constrained methods* | | | | |
| FBP | 1.61 | 38.86 | 6.62 | 107.97 |
| FBP-0 | 2.04 | 38.89 | 8.10 | 107.85 |
| SFBP | 2.14 | 39.28 | 11.10 | 107.48 |
| *general methods* | | | | |
| EBP | 4.84 | 39.14 | 23.30 | 107.48 |
| HNG-E | 1.87 | 38.97 | 7.46 | 107.59 |
| FLWC | 1.53 | 38.86 | 6.05 | 108.04 |

In contrast to the situation for upper bounds, LSAPE based heuristics for GED which aim at the computation of lower bounds [13, 15, 111, 113] as well as methods based on conditional gradient descent [12, 22, 25, 41] where LSAPE occurs as a subproblem crucially depend on the optimality of the computed error-correcting matching. Therefore, these methods cannot use the existing fast LSAPE solvers FBP, FBP-0, and SFBP, unless the triangle inequalities are known to be satisfied. For instance, lower bounds for GED are obtained from the cost $\mathbf{C}(\mathbf{X}^\star)$ of an optimal error-correcting matching for the LSAPE instance $\mathbf{C}$. If $\mathbf{X}^\star$ is not optimal, $\mathbf{C}(\mathbf{X}^\star)$ is in general no valid lower bound.

Table 4.4 shows how different methods for solving LSAPE affect the performance of the algorithm BRANCH suggested in [13, 15], which computes a lower bound for GED. Like BP, given two graphs $G$ and $H$, BRANCH has to solve $1 + |V^G||V^H|$ instances of LSAPE. One of them is of size $(|V^G| + 1) \times (|V^H| + 1)$, the other ones are of size $(\deg^G(u_i) + 1) \times (\deg^H(v_k) + 1)$. The tests were again carried out on the datasets ACYCLIC and MAO, but this time, edit costs that do not satisfy the triangle inequality were used (cf. [1] for more details on the edit costs, especially equation (1) and setting 3 in Table 2). First of all, we can see that the classic approach EBP requires more computational time than other optimized approaches, and that FLWC, FBP, and HNG-E are the fastest methods. Second, we observe that the cost-constrained methods FBP, FBP-0 and SFBP are not able to deal with costs which do not satisfy the triangle inequality: Very often, they compute invalid lower bounds that exceed the exact GED, which we computed using a binary linear programming approach [69]. This is explained by the fact that, if the triangle inequality is not satisfied, optimal error-correcting matchings might well include both insertions and removals. However, those error-correcting matchings are not considered by cost-constrained methods for LSAPE.

**Table 4.4.** Effect of LSAPE solvers on algorithm BRANCH suggested in [13, 15].

| method | ACYCLIC | | MAO | |
|---|---|---|---|---|
| | time in μs | # invalid *LB* | time in μs | # invalid *LB* |
| *cost-constrained methods* | | | | |
| FBP | 1.28 | 4 | 8.30 | 416 |
| FBP-0 | 1.64 | 4 | 10.80 | 416 |
| SFBP | 1.91 | 4 | 12.81 | 416 |
| *general methods* | | | | |
| EBP | 5.34 | 0 | 21.10 | 0 |
| HNG-E | 1.54 | 0 | 7.59 | 0 |
| FLWC | 1.23 | 0 | 8.50 | 0 |

Figure 4.7 shows a case where the cost-constrained methods FBP, FBP-0 and SFBP compute an invalid lower bound. Considering the two graphs $G$ and $H$ extracted from the ACYCLIC dataset, the cost-constrained methods provide a matching $\pi_{\text{FBP}}$ that favors node substitution over removal plus insertion even if removal plus insertion is cheaper. For instance, $\pi_{\text{FBP}}$ substitutes the node $u_7 \in V^G$ by the node $v_9 \in V^H$, although it is cheaper to first delete $u_7$ and then insert $v_9$, as done by the matching $\pi_{\text{FLWC}}$ computed by FLWC. In conclusion, cost-constrained methods do not guarantee valid lower bounds for general edit costs, while our algorithm FLWC does.

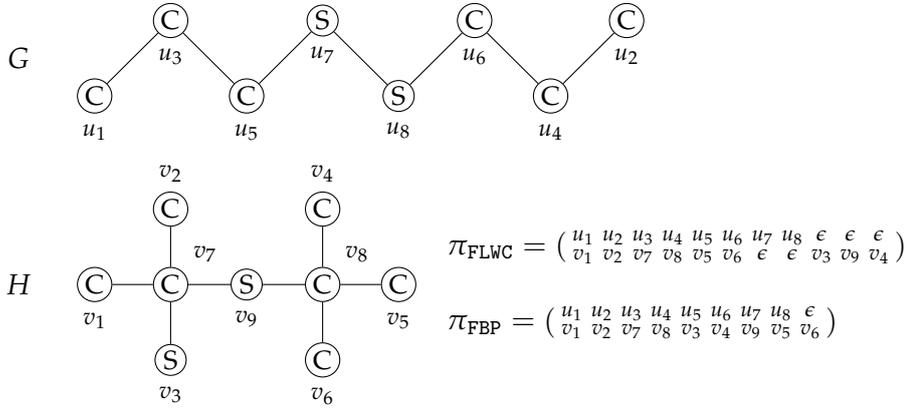

**Figure 4.7.** Example of lower bound computation where FBP, FBP-0, and SFBP do not compute a valid lower bound.

## 4.4 Conclusions and Future Work

In this chapter, we presented FLWC, a new efficient method for solving LSAPE. The technical backbone of our method is a cost-dependent factorization of substitutions into removals and insertions. FLWC runs in $O(\min\{n,m\}^2 \max\{n,m\})$ time and $O(nm)$ space. Its complexities are hence the same as the ones of the most efficient state of the art methods FBP and HNG-E.

The advantage of our method FLWC over FBP is that, unlike FBP, FLWC remains valid for any configuration of the cost matrix and does not require the triangle inequality to hold. Therefore, our method allows to efficiently retrieve a lower bound for the graph edit distance irrespectively of whether or not the edit costs respect the triangle inequality. This is especially important in settings where the edit costs are deduced from calculus and where one does not have an a priori guarantee that the triangle inequality will be satisfied. Situations of this kind occur, for instance, in some quadratic approximations of the graph edit distance problem [22, 25].

One advantage of FLWC over HNG-E is that, although both algorithms have the same time complexity, FLWC is slightly faster in practice. In particular, FLWC is more stable than HNG-E, in the sense that it also allows to quickly solve difficult LSAPE instances with whom HNG-E struggles. A second advantage is that FLWC is much easier to implement: While implementing HNG-E's adaptation of the Hungarian Algorithm to LSAPE requires a thorough knowl-edge of matching theory, `FLWC` can be implemented by slightly transforming the cost matrix and then calling a solver for LSAP. Since LSAP is a very famous combinatorial optimization problem, libraries are available for all major programming languages.

For future work, we are planning to develop an enumeration algorithm that, given an LSAPE instance $\mathbf{C} \in \mathbb{R}^{(n+1) \times (m+1)}$ and constant $K \in \mathbb{N}_{\geq 2}$, efficiently computes not only one, but rather $\min\{K, |\arg\min_{\mathbf{X} \in \Pi_{n,m,\epsilon}} \mathbf{C}(\mathbf{X})|\}$ optimal error-correcting matchings. This technique can then be used for tightening the upper bounds for GED computed by LSAPE based heuristics such as the ones proposed in [10, 11, 13, 15, 32, 50, 83, 111, 113], as the upper bound can be lowered to the minimum over the induced cost of all optimal error-correcting matchings (cf. Chapter 6 for details).

— 5 —

# Exact Algorithms

In this chapter, we provide a systematic overview of algorithms for exactly computing GED. As exactly computing GED is a very hard problem, exact algorithms that scale to large graphs are out of reach. However, efficient exact algorithms are still important; mainly because many objects that are readily modeled as labeled graphs—for instance, some molecular compounds—induce relatively small graphs [87]. Moreover, we show how to speed-up existing solvers using node based tree search for constant, triangular edit costs; and generalize an existing algorithm that uses edge based tree search for exactly computing GED with constant, triangular node edit costs to arbitrary edit costs. Furthermore, we provide a new compact mixed integer linear programming (MIP) formulation for GED.

The results presented in this chapter have previously been presented in the following articles:

- D. B. Blumenthal and J. Gamper, "On the exact computation of the graph edit distance", *Pattern Recognit. Lett.*, 2018, in press. DOI: `10.1016/j.patrec.2018.05.002`
- D. B. Blumenthal and J. Gamper, "Exact computation of graph edit distance for uniform and non-uniform metric edit costs", in *GbRPR 2017*, P. Foggia, C. Liu, and M. Vento, Eds., ser. LNCS, vol. 10310, Cham: Springer, 2017, pp. 211–221. DOI: `10.1007/978-3-319-58961-9_19`

The remainder of this chapter is organized as follows: In Section 5.1, a systematic overview of existing approaches is provided. In Section 5.2, our speed-up of existing node based tree search algorithms for constant, triangular edit costs is presented. In Section 5.3, we show how to generalize the existing edge based tree search algorithm to arbitrary edit costs. In Section 5.4, the new compact MIP formulation is presented. In Section 5.5, the newly proposed techniques are evaluated empirically. Section 5.6 concludes the chapter and points out to future work.

## 5.1 State of the Art

In this section, we provide a systematic overview of the state of the art algorithms for exactly computing GED. In Section 5.1.1, algorithms are presented that compute GED by enumerating all node maps by exploring implicitly constructed search trees. In Section 5.1.2 we present an algorithm that works similarly, but enumerates edge maps instead of node maps. This algorithm requires constant, triangular node edit costs. In Section 5.1.3, we present a recent work that proposes to use parallel depth-first search for speeding-up the previously presented tree search based solvers for GED. In Section 5.1.4, we present algorithms that use mixed integer linear programming (MIP) formulations for computing GED.

### 5.1.1 Node Based Tree Search

The algorithms `A*` [89] and `DFS-GED` [4] compute $GED(G, H)$ by enumerating the set $\Pi(G, H)$ of all node maps between $G$ and $H$. Both algorithms implicitly represent $\Pi(G, H)$ as a tree $T(G, H)$ whose leafs are left-complete partial node maps and whose inner nodes are left-incomplete partial node maps. Before presenting `A*` and `DFS-GED`, we introduce the concept of a partial node map.

**Definition 5.1 ((Left-Complete) Partial Node Map).** Let $G, H \in \mathbb{G}$ be two graphs. A relation $\pi' \subseteq V^G \times (V^H \cup \{\epsilon\})$ is called *partial node map* between $G$ and $H$ if and only if, for all $u \in V^G$ and all $v \in V^H$, the sets $\{v \mid v \in (V^H \cup \{\epsilon\}) \wedge (u, v) \in \pi'\}$ and $\{u \mid u \in V^G \wedge (u, v) \in \pi'\}$ contain at most one node, respectively. We write $\pi'(u) = v$ just in case $(u, v) \in \pi'$, and $\pi'^{-1}(v) = u$ just in case $(u, v) \in \pi'$ and $v \neq \epsilon$. $\Pi'(G, H)$ denotes the set of all partial node maps between $G$ and $H$. For edges $e = (u_1, u_2) \in E^G \cap (\mathrm{supp}(\pi') \times \mathrm{supp}(\pi'))$ and $f = (v_1, v_2) \in E^H \cap (\mathrm{img}(\pi') \times \mathrm{img}(\pi'))$, we introduce the short-hand notations $\pi'(e) := (\pi'(u_1), \pi'(u_2))$ and $\pi'^{-1}(f) :=$ $({\pi'}^{-1}(v_1), {\pi'}^{-1}(v_2))$. A partial node map $\pi'$ is called *left-complete* if and only if $\text{supp}(\pi') = V^G$. Otherwise, it is called *left-incomplete*.

Just as node maps between $G$ and $H$ induce edit paths, partial node maps induce partial edit paths (cf. Table 2.2). Note that, by definition of a partial node map, a partial edit path $P_{\pi'}$ that is induced by a partial node map $\pi' \in \Pi'(G, H)$ never contains node insertions. Hence, the cost of $P_{\pi'}$ is given as follows:

$$c(P_{\pi'}) = \underbrace{\sum_{\substack{u \in \text{supp}(\pi') \\ \pi'(u) \neq \epsilon}} c_V(u, \pi'(u))}_{\text{node substitutions}} + \underbrace{\sum_{\substack{u \in \text{supp}(\pi') \\ \pi'(u) = \epsilon}} c_V(u, \epsilon)}_{\text{node deletions}} \qquad (5.1)$$

$$+ \underbrace{\sum_{\substack{e=(u_1,u_2) \in E^G \\ u_1,u_2 \in \text{supp}(\pi') \\ \pi'(e) \neq \epsilon}} c_E(e, \pi'(e))}_{\text{edge substitutions}} + \underbrace{\sum_{\substack{e=(u_1,u_2) \in E^G \\ u_1,u_2 \in \text{supp}(\pi') \\ \pi'(e) = \epsilon}} c_E(e, \epsilon)}_{\text{edge deletions}}$$

$$+ \underbrace{\sum_{\substack{f=(v_1,v_2) \in E^H \\ v_1,v_2 \in \text{img}(\pi') \\ {\pi'}^{-1}(f) = \epsilon}} c_E(\epsilon, f)}_{\text{edge insertions}}$$

We use the expressions $V^G - \pi' := V^G \setminus \text{supp}(\pi')$, $V^H - \pi' := V^H \setminus \text{img}(\pi')$, $E^G - \pi' := E^G \setminus \{(u_i, u_j) \in E^G \mid u_i, u_j \in \text{supp}(\pi')\}$, and $E^H - \pi' := E^H \setminus \{(v_k, v_l) \in E^H \mid (v_k, v_l) \in \text{img}(\pi') \times \text{img}(\pi')\}$ to denote the nodes and edges of $G$ and $H$ that still have not been mapped by a partial node map $\pi' \in \Pi'(G, H)$. If a partial node map $\pi' \in \Pi'(G, H)$ is left-complete, there is a unique way to extend it to a node map $\pi \in \Pi(G, H)$: We have to map the dummy node $\epsilon$ to all yet unmatched nodes $v \in V^H - \pi'$. The following Proposition 5.1 shows how to extend the cost of a left-complete partial node map $\pi' \in \Pi'(G, H)$ to the cost of the unique node map $\pi \in \Pi(G, H)$ extending $\pi'$.

**Proposition 5.1 (Correctness of `A*` and `DFS-GED` for General Edit Costs).** *Let the operator* `EXTEND-NODE-MAP-COST` $: \Pi'(G, H) \to \mathbb{R}$ *be defined as follows:*

$$\texttt{EXTEND-NODE-MAP-COST}(P_{\pi'}) := c(P_{\pi'}) + \sum_{v \in V^H - \pi'} c_V(\epsilon, v) + \sum_{f \in E^H - \pi'} c_E(\epsilon, f)$$

*Then* $c(P_\pi) = $ `EXTEND-NODE-MAP-COST`$(P_{\pi'})$ *holds for each left-complete partial node map* $\pi' \in \Pi'(G, H)$, *where* $\pi \in \Pi(G, H)$ *is defined as* $\pi := \pi' \cup \{(\epsilon, v) \mid v \in V^H \setminus \text{img}(\pi')\}$.

*Proof.* The proposition follows from the equations (2.1) and (5.1) and the definition of a left-complete partial node map given in Definition 5.1. □

The tree $T(G, H)$ implicitly constructed by `A*` and `DFS-GED` is defined as follows: The root of $T(G, H)$ is the empty partial node map. Inner nodes are left-incomplete partial node maps and leafs are left-complete partial node maps. The levels of $T(G, H)$ correspond to the nodes in $V^G$, which are sorted such that evident nodes will be processed first (cf. [4] and [45] for more details on how the nodes are sorted). For the remainder of this section, we assume w.l.o.g. that $G$'s nodes are sorted as $(u_1, \ldots, u_{|V^G|})$. A partial node map $\pi'$'s level in the search tree is defined as $\text{lev}(\pi') = \max_{u_i \in \text{supp}(\pi')} i$, if $\pi' \neq \emptyset$, and as $\text{lev}(\pi') = 0$, otherwise. Intuitively, $\text{lev}(\pi')$ is the last node in $V^G$ w.r.t. the given ordering which has already been mapped by $\pi$.

If $\text{lev}(\pi') < |V^G|$, $\pi'$ is left-incomplete and hence an inner node in $T(G, H)$. A partial node map $\pi''$ is a child of $\pi'$, if and only if there is a node $v \in (V^H - \pi') \cup \{\epsilon\}$ such that $\pi'' = \pi' \cup \{(u_{\text{lev}(\pi')+1}, v)\}$. In other words, $\pi'$'s children set $CHILDREN(\pi')$ is the set of all partial node maps that extend $\pi'$ to the first yet unmatched node in $V^G$. If $\text{lev}(\pi') = |V^G|$, $\pi'$ is a left-complete partial node map, i.e., a leaf of $T(G, H)$.

`A*` enumerates $\Pi'(G, H)$ by means of the best-first search paradigm [55]. To this purpose, `A*` maintains a set $OPEN \subset \Pi(G, H)$ of pending partial node maps. For each pending partial node map $\pi' \in OPEN$, `A*` maintains the cost $c(P_{\pi'})$ of $\pi'$'s induced partial edit path and a lower bound $LB(\pi')$ for the edit cost of a complete node map which extends a leaf (i.e., left-complete partial node map) in $\pi'$'s down-shadow. This lower bound is defined as

$$LB(\pi') := c(P_{\pi'}) + \text{LSAPE}(\mathbf{C}^{V-\pi'}) + \text{LSAPE}(\mathbf{C}^{E-\pi'}), \quad (5.2)$$

where the LSAPE instances $\mathbf{C}^{V-\pi'} \in \mathbb{R}^{(|V^G-\pi'|+1) \times (|V^H-\pi'|+1)}$ and $\mathbf{C}^{E-\pi'} \in \mathbb{R}^{(|E^G-\pi'|+1) \times (|E^H-\pi'|+1)}$ are defined in terms of the node and edge edit costs between the yet unmatched nodes and edges contained in $V^G - \pi'$, $V^H - \pi'$, $E^G - \pi'$, and $E^H - \pi'$. More precisely, let $(u_{i_r})_{r=1}^{|V^G-\pi'|}$ and $(v_{k_s})_{s=1}^{|V^H-\pi'|}$ be arbitrary, fixed orderings of, respectively, $V^G - \pi'$ and $V^H - \pi'$. The matrix $\mathbf{C}^{V-\pi'}$ is constructed by setting

$$c_{r,s}^{V-\pi'} := c_V(u_{i_r}, v_{k_s}) \qquad (5.3)$$
$$c_{r,|V^H-\pi'|+1}^{V-\pi'} := c_V(u_{i_r}, \epsilon)$$
$$c_{|V^G-\pi'|+1,s}^{V-\pi'} := c_V(\epsilon, v_{k_s})$$

for all $(r,s) \in [|V^G - \pi'|] \times [|V^H - \pi'|]$. The matrix $\mathbf{C}^{E-\pi'}$ is constructed analogously.

Figure 5.1 gives an overview of A*. At initialization, the induced edit cost and the lower bound of the empty partial node map are computed, and *OPEN* is initialized to contain just the empty partial node map (lines 1 to 2). Furthermore, A* maintains an edit cost $c(\pi')$ for each partial node map $\pi' \in OPEN$. If $\pi'$ is left incomplete, $c(\pi')$ equals the cost of its induced partial edit path; otherwise $c(\pi')$ is defined as the cost of the edit path induced by the unique node map extending $\pi'$ (cf. Proposition 5.1). In the main loop (lines 3 to 14), A* first picks and removes a node map $\pi'$ from *OPEN* which minimizes *LB* over *OPEN* (lines 4 to 5). For doing this efficiently, *OPEN* should be implemented as a priority queue. If $\pi'$ is left-complete, the induced edit cost of its unique extension is returned as GED$(G,H)$ (line 7). Otherwise, A* iterates through $\pi'$'s children (lines 9 to 10), whose edit costs are updated and whose lower bounds are recomputed (line 10). If the children are left-complete, their edit costs are extended and their lower bounds are set to the complete edit costs (lines 11 to 13). Subsequently, the children are added to *OPEN* (line 14).

Figure 5.2 gives an overview of DFS-GED. All differences to A* directly stem from the fact that DFS-GED uses depth-first search rather than best-first search for traversing $T(G,H)$. The first difference is that DFS-GED also maintains a global upper bound *UB*, which is initialized by using a fast sub-optimal heuristic, e.g., the one presented in [83] (line 1). The while-loop now terminates once *OPEN* is empty (line 3), and DFS-GED picks a node map $\pi'$ from those node maps that minimize *LB* not over the entire set *OPEN*, but rather over the set *DEEPEST* $\subseteq$ *OPEN*. *DEEPEST* contains the deepest nodes of the current search tree, i.e., the partial node maps in *OPEN* with highest level (lines 4 to 6). For extracting $\pi'$ efficiently, *OPEN* should be implemented as a stack and $\pi'$'s children should be sorted w.r.t. non increasing *LB* before being added to *OPEN*. If $\pi'$ is a left-complete node map, DFS-GED updates

**Input**: Two undirected, labeled graphs $G$ and $H$.
**Output**: The graph edit distance $\text{GED}(G, H)$.

1 initialize $c(\emptyset) \leftarrow 0$ and compute $LB(\emptyset)$;
2 $OPEN \leftarrow \{\emptyset\}$;
3 **while** $true$ **do**
4     $\pi' \leftarrow \arg\min\{LB(\pi'') \mid \pi'' \in OPEN\}$;
5     $OPEN \leftarrow OPEN \setminus \{\pi'\}$;
6     **if** $\text{lev}(\pi') = |V^G|$ **then**
7         **return** $c(\pi')$;
8     **else**
9         **for** $\pi'' \in CHILDREN(\pi')$ **do**
10             update $c(\pi'') \leftarrow c(P_{\pi''})$ and compute $LB(\pi'')$;
11             **if** $\text{lev}(\pi'') = |V^G|$ **then**
12                 $c(\pi'') \leftarrow \texttt{EXTEND-NODE-MAP-COST}(\pi'')$;
13                 $LB(\pi'') \leftarrow c(\pi'')$;
14             $OPEN \leftarrow OPEN \cup \{\pi''\}$;

**Figure 5.1.** The exact algorithm A*.

the global upper bound to the edit cost induced by the unique extension of $\pi'$, if this leads to an improvement (lines 7 to 8). Otherwise, DFS-GED proceeds just like A*, except for the fact that a child $\pi''$ of $\pi'$ is added to $OPEN$ only if $LB(\pi'')$ is smaller than the global upper bound (line 15). After termination of the while-loop, DFS-GED returns $UB$, which now equals $\text{GED}(G, H)$. Note that DFS-GED can easily be turned into an algorithm that quickly computes a, possibly suboptimal, upper bound for $\text{GED}(G; H)$: It suffices to initialize DFS-GED with a time limit and exit the main while-loop starting in line 3 once the time limit has been reached.

### 5.1.2 Edge Based Tree Search for Constant, Triangular Node Edit Costs

While DFS-GED and A* traverse the space of all partial node maps to compute $\text{GED}(G, H)$, the algorithm CSI-GED presented in [52] traverses the space of all valid partial edge maps between $G$ and $H$.

**Definition 5.2 (Partial Edge Map).** A relation $\phi \subseteq \overrightarrow{E^G} \times \overleftrightarrow{E^H} \cup \{\epsilon\}$ is called *partial edge map* between the graphs $G$ and $H$ if and only if $|\{f \mid f \in \overleftrightarrow{E^H} \cup \{\epsilon\} \land (e, f) \in \phi\}| \leq 1$ holds for all $e \in \overrightarrow{E^G}$. The set $\overrightarrow{E^G}$ contains one arbitrarily oriented edge $(u_i, u_j)$ for each undirected edge contained in **Input**: Two undirected, labeled graphs $G$ and $H$.
**Output**: The graph edit distance $\text{GED}(G, H)$.

1 compute initial $UB$, initialize $c(\emptyset) \leftarrow 0$, and compute $LB(\emptyset)$;
2 $OPEN \leftarrow \{\emptyset\}$;
3 **while** $OPEN \neq \emptyset$ **do**
4     $DEEPEST \leftarrow \arg\max\{\text{lev}(\pi') \mid \pi' \in OPEN\}$;
5     $\pi' \leftarrow \arg\min\{LB(\pi'') \mid \pi'' \in DEEPEST\}$;
6     $OPEN \leftarrow OPEN \setminus \{\pi'\}$;
7     **if** $\text{lev}(\pi') = |V^G|$ **then**
8       **if** $c(\pi) < UB$ **then** $UB \leftarrow c(\pi)$;
9     **else**
10       **for** $\pi'' \in \text{CHILDREN}(\pi')$ **do**
11         update $c(\pi'') \leftarrow c(P_{\pi''})$ and compute $LB(\pi'')$;
12         **if** $\text{lev}(\pi'') = |V^G|$ **then**
13           $c(\pi'') \leftarrow \text{EXTEND-NODE-MAP-COST}(\pi'')$;
14           $LB(\pi'') \leftarrow c(\pi'')$;
15         **if** $LB(\pi'') < UB$ **then** $OPEN \leftarrow OPEN \cup \{\pi''\}$;

16 **return** $UB$;

**Figure 5.2.** The exact algorithm DFS-GED.

$E^G$, and $\overleftrightarrow{E^H}$ contains two directed edges $(v_k, v_l)$ and $(v_l, v_k)$ for each undirected edge contained in $E^H$. We write $\phi(u_i, u_j) = (v_k, v_l)$ just in case if $((u_i, u_j), (v_k, v_l)) \in \phi$.

A partial edge map $\phi$ induces a relation $\pi_\phi$ on $V^G \times V^H$: If $\phi(u_i, u_j) = (v_k, v_l)$, then $(u_i, v_k) \in \pi_\phi$ and $(u_j, v_l) \in \pi_\phi$. The main idea of CSI-GED is to enumerate all those edge maps for which this relation is partial node map.

**Definition 5.3 (Valid Partial Edge Map).** A partial edge map $\phi$ between graphs $G$ and $H$ is called *valid*, if and only if $\pi_\phi$ is a partial node map. The set of all valid edge maps between $G$ and $H$ is denoted as $\Phi(G, H)$.

Together with its induced partial node map $\pi_\phi$, a valid partial edge map $\phi$ corresponds to a partial edit path $P_\phi$ between $G$ and $H$ that contains only node substitutions, edge substitutions, and edge deletions. The cost of $P_\phi$ is hence given as follows:

$$c(P_\phi) = \underbrace{\sum_{u \in \text{supp}(\pi_\phi)} c_V(u, \pi_\phi(u))}_{\text{node substitutions}} \tag{5.4}$$

$$+ \underbrace{\sum_{\substack{e \in \mathrm{supp}(\phi) \\ \phi(e) \neq \epsilon}} c_E(e, \phi(e))}_{\text{edge substitutions}} + \underbrace{\sum_{\substack{e \in \mathrm{supp}(\phi) \\ \phi(e) = \epsilon}} c_E(e, \epsilon)}_{\text{edge deletions}}$$

We use the expressions $V^G - \phi := V^G \setminus \mathrm{supp}(\pi_\phi)$, $V^H - \phi := V^H \setminus \mathrm{img}(\pi_\phi)$, $E^G - \phi := E^G \setminus \{(u_i, u_j) \in E^G \mid (u_i, u_j) \in \mathrm{supp}(\phi) \vee (u_j, u_i) \in \mathrm{supp}(\phi)\}$, and $E^H - \phi := E^H \setminus \{(v_k, v_l) \in E^H \mid (v_k, v_l) \in \mathrm{img}(\phi) \vee (v_l, v_k) \in \mathrm{img}(\phi)\}$ to denote those nodes and edges of $G$ and $H$ that still have not been mapped by a partial edge map $\phi \in \Phi(G, H)$. A valid partial edge map $\phi$ is called *left-complete*, if and only if $\mathrm{supp}(\phi) = \overrightarrow{E^G}$ and hence $E^G - \phi = \emptyset$. Otherwise, it is called *left-incomplete*. $\overline{\Phi}(G, H)$ denotes the set of all left-complete partial edge maps between $G$ and $H$. The following Proposition 5.2 constitutes the backbone of CSI-GED.

**Proposition 5.2 (Correctness of CSI-GED for Constant, Triangular Node Edit Costs).** *Let the operator* EXTEND-EDGE-MAP-COST-1 $: \Phi(G, H) \to \mathbb{R}$ *be defined as follows:*

$$\mathtt{EXTEND\text{-}EDGE\text{-}MAP\text{-}COST\text{-}1}(P_\phi) := c(P_\phi) + \sum_{f \in E^H - \phi} c_E(\epsilon, f)$$
$$+ \Gamma(\ell_V^G[\![V^G - \phi]\!], \ell_V^H[\![V^H - \phi]\!], c_V^{\mathrm{sub}}, c_V^{\mathrm{del}}, c_V^{\mathrm{ins}})$$

*Then the equation* $\mathrm{GED}(G, H) = \min_{\phi \in \overline{\Phi}(G, H)} \mathtt{EXTEND\text{-}EDGE\text{-}MAP\text{-}COST\text{-}1}(P_\phi)$ *holds for constant, triangular node edit cost functions $c_V$.*

*Proof.* Cf. Theorem 1 in [52] for the proof of this proposition. □

CSI-GED enumerates $\Phi(G, H)$ by implicitly constructing a tree $T(G, H)$ whose leafs are left-complete valid partial edge maps and whose inner nodes are left-incomplete valid partial edge maps. The root of $T(G, H)$ is the empty edge map, and the tree's levels correspond to the indices of the edges $e_r \in \overrightarrow{E^G}$, which are sorted in a fixed order $(e_1, \ldots, e_{|E^G|})$ (cf. [52] for details on how the edges are sorted). We define a valid partial edge map $\phi$'s level in the search tree as $\mathrm{lev}(\phi) = \max\{r \mid e_r \in \mathrm{supp}(\phi)\}$, if $\phi \neq \emptyset$, and as $\mathrm{lev}(\phi) = 0$, otherwise. Intuitively, $\mathrm{lev}(\phi)$ is the index of the last edge in $\overrightarrow{E^G}$ w.r.t. the given ordering which has already been mapped by $\phi$.

If $\mathrm{lev}(\phi) < |\overrightarrow{E^G}|$, the valid partial edge map $\phi$ is left-incomplete and hence an inner node in $T(G, H)$. A partial edge map $\phi'$ is a child of $\phi$, if and

**Input**: Two undirected, labeled graphs $G$ and $H$.
**Output**: The graph edit distance $\text{GED}(G, H)$.

1 compute initial $UB$, initialize $\mathbf{C}^E$ and $c(\emptyset) \leftarrow 0$, and compute $LB(\emptyset)$;
2 $OPEN \leftarrow \{\emptyset\}$;
3 **while** $OPEN \neq \emptyset$ **do**
4 $\quad DEEPEST \leftarrow \arg\max\{\text{lev}(\phi) \mid \phi \in OPEN\}$;
5 $\quad \phi \leftarrow \arg\min\{c^E_{e_{\text{lev}(\phi')}, \phi'(e_{\text{lev}(\phi')})} \mid \phi' \in DEEPEST\}$;
6 $\quad OPEN \leftarrow OPEN \setminus \{\phi\}$;
7 $\quad$ **if** $\text{lev}(\phi') = |E^G|$ **then**
8 $\quad\quad$ **if** $c(\phi) < UB$ **then** $UB \leftarrow c(\phi)$;
9 $\quad$ **else**
10 $\quad\quad$ **for** $\phi' \in CHILDREN(\phi)$ **do**
11 $\quad\quad\quad$ update $c(\phi') \leftarrow c(P_{\phi'})$ and compute $LB(\phi')$;
12 $\quad\quad\quad$ **if** $\text{lev}(\phi') = |E^G|$ **then**
13 $\quad\quad\quad\quad$ $c(\phi') \leftarrow \texttt{EXTEND-EDGE-MAP-COST-1}(\phi')$;
14 $\quad\quad\quad\quad$ $LB(\phi') \leftarrow c(\phi')$;
15 $\quad\quad\quad$ **if** $LB(\phi') < UB$ **then** $OPEN \leftarrow OPEN \cup \{\phi'\}$;

16 **return** $UB$;

**Figure 5.3.** The exact algorithm `CSI-GED`.

only if $\phi'$ is valid and there is an edge $e \in (\overleftrightarrow{E^H} \setminus \text{img}(\phi)) \cup \{\epsilon\}$ such that $\phi' = \phi \cup \{(e_{\text{lev}(\pi)+1}, e)\}$. In other words, $\phi$'s children set $CHILDREN(\phi)$ is the set of all valid partial edge maps that extend $\phi$ to the first yet unmatched edge in $\overrightarrow{E^G}$. If $\text{lev}(\phi) = |\overrightarrow{E^G}|$, $\phi$ is a left-complete and hence a leaf in $T(G, H)$.

Figure 5.3 gives an overview of the algorithm `CSI-GED`, which uses depth-first search to traverse $T(G, H)$. Although `CSI-GED` was originally presented as a recursive algorithm [52], we here show the iterative version, in order to increase its comparability to `A*` and `DFS-GED`. We see that `CSI-GED`'s structure is very similar to the structure of `DFS-GED`, which does depth-first search on partial node maps. Like `DFS-GED`, for each pending node $\phi$ in the search tree, `CSI-GED` maintains an edit cost $c(\phi)$ and a lower bound $LB(\phi)$ for the edit cost of a leaf in the $\phi$'s down shadow. If $\phi$ is an inner node (i.e., left-incomplete), the edit cost $c(\phi)$ is set to $c(P_\phi)$, otherwise it is set to `EXTEND-EDGE-MAP-COST-1`$(\phi)$. For the definition of the lower bound $LB(\phi)$ employed by `CSI-GED`, we refer to [52]. The only structural difference between `DFS-GED` and `CSI-GED` is that, in line 5, `CSI-GED` does not use the lower bound $LB$ for deciding which valid edge map to pick from the set of pending edge maps $OPEN$. Instead, `CSI-GED` uses the matrix $\mathbf{C}^E \in \mathbb{R}^{|E^G| \times (2|E^H|+1)}$, whose entry $c_{e,f}^E$ is an estimate of the graph edit distance under the constraint that the edge $e \in \overrightarrow{E^G}$ is mapped to the edge $f \in \overleftrightarrow{E^H} \cup \{\epsilon\}$ (cf. [52] for the definition of $\mathbf{C}^E$). By maintaining a time limit and exiting the main while-loop starting in line 3 once the time limit has been reached, just like `DFS-GED`, `CSI-GED` can be turned into an algorithm that quickly computes suboptimal upper bounds for GED.

### 5.1.3 Parallelized Depth-First Search

In [3], it is proposed to use parallelized depth-first search for speeding-up the computation of GED. The resulting algorithm `P-DFS-GED` can be viewed as a combination of `A*` and `DFS-GED`. In a first step, `P-DFS-GED` runs `A*` until the queue *OPEN* has reached size $N$ and initializes local queues $OPEN_t := \emptyset$ for all slave threads $t \in [T-1]$, where the number of threads $T \in \mathbb{N}_{\geq 2}$ and the queue size $N \in \mathbb{N}_{\geq T-1}$ are parameters of `P-DFS-GED`. Subsequently, `P-DFS-GED` sorts the partial node maps $\pi'$ contained in *OPEN* w.r.t. non-decreasing lower bound $LB(\pi')$ (cf. equation (5.2) above), and alternately distributes them over the local queues $OPEN_t$.

In each slave thread $t \in [T-1]$, `P-DFS-GED` then runs `DFS-GED` from the local queue $OPEN_t$, maintaining a load

$$\omega_t := \sum_{\pi' \in OPEN_t} |V^G - \pi'|,$$

which is defined as the total number of nodes in $G$ that still have not been assigned by the partial node maps contained in $OPEN_t$. All slave threads share and update a global upper bound *UB*.

When a slave thread $t \in [T-1]$ becomes idle, i.e., when $OPEN_t = \emptyset$, the master thread locks a slave thread $t' := \arg\max_{t'' \in [T-1]} \omega_{t''}$ with maximum load, sorts the partial node maps $\pi'$ contained in $OPEN_{t'}$ w.r.t. non-decreasing $LB(\pi')$, and then alternately distributes them between $OPEN_t$ and $OPEN_{t'}$. Once all slave threads have become idle, `P-DFS-GED` returns *UB*, which now equals $GED(G, H)$. Like `DFS-GED` and `CSI-GED`, `P-DFS-GED` can be turned into an algorithm that quickly computes suboptimal upper bounds for GED, if it is set up to return *UB* as soon as a given time limit has been reached. Note that although `P-DFS-GED` is designed as a parallelization of the node based depth-first search `DFS-GED`, it can straightforwardly be adapted to parallelize the edge based depth-first search `CSI-GED`.

$$
\begin{aligned}
\min \quad & \sum_{u_i \in V^G} \sum_{v_k \in V^H} c_V(u_i, v_k) x_{i,k}^{\text{sub}} + \sum_{u_i \in V^G} c_V(u_i, \epsilon) x_i^{\text{del}} + \sum_{v_k \in V^H} c_V(\epsilon, v_k) x_k^{\text{ins}} \\
& + \sum_{(u_i, u_j) \in E^G} \sum_{(v_k, v_l) \in E^H} c_E((u_i, u_j), (v_k, v_l)) y_{i,j,k,l}^{\text{sub}} \\
& + \sum_{(u_i, u_j) \in E^G} c_E((u_i, u_j), \epsilon) y_{i,j}^{\text{del}} + \sum_{(v_k, v_l) \in E^H} c_E(\epsilon, (v_k, v_l)) y_{k,l}^{\text{ins}} \\
\text{s.t.} \quad & x_i^{\text{del}} + \sum_{v_k \in V^H} x_{i,k}^{\text{sub}} = 1 && \forall u_i \in V^G \\
& x_k^{\text{ins}} + \sum_{u_i \in V^G} x_{i,k}^{\text{sub}} = 1 && \forall v_k \in V^H \\
& y_{i,j}^{\text{del}} + \sum_{(v_k, v_l) \in E^H} y_{i,j,k,l}^{\text{sub}} = 1 && \forall (u_i, u_j) \in E^G \\
& y_{k,l}^{\text{ins}} + \sum_{(u_i, u_j) \in E^G} y_{i,j,k,l}^{\text{sub}} = 1 && \forall (v_k, v_l) \in E^H \\
& y_{i,j,k,l}^{\text{sub}} - x_{i,k}^{\text{sub}} x_{j,l}^{\text{sub}} - x_{i,l}^{\text{sub}} x_{j,k}^{\text{sub}} = 0 && \forall ((v_k, v_l), (v_k, v_l)) \in E^G \times E^H \\
& \mathbf{x}^{\text{sub}} \in \{0,1\}^{|V^G| \times |V^H|}, \mathbf{x}^{\text{del}} \in \{0,1\}^{|V^G|}, \mathbf{x}^{\text{ins}} \in \{0,1\}^{|V^H|} \\
& \mathbf{y}^{\text{sub}} \in \{0,1\}^{|E^G| \times |E^H|}, \mathbf{y}^{\text{del}} \in \{0,1\}^{|E^G|}, \mathbf{y}^{\text{ins}} \in \{0,1\}^{|E^H|}
\end{aligned}
$$

**Figure 5.4.** A quadratic programming formulation of GED.

### 5.1.4 MIP Formulations

Recall the alternative Definition 2.6 of GED, which defines the problem of computing GED as a minimization problem over the set of all node maps between two graphs $G$ and $H$. This definition can straightforwardly be transformed into the quadratic programming formulation of GED detailed in Figure 5.4 [25, 82]. The binary decision variables $x_{i,k}^{\text{sub}}$, $x_i^{\text{del}}$, and $x_k^{\text{ins}}$ indicate, respectively, whether the node $u_i \in V^G$ is to be substituted by the node $v_k \in V^H$, whether $u_i$ is to be deleted, and whether $v_k$ is to be inserted. Analogously, the binary decision variables $y_{i,j,k,l}^{\text{sub}}$, $y_{i,j}^{\text{del}}$, and $y_{k,l}^{\text{ins}}$ indicate, respectively, whether the edge $(u_i, u_j) \in E^G$ is to be substituted by the edge $(v_k, v_l) \in E^H$, whether $(u_i, u_j)$ is to be deleted, and whether $(v_k, v_l)$ is to be inserted. The quadratic constraint $y_{i,j,k,l}^{\text{sub}} - x_{i,k}^{\text{sub}} x_{j,l}^{\text{sub}} - x_{i,l}^{\text{sub}} x_{j,k}^{\text{sub}} = 0$ ensures that $(u_i, u_j)$ is substituted by $(v_k, v_l)$ if and only if the node map $\pi$ encoded by $\mathbf{x}^{\text{sub}}$, $\mathbf{x}^{\text{del}}$, and $\mathbf{x}^{\text{ins}}$ satisfies $\pi(u_i, u_j) = (v_k, v_l)$.

$$\min \sum_{u_i \in V^G} \sum_{v_k \in V^H} c_V(u_i, v_k) x_{i,k}^{\text{sub}} + \sum_{u_i \in V^G} c_V(u_i, \epsilon) x_i^{\text{del}} + \sum_{v_k \in V^H} c_V(\epsilon, v_k) x_k^{\text{ins}}$$
$$+ \sum_{(u_i,u_j) \in E^G} \sum_{(v_k,v_l) \in E^H} c_E((u_i, u_j), (v_k, v_l)) y_{i,j,k,l}^{\text{sub}}$$
$$+ \sum_{(u_i,u_j) \in E^G} c_E((u_i, u_j), \epsilon) y_{i,j}^{\text{del}} + \sum_{(v_k,v_l) \in E^H} c_E(\epsilon, (v_k, v_l)) y_{k,l}^{\text{ins}}$$

$$\text{s.t.} \quad x_i^{\text{del}} + \sum_{v_k \in V^H} x_{i,k}^{\text{sub}} = 1 \qquad \forall u_i \in V^G$$
$$x_k^{\text{ins}} + \sum_{u_i \in V^G} x_{i,k}^{\text{sub}} = 1 \qquad \forall v_k \in V^H$$
$$y_{i,j}^{\text{del}} + \sum_{(v_k,v_l) \in E^H} y_{i,j,k,l}^{\text{sub}} = 1 \qquad \forall (u_i, u_j) \in E^G$$
$$y_{k,l}^{\text{ins}} + \sum_{(u_i,u_j) \in E^G} y_{i,j,k,l}^{\text{sub}} = 1 \qquad \forall (v_k, v_l) \in E^H$$
$$y_{i,j,k,l}^{\text{sub}} - x_{i,k}^{\text{sub}} - x_{i,l}^{\text{sub}} \leq 0 \qquad \forall ((v_k, v_l), (v_k, v_l)) \in E^G \times E^H$$
$$y_{i,j,k,l}^{\text{sub}} - x_{j,l}^{\text{sub}} - x_{j,k}^{\text{sub}} \leq 0 \qquad \forall ((v_k, v_l), (v_k, v_l)) \in E^G \times E^H$$
$$\mathbf{x}^{\text{sub}} \in \{0,1\}^{|V^G| \times |V^H|}, \mathbf{x}^{\text{del}} \in \{0,1\}^{|V^G|}, \mathbf{x}^{\text{ins}} \in \{0,1\}^{|V^H|}$$
$$\mathbf{y}^{\text{sub}} \in \{0,1\}^{|E^G| \times |E^H|}, \mathbf{y}^{\text{del}} \in \{0,1\}^{|E^G|}, \mathbf{y}^{\text{ins}} \in \{0,1\}^{|E^H|}$$

**Figure 5.5.** The MIP formulation F-1.

MIP based solvers for GED compute GED by feeding linearizations of the quadratic program shown in Figure 5.4 into MIP solvers such as Gurobi Optimization [54] or IBM CPLEX [58]. The integer linear program F-1 suggested in [68, 69] and displayed in Figure 5.5 is a straightforward linearization of the quadratic programming formulation shown in Figure 5.4. F-1 has $O(|E^G||E^H|)$ binary variables and $O(|E^G||E^H|)$ constraints.

The linearization F-2 suggested in [69] and displayed in Figure 5.7 improves F-1 by reducing the number of variables and constraints. It uses the fact that the node and edge substitution variables $\mathbf{x}^{\text{sub}}$ and $\mathbf{y}^{\text{sub}}$ implicitly encode the node and edge deletion and insertion variables $\mathbf{x}^{\text{del}}$, $\mathbf{x}^{\text{ins}}$, $\mathbf{y}^{\text{del}}$, and $\mathbf{y}^{\text{ins}}$. F-2 uses modified edit costs $c'_V$ and $c'_E$, which are defined as

$$c'_V(u_i, v_k) := c_V(u_i, v_k) - c_V(u_i, \epsilon) - c_V(\epsilon, v_k)$$
$$c'_E((u_i, u_j), (v_k, v_l)) := c_E((u_i, u_j), (v_k, v_l)) - c_E((u_i, u_j), \epsilon) - c_E(\epsilon, (v_k, v_l));$$

$$\min \sum_{u_i V^G} \sum_{v_k \in V^H} c'_V(u_i, v_k) x^{\text{sub}}_{i,k} + \sum_{(u_i,u_j) \in E^G} \sum_{(v_k,v_l) \in E^H} c'_E((u_i,u_j),(v_k,v_l)) y^{\text{sub}}_{i,j,k,l} + C$$

$$\text{s.t.} \quad \sum_{v_k \in V^H} x^{\text{sub}}_{i,k} \leq 1 \qquad \forall u_i \in V^G$$

$$\sum_{u_i \in V^G} x^{\text{sub}}_{i,k} \leq 1 \qquad \forall v_k \in V^H$$

$$\sum_{(v_k,v_l) \in E^H} y^{\text{sub}}_{i,j,k,l} - x^{\text{sub}}_{i,k} - x^{\text{sub}}_{j,k} \leq 0 \qquad \forall (v_k, (u_i, u_j)) \in V^H \times E^G$$

$$\mathbf{x}^{\text{sub}} \in \{0,1\}^{|V^G| \times |V^H|}, \mathbf{y}^{\text{sub}} \in \{0,1\}^{|E^G| \times |E^H|}$$

**Figure 5.6.** The MIP formulation F-2.

$$\min \sum_{u_i V^G} \sum_{v_k \in V^H} c'_V(u_i, v_k) x^{\text{sub}}_{i,k}$$
$$+ \sum_{(u_i,u_j) \in E^G} \sum_{(v_k,v_l) \in E^H} c'_E((u_i,u_j),(v_k,v_l))(y^{\text{sub}}_{i,j,k,l} + y^{\text{sub}}_{i,j,l,k}) + C$$

$$\text{s.t.} \quad \sum_{v_k \in V^H} x^{\text{sub}}_{i,k} \leq 1 \qquad \forall u_i \in V^G$$

$$\sum_{u_i \in V^G} x^{\text{sub}}_{i,k} \leq 1 \qquad \forall v_k \in V^H$$

$$\sum_{(u_i,u_j) \in E^G} \sum_{(v_k,v_l) \in E^H} y_{i,j,k,l} + y_{i,j,l,k} - d_{i,k} x^{\text{sub}}_{i,k} \leq 0 \;\forall (u_i, v_k) \in V^G \times V^H$$

$$\mathbf{x}^{\text{sub}} \in \{0,1\}^{|V^G| \times |V^H|}, \mathbf{y}^{\text{sub}} \in \{0,1\}^{|E^G| \times (|E^H|+|E^H|)}$$

**Figure 5.7.** The MIP formulation F-3.

as well as a constant $C$ defined as $C := \sum_{u_i \in V^G} c_V(u_i, \epsilon) + \sum_{v_k \in V^H} c_V(\epsilon, v_k) + \sum_{(u_i,u_j) \in E^G} c_E((u_i, u_j), \epsilon) + \sum_{(v_k,v_l) \in E^H} c_E(\epsilon, (v_k, v_l))$. F-2 has $O(|V^H||E^G|)$ constraints and $O(|E^G||E^H|)$ binary variables.

The linearization F-3 [42] displayed in Figure 5.7 further improves F-2 for dense graphs. The difference w.r.t. F-2 is that, for each pair of edges $((u_i, u_j), (v_k, v_l)) \in E^G \times E^H$, F-3 contains *two* binary decision variables $y^{\text{sub}}_{i,j,k,l}$ and $y^{\text{sub}}_{i,j,l,k}$. This doubles the number of edge variables but allows to decrease the number of constraints. The variable $y^{\text{sub}}_{i,j,k,l}$ indicates whether the node map $\pi$ encoded by $\mathbf{x}^{\text{sub}}$ satisfies $\pi(u_i) = v_k$ and $\pi(u_j) = v_l$, while $y^{\text{sub}}_{i,j,l,k}$ indicates whether $\pi$ satisfies $\pi(u_i) = v_l$ and $\pi(u_j) = v_k$. That is, both $y^{\text{sub}}_{i,j,k,l}$ and $y^{\text{sub}}_{i,j,l,k}$ equal 1 only if $\pi(u_i, u_j) = (v_k, v_l)$. Moreover, F-3 uses constants $d_{i,k}$, which are defined as $d_{i,k} := \min\{\deg^G(u_i), \deg^H(v_k)\}$ for all $(u_i, v_k) \in V^G \times V^H$. F-3 has $O(|V^G||V^H|)$ constraints and $O(|E^G||E^H|)$ binary variables.

The linearization ADJ-IP suggested in [60] and displayed in Figure 5.8 requires the edge edit costs $c_E$ to be constant and symmetric. Furthermore, the linearization ADJ-IP is designed for graphs without edge labels, as it sets all edge substitution costs to 0. If used with graphs whose edges are labeled, it ignores all edit costs induced by substituting an edge $(u_i, u_j) \in E^G$ by an edge $(v_k, v_l) \in E^H$ with $\ell^G_E(u_i, u_j) \neq \ell^H_E(v_k, v_l)$ and hence only yields a lower bound for GED$(G, H)$ rather than the exact edit distance. ADJ-IP has $O((|V^G| + |V^H|)^2)$ constraints and $O((|V^G| + |V^H|)^2)$ binary variables.

## 5.2 Speed-Up of Node Based Tree Search

One bottleneck for the node based tree searches A* and DFS-GED is that the lower bound $LB(\pi')$ defined in equation (5.2) has to be recomputed for each partial node map $\pi'$ which is added to *OPEN* (line 10 in Figure 5.1 and line 11 in Figure 5.2). Computing $LB(\pi')$ requires solving the LSAPE instances $\mathbf{C}^{V-\pi'}$ and $\mathbf{C}^{E-\pi'}$, which are of sizes $O(|V^G|) \times O(|V^H|)$ and $O(|E^G|) \times O(|E^H|)$, respectively. As detailed in Chapter 4, this requires $O(\min\{|V^G|, |V^H|\}^2 \max\{|V^G|, |V^H|\} + \min\{|E^G|, |E^H|\}^2 \max\{|E^G|, |E^H|\})$ time.

Our speed-up for uniform edit costs builds upon the fact that, if the edit costs are constant and triangular, the LSAPE instances $\mathbf{C}^{V-\pi'}$ and $\mathbf{C}^{E-\pi'}$ can be solved via multiset intersection.

**Proposition 5.3 (Correctness of A* and DFS-GED for Constant, Triangular Edit Costs).** *Let $\pi' \in \Pi'(G, H)$ be a partial node map and the matrices $\mathbf{C}^{V-\pi'} \in \mathbb{R}^{(|V^G-\pi'|+1) \times (|V^H-\pi'|+1)}$ and $\mathbf{C}^{E-\pi'} \in \mathbb{R}^{(|E^G-\pi'|+1) \times (|E^H-\pi'|+1)}$ be defined as specified in equation (5.3). Then the equation*

$$\text{LSAPE}(\mathbf{C}^{V-\pi'}) = \Gamma(\ell^G_V[\![V^G - \pi']\!], \ell^H_V[\![V^H - \pi']\!], c^{\text{sub}}_V, c^{\text{del}}_V, c^{\text{ins}}_V)$$

$$\min \sum_{u_i \in V^G} \sum_{v_k \in V^H} [c_V(u_i, v_k)x_{i,k} + \frac{c_E^{\text{del}}}{2}(s_{i,k} + t_{i,k})]$$

$$+ \sum_{u_i \in V^G} \sum_{u_j \in V^G} [c_V(u_i, \epsilon)x_{i,|V^H|+j} + \frac{c_E^{\text{del}}}{2}(s_{i,|V^H|+j} + t_{i,|V^H|+j})]$$

$$+ \sum_{v_l \in V^H} \sum_{v_k \in V^H} [c_V(\epsilon, v_k)x_{|V^G|+l,k} + \frac{c_E^{\text{del}}}{2}(s_{|V^G|+l,k} + t_{|V^G|+l,k})]$$

$$\text{s. t.} \quad \sum_{u_j \in V^G} x_{i,|V^H|+j} + \sum_{v_k \in V^H} x_{i,k} = 1 \quad \forall u_i \in V^G$$

$$\sum_{v_k \in V^H} x_{|V^G|+k,|V^H|+i} + \sum_{u_j \in V^G} x_{j,|V^H|+i} = 1 \quad \forall u_i \in V^G$$

$$\sum_{v_l \in V^H} x_{|V^G|+l,k} + \sum_{u_i \in V^G} x_{i,k} = 1 \quad \forall v_k \in V^H$$

$$\sum_{u_i \in V^G} x_{|V^G|+k,|V^H|+i} + \sum_{v_l \in V^H} x_{|V^G|+k,l} = 1 \quad \forall v_k \in V^H$$

$$s_{i,k} - t_{i,k} + \sum_{u_j \in N^G(u_i)} x_{j,k} - \sum_{v_l \in N^H(v_k)} x_{i,l} = 0 \quad \forall (u_i, v_k) \in V^G \times V^H$$

$$s_{i,|V^H|+j} - t_{i,|V^H|+j} + \sum_{u_r \in N^G(u_i)} x_{r,|V^H|+j} = 0 \quad \forall (u_i, u_j) \in V^G \times V^G$$

$$s_{|V^G|+l,k} - t_{|V^G|+l,k} - \sum_{v_s \in N^H(v_k)} x_{|V^G|+l,s} = 0 \quad \forall (v_l, v_k) \in V^H \times V^H$$

$$\mathbf{x}, \mathbf{s}, \mathbf{t} \in \{0,1\}^{(|V^G|+|V^H|) \times (|V^H|+|V^G|)}$$

**Figure 5.8.** The MIP formulation `ADJ-IP`.

*holds for constant, triangular node edit cost functions $c_V$; and the equation*

$$\text{LSAPE}(\mathbf{C}^{E-\pi'}) = \Gamma(\ell_E^G[\![E^G - \pi']\!], \ell_E^H[\![E^H - \pi']\!], c_E^{\text{sub}}, c_E^{\text{del}}, c_E^{\text{ins}})$$

*holds for constant, triangular edge edit cost functions $c_E$.*

*Proof.* For proving the first part of the proposition, we assume w. l. o. g. that $|V^G - \pi'| \leq |V^H - \pi'|$. Then $\Gamma(\ell_V^G[\![V^G - \pi']\!], \ell_V^H[\![V^H - \pi']\!], c_V^{\text{sub}}, c_V^{\text{del}}, c_V^{\text{ins}}) = c_V^{\text{sub}}(|V^G - \pi'| - |\ell_V^G[\![V^G - \pi']\!] \cap \ell_V^H[\![V^H - \pi']\!]|) + c_V^{\text{ins}}(|V^H - \pi'| - |V^G - \pi'|)$ holds by definition of $\Gamma$. Furthermore, Proposition 4.2 implies that, if $c_V$ is triangular, an optimal error-correcting matching for $\mathbf{C}^{V-\pi'}$ contains exactly $|V^H - \pi'| - |V^G - \pi'|$ insertions and exactly $|V^G - \pi'|$ substitutions. If $c_V$ is constant, each insertion incurs a cost of $c_V^{\text{ins}}$, exactly $|\ell_V^G[\![V^G - \pi']\!] \cap \ell_V^H[\![V^H - \pi']\!]|$ substitutions incur a cost of 0, and the remaining substitutions incur a cost of $c_V^{\text{sub}}$. These observations prove the first part of the proposition. The second part can be shown analogously. $\square$

It has been shown that the intersection of *sorted* multisets can be computed in linear time [111]. Together with Proposition 5.3, this immediately implies that, if the edit costs are constant and triangular, $LB(\pi')$ can be computed in $O(\max\{|V^G|, |V^H|\} \log \max\{|V^G|, |V^H|\} + \max\{|E^G|, |E^H|\} \log \max\{|E^G|, |E^H|\})$ time: We first sort the labels of the nodes and the edges that have not been assigned by $\pi'$ and then compute the intersection sizes of the resulting sorted multisets.

In order to further reduce the complexity of the computation of $LB(\pi')$, we proceed as follows: When initializing DFS-GED, we *once* sort the multisets $\ell_V^G[\![V^G]\!]$, $\ell_V^H[\![V^H]\!]$, $\ell_E^G[\![E^G]\!]$, and $\ell_E^H[\![E^H]\!]$, which contain the labels of *all* nodes and edges. For each $\{\{\alpha_s\}\}$ of the resulting sorted multisets and each partial node map $\pi'$, we maintain a boolean vector that indicates whether the node or edge with label $\alpha_s$ is still unassigned by $\pi'$. This vector can be updated in constant additional time when updating the cost $c(\pi')$ of the partial edit path induced by $\pi'$. For each partial node map $\pi'$, $LB(\pi')$ can then be computed in $O(\max\{|V^G|, |V^H|\} + \max\{|E^G|, |E^H|\})$ time by using a variation of the algorithm for multiset intersection presented in [111].

## 5.3 Generalization of Edge Based Tree Search

For generalizing CSI-GED to general node edit costs, we prove the following generalized version of Proposition 5.2.

**Proposition 5.4 (Correctness of CSI-GED for General Edit Costs).** *Let* $(u_{i_r})_{r=1}^{|V^G - \phi|}$ *and* $(v_{k_s})_{s=1}^{|V^H - \phi|}$ *be arbitrary, fixed orderings of, respectively, $V^G - \phi$ and $V^H - \phi$, and the operator* EXTEND-EDGE-MAP-COST-2 $: \Phi(G, H) \to \mathbb{R}$ *be defined as*

$$\text{EXTEND-EDGE-MAP-COST-2}(P_\phi) := c(P_\phi) + \sum_{f \in E^H - \phi} c_E(\epsilon, f) + \text{LSAPE}(\mathbf{C}^{V-\phi}),$$

*where the LSAPE instance* $\mathbf{C}^{V-\phi} \in \mathbb{R}^{(|V^G-\phi|+1) \times (|V^H-\phi|+1)}$ *is constructed by setting*

$$c_{r,s}^{V-\phi} := c_V(u_{i_r}, v_{k_s})$$

$$c_{r,|V^H-\phi|+1}^{V-\phi} := c_V(u_{i_r}, \epsilon)$$
$$c_{|V^G-\phi|+1,s}^{V-\phi} := c_V(\epsilon, v_{k_s})$$

*for all* $(r,s) \in [|V^G - \phi|] \times [|V^H - \phi|]$. *Then it holds that* $\text{GED}(G,H) = \min_{\phi \in \overline{\Phi}(G,H)} \texttt{EXTEND-EDGE-MAP-COST-2}(P_\phi)$.

*Proof.* We show the following two statements, which, together with the alternative Definition 2.6 of GED, prove the proposition:

1. For each node map $\pi \in \Pi(G,H)$, there is a left-complete valid partial edge map $\phi \in \overline{\Phi}(G,H)$ such that $c(P_\pi) \geq \texttt{EXTEND-EDGE-MAP-COST-2}(P_\phi)$.
2. For each left-complete valid partial edge map $\phi \in \overline{\Phi}(G,H)$, there is a node map $\pi \in \Pi(G,H)$ such that $c(P_\pi) \leq \texttt{EXTEND-EDGE-MAP-COST-2}(P_\phi)$.

For proving the first statement, let $\pi \in \Pi(G,H)$ be a node map and $(u_i, u_j) \in \overrightarrow{E^G}$. We construct a left-complete partial node map $\phi$ by setting $\pi(u_i, u_j) := (\pi(u_i), \pi(u_j))$, if the edge $(u_i, u_j)$ is preserved under $\pi$, i.e., if we have $(\pi(u_i), \pi(u_j)) \in E^H$. Otherwise, we define $\pi(u_i, u_j) := \epsilon$. By construction, $\pi_\phi$ equals the restriction of $\pi$ to $\text{supp}(\phi)$. Since $\pi$ is a node map, this implies that $\pi_\phi$ is a partial node map and hence that $\phi$ is valid. We therefore have $\phi \in \overline{\Phi}(G,H)$.

By construction of $\phi$, all edit operations contained in the partial edit path $P_\phi$ induced by $\phi$ are also contained in the partial edit path $P_\pi$ induced by $\pi$. More precisely, $P_\phi$ consists of all edge substitutions and deletions contained in $P_\pi$ together with all node substitutions between $\text{supp}(\pi_\phi)$ and $\text{img}(\pi_\phi)$. This implies the equation

$$c(P_\pi) = c(P_\phi) + \underbrace{\sum_{f \in E^H - \phi} c_E(\epsilon, f) + \sum_{u \in V^G - \phi} c_V(u, \pi(u)) + \sum_{\substack{v \in V^H - \phi \\ \pi^{-1}(v) = \epsilon}} c_V(\epsilon, v)}_{=:A},$$

which proves the first statement, since $\pi' := \pi \cap (((V^G - \phi) \cup \{\epsilon\}) \times ((V^H - \phi) \cup \{\epsilon\}))$ is a feasible LSAPE solution for $\mathbf{C}^{V-\phi}$ with $\mathbf{C}^{V-\phi}(\pi') = A$.

For proving the second statement, let $\phi \in \overline{\Phi}(G,H)$ be a left-complete valid partial edge map and $\pi'$ be an optimal LSAPE solution for $\mathbf{C}^{V-\phi}$. We construct a node map $\pi \in \Pi(G,H)$ by setting $\pi := \pi_\phi \cup \pi'$. By construction, we have $\sum_{(u,v) \in \pi} c_V(u,v) = \sum_{u \in \text{supp}(\pi_\phi)} c_V(u, \pi_\phi(u)) + \text{LSAPE}(\mathbf{C}^{V-\phi})$. Furthermore, we know that, if $\pi(u_i, u_j) \neq \epsilon$ for some $(u_i, u_j) \in \overrightarrow{E^G}$, then $(\pi(u_i), \pi(u_j)) \in E^H$. In other words, all edge substitutions that are contained in the edit path $P_\phi$ induced by $\phi$ are also contained in the edit path $P_\pi$ induced by $\pi$. Let $E^{\pi-\phi} := \{(e,f) \mid (e,f) \in (E^G - \phi) \times (E^H - \phi) \wedge \pi(e) = f\} \subseteq E^G \times E^H$ be the set of all pairs of edges that are substituted by $P_\pi$ but deleted and reinserted by $P_\phi$. Then it holds that

$$c(P_\pi) = \texttt{EXTEND-EDGE-MAP-COST-2}(P_\phi) + \sum_{(e,f) \in E^{\pi-\phi}} \Delta_{e,f},$$

where $\Delta_{e,f} := c_E(e,f) + c_E(e,\epsilon) + c_E(\epsilon,f)$. As argued in the proof of Theorem 3.4, we can assume w.l.o.g. that edge substitutions are strongly irreducible. This implies $\Delta_{e,f} \leq 0$ for all $(e,f) \in E^{\pi-\phi}$, and hence finishes the proof of the second statement. □

Proposition 5.4 indicates how to extend `CSI-GED` to general edit costs: We just have to call `EXTEND-EDGE-MAP-COST-2` (which requires solving an instance of LSAPE) instead of `EXTEND-EDGE-MAP-COST-1` (which requires computing the size of a multiset intersection) in line 13 of Figure 5.3. Since computing the size of a multiset intersection requires linear time, whereas solving an instance of LSAPE needs cubic time, this comes at the price of an increased runtime. However, the increase is very moderate, as `EXTEND-EDGE-MAP-COST-2` is called only at the leafs of `CSI-GED`'s search tree.

## 5.4 A Compact MIP Formulation

Figure 5.9 shows the new, compact linearization `COMPACT-MIP`. `COMPACT-MIP` has $O(|V^G||V^H|)$ constraints, $O(|V^G||V^H|)$ binary variables, and $O(|V^G||V^H|)$ continuous variables. It is hence the smallest available MIP formulation of GED. Moreover, unlike the linearization `ADJ-IP` presented above (cf. Figure 5.8) `COMPACT-MIP` works for general graphs and edit cost.

`COMPACT-MIP` has fewer variables and constraints, because is does not contain edge variables. Instead, it contains continuous variables $\mathbf{z}^{\text{sub}}$, $\mathbf{z}^{\text{del}}$, $\mathbf{z}^{\text{ins}}$, which, at the optimum, contain the edit costs which are induced by the node assignment $\pi$ encoded by optimal binary node variables $\mathbf{x}^{\text{sub}}$, $\mathbf{x}^{\text{del}}$, and $\mathbf{x}^{\text{ins}}$. Moreover, `COMPACT-MIP` uses constants $a_{i,k}^{\text{sub}}$, $a_i^{\text{del}}$, and $a_k^{\text{ins}}$, which are defined as

$$a_{i,k}^{\text{sub}} := \Big[ \sum_{u_j \in N^G(u_i)} \sum_{v_l \in N^H(v_k)} c_E((u_i,u_j),(v_k,v_l))$$

$$+ \sum_{u_j \in N^G(u_i)} (|V^H| - \deg^H(v_k) + 1) c_E((u_i, u_j), \epsilon)$$

$$+ \sum_{v_l \in N^H(v_k)} (|V^G| - \deg^G(u_i) + 1) c_E(\epsilon, (v_k, v_l)) \Big] / 2$$

$$a_i^{\text{del}} := \Big[ \sum_{u_j \in N^G(u_i)} (|V^H| + 1) c_E((u_i, u_j), \epsilon) \Big] / 2$$

$$a_k^{\text{ins}} := \Big[ \sum_{v_l \in N^H(v_k)} (|V^G| + 1) c_E(\epsilon, (v_k, v_l)) \Big] / 2,$$

for all $(u_i, v_k) \in V^G \times V^H$. Theorem 5.1 below shows that COMPACT-MIP is indeed a MIP formulation of GED.

**Theorem 5.1 (Correctness of COMPACT-MIP).** *For all G and H, it holds that the minimum of the MIP COMPACT-MIP shown in Figure 5.9 equals* $\text{GED}(G, H)$.

*Proof.* Let $\mathbf{x}^{\text{sub}} \in \{0,1\}^{|V^G| \times |V^H|}$, $\mathbf{x}^{\text{del}} \in \{0,1\}^{|V^G|}$, $\mathbf{x}^{\text{ins}} \in \{0,1\}^{|V^H|}$, and $\mathbf{x} := (\mathbf{x}^{\text{sub}}, \mathbf{x}^{\text{del}}, \mathbf{x}^{\text{ins}})$. For all $(u_i, v_k) \in V^G \times V^H$, we define the linear expressions $c_{i,k}^{\text{sub}}(\mathbf{x})$, $c_i^{\text{del}}(\mathbf{x})$, and $c_k^{\text{ins}}(\mathbf{x})$ as follows:

$$c_{i,k}^{\text{sub}}(\mathbf{x}) := c_V(u_i, v_k) + \Big[ \sum_{u_j \in N^G(u_i)} \sum_{v_l \in N^H(v_k)} c_E((u_i, u_j), (v_k, v_l)) x_{j,l}^{\text{sub}}$$
$$+ \sum_{u_j \in N^G(u_i)} \sum_{v_l \notin N^H(v_k)} c_E((u_i, u_j), \epsilon) x_{j,l}^{\text{sub}}$$
$$+ \sum_{u_j \notin N^G(u_i)} \sum_{v_l \in N^H(v_k)} c_E(\epsilon, (v_k, v_l)) x_{j,l}^{\text{sub}}$$
$$+ \sum_{u_j \in N^G(u_i)} c_E((u_i, u_j), \epsilon) x_j^{\text{del}} + \sum_{v_l \in N^H(v_k)} c_E(\epsilon, (v_k, v_l)) x_l^{\text{ins}} \Big] / 2$$

$$c_i^{\text{del}}(\mathbf{x}) := c_V(u_i, \epsilon) + \Big[ \sum_{u_j \in N^G(u_i)} \sum_{v_l \in V^H} c_E((u_i, u_j), \epsilon) x_{j,l}^{\text{sub}}$$
$$+ \sum_{u_j \in N^G(u_i)} c_E((u_i, u_j), \epsilon) x_j^{\text{del}} \Big] / 2$$

$$c_k^{\text{ins}}(\mathbf{x}) := c_V(\epsilon, v_k) + \Big[ \sum_{u_j \in V^G} \sum_{v_l \in N^H(v_k)} c_E(\epsilon, (v_k, v_l)) x_{j,l}^{\text{sub}}$$
$$+ \sum_{v_l \in N^H(v_k)} c_E(\epsilon, (v_k, v_l)) x_l^{\text{ins}} \Big] / 2$$

The expressions $c_{i,k}^{\text{sub}}(\mathbf{x})$ equals the overall edit cost induced by substituting node $u_i$ by node $v_k$, given all other node edit operations encoded by $\mathbf{x}$. Similarly, $c_i^{\text{del}}(\mathbf{x})$ and $c_k^{\text{ins}}(\mathbf{x})$ equal the overall edit costs induced by deleting the node $u_i$ and inserting the node $v_k$, respectively. Using these expression,

$$\min \sum_{u_i \in V^G} \sum_{v_k \in V^H} z_{i,k}^{\text{sub}} + \sum_{u_i \in V^G} z_i^{\text{del}} + \sum_{v_k \in V^H} z_k^{\text{ins}}$$

$$\text{s.t.} \quad x_i^{\text{del}} + \sum_{v_k \in V^H} x_{i,k}^{\text{sub}} = 1 \quad \forall u_i \in V^G$$

$$x_k^{\text{ins}} + \sum_{u_i \in V^G} x_{i,k}^{\text{sub}} = 1 \quad \forall v_k \in V^H$$

$$\Big[ \sum_{u_j \in N^G(u_i)} \sum_{v_l \in N^H(v_k)} c_E((u_i, u_j), (v_k, v_l)) x_{j,l}^{\text{sub}}$$

$$+ \sum_{u_j \in N^G(u_i)} \sum_{v_l \notin N^H(v_k)} c_E((u_i, u_j), \epsilon) x_{j,l}^{\text{sub}}$$

$$+ \sum_{u_j \notin N^G(u_i)} \sum_{v_l \in N^H(v_k)} c_E(\epsilon, (v_k, v_l)) x_{j,l}^{\text{sub}}$$

$$+ \sum_{u_j \in N^G(u_i)} c_E((u_i, u_j), \epsilon) x_j^{\text{del}}$$

$$+ \sum_{v_l \in N^H(v_k)} c_E(\epsilon, (v_k, v_l)) x_l^{\text{ins}} \Big] / 2$$

$$+ c_V(u_i, v_k) - a_{i,k}^{\text{sub}}(1 - x_{i,k}^{\text{sub}}) - z_{i,k}^{\text{sub}} \le 0 \; \forall (u_i, v_k) \in V^G \times V^H$$

$$\Big[ \sum_{u_j \in N^G(u_i)} \sum_{v_l \in V^H} c_E((u_i, u_j), \epsilon) x_{j,l}^{\text{sub}}$$

$$+ \sum_{u_j \in N^G(u_i)} c_E((u_i, u_j), \epsilon) x_j^{\text{del}} \Big] / 2$$

$$+ c_V(u_i, \epsilon) - a_i^{\text{del}}(1 - x_i^{\text{del}}) - z_i^{\text{del}} \le 0 \quad \forall u_i \in V^G$$

$$\Big[ \sum_{u_j \in V^G} \sum_{v_l \in N^H(v_k)} c_E(\epsilon, (v_k, v_l)) x_{j,l}^{\text{sub}}$$

$$+ \sum_{v_l \in N^H(v_k)} c_E(\epsilon, (v_k, v_l)) x_l^{\text{ins}} \Big] / 2$$

$$+ c_V(\epsilon, v_k) - a_k^{\text{ins}}(1 - x_k^{\text{ins}}) - z_k^{\text{ins}} \le 0 \quad \forall v_k \in V^H$$

$$\mathbf{x}^{\text{sub}} \in \{0,1\}^{|V^G| \times |V^H|}, \mathbf{x}^{\text{del}} \in \{0,1\}^{|V^G|}, \mathbf{x}^{\text{ins}} \in \{0,1\}^{|V^H|}$$

$$\mathbf{z}^{\text{sub}} \in \mathbb{R}_{\ge 0}^{|V^G| \times |V^H|}, \mathbf{z}^{\text{del}} \in \mathbb{R}_{\ge 0}^{|V^G|}, \mathbf{z}^{\text{ins}} \in \mathbb{R}_{\ge 0}^{|E^H|}$$

**Figure 5.9.** The MIP formulation `COMPACT-MIP`.

the quadratic programming formulation of GED shown in Figure 5.4 can be rewritten as the problem to minimize the objective function

$$\sum_{u_i \in V^G} \sum_{v_k \in V^H} x_{i,k}^{\text{sub}} c_{i,k}^{\text{sub}}(\mathbf{x}) + \sum_{u_i \in V^G} x_i^{\text{del}} c_i^{\text{del}}(\mathbf{x}) + \sum_{v_k \in V^H} x_k^{\text{ins}} c_k^{\text{ins}}(\mathbf{x})$$

over all $\mathbf{x}^{\text{sub}} \in \{0,1\}^{|V^G| \times |V^H|}$, $\mathbf{x}^{\text{del}} \in \{0,1\}^{|V^G|}$, and $\mathbf{x}^{\text{ins}} \in \{0,1\}^{|V^H|}$ that respect the constraints $x_i^{\text{del}} + \sum_{v_k \in V^H} x_{i,k}^{\text{sub}} = 1$ for all $u_i \in V^G$ and the constraints $x_k^{\text{ins}} + \sum_{u_i \in V^G} x_{i,k}^{\text{sub}} = 1$ for all $v_k \in V^H$.

The theorem hence follows, if we can show that, at the optimum, the equations $z_{i,k}^{\text{sub}} = x_{i,k}^{\text{sub}} c_{i,k}^{\text{sub}}(\mathbf{x})$, $z_i^{\text{del}} = x_i^{\text{del}} c_i^{\text{del}}(\mathbf{x})$, and $z_k^{\text{ins}} = x_k^{\text{ins}} c_k^{\text{ins}}(\mathbf{x})$ hold for all $(u_i, v_k) \in V^G \times V^H$. For showing that this is the case, we rewrite the last three sets of constraints of `COMPACT-MIP` as follows:

$$c_{i,k}^{\text{sub}}(\mathbf{x}) - a_{i,k}^{\text{sub}}(1 - x_{i,k}^{\text{sub}}) \leq z_{i,k}^{\text{sub}} \quad \forall (u_i, v_k) \in V^G \times V^H \quad (5.5)$$

$$c_i^{\text{del}}(\mathbf{x}) - a_i^{\text{del}}(1 - x_i^{\text{del}}) \leq z_i^{\text{del}} \quad \forall u_i \in V^G \quad (5.6)$$

$$c_k^{\text{ins}}(\mathbf{x}) - a_k^{\text{ins}}(1 - x_k^{\text{ins}}) \leq z_k^{\text{ins}} \quad \forall v_k \in V^H \quad (5.7)$$

For establishing $z_{i,k}^{\text{sub}} = x_{i,k}^{\text{sub}} c_{i,k}^{\text{sub}}(\mathbf{x})$, we distinguish the cases $x_{i,k}^{\text{sub}} = 1$ and $x_{i,k}^{\text{sub}} = 0$. In the first case, (5.5) implies $z_{i,k}^{\text{sub}} \geq c_{i,k}^{\text{sub}}(\mathbf{x}) = x_{i,k}^{\text{sub}} c_{i,k}^{\text{sub}}(\mathbf{x})$ and hence $z_{i,k}^{\text{sub}} = x_{i,k}^{\text{sub}} c_{i,k}^{\text{sub}}(\mathbf{x})$ at the optimum. In the second case, the left-hand side of (5.5) is negative, as $c_{i,k}^{\text{sub}}(\mathbf{x}) \leq a_i^{\text{del}}$ holds for all $\mathbf{x}$. This implies that, at the optimum, we have $z_{i,k}^{\text{sub}} = 0 = x_{i,k}^{\text{sub}} c_{i,k}^{\text{sub}}(\mathbf{x})$, as required. The equations $z_i^{\text{del}} = x_i^{\text{del}} c_i^{\text{del}}(\mathbf{x})$ and $z_k^{\text{ins}} = x_k^{\text{ins}} c_k^{\text{ins}}(\mathbf{x})$ can be established analogously. □

## 5.5 Empirical Evaluation

We carried out extensive experiments to evaluate the exact GED algorithms presented in this chapter. In Section 5.5.1 we describe the setup of the experiments; in Section 5.5.2, we report the results; and in Section 5.5.3, we concisely summarize the most important experimental findings.

### 5.5.1 Setup and Datasets

**Datasets and Edit Costs.** We tested on the datasets PROTEIN, GREC, and LETTER (H). For all three datasets, we ran tests on uniform and non-uniform edit costs, where the non-uniform edit costs were computed as suggested in [2, 84] (cf. Section 2.4).

**Compared Methods.** In the experiments, we compared the performances of `A*`, `DFS-GED`, `CSI-GED`, `F-2`, and `COMPACT-MIP`. We did not include the parallelization technique `P-DFS-GED` in our experiments. Moreover, we excluded `F-1` (because, in [69], it is reported to always perform worse than `F-2`), `F-3` (because it is designed for dense graphs but we tested on sparse graphs), and `ADJ-IP` (because it does not compute the exact GED on datasets with edge labels).

**Protocol and Test Metrics.** We used the test protocol suggested in [52]: For all datasets $\mathcal{D}$ and all $i \in \{3, 6, \ldots, \max_{G \in \mathcal{D}} |G|\}$, we defined a size-constrained test group $\mathcal{G}_i$ that contains at most four randomly selected graphs $G \in \mathcal{D}$ satisfying $|V^G| = i \pm 1$. For each tested algorithm `ALG` and each test group $\mathcal{G}_i$, all pairwise comparisons between graphs contained in $\mathcal{G}_i$ were carried out. We set a time limit of 1000 seconds for each individual GED computation and recorded the following metrics:

- timeouts(`ALG`, $i$): The percentage of pairwise comparisons between graphs in $\mathcal{G}_i$ where `ALG` did not finish within 1000 seconds.
- runtime(`ALG`, $i$): `ALG`'s average runtime across all pairwise comparisons between graphs in $\mathcal{G}_i$.

**Implementation and Hardware Specifications.** All algorithms were implemented in C++ and employ the same data structures and subroutines. For implementing the IP based algorithms `F-2` and `COMPACT-MIP`, we used Gurobi Optimization [54], which was set up to run in one thread. All tests were carried out on a machine with two Intel Xeon E5-2667 v3 processors with 8 cores each and 98 GB of main memory running GNU/Linux.

### 5.5.2 Results of the Experiments

**Effects of Speed-up and Generalization.** Figure 5.10 and Table 5.1 show the effects of the speed-up of `DFS-GED` for constant triangular edit costs and the generalization of `CSI-GED` to arbitrary edit costs presented in Section 5.2 and Section 5.3, respectively. Our algorithms are starred. The values in Table 5.1 are obtained by averaging the runtimes across the test groups; bold values highlight the best performing algorithm. For enabling comparisons between the different versions of `CSI-GED` and `DFS-GED`, the tests were carried out for uniform edit costs. We only show the results for the test groups $\mathcal{G}_3$, $\mathcal{G}_6$, and $\mathcal{G}_9$, since on the test groups containing larger graphs, `CSI-GED` and `DFS-GED` were not competitive (cf. Figures 5.11 to 5.12 and Tables 5.2 to 5.3 below). We see that the difference between the runtime of the original uniform version of `CSI-GED` and our generalization to arbitrary edit costs is small: On LETTER (H), the uniform version was on average 1.44 times faster than the general version; on GREC, it was around 1.03 times faster; and on PROTEIN, it was around 1.01 times faster. On the other hand, our speed-up accelerated the original general version of `DFS-GED` by a factor of 1.92 on LETTER (H), by a factor of 4.33 on GREC, and by a factor of 10.67 on PROTEIN. These results mirror the fact that our generalization of `CSI-GED` increases the runtime only at the leafs of the search tree, while our speed-up of `DFS-GED` accelerates the computations that have to be carried out at each inner node.

**Table 5.1.** Averaged effects of speed-up and generalization on `DFS-GED` and `CSI-GED`.

| algorithm | LETTER (H) | GREC | PROTEIN |
|---|---|---|---|
| | avg. runtime across $\mathcal{G}_i$, $i \leq 9$, in seconds | | |
| `CSI-GED` (original) | **$81.73 \cdot 10^{-5}$** | **$89.73 \cdot 10^{-3}$** | 23.12 |
| `CSI-GED` (generalization)$^\star$ | $11.80 \cdot 10^{-4}$ | $92.31 \cdot 10^{-3}$ | 23.28 |
| `DFS-GED` (original) | $31.26 \cdot 10^{-4}$ | $51.65 \cdot 10^{-2}$ | $92.42 \cdot 10^{-2}$ |
| `DFS-GED` (speed-up)$^\star$ | $16.25 \cdot 10^{-4}$ | $11.92 \cdot 10^{-2}$ | **$86.61 \cdot 10^{-3}$** |

**Results for Non-Uniform Edit Costs.**   Figure 5.11 and Table 5.2 show the results for non-uniform edit costs. The original version of `CSI-GED` and our speed-up of `DFS-GED` were not included in the experiments, as they only work for uniform edit costs. Across all datasets, all tested algorithms managed to successfully compute GED within 1000 seconds on the test groups $\mathcal{G}_3$, $\mathcal{G}_6$, and $\mathcal{G}_9$, which contain graphs with up to 10 nodes. On these test groups, our generalization of `CSI-GED` was clearly the best algorithm, followed by `DFS-GED`: On LETTER (H), `CSI-GED` was on average around 3.23 times faster than `DFS-GED`; on GREC, it was around 7.18 times faster; and on PROTEIN, it was around 13.88 times faster. On the test groups that contain larger graphs, `F-2` performed best. Most notably, `F-2` clearly outperformed `COMPACT-MIP`, although the IP employed by the latter is smaller than the one used by the former. Across all datasets, `F-2` managed to compute ged between graphs with up to 16 nodes within 1000 seconds, but struggled on larger graphs. For instance, on PROTEIN, `F-2` (like all other tested algorithms) always failed on all test groups $\mathcal{G}_i$ with $i \geq 21$. We therefore do not display the outcomes for the test groups $\mathcal{G}_i$ with $i > 21$ of the PROTEIN dataset.

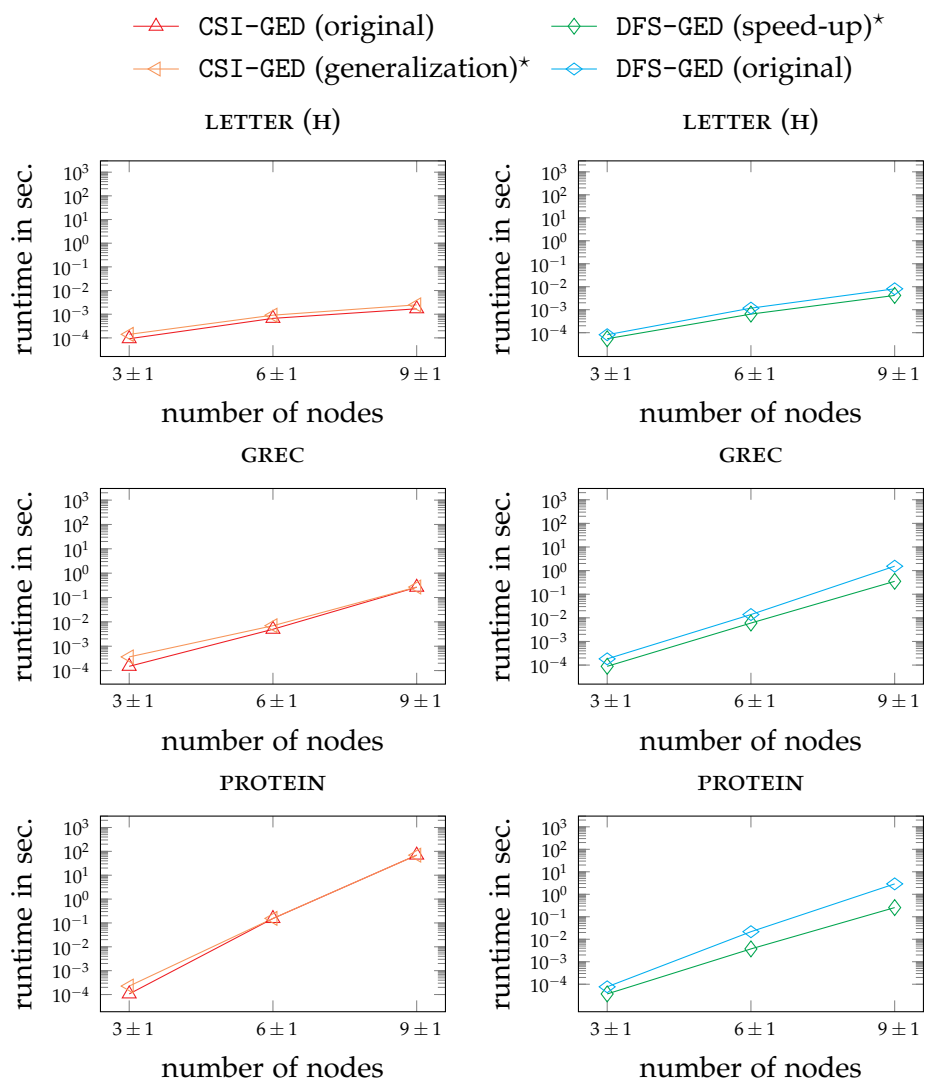

**Figure 5.10.** Effects of speed-up and generalization on DFS-GED and CSI-GED.

**Results for Uniform Edit Costs.** Figure 5.12 and Table 5.3 show the outcomes for uniform edit costs. We do not display the results for the general versions of CSI-GED and DFS-GED, since they were outperformed by the uniform counterparts (cf. Figure 5.10 and Table 5.1 above). Just as for non-uniform edit costs, F-2 was the best algorithm for graphs with more than 10 nodes: Across all datasets, it managed to compute ged between graphs with up to 16 nodes. However, F-2 (like all other tested algorithms) again struggled with substantially larger graphs, as it always failed to compute ged within 1000 seconds on the test groups $\mathcal{G}_i$ with $i \geq 24$ of the PROTEIN dataset. Again, we therefore do not display the results for the test groups $\mathcal{G}_i$ with $i > 24$ of the PROTEIN dataset. For small graphs with up to 10 nodes contained in the test groups $\mathcal{G}_3$, $\mathcal{G}_6$, and $\mathcal{G}_9$, our speed-up of DFS-GED performed best: On LETTER (H) and GREC, only CSI-GED performed better than DFS-GED (factor 1.99 on LETTER (H) and factor 1.33 on GREC). However, on the test groups $\mathcal{G}_3$, $\mathcal{G}_6$, and $\mathcal{G}_9$ of the PROTEIN dataset, the speed up version of DFS-GED was around 266.94 times faster than CSI-GED and around 1.50 times faster that the best competitor F-2, which performed worse than DFS-GED also on the small graphs of the datasets LETTER (H) and GREC.

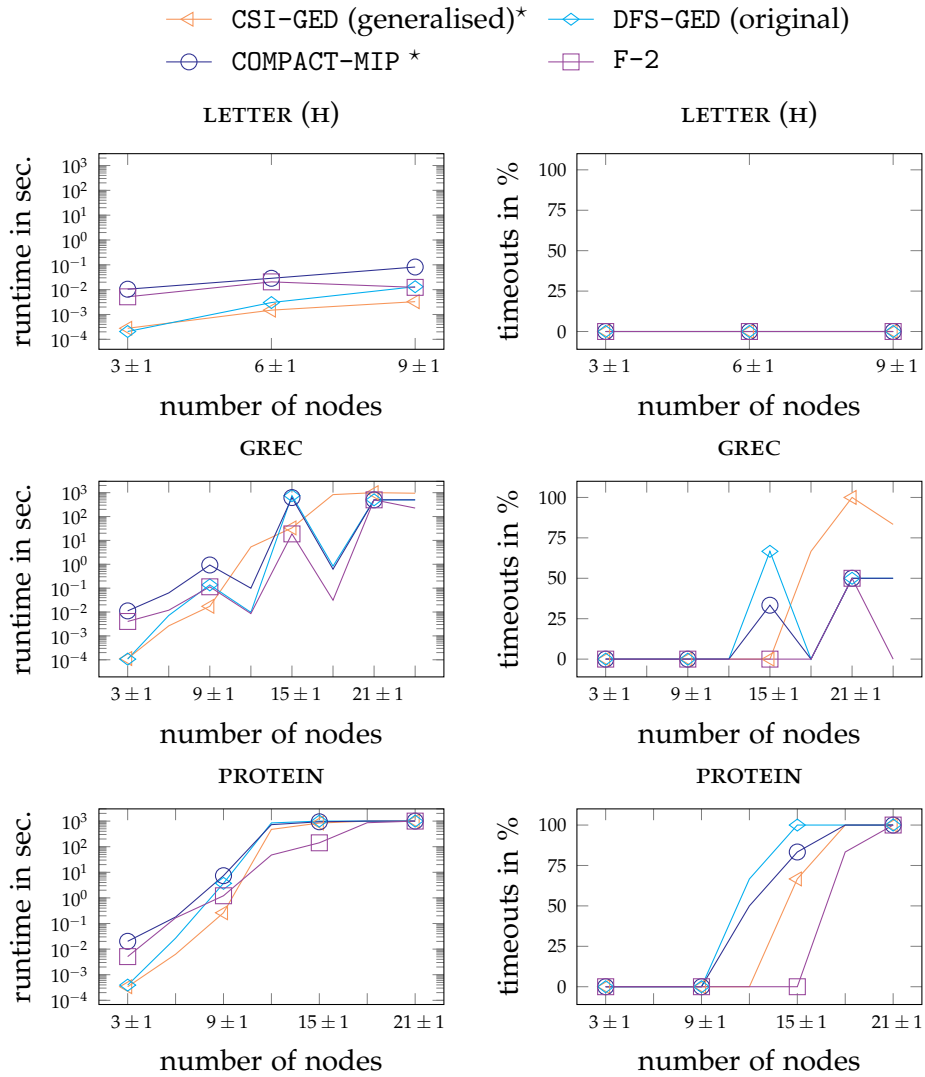

**Figure 5.11.** Results for non-uniform edit costs.

**Table 5.2.** Averaged results for non-uniform edit costs.

| algorithm | LETTER (H) | GREC | PROTEIN |
|---|---|---|---|
| | avg. runtime across $\mathcal{G}_i$, $i \leq 9$, in seconds | | |
| CSI-GED (generalization)* | **$16.87 \cdot 10^{-4}$** | **$66.95 \cdot 10^{-4}$** | **$89.72 \cdot 10^{-3}$** |
| DFS-GED (original) | $54.64 \cdot 10^{-4}$ | $48.06 \cdot 10^{-3}$ | $12.45 \cdot 10^{-1}$ |
| COMPACT-MIP * | $40.26 \cdot 10^{-3}$ | $33.60 \cdot 10^{-2}$ | $25.26 \cdot 10^{-1}$ |
| F-2 | $12.71 \cdot 10^{-3}$ | $43.32 \cdot 10^{-3}$ | $42.12 \cdot 10^{-2}$ |
| | avg. runtime across $\mathcal{G}_i$, $i > 9$, in seconds | | |
| CSI-GED (generalization)* | — | $56.29 \cdot 10^{1}$ | $82.85 \cdot 10^{1}$ |
| DFS-GED (original) | — | $35.06 \cdot 10^{1}$ | $96.24 \cdot 10^{1}$ |
| COMPACT-MIP * | — | $32.51$ | $91.67 \cdot 10^{1}$ |
| F-2 | — | **$14.91 \cdot 10^{1}$** | **$51.37 \cdot 10^{1}$** |
| | avg. timeouts across $\mathcal{G}_i$, $i > 9$, in % | | |
| CSI-GED (generalization)* | — | 50.00 | 66.67 |
| DFS-GED (original) | — | 33.33 | 91.67 |
| COMPACT-MIP * | — | 26.67 | 83.33 |
| F-2 | — | **10.00** | **45.83** |

### 5.5.3 Upshot of the Results

The main outcome of the experiments is that, for sparse, small graphs with up to 10 nodes, our generalization of CSI-GED is the best algorithm for non-uniform edit costs and our speed-up of DFS-GED is the best algorithm for uniform edit costs. If the number of nodes increases, the IP-based approach F-2 is the most performant algorithm. Furthermore, the experiments show that generalizing CSI-GED to arbitrary edit costs only slightly increases its runtime, while the speed-up of DFS-GED for constant, triangular edit costs accelerates the algorithm significantly. Note that, since the test datasets contain only sparse graphs with similar topological properties (cf. Table 2.4), these findings do not necessarily generalize to denser graphs.

Our tests also confirm the result that the standard approach A* is not competitive, as reported in [4, 52, 69]: Even on small graphs, A* very often failed because it ran out of memory; and if it did not fail, it was almost always the slowest algorithm. In order not to overload the plots, we hence do not show the results for A*. Furthermore, the experiments showed that the memory demands of all algorithms except for A* are negligible.

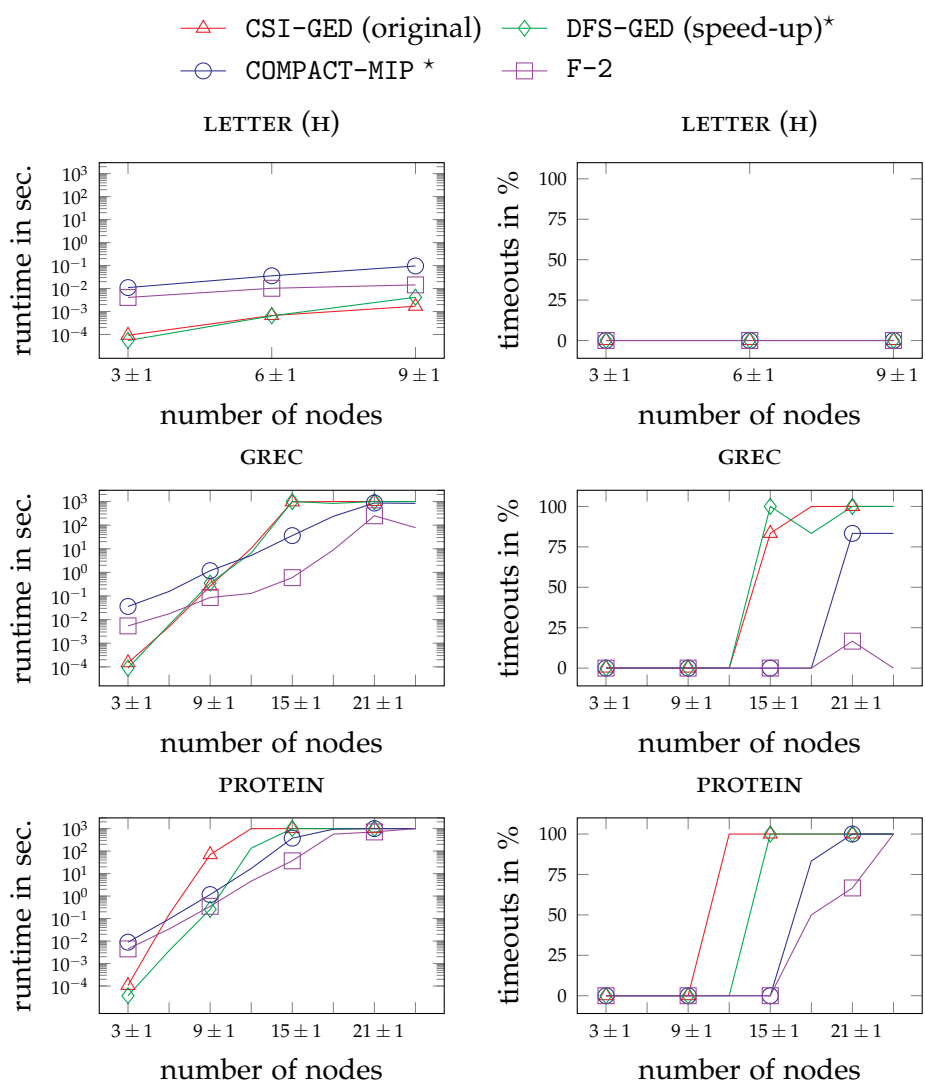

**Figure 5.12.** Results for uniform edit costs.

## 5.6 Conclusions and Future Work

In this chapter, we provided an overview of state of the art algorithms for the exact computation of GED, suggested three new techniques, and reported the results of experiments that tested available exact single-thread algorithms. One negative takeaway message of these experiments is that, even on sparse graphs such as the ones contained in the test datasets, no currently available single-thread algorithm manages to reliably compute GED within reasonable time between graphs with more than 16 nodes. However, there is at least one good reason to believe that there is room for improvement: F-2, i.e., the algorithm which among all available single-thread algorithms can cope with the largest graphs, uses a generic IP solver. Therefore, it should in principle be possible to design a specialized exact GED algorithm which is at least as but, hopefully, more performant than F-2.

Table 5.3. Averaged results for uniform edit costs.

| algorithm | LETTER (H) | GREC | PROTEIN |
|---|---|---|---|
| | avg. runtime across $\mathcal{G}_i$, $i \leq 9$, in seconds | | |
| CSI-GED (original) | **$81.73 \cdot 10^{-5}$** | **$89.73 \cdot 10^{-3}$** | 23.12 |
| DFS-GED (speed-up)* | $16.25 \cdot 10^{-4}$ | $11.93 \cdot 10^{-2}$ | **$86.61 \cdot 10^{-3}$** |
| COMPACT-MIP * | $47.56 \cdot 10^{-3}$ | $45.81 \cdot 10^{-2}$ | $42.34 \cdot 10^{-2}$ |
| F-2 | $95.75 \cdot 10^{-4}$ | $36.58 \cdot 10^{-3}$ | $12.97 \cdot 10^{-2}$ |
| | avg. runtime across $\mathcal{G}_i$, $i > 9$, in seconds | | |
| CSI-GED (original) | — | $79.58 \cdot 10^{1}$ | $10.00 \cdot 10^{2}$ |
| DFS-GED (speed-up)* | — | $76.80 \cdot 10^{1}$ | $82.71 \cdot 10^{1}$ |
| COMPACT-MIP * | — | $39.61 \cdot 10^{1}$ | $6.70 \cdot 10^{1}$ |
| F-2 | — | **67.58** | **$46.39 \cdot 10^{1}$** |
| | avg. timeouts across $\mathcal{G}_i$, $i > 9$, in % | | |
| CSI-GED (original) | — | 76.67 | 100.00 |
| DFS-GED (speed-up)* | — | 76.67 | 80.00 |
| COMPACT-MIP * | — | 33.33 | 56.57 |
| F-2 | — | **3.33** | **43.33** |

A second, comparative look at the available specialized algorithms A*, DFS-GED, and CSI-GED provides us with many possible starting points for the design of more efficient exact algorithms. For instance, one could try to make DFS-GED or CSI-GED employ more sophisticated lower bounds such as the ones proposed in [13, 15] and discussed in Chapter 6 below for pruning unpromising node or edge maps. One could try to develop new heuristics for determining the order in which the deepest node or edge maps of a depth-first search tree are processed. One could try out an edge based best-first search. Or else, one could use an entirely different search paradigm such as the hybrid best-first search suggested in [5] for traversing the space of all partial node maps or the space of all valid partial edge maps.

# — 6 —

# Heuristic Algorithms

Because of the hardness of exactly computing GED or approximating it within provable approximation ratios and since, very often, one is actually not interested in the exact value of GED but rather wants to find the graph $H$ that is most similar to a given query graph $G$, during the past years, a huge variety of heuristic algorithms for GED have been proposed. These heuristics approximate GED via lower or upper bounds, using techniques such as transformations to the linear sum assignment problem with error-correction (LSAPE), linear programming, and local search. Since no theoretical guarantees can be provided for the produced bounds, the heuristics are always evaluated empirically. The most frequently used evaluation criteria are the runtime behavior of the heuristics, the tightness of the produced upper bounds, and the performance of pattern recognition frameworks that use the bounds produced by the heuristics as underlying distance measures.

In this chapter, we provide a systematic overview of the most important heuristics. The fist column of Table 6.1 shows the suggested taxonomy. Whenever possible, we model the compared heuristics as instantiations of one of the following three paradigms: `LSAPE-GED`, `LP-GED`, and `LS-GED`. Instantiations of `LSAPE-GED` use transformations to LSAPE for heuristically computing GED. All instantiation of `LSAPE-GED` produce upper bounds, some also yield lower bounds. Instantiations of `LP-GED` compute lower and upper bounds for GED by employing linear programming (LP) relaxations of mixed integer programming (MIP) formulations of GED. And instantiations of the paradigm `LS-GED` improve initially computed or randomly generated upper bounds by using variants of local search.

Moreover, we present the new heuristics `BRANCH`, `BRANCH-FAST`,

| paradigm | new heuristics proposed in this chapter |
|---|---|
| `LSAPE-GED` | `BRANCH, BRANCH-FAST, RING, RING-ML, MULTI-SOL` |
| `LP-GED` | — |
| `LS-GED` | `K-REFINE, RANDPOST` |
| miscellaneous heuristics | `BRANCH-TIGHT` |

**Table 6.1.** Newly proposed heuristics and suggested taxonomy for heuristic computation of GED.

`BRANCH-TIGHT`, `RING`, `RING-ML`, `MULTI-SOL`, `K-REFINE`, and `RANDPOST`. The second column of Table 6.1 locates them within the suggested taxonomy. The algorithms `BRANCH` and `BRANCH-FAST` instantiate the paradigm `LSAPE-GED` to quickly compute lower and upper bounds for GED. Both of them exhibit excellent tradeoffs between runtime behavior and quality of the produced bounds. `BRANCH-TIGHT` is an anytime algorithm that iteratively improves the lower bounds produced by `BRANCH`. It performs particularly well on datasets where editing edges is more expensive than editing nodes. `RING` and `RING-ML` are instantiations of `LSAPE-GED` that are designed to yield tight upper bounds. Both algorithms decompose the input graphs into rings whose centers or roots correspond the graphs' nodes (cf. Definition 6.2 for the definition of rooted rings). Subsequently, this decomposition is used to populate instances of LSAPE. The post-processing technique `MULTI-SOL` tightens the upper bounds computed by all instantiations of `LSAPE-GED`. It enumerates several optimal LSAPE solutions and returns the smallest upper bound that is induced by one of the solutions. The heuristic `K-REFINE` improves and generalizes the existing local search algorithm `K-REFINE`. On small graphs, `K-REFINE` is among the most accurate GED heuristics and, at the same time, clearly outperforms similarly accurate algorithms in terms of runtime; on larger graphs, it yields an excellent tradeoff between runtime and quality of the produced upper bounds. The framework `RANDPOST` stochastically generates initial solutions for instantiations of `LS-GED`, such that the local search is intensified in promising regions of the set of all node maps. `RANDPOST` is particularly efficient on larger graphs.

Finally, we carried out extensive experiments to test the heuristics presented in this chapter. We also addressed the following meta-questions Q-1 and Q-2, which, to the best of our knowledge, have not been explicitly discussed in the literature.

Q-1 Is it indeed beneficial to use GED as a guidance for the design of graph distance measures, if these distance measures are to be used within pattern recognition frameworks?

Q-2 Do graph distance measures defined by upper bounds for GED or graph distance measures defined by lower bounds for GED perform better when used within pattern recognition frameworks?

The results presented in this chapter have previously been presented in the following articles:

- D. B. Blumenthal, S. Bougleux, J. Gamper, and L. Brun, "Ring based approximation of graph edit distance", in *S+SSPR 2018*, X. Bai, E. Hancock, T. Ho, R. Wilson, B. Biggio, and A. Robles-Kelly, Eds., ser. LNCS, vol. 11004, Cham: Springer, 2018, pp. 293–303. DOI: 10.1007/978-3-319-97785-0_28
- D. B. Blumenthal and J. Gamper, "Improved lower bounds for graph edit distance", *IEEE Trans. Knowl. Data Eng.*, vol. 30, no. 3, pp. 503–516, 2018. DOI: 10.1109/TKDE.2017.2772243
- D. B. Blumenthal and J. Gamper, "Correcting and speeding-up bounds for non-uniform graph edit distance", in *ICDE 2017*, IEEE Computer Society, 2017, pp. 131–134. DOI: 10.1109/ICDE.2017.57
- D. B. Blumenthal, N. Boria, J. Gamper, S. Bougleux, and L. Brun, "Comparing heuristics for graph edit distance computation", *VLDB J.*, 2019, in press. DOI: 10.1007/s00778-019-00544-1
- D. B. Blumenthal, S. Bougleux, J. Gamper, and L. Brun, "Upper bounding GED via transformations to LSAPE based on rings and machine learning", 2019. arXiv: 1907.00203 [cs.DS]
- N. Boria, D. B. Blumenthal, S. Bougleux, and L. Brun, "Improved local search for graph edit distance", 2019. arXiv: 1907.02929 [cs.DS]

The remainder of this chapter is organized as follows: In Section 6.1, a systematic overview of existing approaches is provided. In the Sections 6.2 to 6.7, the new heuristics BRANCH and BRANCH-FAST (Section 6.2), BRANCH-TIGHT (Section 6.3), RING and RING-ML (Section 6.4), MULTI-SOL (Section 6.5), K-REFINE (Section 6.6), and RANDPOST (Section 6.7) are presented. In Section 6.8, the presented heuristics are evaluated empirically. Section 6.9 concludes the chapter and points out to future work.

## 6.1 State of the Art

In this section, we provide an overview of the most important existing heuristics for upper or lower bounding GED. Each heuristic is denoted by a name written in typewriter font. Whenever possible, this name is taken from the original publication. If no original name is available, we invented a name which reflects the main technical ingredient of the heuristic.

Some of the presented heuristics are designed for special edit cost functions. For instance, some of them require constant edge edit costs, i.e., expect that there are constants $c_E^{\text{sub}}, c_E^{\text{del}}, c_E^{\text{ins}} \in \mathbb{R}$ such that $c_E(\beta, \beta') = c_E^{\text{sub}}$, $c_E(\beta, \epsilon) = c_E^{\text{del}}$, and $c_V(\epsilon, \beta') = c_E^{\text{ins}}$ holds for all $(\beta, \beta') \in \Sigma_E \times \Sigma_E$ with $\beta \neq \beta'$. In order to be able to run a heuristic designed for constant edge edit costs on datasets where this constraint is not satisfied, one can compute constants $c_E^{\text{sub}} := \min\{c_E(\beta, \beta') \mid (\beta, \beta') \in \ell_E^G[E^G] \times \ell_E^H[E^H] \wedge \beta \neq \beta'\}$, $c_E^{\text{del}} := \min\{c_E(\beta, \epsilon) \mid \beta \in \ell_E^G[E^G]\}$, and $c_E^{\text{del}} := \min\{c_E(\epsilon, \beta') \mid \beta' \in \ell_E^H[E^H]\}$ before running the heuristic on graphs $G$ and $H$. The same technique can be used for enforcing other cost constraints such as uniformity.

In the Sections 6.1.1 to 6.1.3, we introduce the paradigms LSAPE-GED, LP-GED, and LS-GED and present heuristics that can be modeled as their instantiations or extensions. In Section 6.1.4, we present miscellaneous heuristics that cannot be subsumed under any of the three paradigms.

### 6.1.1 LSAPE Based Heuristics

Instantiations of the paradigm LSAPE-GED use transformations from GED to LSAPE (cf. Definition 2.26 and Chapter 4 above) to compute upper and, possibly, lower bounds for GED, as well as node maps that induce the upper bounds. More precisely, they proceed as described in Figure 6.1: In a first step, the input graphs $G$ and $H$ and the edit cost functions are used to construct an LSAPE instance $\mathbf{C}$ of size $(|V^G|+1) \times (|V^H|+1)$ such that optimal LSAPE solutions for $\mathbf{C}$ induce cheap edit paths between $G$ and $H$ (line 1). This construction phase is where different instantiations of LSAPE-GED vary from each other. Subsequently, the LSAPE instance $\mathbf{C}$ is solved — either optimally or greedily — and the cost $c(P_\pi)$ of the edit path induced by the obtained LSAPE solution $\pi$ is interpreted as an upper bound for $\text{GED}(G, H)$ (line 2).

**Input**: Two undirected, labeled graphs $G$ and $H$, node and edge edit cost functions $c_V$ and $c_E$.
**Output**: An upper bound $UB$ and, possibly, a lower bound $LB$ for $\text{GED}(G, H)$, as well as a node map $\pi \in \Pi(G, H)$ with $c(P_\pi) = UB$.

1 use information encoded in $G$, $H$, $c_V$, and $c_E$ to construct LSAPE instance $\mathbf{C} \in \mathbb{R}^{(|V^G|+1) \times (|V^H|+1)}$;
2 use optimal or greedy solver to compute cheap LSAPE solution $\pi \in \Pi_{|V^G|,|V^H|,\epsilon}$;
3 set upper bound to $UB := c(P_\pi)$;
4 **if** *line 1 ensures $\xi(G, H, c_V, c_E) \text{LSAPE}(\mathbf{C}) \leq \text{GED}(G, H)$* **then**
5    **if** *optimal solver was used in line 2* **then**
6       set lower bound to $LB := \xi(G, H, c_V, c_E)\mathbf{C}(\pi)$;
7       **return** $LB$, $UB$, and $\pi$;
8    **else**
9       **return** $UB$ and $\pi$;
10 **return** $UB$ and $\pi$;

**Figure 6.1.** The paradigm LSAPE-GED.

If the protocol for constructing the LSAPE instance $\mathbf{C}$ ensures that one can define a scaling function $\xi(G, H, c_V, c_E)$ such that

$$\xi(G, H, c_V, c_E) \text{LSAPE}(\mathbf{C}) \leq \text{GED}(G, H) \quad (6.1)$$

holds for all graphs $G, H \in \mathbb{G}$ and all edit cost functions $c_V$ and $c_E$ and an optimal LSAPE solver was used to compute $\pi$, $\xi(G, H, c_V, c_E)\mathbf{C}(\pi)$ is returned as a lower bound for GED along with the node map $\pi$ and its induced upper bound (lines 4 to 7). Otherwise, only the upper bound and the node map are returned (lines 9 to 10).

Assume that an instantiation of LSAPE-GED constructs its LSAPE instance $\mathbf{C}$ in $O(\omega)$ time. As pointed out in Chapter 4, optimally solving $\mathbf{C}$ requires $O(\min\{|V^G|, |V^H|\}^2 \max\{|V^G|, |V^H|\})$ time, while the complexity of greedily computing a cheap suboptimal solution is $O|V^G||V^H|)$. The induced cost of the obtained node map $\pi$ can be computed in $O(\max\{|E^G|, |E^H|\})$ time. The heuristic's overall runtime complexity is hence $O(\omega + \min\{|V^G|, |V^H|\}^2 \max\{|V^G|, |V^H|\} + \max\{|E^G|, |E^H|\})$ if an optimal solver is used in line 2, and $O(\omega + |V^G||V^H| + \max\{|E^G|, |E^H|\})$ if $\mathbf{C}$ is solved greedily.

#### 6.1.1.1 The Algorithm NODE

The algorithm NODE [60] is a very simple instantiation of LSAPE: It completely ignores the edges of the input graphs $G$ and $H$ and just defines $\mathbf{C}$ as the node edit cost matrix between $G$ and $H$. In other words, it sets

$$c_{i,k} := c_V(u_i, v_k)$$
$$c_{i,|V^H|+1} := c_V(u_i, \epsilon)$$
$$c_{|V^G|+1,k} := c_V(\epsilon, v_k)$$

for all $(i, k) \in [|V^G|] \times [|V^H|]$.

The time complexity of constructing $\mathbf{C}$ is $O(|V^G||V^H|)$. As inequality (6.1) with $\xi :\equiv 1$ holds for all graphs $G, H \in \mathbb{G}$ and all edit cost functions $c_V$ and $c_E$, NODE computes both an upper and a lower bound for GED.

#### 6.1.1.2 The Algorithm BP

Unlike NODE, the algorithm BP [83] also considers edges. Informally, this is done by adding to $c_{i,k}$ as defined by NODE the optimal cost of transforming the edges that are incident with $u_i$ in $G$ into the edges that are incident with $v_k$ in $H$.

Formally, for each $(i, k) \in [|V^G|] \times [|V^H|]$, an auxiliary LSAPE instance $\mathbf{C}^{i,k} \in \mathbb{R}^{(\deg^G(u_i)+1) \times (\deg^H(v_k)+1)}$ is constructed. Let $(u_{i_j})_{j=1}^{\deg^G(u_i)}$ be an enumeration of $u_i$'s neighborhood $N^G(u_i)$, and $(v_{k_l})_{l=1}^{\deg^H(v_k)}$ be an enumeration of $v_k$'s neighborhood $N^H(v_k)$. BP sets

$$c_{j,l}^{i,k} := c_E((u_i, u_{i_j}), (v_k, v_{k_l}))$$
$$c_{j,\deg^H(v^k)+1}^{i,k} := c_E((u_i, u_{i_j}), \epsilon)$$
$$c_{\deg^G(u_i),l}^{i,k} := c_E(\epsilon, (v_k, v_{k_l}))$$

for all $(j, l) \in [\deg^G(u_i)] \times [\deg^H(v_k)]$, and optimally solves the resulting LSAPE instance. Once this has been done for all $(i, k) \in [|V^G|] \times [|V^H|]$, the final LSAPE instance $\mathbf{C}$ is constructed by setting

$$c_{i,k} := c_V(u_i, v_k) + \text{LSAPE}(\mathbf{C}^{i,k})$$
$$c_{i,|V^H|+1} := c_V(u_i, \epsilon) + \sum_{j=1}^{\deg^G(u_i)} c_E((u_i, u_{i_j}), \epsilon)$$

$$c_{|V^G|+1,k} := c_V(\epsilon, v_k) + \sum_{l=1}^{\deg^H(v_k)} c_E(\epsilon, (v_k, v_{k_l}))$$

for all $(i,k) \in [|V^G|] \times [|V^H|]$.

BP requires $O(|V^G||V^H|\Delta_{\min}^{G,H^2}\Delta_{\max}^{G,H})$ time for constructing **C**. This construction does not guarantee that inequality (6.1) holds, which implies that BP only returns an upper bound for GED.

#### 6.1.1.3 The Algorithm STAR

The algorithm STAR [111] considers the neighbors of the nodes $u_i \in V^G$ and $v_k \in V^H$ when populating the cell $c_{i,k}$ of its LSAPE instance **C**. It requires uniform edit cost functions $c_V$ and $c_E$ and ignores the edge labels of the input graphs. Let $C$ be the constant such that $c_V(\alpha, \alpha') = c_E(\beta, \beta') = C$ holds for all $(\alpha, \alpha') \in (\Sigma_V \cup \{\epsilon\}) \times (\Sigma_V \cup \{\epsilon\})$ with $\alpha \neq \alpha'$ and all $(\beta, \beta') \in (\Sigma_E \cup \{\epsilon\}) \times (\Sigma_E \cup \{\epsilon\})$ with $\beta \neq \beta'$. STAR then defines its LSAPE instance **C** by setting

$$c_{i,k} := \Gamma(\ell_V^G[\![N^G(u_i)]\!], \ell_V^H[\![N^H(v_k)]\!], C, C, C)$$
$$+ C \cdot [\delta_{\ell_V^G(u_i) \neq \ell_V^H(v_k)} + \Delta_{\max}^{i,k} - \Delta_{\min}^{i,k}]$$
$$c_{i,|V^H|+1} := C \cdot [1 + 2\deg^G(u_i)]$$
$$c_{|V^G|+1,k} := C \cdot [1 + 2\deg^H(v_k)]$$

for all $(i,k) \in [|V^G|] \times [|V^H|]$, where $\Delta_{\min}^{i,k} := \min\{\deg^G(u_i), \deg^H(v_k)\}$ and $\Delta_{\min}^{i,k} := \max\{\deg^G(u_i), \deg^H(v_k)\}$.

STAR requires $O(\max\{|V^G|, |V^H|\}\Delta_{\max}^{G,H} \log(\Delta_{\max}^{G,H}) + |V^G||V^H|\Delta_{\min}^{G,H})$ time for constructing its LSAPE instance **C**. Furthermore, inequality (6.1) holds if the scaling function $\xi$ is defined as $\xi(G, H, c_V, c_E) := 1/\max\{4, \Delta_{\max}^{G,H} + 1\}$. STAR hence returns both a lower and an upper bound for GED.

#### 6.1.1.4 The Algorithm BRANCH-CONST

The algorithm BRANCH-CONST suggested in [113, 114] is designed for constant edge edit costs $c_E$, i.e., settings where all edge substitutions, deletions, and insertions incur costs $c_E^{\text{sub}}$, $c_E^{\text{del}}$, and $c_E^{\text{ins}}$, respectively. BRANCH-CONST defines its LSAPE instance **C** by setting

$$c_{i,k} := c_V(u_i, v_k) + 0.5\Gamma(\ell_E^G[\![E^G(u_i)]\!], \ell_E^H[\![E^H(v_k)]\!], c_E^{\text{sub}}, c_E^{\text{del}}, c_E^{\text{ins}})$$

$$c_{i,|V^H|+1} := c_V(u_i, \epsilon) + 0.5 \deg^G(u_i) c_E^{\text{del}}$$
$$c_{|V^G|+1,k} := c_V(\epsilon, v_k) + 0.5 \deg^H(v_k) c_E^{\text{ins}}$$

for all $(i,k) \in [|V^G|] \times [|V^H|]$.

Like STAR, BRANCH-CONST requires $O(\max\{|V^G|, |V^H|\} \Delta_{\max}^{G,H} \log(\Delta_{\max}^{G,H}) + |V^G||V^H|\Delta_{\min}^{G,H})$ time for constructing its LSAPE instance **C**. As inequality (6.1) with $\zeta :\equiv 1$ holds for each input, BRANCH-CONST returns an upper and a lower bound for GED.

#### 6.1.1.5 The Algorithm SUBGRAPH

The algorithm SUBGRAPH [32] considers more global information than the previously presented heuristics for constructing its LSAPE instance **C**. Given a constant $K \in \mathbb{N}_{\geq 1}$, SUBGRAPH constructs graphlets $G_i := G[\bigcup_{s=0}^K N_s^G(u_i)]$ and $H_k := H[\bigcup_{s=0}^K N_s^H(v_k)]$ for all $(i,k) \in [|V^G|] \times [|V^H|]$, i.e., associates all nodes in the input graphs to the subgraphs which are induced by the sets of all nodes that are at distance at most $K$. For graphlets $G_i$ and $H_k$, SUBGRAPH defines $\text{GED}_{i,k}(G_i, H_k) := \min\{c(P_\pi) \mid \pi \in \Pi(G_i, H_i) \wedge \pi(u_i) = v_k\}$ as the edit distance under the restriction that $G^i$'s root node $u_i$ be mapped to $H^k$'s root node $v_k$. SUBGRAPH then constructs its LSAPE instance **C** by setting

$$c_{i,k} := \text{GED}_{i,k}(G_i, H_k)$$
$$c_{i,|V^H|+1} := \text{GED}(G_i, \mathcal{E})$$
$$c_{|V^G|+1,k} := \text{GED}(\mathcal{E}, H_k)$$

for all $(i,k) \in [|V^G|] \times [|V^H|]$, where $\mathcal{E}$ denotes the empty graph.

The time complexity of SUBGRAPH's construction phase of its LSAPE instance **C** is exponential in $\Delta_{\max}^{G,H}$. This implies that, unless $\max \deg(G)$ and $\max \deg(H)$ are bounded by a constant, SUBGRAPH does not run in polynomial time. SUBGRAPH only computes an upper bound for GED.

#### 6.1.1.6 The Algorithm WALKS

The algorithm WALKS [50] requires constant and symmetric edit cost functions $c_V$ and $c_E$ and aims at computing a tight upper bound for GED by associating each node in the input graphs to the set of walks of size $K$ that start at this node. Given constant $K \in \mathbb{N}_{\geq 1}$, a node $u_i \in V^G$, and a node $v_k \in V^H$, WALKS defines $\mathcal{W}_i^G$ and $\mathcal{W}_k^H$ as, respectively, the sets of walks of size $K$ that start at $u_i$ and $v_k$. Walks $W \in \mathcal{W}_i^G$ and $W' \in \mathcal{W}_k^H$ are called *similar* if they encode the same sequences of node and edge labels. Otherwise, $W$ and $W'$ are called *different*.

WALKS now computes the matrix products $\mathbf{W}_G^K$, $\mathbf{W}_H^K$, and $\mathbf{W}_\times^K$, where $\mathbf{W}_G$ is the adjacency matrix of $G$, $\mathbf{W}_H$ is the adjacency matrix of $H$, and $\mathbf{W}_\times$ is the adjacency matrix of the direct product graph $G \times H$ of $G$ and $H$. $G \times H$ contains a node $(u_i, v_k)$ for each $(i,k) \in [|V^G|] \times [|V^H|]$ such that $\ell_V^G(u_i) = \ell_V^H(v_k)$. Two nodes $(u_i, v_k)$ and $(u_j, v_l)$ of the product graph $G \times H$ are connected by an edge if and only if $(u_i, u_j) \in E^G$, $(v_k, v_l) \in E^H$, and $\ell_E^G(u_i, u_j) = \ell_E^H(v_k, v_l)$.

With the help of $\mathbf{W}_G^K$, $\mathbf{W}_H^K$, and $\mathbf{W}_\times^K$, for each node label $\alpha \in \Sigma_V$, WALKS computes an estimate $\widehat{h}_{i\setminus k}(\alpha)$ of the number of walks $W \in \mathcal{W}_i^G$ that end at a node with label $\alpha$ and must be substituted by a different walk $W' \in \mathcal{W}_k^H$. Analogously, $\widehat{h}_{k\setminus i}(\alpha)$ is computed as an estimate of the number of walks $W' \in \mathcal{W}_k^H$ that end at a node with label $\alpha$ and must be substituted by a different walk $W \in \mathcal{W}_i^G$. Moreover, WALKS computes an estimate $\widehat{r}_{i\setminus k} := \sum_{\alpha \in \Sigma_V} \widehat{h}_{i\setminus k}(\alpha) - \min\{\widehat{h}_{i\setminus k}(\alpha), \widehat{h}_{k\setminus i}(\alpha)\}$ of the number of walks in $W \in \mathcal{W}_i^G$ that must be substituted by different a different walk $W' \in \mathcal{W}_k^H$ that does not end at the same node label, and an estimate $\widehat{r}_{k\setminus i} := \sum_{\alpha \in \Sigma_V} \widehat{h}_{k\setminus i}(\alpha) - \min\{\widehat{h}_{i\setminus k}(\alpha), \widehat{h}_{k\setminus i}(\alpha)\}$ of the number of walks in $W' \in \mathcal{W}_k^H$ that must be substituted by a different walk $W \in \mathcal{W}_i^G$ that does not end at the same node label. With these ingredients, WALKS constructs its LSAPE instance $\mathbf{C}$ by setting

$$\begin{aligned}
c_{i,k} &:= [(\delta_{\ell_V^G(u_i) \neq \ell_V^H(v_k)} + K - 1)c_V^{\text{sub}} + Kc_E^{\text{sub}}] \cdot \sum_{\alpha \in \Sigma_V} \min\{\widehat{h}_{i\setminus k}(\alpha), \widehat{h}_{k\setminus i}(\alpha)\} \\
&\quad + [(\delta_{\ell_V^G(u_i) \neq \ell_V^H(v_k)} + K)c_V^{\text{sub}} + Kc_E^{\text{sub}}] \cdot \min\{\widehat{r}_{i\setminus k}, \widehat{r}_{k\setminus i}\} \\
&\quad + [(\delta_{\ell_V^G(u_i) \neq \ell_V^H(v_k)} + K)c_V^{\text{del}} + Kc_E^{\text{del}}] \cdot |\widehat{r}_{i\setminus k} - \widehat{r}_{k\setminus i}| \\
c_{i,|V^H|+1} &:= [(\delta_{\ell_V^G(u_i) \neq \ell_V^H(v_k)} + K)c_V^{\text{del}} + Kc_E^{\text{del}}] \cdot |\mathcal{W}_i^G| \\
c_{|V^G|+1,k} &:= [(\delta_{\ell_V^G(u_i) \neq \ell_V^H(v_k)} + K)c_V^{\text{del}} + Kc_E^{\text{del}}] \cdot |\mathcal{W}_k^H|
\end{aligned}$$

for all $(i,k) \in [|V^G|] \times [|V^H|]$.

WALKS requires $O((|V^G||V^H|)^\omega)$ time for computing its LSAPE instance $\mathbf{C}$, where $O(n^\omega)$ is the complexity of multiplying two matrices with $n$ rows and $n$ columns. The asymptotically fastest matrix multiplication algorithms achieve $\omega < 2.38$ [65]; the fastest practically useful matrix multiplication algorithm runs in $O(n^{\log_2(7)}) \approx O(n^{2.81})$ time [105]. WALKS only computes an upper bound for GED.

#### 6.1.1.7 The Extension CENTRALITIES

Assume that an LSAPE instance $\mathbf{C} \in \mathbb{R}^{(|V^G|+1) \times (|V^H|+1)}$ has been constructed by one of the instantiations of LSAPE-GED presented above. In [40, 88], it is suggested to define a node centrality measure $\phi$ that maps central nodes to large and non-central nodes to small non-negative reals. Suggested centrality measures are, for instance, the degrees, the eigenvector centralities [17], and the pagerank centralities [27] of the nodes of the input graphs.

With the help of $\phi$, the upper bound for GED induced by $\mathbf{C}$ can be improved. To this purpose, a second LSAPE instance $\mathbf{C}' \in \mathbb{R}^{(|V^G|+1) \times (|V^H|+1)}$ is constructed by setting

$$c'_{i,k} := (1-\gamma) \cdot c_{i,k} + \gamma \cdot |\phi(u_i) - \phi(v_k)|$$
$$c'_{i,|V^H|+1} := (1-\gamma) \cdot c_{i,|V^H|+1} + \gamma \cdot \phi(u_i)$$
$$c'_{|V^G|+1,k} := (1-\gamma) \cdot c_{|V^G|+1,k} + \gamma \cdot \phi(v_k)$$

for all $(i,k) \in [|V^G|] \times [|V^H|]$, where $0 \leq \gamma \leq 1$ is a meta-parameter. Subsequently, two cheap or optimal LSAPE solutions $\pi, \pi' \in \Pi(|V^G|, |V^H|)$ for $\mathbf{C}$ and $\mathbf{C}'$ are computed, and the returned upper bound for GED is improved from $UB := c(P_\pi)$ to $UB := \min\{c(P_\pi), c(P_{\pi'})\}$ (cf. line 3 of Figure 6.1).[1]

### 6.1.2 LP Based Heuristics

Heuristics that use linear programs (LP) for upper and lower bounding GED proceed as described in Figure 6.2: In a first step, the quadratic program shown in Figure 5.4 is linearized to obtain a (mixed) integer linear programming (MIP) formulation $\widehat{F}$ (line 1). This linearization phase is where different instantiations of LP-GED vary from each other; existing linearizations are described in Section 5.1.4 and Section 5.4 above. Next, an LP $F$ is defined as the continuous relaxation of $\widehat{F}$, $F$ is solved, and the lower bound $LB$ is set to the cost of the optimal solution of $F$.

---

[1] In the original publications, this technique is suggested for the LSAPE instance produced by BP (cf. Section 6.1.1.2). It can, however, be employed in combination with the LSAPE instances produced by any instantiation of LSAPE-GED.

**Input**: Graphs $G$ and $H$, node edit costs $c_V$, edge edit costs $c_E$.
**Output**: An upper bound $UB$ and a lower bound $LB$ for $\text{GED}(G, H)$, as well as a node map $\pi^\star \in \Pi(G, H)$ with $c(P_{\pi^\star}) = UB$.

1. construct linearization $\widehat{F}$ of the quadratic programming formulation detailed in Figure 5.4;
2. relax all integrality constraints of $\widehat{F}$ to obtain LP $F$;
3. solve $F$ and set $LB$ to the obtained minimum;
4. use optimal continuous solution to construct projection problem $\mathbf{C} \in [0,1]^{(|V^G|+1)\times(|V^H|+1)}$;
5. compute $\pi^\star \in \arg\min_{\pi \in \Pi(|V^G|,|V^H|)} \mathbf{C}(\pi)$;
6. set $UB := c(P_{\pi^\star})$;
7. **return** $LB$, $UB$, and $\pi^\star$;

**Figure 6.2.** The paradigm `LP-GED`.

In the literature, LP based heuristics for GED are usually described as algorithms that only yield lower bounds. However, they can straightforwardly be extended to also compute node maps with their induced upper bounds. To that purpose, after solving the LP $F$, an LSAPE instance $\mathbf{C} \in \mathbb{R}^{(|V^G|+1)\times(|V^H|+1)}$ is constructed, whose optimal solutions $\pi^\star \in \arg\min_{\pi \in \Pi(|V^G|,|V^H|)} \mathbf{C}(\pi)$ can be viewed as projections of the previously computed optimal and possibly continuous solution for $F$ to the discrete domain (line 4). Subsequently, an optimal solution $\pi^\star$ for $\mathbf{C}$ is computed (line 5), the upper bound $UB$ is set to its induced edit cost (line 6), and $LB$, $UB$, and $\pi^\star$ are returned (line 7).

In theory, the LP $F$ can be solved in $O(\text{var}(F)^{3.5}\,\text{enc}(F))$ time, where $\text{var}(F)$ is the number of variables contained in $F$ and $\text{enc}(F)$ is the number of bits needed to encode $F$ [61]. However, popular LP solvers such as IBM CPLEX or Gurobi Optimization often use asymptotically slower algorithms that perform better in practice. Solving the projection problem $\mathbf{C}$ requires $O(\min\{|V^G|,|V^H|\}^2 \max\{|V^G|,|V^H|\})$ time.

### 6.1.3 Local Search Based Heuristics

Figure 6.3 shows how to compute node maps and their induced upper bounds for GED via variants of local search. In a first step, an initial node map $\pi \in \Pi(G, H)$ is generated randomly or constructed, for instance, by calling one of the instantiations of `LSAPE-GED` presented above (line 1). Subsequently, a variant of local search is run which, starting at $\pi$, produces an improved

**Input**: Graphs $G$ and $H$, node edit costs $c_V$, edge edit costs $c_E$.
**Output**: An upper bound $UB$ for $\text{GED}(G, H)$, as well as a node map
$\pi' \in \Pi(G, H)$ with $c(P_{\pi'}) = UB$.
1 compute or randomly construct initial node map $\pi \in \Pi(G, H)$;
2 use information encoded in $G$, $H$, $c_V$, and $c_E$ to construct node map
$\pi' \in \Pi(G, H)$ with $c(P_{\pi'}) \leq c(P_\pi)$ via local search starting at $\pi$;
3 set upper bound to $UB := c(P_{\pi'})$;
4 **return** $UB$ and $\pi'$;

**Figure 6.3.** The paradigm LS-GED.

node map $\pi' \in \Pi(G, H)$ with $c(P_{\pi'}) \leq c(P_\pi)$ (line 2). This refinement phase is where different instantiations of LS-GED vary from each other. Once $\pi'$ has been computed, $UB := c(P_{\pi'})$ and $\pi'$ are returned (lines 3 to 4).

#### 6.1.3.1 The Algorithm REFINE

Given an initial node map $\pi \in \Pi(G, H)$, the algorithm REFINE [111] proceeds as follows: Let $((u_s, v_s))_{s=1}^{|\pi|}$ be an arbitrary ordering of the node assignments contained in $\pi$ and let $G_\pi := (V_\pi^G \cup V_\pi^H, A_\pi)$ be an auxiliary directed bipartite graph, where $V_\pi^G := \{u_s \mid s \in [|\pi| + 1]\}$, $V_\pi^H := \{v_s \mid s \in [|\pi|]\}$, and $A_\pi := \pi \cup \{(v_s, u_{s'}) \mid (s, s') \in [|\pi|] \times [|\pi|] \wedge s \neq s'\}$. In other words, $G_\pi$ contains a forward arc for each assignment contained in $\pi$ and backward arcs between nodes in $V_\pi^G$ and $V_\pi^H$ that are not assigned to each other by $\pi$. A directed cycle $C \subseteq A_\pi$ in $G_\pi$ with $|C| = 4$ is called 2-swap: If, in the node map $\pi$, the two node assignments that correspond to $C$'s forward arcs are replaced by those that correspond to $C$'s backward arcs, there are exactly two nodes from $V_\pi^G$ and exactly two nodes from $V_\pi^H$ whose assignments are swapped. Since there is a one-to-one correspondence between 2-swaps and two-element subsets of forward arcs contained in $G_\pi$, there are exactly $\binom{|\pi|}{2} = O((|V^G| + |V^H|)^2)$ 2-swaps.

For each 2-swap $C = \{(u_s, v_s), (v_s, u_{s'}), (u_{s'}, v_{s'}), (v_{s'}, u_s)\}$, REFINE checks if the swapped node map $\pi' := (\pi \setminus \{(u_s, v_s), (u_{s'}, v_{s'})\}) \cup \{(u_s, v_{s'}), (u_{s'}, v_s)\}$ induces a smaller upper bound than $\pi$. If, at the end of the for-loop, a node map $\pi'$ has been found that improves the upper bound, $\pi$ is updated to the node map that yields the largest improvement and the process iterates. Otherwise, the output node map $\pi'$ is set to $\pi$ and REFINE terminates.

For checking if a 2-swap $C$ improves the induced upper bound, it suffices to consider the edges that are incident with the nodes involved in the swap (cf. Section 6.6 below). Therefore, one iteration of `REFINE` runs in $O((|V^G| + |V^H|)^2 \Delta_{\max}^{G,H})$ time, where $\Delta_{\max}^{G,H} := \max\{\max \deg(G), \max \deg(H)\}$. Since the induced upper bound improves in each iteration, this gives an overall runtime complexity of $O(UB(|V^G| + |V^H|)^2 \Delta_{\max}^{G,H})$ for integral edit costs, where $UB$ is the initial upper bound.

#### 6.1.3.2 The Algorithm `BP-BEAM`

Given an initial node map $\pi \in \Pi(G, H)$ and a constant $K \in \mathbb{N}_{\geq 1}$, the algorithm `BP-BEAM` [92] starts by producing a random ordering $((u_s, v_s))_{s=1}^{|\pi|}$ of the node assignments contained in $\pi$. `BP-BEAM` now constructs an output node map $\pi'$ with $c(P_{\pi'}) \leq c(P_\pi)$ by partially traversing an implicitly constructed tree $T$ via beam search with beam size $K$. The nodes of $T$ are tuples $(\pi'', c(P_{\pi''}), s)$, where $\pi'' \in \Pi(G, H)$ is an ordered node map, $c(P_{\pi''})$ is its induced edit cost, and $s \in [|\pi|]$ is the depth of the tree node in $T$. Tree nodes $(\pi'', c(P_{\pi''}), s)$ with $s = |\pi|$ are leafs, and the children of an inner node $(\pi'', c(P_{\pi''}), s)$ are $\{(\text{swap}(\pi'', s, s'), c(P_{\text{swap}(\pi'', s, s')}), s + 1) \mid s' \in \{s, \ldots, |\pi|\}\}$. Here, $\text{swap}(\pi'', s, s')$ is the ordered node map obtained from $\pi''$ by swapping the assignments $(u_s, v_s)$ and $(u_{s'}, v_{s'})$, i.e., setting $v_s := v_{s'}$ and $v_{s'} := v_s$.

At initialization, `BP-BEAM` sets the output node map $\pi'$ to the initial node map $\pi$. Furthermore, `BP-BEAM` maintains a priority queue $q$ of tree nodes which is initialized as $q := \{(\pi, c(P_\pi), 1)\}$ and sorted w.r.t. non-decreasing induced edit cost of the contained node maps. As long as $q$ is non-empty, `BP-BEAM` extracts the top node $(\pi'', c(P_{\pi''}), s)$ from $q$ and updates the output node map $\pi'$ to $\pi''$ if $c(P_{\pi''}) < c(P_{\pi'})$. If $s < |\pi|$, i.e., if the extracted tree node is no leaf, `BP-BEAM` adds all of its children to the priority queue $q$ and subsequently discards all but the first $K$ tree nodes contained in $q$. Once $q$ is empty, the cheapest encountered node map $\pi'$ is returned.

By construction of $T$, we know that at most $1 + K(|\pi| - 1) = O(|V^G| + |V^H|)$ tree nodes are extracted from $q$. For each extracted inner node, `BP-BEAM` constructs all children, which requires $O((|V^G| + |V^H|)\Delta_{\max}^{G,H})$ time, and subsequently sort $q$, which requires $O((|V^G| + |V^H|) \log(|V^G| + |V^H|))$ time. `BP-BEAM` hence runs in $O((|V^G| + |V^H|)^2(\Delta_{\max}^{G,H} + \log(|V^G| + |V^H|)))$ time.

#### 6.1.3.3 The Algorithm `IBP-BEAM`

Since the size of the priority queue $q$ is restricted to $K$, which parts of the search tree $T$ are visited by `BP-BEAM` crucially depends on the ordering of the initial node map $\pi$. Therefore, `BP-BEAM` can be improved by considering not one but several initial orderings. The algorithm `IBP-BEAM` suggested in [44] does exactly this. That is, given a constant number of iterations $I \in \mathbb{N}_{\geq 1}$, `IBP-BEAM` runs `BP-BEAM` with $I$ different randomly created orderings of the initial node map $\pi$ and then returns the cheapest node map $\pi'$ encountered in one of the iterations. Therefore, `IBP-BEAM` runs in $O(I(|V^G|+|V^H|)^2(\Delta_{\max}^{G,H} + \log(|V^G|+|V^H|)))$ time.

#### 6.1.3.4 The Algorithm `IPFP`

The algorithm `IPFP` [67] can be seen as an adaptation of the seminal Frank-Wolfe algorithm [48] to cases where an integer solution is required.

Its adaptation to the case of GED, first suggested in [25], implicitly constructs a matrix $\mathbf{D} \in \mathbb{R}^{((|V^G|+1)\cdot(|V^H|+1))\times((|V^G|+1)\cdot(|V^H|+1))}$ such that $\min_{\mathbf{X}\in\Pi(G,H)} \text{vec}(\mathbf{X})^\top \mathbf{D}\,\text{vec}(\mathbf{X}) = \text{GED}$. Several ways to construct such a matrix $\mathbf{D}$ are discussed in depth in Section 3.1.2 and Section 3.3 above.

Let the cost function $Q$ be defined as $Q(\mathbf{X}, \mathbf{D}) := \text{vec}(\mathbf{X})^\top \mathbf{D}\,\text{vec}(\mathbf{X})$. Starting from an initial node map $\mathbf{X}_0 \in \Pi(G, H)$ with induced upper bound $UB := Q(\mathbf{X}_0, \mathbf{D})$, the algorithm converges to a, possibly fractional, local minimum for GED by repeating the five following steps:

1. Populate LSAPE instance $\mathbf{C}_k := \mathbf{D}\,\text{vec}(\mathbf{X}_k)$.
2. Compute $\mathbf{B}_{k+1} \in \arg\min_{\mathbf{B}\in\Pi(G,H)} \mathbf{C}_k(\mathbf{B})$.
3. Set $UB := \min\{UB, Q(\mathbf{B}_{k+1}, \mathbf{D})\}$.
4. Compute $\alpha_{k+1} := \min_{\alpha\in[0,1]} Q(\mathbf{X}_k + \alpha \cdot (\mathbf{B}_{k+1} - \mathbf{X}_k), \mathbf{D})$.
5. Set $\mathbf{X}_{k+1} := \mathbf{X}_k + \alpha_{k+1}(\mathbf{B}_{k+1} - \mathbf{X}_k)$.

The algorithm iterates until $|Q(\mathbf{X}_k, \mathbf{D}) - \mathbf{C}_k(\mathbf{B}_{k+1})|/Q(\mathbf{X}_k, \mathbf{D})$ is smaller than a convergence threshold $\varepsilon$ or a maximal number of iterations $I$ has been reached. Subsequently, the possibly fractional local optimum $\mathbf{X}_{k+1}$ is projected to the closest integral solution $\widehat{\mathbf{X}}$, and the upper bound $UB := \min\{UB, Q(\widehat{\mathbf{X}}, \mathbf{D})\}$ is returned.

Populating the LSAPE instance $\mathbf{C}_k$ in step 1 requires $O(k|V^G||V^H|\max\{|V^G|, |V^H|\})$ time. Solving the LSAPE instance in step 2 requires $O(\min\{|V^G|,|V^H|\}^2 \max\{|V^G|,|V^H|\})$ time. Updating the upper bound in step 3 requires $O(\max\{|V^G|,|V^H|\}^2)$ time. Determining the optimal step width $\alpha_{k+1}$ in step 4 can be done analytically in $O(|V^G||V^H|)$ time. And projecting the final fractional solution $\mathbf{X}_{k+1}$ to the integral solution $\hat{\mathbf{X}}$ requires $O(\min\{|V^G|,|V^H|\}^2 \max\{|V^G|,|V^H|\})$ time. IPFP's overall runtime complexity is hence $O(I^2|V^G||V^H|\max\{|V^G|,|V^H|\})$.

Slightly different versions of IPFP that use LSAP instead of LSAPE as a linear model have been presented in [22] and [12] and in Section 3.1.2 and Section 3.3 above. As already mentioned there, the main advantage of these versions w.r.t. the one presented in this section is that they are easier to implement: Unlike LSAPE, LSAP is a standard combinatorial optimization problem and libraries for solving it are available for all major programming languages. The drawback of the version presented in [22] is that is uses a significantly larger quadratic matrix $\mathbf{D}$, while the drawback of the version presented in [12] and Section 3.3 is that it can be used only for quasimetric edit cost functions.

#### 6.1.3.5 The Extension `MULTI-START`

`MULTI-START` was suggested in [41] as an extension to the IPFP algorithm. However, it can be employed to improve any instantiation of the paradigm `LS-GED`. While instantiations of `LS-GED` compute locally optimal node maps, the quality of the local optimum highly depends on the initialization of the method, which is a general drawback of local search methods. Hence, the `MULTI-START` extension to the framework simply proposes to use $K$ different initial solutions, runs the `LS-GED` framework on each of them (possibly in parallel), and return the best among the $K$ computed local optima.

In order to further reduce the computing time of `MULTI-START` when parallelization is available, it was suggested in [19] to run in parallel more local searches than the number of desired local optima and to stop the whole process when the number of local searches that have converged has reached the number of desired local optima. In this context, the framework runs with two parameters: $K$ is the number of initial solutions, and $0 < \rho \leq 1$ is defined such that $\lceil \rho \cdot K \rceil$ is the number of desired computed local optima.

### 6.1.4 Miscellaneous Heuristics

### 6.1.5 The Algorithm `HED`

Given two input graphs $G$ and $H$, like `BP`, the algorithm `HED` [46] starts by constructing auxiliary LSAPE instances $\mathbf{C}^{i,k} \in \mathbb{R}^{(\deg^G(u_i)+1) \times (\deg^H(v_k)+1)}$ for all $(i,k) \in [|V^G|] \times [|V^H|]$ (cf. Section 6.1.1.2 above for details). Subsequently, `BP` constructs an LSAPE instance $\mathbf{C} \in \mathbb{R}^{(|V^G|+1) \times (|V^H|+1)}$ by setting

$$c_{i,k} := c_V(u_i, v_k) + 0.5 \cdot \mathbf{C}^{i,k}(\pi^{i,k})$$

$$c_{i,|V^H|+1} := c_V(u_i, \epsilon) + 0.5 \cdot \sum_{j=1}^{\deg^G(u_i)} c_E((u_i, u_{i_j}), \epsilon)$$

$$c_{|V^G|+1,k} := c_V(\epsilon, v_k) + 0.5 \cdot \sum_{l=1}^{\deg^H(v_k)} c_E(\epsilon, (v_k, v_{k_l}))$$

for all $(i,k) \in [|V^G|] \times [|V^H|]$, where $(u_{i_j})_{j=1}^{\deg^G(u_i)}$ is an enumeration of $u_i$'s neighborhood $N^G(u_i)$, and $(v_{k_l})_{l=1}^{\deg^H(v_k)}$ is an enumeration of $v_k$'s neighborhood $N^H(v_k)$.

Once $\mathbf{C}$ has been constructed, `HED` computes a lower bound

$$LB := 0.5 \cdot \sum_{i=1}^{|V^G|} \min_{k \in [|V^H|+1]} c_{i,k} + 0.5 \cdot \sum_{k=1}^{|V^H|} \min_{i \in [|V^G|+1]} c_{i,k}$$

for GED by summing the minima of $\mathbf{C}$'s rows and columns. Note that, in general, $LB$ does not correspond to a feasible LSAPE solution, because of which `HED` does not compute node maps and upper bounds for GED. Furthermore, it holds that $LB \leq \text{LSAPE}(\mathbf{C})$, which implies that the lower bound computed by `HED` is never tighter than the lower bound computed by `BRANCH`.

As detailed in Section 6.1.1.2, the LSAPE instance $\mathbf{C}$ can be constructed in $O(|V^G||V^H|\Delta_{\min}^{G,H^2}\Delta_{\max}^{G,H})$ time. This implies that the overall runtime complexity of `HED` is $O(|V^G||V^H|\Delta_{\min}^{G,H^2}\Delta_{\max}^{G,H})$.

### 6.1.6 The Algorithm `SA`

The algorithm `SA` [94] uses simulated annealing to improve the upper bound computed by an instantiation of the paradigm `LSAPE-GED` discussed in Sec-tion 6.1.1 above.² SA is hence similar to the local search based heuristics presented in Section 6.1.3. The difference is that, instead of varying an initial node map, SA varies the processing order for greedily computing a cheap solution for an initially computed LSAPE instance.

Assume w. l. o. g. that $G$ and $H$ are two input graphs with $|V^G| \geq |V^H|$. SA starts by running an instantiation of LSAPE-GED to obtain an initial node map $\pi \in \Pi(G, H)$, an LSAPE instance $\mathbf{C} \in \mathbb{R}^{(|V^G|+1) \times (|V^H|+1)}$, and, possibly, a lower bound $LB$. If the employed LSAPE-GED instantiation does not yield a lower bound, $LB$ can be computed with any other method that produces a lower bound.

Given a maximal number of iterations $N$ and start and end probabilities $p_1$ and $p_N$ with $1 > p_1 \geq p_N > 0$ for accepting an unimproved node map, SA initializes an ordering $\sigma : [|V^G|] \to [|V^G|]$ of the first $|V^G|$ rows of $\mathbf{C}$ by setting $\sigma(i) := i$ for all $i \in [|V^G|]$, computes a cooling factor $a := (\log(p_1)/\log(p_I))^{1/(N-1)}$ such that $p_1^{a^{-(N-1)}} = p_N$, and sets the current acceptance probability to $p := p_1$. Furthermore, SA maintains the best encountered node map $\pi'$ and the current node map $\pi''$ which are initialized as $\pi' := \pi'' := \pi$, as well as a counter $r$ initialized as $r := 0$ which counts the number of consecutive iterations without improvement of the best encountered node map.

As long as the maximal number of iterations $N$ has not been reached and the upper bound $c(P_{\pi'})$ of the best encountered node map is greater than $LB$, SA does the following: First, a candidate row ordering $\sigma'$ is obtained from the current ordering $\sigma$ by setting $\sigma'(1) := \sigma(i)$, $\sigma'(j) := \sigma(j-1)$ for all $j \in [i] \setminus \{1\}$, and $\sigma'(j) := \sigma(j)$ for all $j \in [|V^G|] \setminus [i]$, where $i \in [|V^G|]$ is a randomly selected row of $\mathbf{C}$. Next, a candidate node map $\pi'''$ is computed by greedily assigning the $\sigma'$-ordered rows of $\mathbf{C}$ to the cheapest unassigned columns. If the upper bound induced by $\pi'''$ is cheaper than the upper bound of the current node map $\pi''$, the current node map $\pi''$ and the current ordering $\sigma$ are updated to $\sigma'$ and $\pi'''$, respectively. Otherwise, they are updated with a probability that is proportional to the current acceptance probability $p$ and inversely proportional to the deterioration $c(P_{\pi'''}) - c(P_{\pi''})$ of the induced upper bound.

---

²In [94], SA is presented as a technique for improving the upper bound computed by the LSAPE-GED instantiation BP. Since SA can be used with any instantiation of LSAPE-GED, we here present a more general version.

After updating the current node map $\pi''$ and the current ordering $\sigma$, SA checks if the upper bound induced by $\pi''$ is tighter than the upper bound induced by the best encountered node map $\pi'$. If this is the case, $\pi'$ is updated to $\pi''$ and the number $r$ of consecutive iterations without improvement is reset to 0. Otherwise, $r$ is incremented and the current ordering $\sigma$ is reshuffled randomly with probability $r/N$. Finally, the current acceptance probability $p$ is set to $p_1^{q^{-s}}$, where $s$ is the number of the current iteration, and SA iterates. After exiting the main loop, SA returns the node map $\pi'$ and its induced upper bound $UB := c(P_{\pi'})$.

The dominant operations in one iteration of SA are the greedy computation of the candidate node map $\pi'''$ and the computation of its induced upper bound. One iteration of SA hence runs in $O(|V^G||V^H| + \max\{E^G, E^H\})$ time. This implies that SA's overall runtime complexity is $O(\omega + N \cdot (|V^G||V^H| + \max\{E^G, E^H\}))$, where $O(\omega)$ is the runtime required for computing the initial upper and lower bounds as well as the LSAPE instance $\mathbf{C}$.

### 6.1.7 The Algorithm BRANCH-COMPACT

The algorithm BRANCH-COMPACT [113] computes a lower bound for GED with uniform edit cost functions $c_V$ and $c_E$. Recall that $c_V$ and $c_E$ are uniform if there is a constant $c \in \mathbb{R}_{>0}$ such that $c_V(\alpha, \alpha') = c_E(\beta, \beta') = c$ holds for all node labels $(\alpha, \alpha') \in (\Sigma_V \cup \{\epsilon\}) \times (\Sigma_V \cup \{\epsilon\})$ with $\alpha \neq \alpha'$ and all edge labels $(\beta, \beta') \in (\Sigma_E \cup \{\epsilon\}) \times (\Sigma_E \cup \{\epsilon\})$ with $\beta \neq \beta'$. BRANCH-COMPACT computes neither node maps nor upper bounds.

Given input graphs $G$ and $H$, BRANCH-COMPACT starts by constructing branches $\mathcal{B}_i^G := (\ell_V^G(u_i), \ell_E^G[\![E^G(u_i)]\!])$ and $\mathcal{B}_k^H := (\ell_V^H(v_k), \ell_E^H[\![E^H(v_k)]\!])$ for all $u_i \in V^G$ and all $v_k \in V^H$. Subsequently, BRANCH-COMPACT sorts the branches in non-decreasing lexicographical order, i. e., computes orderings $\sigma^G : [|V^G|] \to [|V^G|]$ and $\sigma^H : [|V^H|] \to [|V^H|]$ such that $\mathcal{B}_{\sigma^G(i)}^G \preceq_L \mathcal{B}_{\sigma^G(i+1)}^G$ holds for all $i \in [|V^G|-1]$ and $\mathcal{B}_{\sigma^H(k)}^H \preceq_L \mathcal{B}_{\sigma^H(k+1)}^H$ holds for all $k \in [|V^H|-1]$.

BRANCH-COMPACT now performs a first parallel linear scan over the sorted sequences of branches $(\mathcal{B}_{\sigma^G(i)}^G)_{i=1}^{|V^G|}$ and $(\mathcal{B}_{\sigma^H(k)}^H)_{k=1}^{|V^H|}$ to delete a maximal number of pairs of branches $(\mathcal{B}_{\sigma^G(i)}^G, \mathcal{B}_{\sigma^H(k)}^H)$ with $\mathcal{B}_{\sigma^G(i)}^G = \mathcal{B}_{\sigma^H(k)}^H$. Subsequently, BRANCH-COMPACT initializes its lower bound as $LB := 0$ and performs a second parallel linear scan over the remaining branches. In this scan, a maximal number of pairs of branches $(\mathcal{B}_{\sigma^G(i)}^G, \mathcal{B}_{\sigma^H(k)}^H)$ with $\ell_V^G(u_{\sigma^G(i)}) = \ell_V^H(v_{\sigma^H(k)})$ is deleted and $LB$ is incremented by $c/2$ for each deleted pair of branches.

Finally, *LB* is set to $LB := LB + c(\max\{|V^G|, |V^H|\} - D)$, where $D$ is the number of pairs of branches that have been deleted during the two scans.

Once the branches of the input graphs have been sorted, BRANCH-COMPACT requires $O(\max\{|V^G|, |V^H|\})$ time for computing its lower bound. For sorting the branches, BRANCH-COMPACT first has to sort the edge label multisets $\ell_E^G[\![E^G(u_i)]\!]$ and $\ell_E^H[\![E^H(v_k)]\!]$ for all $u_i \in V^G$ and all $v_k \in V^H$. This requires $O(\max\{|V^G|, |V^H|\} \Delta_{\max}^{G,H} \log(\Delta_{\max}^{G,H}))$ time. BRANCH-COMPACT's overall runtime complexity is hence $O(\max\{|V^G|, |V^H|\}(\Delta_{\max}^{G,H} \log(\Delta_{\max}^{G,H}) + \log(\max\{|V^G|, |V^H|\})))$.

### 6.1.8 The Algorithm PARTITION

Like BRANCH-COMPACT, the algorithm PARTITION [113] computes a lower bound for GED with uniform edit costs. Given input graphs $G$ and $H$ and a constant $K \in \mathbb{R}_{\geq 1}$, PARTITION starts by initializing a collection $\mathcal{S} := \emptyset$ of $K'$-sized substructures of $G$ that are not subgraph-isomorphic to $H$, where $K' \in [K]$ and a $K'$-sized substructure of $G$ is a connected subgraph of $G$ that is composed of $K'$ elements (nodes or edges). For instance, 1-sized substructures are single nodes or edges, 2-sized substructures are nodes together with an incident edge, and 3-sized substructures are nodes together with two incident edges or edges together with their terminal nodes.

Starting with $K' := 1$, PARTITION now consecutively checks for each $K'$-sized substructure $S^G \subseteq G$ of $G$ if there is a $K'$-sized substructure of $H$ which is isomorphic to $S^G$. If this is not the case $S^G$ is added to $\mathcal{S}$ and deleted from $G$. Once all $K'$-sized substructure have been considered, $K'$ is incremented and the process iterates if $K' \leq K$. Otherwise, PARTITION returns the lower bound $LB := c|\mathcal{S}|$.

Since $G$ and $H$ have, respectively, $O(|E^G|)$ and $O(|E^H|)$ substructures of sizes 1, 2, and 3, PARTITION with $K \leq 3$ runs in $O(|E^G||E^H|)$ time. Determining non-isomorphic substructures of size $K > 3$ cannot be done naively but requires to call subgraph isomorphism verification algorithms such as the one proposed in [37]. In the worst case, these algorithms require super-polynomial time (e.g. $O(|V^G|!|V^G|)$ for the algorithm proposed in [37]), but are usually fast in practice for small $K$.

### 6.1.9 The Algorithm HYBRID

The algorithm HYBRID [113] improves the lower bounds of the algorithms BRANCH-CONST and PARTITION presented in Section 6.1.1.4 and Section 6.1.8. Given input graphs $G$ and $H$ and a constant $K \in \mathbb{R}_{\geq 1}$, HYBRID first runs PARTITION with the maximal size of the considered substructures set to $K$ to obtain a collection $\mathcal{S}$ of substructures $S^G \subseteq G$ of $G$ that are not subgraph-isomorphic to $H$.

Let $\mathcal{C}(\mathcal{S}) := \times_{S^G \in \mathcal{S}} S^G$ be the set of all configurations of nodes or edges that appear in the non-isomorphic substructures. For each configuration $a := (a_s)_{s=1}^{|\mathcal{S}|} \in \mathcal{C}(\mathcal{S})$, HYBRID creates a modified graph $G_a$, where all nodes or edges $a_s$ contained in the configuration $a$ are labeled with a special wildcard label $\gamma$. Subsequently, HYBRID runs a variant of BRANCH-CONST on the graphs $G_a$ and $H$, which edits $\gamma$-labeled nodes and edges for free. Finally, HYBRID returns the lower bound $LB := |\mathcal{S}| + \min\{LB_a \mid a \in \mathcal{C}(\mathcal{S})\}$, where $LB_a$ denotes the lower bound returned by the wildcard version of BRANCH-CONST if run on the graphs $G_a$ and $H$. This lower bound is guaranteed to be at least as tight as the lower bounds computed by PARTITION and BRANCH-CONST.

Let $O(\omega_1)$ be the runtime complexity of PARTITION with the maximal size of the considered substructures set to $K$ and $O(\omega_2)$ be the runtime complexity of BRANCH-CONST. Then HYBRID runs in $O(\omega_1 + \omega_2|\mathcal{C}(\mathcal{S})|)$ time. Note that $|\mathcal{C}(\mathcal{S})|$ can get huge. For instance, assume that PARTITION manages to completely partition $G$ into non-isomorphic substructures of size 2. Then it holds that $|\mathcal{C}(\mathcal{S})| = \prod_{S^G \in \mathcal{S}} |S^G| = 2^{|V^G|}$. HYBRID's runtime complexity is hence $\Omega(\omega_1 + \omega_2 2^{|V^G|})$ and thus not polynomially bounded.

## 6.2 Two New LSAPE Based Lower and Upper Bounds

In this section, we present the algorithms BRANCH (Section 6.2.1) and BRANCH-FAST (Section 6.2.2). Both algorithms instantiate the paradigm LSAPE-GED and compute both lower and upper bounds for GED. BRANCH and BRANCH-FAST can both be viewed as generalizations of BRANCH-CONST to arbitrary edit costs. The lower bound computed by BRANCH is always at least as tight as the lower bound produced by BRANCH-FAST; and both lower bounds are pseudo-metrics if the underlying edit cost functions satisfy certain constraints.

### 6.2.1 The Algorithm BRANCH

BRANCH is very similar to the algorithm BP presented in Section 6.1.1.2, but also computes a lower and not only an upper bound. In fact, in [93], it is claimed that BP can be extended to also compute a lower bound. Personal communication with the authors of [93] established that their intended extension is equivalent with BRANCH as presented in this section. However, the presentation provided in [93] is incorrect: if the lower bound is computed as detailed in [93], it can happen that it exceeds the exact GED. For an example where this happens, cf. [13, 15]. Since the presentation given in [93] does not mirror the authors' intentions, we here do not restate this counterexample.

Like BP, for each $(i,k) \in [|V^G|] \times [|V^H|]$, BRANCH constructs an auxiliary LSAPE instance $\mathbf{C}^{i,k} \in \mathbb{R}^{(\deg^G(u_i)+1) \times (\deg^H(v_k)+1)}$ by setting

$$c^{i,k}_{j,l} := c_E((u_i, u_{i_j}), (v_k, v_{k_l}))$$
$$c^{i,k}_{j,\deg^H(v^k)+1} := c_E((u_i, u_{i_j}), \epsilon)$$
$$c^{i,k}_{\deg^G(u_i),l} := c_E(\epsilon, (v_k, v_{k_l}))$$

for all $(j,l) \in [\deg^G(u_i)] \times [\deg^H(v_k)]$, where $(u_{i_j})_{j=1}^{\deg^G(u_i)}$ and $(v_{k_l})_{l=1}^{\deg^H(v_k)}$ are enumerations of $u_i$'s neighborhood $N^G(u_i)$ and $v_k$'s neighborhood $N^H(v_k)$, respectively. Subsequently, BRANCH optimally solves the LSAPE instances $\mathbf{C}^{i,k}$ for all $(i,k) \in [|V^G|] \times [|V^H|]$; and constructs its final LSAPE instance $\mathbf{C}$ by setting

$$c_{i,k} := c_V(u_i, v_k) + 0.5 \cdot \text{LSAPE}(\mathbf{C}^{i,k})$$
$$c_{i,|V^H|+1} := c_V(u_i, \epsilon) + 0.5 \cdot \sum_{j=1}^{\deg^G(u_i)} c_E((u_i, u_{i_j}), \epsilon)$$
$$c_{|V^G|+1,k} := c_V(\epsilon, v_k) + 0.5 \cdot \sum_{l=1}^{\deg^H(v_k)} c_E(\epsilon, (v_k, v_{k_l}))$$

for all $(i,k) \in [|V^G|] \times [|V^H|]$.

The only difference between BP and BRANCH is hence that BRANCH includes the factor 0.5 in the construction of $\mathbf{C}$. Therefore, like BP, BRANCH requires $O(|V^G||V^H|\Delta_{\min}^{G,H^2}\Delta_{\max}^{G,H})$ time for constructing $\mathbf{C}$. The following Theorem 6.1 shows that, if $\mathbf{C}$ is constructed as done by BRANCH, inequality (6.1) with $\xi := 1$ holds for each input, and hence implies that BRANCH computes a lower bound for GED.

**Theorem 6.1 (Correctness of BRANCH).** *If the LSAPE instance $\mathbf{C} \in \mathbb{R}^{(|V^G|+1) \times (|V^H|+1)}$ is constructed as done by BRANCH, then $\mathrm{LSAPE}(\mathbf{C}) \leq \mathrm{GED}(G, H)$ holds for all graphs G and H and all edit cost functions $c_V$ and $c_E$.*

*Proof.* Let $\pi \in \Pi(G, H)$ be a node map with induced edit cost $c(P_\pi) = \mathrm{GED}(G, H)$. Furthermore, let $SUB(\pi) := \{(u_i, v_k) \in V^G \times V^H \mid \pi(u_i) = v_k\}$, $DEL(\pi) := \{u_i \in V^G \mid \pi(u_i) = \epsilon\}$, and $INS(\pi) := \{v_k \in V^H \mid \pi^{-1}(v_k) = \epsilon\}$. For all $(u_i, v_k) \in SUB(\pi)$, we define

$$c_E^{\mathrm{sub}}(u_i, v_k) := \sum_{\substack{u_j \in N^G(u_i) \\ \pi(u_j) \in N^H(v_k)}} c_E((u_i, u_j), (v_k, \pi(u_j))) \\ + \sum_{\substack{u_j \in N^G(u_i) \\ \pi(u_j) \notin N^H(v_k)}} c_E((u_i, u_j), \epsilon) + \sum_{\substack{v_l \in N^H(v_k) \\ \pi^{-1}(v_l) \notin N^G(u_i)}} c_E(\epsilon, (v_k, v_l)),$$

and rearrange the terms in equation (2.1) to write $c(P_\pi)$ as follows:

$$c(P_\pi) = \sum_{(u_i, v_k) \in SUB(\pi)} \underbrace{c_V(u_i, v_k) + 0.5 \cdot c_E^{\mathrm{sub}}(u_i, v_k)}_{=: c^{\mathrm{sub}}(u_i, v_k)} \\ + \sum_{u_i \in DEL(\pi)} \underbrace{c_V(u_i, \epsilon) + 0.5 \cdot \sum_{u_j \in N^G(u_i)} c_E((u_i, u_j), \epsilon)}_{=: c^{\mathrm{del}}(u_i)} \\ + \sum_{v_k \in INS(\pi)} \underbrace{c_V(\epsilon, v_k) + 0.5 \cdot \sum_{v_l \in N^H(v_k)} c_E(\epsilon, (v_k, v_l))}_{=: c^{\mathrm{ins}}(v_k)}$$

By construction of BRANCH's LSAPE instance $\mathbf{C}$, $c^{\mathrm{del}}(u_i) = c_{i, |V^H|+1}$ holds for all $u_i \in DEL(\pi)$, and $c^{\mathrm{ins}}(v_k) = c_{|V^G|+1, k}$ holds for all $v_k \in INS(\pi)$. Hence, the theorem follows if we can show that

$$c^{\mathrm{sub}}(u_i, v_k) \geq c_{i,k} \qquad (6.2)$$

holds for all $(u_i, v_k) \in SUB(\pi)$. Let $(u_i, v_k) \in SUB(\pi)$, and $(u_{i_j})_{j=1}^{\deg^G(u_i)}$ and $(v_{k_l})_{l=1}^{\deg^H(v_k)}$ be enumerations of $N^G(u_i)$ and $N^H(v_k)$, respectively. We define an error-correcting matching $\pi^{i,k} \in \Pi_{\deg^G(u_i), \deg^H(v_k), \epsilon}$ as $\pi^{i,k} := \{(j, l) \in [\deg^G(u_i)] \times [\deg^H(v_k)] \mid \pi(u_{i_j}) = v_{k_l}\} \cup \{(j, \deg^H(v_k) + 1) \in [\deg^G(u_i)] \times \{\deg^H(v_k) + 1\} \mid \pi(u_{i_j}) \notin N^H(v_k)\} \cup \{(\deg^G(u_i) + 1, l) \in \{\deg^G(u_i) + 1\} \times [\deg^H(v_k)] \mid \pi^{-1}(v_{k_l}) \notin N^G(u_i)\}$. Then it holds that

$$c_E^{\mathrm{sub}}(u_i, v_k) = \mathbf{C}^{i,k}(\pi^{i,k}) \geq \mathrm{LSAPE}(\mathbf{C}^{i,k}),$$

which implies equation (6.2) and hence proves the theorem. □

Given graphs $G$ and $H$ and edit cost functions $c_V$ and $c_E$, let the branch edit distance $\text{BED}(G, H)$ be defined as the lower bound for $\text{GED}(G, H)$ computed by `BRANCH`. Instead of viewing BED as a proxy for GED, it can be viewed as computationally tractable graph distance measures. If regarded in this way, it is highly desirable that BED indeed behaves like a distance: Given a domain $\mathbb{G}$ of graphs on label sets $\Sigma_V$ and $\Sigma_E$ and metric edit cost functions $c_V$ and $c_E$, we ideally want BED to be a metric. Unfortunately, this desideration is unrealistic, because the existence of a polynomially computable mapping $\chi$ that satisfies $\chi(G, H) = 0$ if and only if $G \simeq H$ implies that the problem of deciding if two graphs are isomorphic is in $P$. Whether this is the case, is still open [76]. The best we can hope for is thus that BED is a pseudo-metric, i.e., that is is non-negative, symmetric, fulfills the triangle inequality, and respects the constraint $\text{BED}(G, H) = 0$ for all $G$ and $H$ with $G \simeq H$. The following Proposition 6.1 shows that this is indeed the case.

**Proposition 6.1 (Pseudo-Metricity of `BRANCH`'s Lower Bound).** *For metric edit cost functions $c_V$ and $c_E$, BED is a pseudo-metric on the domain $\mathbb{G}$ of graphs on label sets $\Sigma_V$ and $\Sigma_E$.*

*Proof.* Since $c_V$ and $c_E$ are metrics, the LSAPE instance $\mathbf{C}$ employed by `BRANCH` is symmetric and non-negative. This implies symmetricity and non-negativity for BED. For checking the constraint $\text{BED}(G, H) = 0$ for all $G, H \in \mathbb{G}$ with $G \simeq H$, consider two isomorphic graphs $G$ and $H$ and a node map $\pi \in \Pi(G, H)$ with $c(P_\pi) = 0$. By construction of $\mathbf{C}$, it holds that $\mathbf{C}(\pi) = 0$ and thus $\text{BED}(G, H) = 0$.

It remains to check that BED satisfies the triangle inequality $\text{BED}(G, F) \leq \text{BED}(G, H) + \text{BED}(H, F)$ for all graphs $G, H, F \in \mathbb{G}$. Let $\mathbf{C}^{G,H}$, $\mathbf{C}^{H,F}$, and $\mathbf{C}^{G,F}$ be the LSAPE instances constructed by `BRANCH` to compute $\text{BED}(G, H)$, $\text{BED}(H, F)$ and $\text{BED}(G, F)$, respectively. Furthermore, let $\pi_{G,H} \in \Pi_{|V^G|,|V^H|,\epsilon}$ and $\pi_{H,F} \in \Pi_{|V^H|,|V^F|,\epsilon}$ be optimal error-correcting matchings for $\mathbf{C}^{G,H}$ and $\mathbf{C}^{H,F}$, respectively. We define an error-correcting matching $\pi_{G,F} \in \Pi_{|V^G|,|V^F|,\epsilon}$ as follows:

$$\begin{aligned} \pi^{G,F} := & \{(\pi_{G,H}^{-1}(k), \pi_{H,F}(k)) \mid k \in [|V^H|]\} \\ & \cup \{(i, |V^F| + 1) \mid i \in \pi_{G,H}^{-1}[\{|V^H| + 1\}]\} \\ & \cup \{(|V^G| + 1, r) \mid r \in \pi_{H,F}[\{|V^H| + 1\}]\} \end{aligned}$$

By construction of $\pi_{G,F}$, we have

$$\begin{aligned}
\mathbf{C}^{G,F}(\pi_{G,F}) &= \sum_{k\in[|V^H|]} c^{G,F}_{\pi^{-1}_{G,H}(k),\pi_{H,F}(k)} \\
&\quad + \sum_{i\in\pi^{-1}_{G,H}[\{|V^H|+1\}]} c^{G,F}_{i,|V^F|+1} + \sum_{r\in\pi_{H,F}[\{|V^H|+1\}]} c^{G,F}_{|V^G|+1,r} \\
&= \sum_{k\in[|V^H|]} c^{G,F}_{\pi^{-1}_{G,H}(k),\pi_{H,F}(k)} \\
&\quad + \sum_{i\in\pi^{-1}_{G,H}[\{|V^H|+1\}]} c^{G,H}_{i,|V^H|+1} + \sum_{r\in\pi_{H,F}[\{|V^H|+1\}]} c^{H,F}_{|V^H|+1,r} \\
&\leq \sum_{k\in[|V^H|]} c^{G,H}_{\pi^{-1}_{G,H}(k),k} + \sum_{i\in\pi^{-1}_{G,H}[\{|V^H|+1\}]} c^{G,H}_{i,|V^H|+1} \\
&\quad + \sum_{k\in[|V^H|]} c^{H,F}_{k,\pi_{H,F}(k)} + \sum_{r\in\pi_{H,F}[\{|V^H|+1\}]} c^{H,F}_{|V^H|+1,r} \\
&= \mathbf{C}^{G,H}(\pi_{G,H}) + \mathbf{C}^{H,F}(\pi_{H,F}) = \mathrm{BED}(G,H) + \mathrm{BED}(H,F),
\end{aligned}$$

where the second equality follows from the definitions of $\mathbf{C}^{G,H}$, $\mathbf{C}^{H,F}$, and $\mathbf{C}^{G,F}$, the inequality follows from the metricity of $c_V$ and $c_E$, and the last equality follows from the optimality of $\pi_{G,H} \in \Pi_{|V^G|,|V^H|,\epsilon}$ and $\pi_{H,F} \in \Pi_{|V^H|,|V^F|,\epsilon}$. Since $\mathrm{BED}(G,F) \leq \mathbf{C}^{G,F}(\pi_{G,F})$, this implies that BED satisfies the triangle inequality and hence finishes the proof of the proposition. □

### 6.2.2  The Algorithm BRANCH-FAST

The algorithm BRANCH-FAST speeds-up BRANCH at the cost of producing a looser lower bound. For all $(i,k) \in [|V^G|] \times [|V^H|]$, BRANCH-FAST computes the minimal deletion cost $c^i_{\min} := \min\{c_E(e,\epsilon) \mid e \in E^G(u_i)\}$, the minimal insertion cost $c^k_{\min} := \min\{c_E(\epsilon,f) \mid f \in E^H(v_k)\}$, as well as the minimal substitution cost $c^{i,k}_{\min} := \min\{c_E(e,f) \mid (e,f) \in E^G(u_i) \times E^H(v_k) \wedge \ell^G_E(e) \neq \ell^H_E(f)\}$ for the sets $E^G(u_i)$ and $E^H(v_k)$ of edges that are incident to $u_i$ in $G$ and to $v_k$ in $H$, respectively. With these ingredients, BRANCH-FAST constructs its LSAPE instance $\mathbf{C}$ by setting

$$c_{i,k} := c_V(u_i,v_k) + 0.5 \cdot \Gamma(\ell^G_E[\![E^G(u_i)]\!], \ell^H_E[\![E^H(v_k)]\!], c^{i,k}_{\min}, c^i_{\min}, c^k_{\min})$$
$$c_{i,|V^H|+1} := c_V(u_i,\epsilon) + 0.5 \cdot \deg^G(u_i) c^i_{\min}$$
$$c_{|V^G|+1,k} := c_V(\epsilon,v_k) + 0.5 \cdot \deg^H(v_k) c^k_{\min}$$

for all $(i,k) \in [|V^G|] \times [|V^H|]$.

By sorting all sets of incident edge labels before populating $\mathbf{C}$, BRANCH-FAST can reduce the time complexity of constructing $\mathbf{C}$ to $O(\max\{|V^G|, |V^H|\}\Delta_{\max}^{G,H}\log(\Delta_{\max}^{G,H}) + |V^G||V^H|\Delta_{\min}^{G,H}\Delta_{\max}^{G,H})$. The following Theorem 6.2 shows that BRANCH-FAST computes a lower bound for GED that is never tighter than the lower bound produced by BRANCH. By Theorem 6.1, this means that inequality (6.1) with $\xi :\equiv 1$ holds for each input, which, in turn, implies that BRANCH-FAST returns both an upper and a lower bound for GED.

**Theorem 6.2 (Correctness of BRANCH-FAST).** *If the LSAPE instances $\mathbf{C} \in \mathbb{R}^{(|V^G|+1)\times(|V^H|+1)}$ and $\mathbf{C}' \in \mathbb{R}^{(|V^G|+1)\times(|V^H|+1)}$ are constructed as done by BRANCH-FAST and BRANCH, respectively, then $\mathrm{LSAPE}(\mathbf{C}) \leq \mathrm{LSAPE}(\mathbf{C}')$ holds for all graphs G and H and all edit cost functions $c_V$ and $c_E$.*

*Proof.* For proving the theorem, we show that the inequalities

$$c_{i,k} \leq c'_{i,k} \tag{6.3}$$
$$c_{i,|V^H|+1} \leq c'_{i,|V^H|+1} \tag{6.4}$$
$$c_{|V^G|+1,k} \leq c'_{|V^G|+1,k} \tag{6.5}$$

hold for all $(i,k) \in [|V^G|] \times [|V^H|]$.

The inequalities (6.4) and (6.5) immediately follow from the definitions of $\mathbf{C}$ and $\mathbf{C}'$. To show that inequality (6.3) holds, we fix $(i,k) \in [|V^G|] \times [|V^H|]$ and prove the inequality

$$\Gamma(\ell_E^G[\![E^G(u_i)]\!], \ell_E^H[\![E^H(v_k)]\!], c_{\min}^{i,k}, c_{\min}^i, c_{\min}^k) \leq \mathrm{LSAPE}(\mathbf{C}^{i,k}), \tag{6.6}$$

where $\mathbf{C}^{i,k} \in \mathbb{R}^{(\deg^G(u_i)+1)\times(\deg^H(v_k)+1)}$ is the auxiliary LSAPE instance employed by BRANCH to populate the cell $c'_{i,k}$. Recall that, by the proof of Theorem 3.4, we can assume w. l. o. g. that the edge edit costs $c_E$ are triangular. Therefore, Proposition 4.2 implies that an an optimal error-correcting matching for $\mathbf{C}^{i,k}$ contains exactly $\min\{|\ell_E^G[\![E^G(u_i)]\!]|, |\ell_E^H[\![E^H(v_k)]\!]|\} - |\ell_E^G[\![E^G(u_i)]\!] \cap \ell_E^H[\![E^H(v_k)]\!]|$ edge substitutions, exactly $\max\{|\ell_E^G[\![E^G(u_i)]\!]| - |\ell_E^H[\![E^H(v_k)]\!]|, 0\}$ edge deletions, and exactly $\max\{|\ell_E^H[\![E^H(v_k)]\!]| - |\ell_E^G[\![E^G(u_i)]\!]|, 0\}$ edge insertions. All edge substitutions, deletions, and insertions have costs not larger than $c_{\min}^{i,k}$, $c_{\min}^i$, and $c_{\min}^k$, respectively. Plugging these observations into Definition 2.16 of the operator $\Gamma$ yields the desired inequality (6.6). □

Proposition 6.2 below shows that BRANCH-FAST, BRANCH, and BRANCH-CONST are equivalent for constant, triangular edge edit cost functions $c_E$. By Proposition 6.1, this immediately implies the following corollary.

**Corollary 6.1 (Pseudo-Metricity of BRANCH-FAST's and BRANCH-CONST's Lower Bounds).** *For metric node edit cost functions $c_V$ and metric, constant edge edit cost functions $c_E$, the lower bounds computed by BRANCH and BRANCH-CONST are pseudo-metrics on the domain $\mathbb{G}$ of graphs on label sets $\Sigma_V$ and $\Sigma_E$.*

**Proposition 6.2 (Equivalence of BRANCH, BRANCH-FAST, and BRANCH-CONST).** *For constant, triangular edge edit cost functions $c_E$, the lower bounds computed by BRANCH, BRANCH-FAST, and BRANCH-CONST coincide for all graphs $G, H \in \mathbb{G}$.*

*Proof.* The proposition follows from the arguments provided in the proof of Theorem 6.1 above and the fact that, if the edge edit cost function $c_E$ is constant, we have $c_{\min}^{i,k} = c_E^{\text{sub}}$, $c_{\min}^i = c_E^{\text{del}}$, and $c_{\min}^k = c_E^{\text{ins}}$, for all $(i,k) \in [|V^G|] \times [|V^H|]$. □

## 6.3 An Anytime Algorithm for Tight Lower Bounds

In this section, we present BRANCH-TIGHT, an anytime algorithm that improves the lower bound computed by BRANCH. We first present the algorithm and characterize it in terms of runtime complexity (Section 6.3.1), and then prove its correctness (Section 6.3.2).

### 6.3.1 Presentation and Complexity of BRANCH-TIGHT

Figure 6.4 gives an overview of the algorithm. Given graphs $G$ and $H$, BRANCH-TIGHT starts by enforcing $|V^G| = |V^H| =: N$ (lines 1 to 5). If the edit cost functions $c_V$ and $c_E$ are quasimetric, this is done by adding $\max\{|V^G|, |V^H|\} - |V^G|$ isolated dummy nodes to $G$ and adding $\max\{|V^G|, |V^H|\} - |V^H|$ isolated dummy nodes to $H$. Otherwise, $|V^H|$ isolated dummy nodes are added to $G$ and $|V^G|$ isolated dummy nodes are added to $H$. In the next step, all missing edges in $G$ and $H$ are replaced by dummy edges. By Lemma 6.1 below, both of these preprocessing operations leave $\text{GED}(G, H)$ invariant. Moreover, they ensure that there is an optimal node map between $G$ and $H$ that contains only node and edge substitutions.

After preprocessing the input graphs, BRANCH-TIGHT runs an anytime algorithm that, given a maximal number of iterations $I$, computes lower bounds $(LB_s)_{r=1}^I$ and upper bounds $(UB_s)_{r=1}^I$ for GED such that $LB_1$ equals the lower bound computed by the algorithm BRANCH presented in Section 6.2.1 above and $LB_{r+1} \geq LB_s$ holds for all $r \in [I-1]$ (lines 8 to 15). Once $I$ or a given time

**Input**: Two graphs $G$ and $H$, number of iterations $I \geq 2$, convergence threshold $\varepsilon \geq 0$, time limit $t > 0$.
**Output**: Lower and upper bounds $LB$ and $UB$ for $\mathrm{GED}(G, H)$, as well as a node map $\pi \in \Pi(G, H)$ with $c(P_\pi) = UB$.

1 **if** *edit cost functions $c_V$ and $c_E$ are quasimetric* **then**
2     $N := \max\{|V^G|, |V^H|\}$;
3 **else**
4     $N := |V^G| + |V^H|$;
5 add $N - |V^G|$ and $N - |V^H|$ isolated dummy nodes to $G$ and $H$, respectively;
6 replace all missing edges in $G$ and $H$ by dummy edges;
7 set $I' := I$;
8 **for** $r \in [I]$ **do**
9     construct LSAP instance $\mathbf{C}^r \in \mathbb{R}^{N \times N}$ according to equations (6.7) to (6.8);
10     compute optimal perfect matching $\pi^r \in \Pi_{N,N}$ for $\mathbf{C}^r$;
11     set $LB_r := \mathrm{LSAP}(\mathbf{C}^r)$ and $UB_r := c(P_{\pi^r})$;
12     **if** $r \geq 2$ **then**
13        **if** $(LB_r - LB_{r-1})/LB_{r-1} < \varepsilon$ *or time limit $t$ expired* **then**
14           set $I' := r$;
15           **break**;

16 set $r^\star := \arg\min_{r \in [I']} UB_r$, $\pi := \pi_{r^\star}$, and $UB := c(P_\pi)$;
17 **return** $LB := LB_{I'}$, $UB$, and $\pi$;

**Figure 6.4.** The anytime algorithm `BRANCH-TIGHT`.

limit has been reached or the lower bound has converged, `BRANCH-TIGHT` returns the last lower bound $LB := LB_{I'}$ and the best encountered upper bound along with the corresponding node map, where $I' \leq I$ is the actual number of iterations.

For each $(u_i, v_k) \in V^G \times V^H$ and each iteration $r \in [I]$, `BRANCH-TIGHT` constructs and solves LSAP instances $\mathbf{C}^{i,k,r} \in \mathbb{R}^{(N-1) \times (N-1)}$ defined as

$$c_{j,l}^{i,k,r} := \begin{cases} 0.5 \cdot c_E((u_i, u_j), (v_k, v_l)) & \text{if } r = 1 \\ c_{j,l}^{i,k,r-1} - s_{j,l}^{i,k,r-1} - \frac{s_{i,k}^{r-1}}{N-1} + s_{i,k}^{j,l,r-1} + \frac{s_{j,l}^{r-1}}{N-1} & \text{else} \end{cases} \quad (6.7)$$

for all $(u_j, v_l) \in V^G \setminus \{u_i\} \times V^H \setminus \{v_k\}$. Here, $s_{j,l}^{r,i,k}$ is the slack of the variable $x_{j,l}$ in an optimal LSAP solution of the LSAP instance $\mathbf{C}^{i,k,r}$, and $s_{i,k}^r$ is the slack of the variable $x_{i,k}$ in an optimal solution of the LSAP instance $\mathbf{C}^r \in \mathbb{R}^{N \times N}$, which, in turn, is constructed by setting

$$c_{i,k}^r := c_V(u_i, v_k) + \mathrm{LSAP}(\mathbf{C}^{i,k,r}) \quad (6.8)$$

for all $(u_i, v_k) \in V^G \times V^H$. After constructing $\mathbf{C}^r$, an optimal solution $\pi^r$ for $\mathbf{C}^r$ is computed, $LB_r$ is set to $\mathrm{LSAP}(\mathbf{C}^r)$, and $UB_r$ is set to the cost of the edit path induced by $\pi^r$. Subsequently, $r$ is incremented and the process iterates. Theorem 6.3 below shows that this process indeed yields a sequence $(LB_r)_{r=1}^{I'}$ of lower bounds that satisfies $LB_1 = \text{BED}(G,H)$ and $\text{GED}(G,H) \geq LB_r \geq LB_{r-1}$, for all $r \geq 2$.

BRANCH-TIGHT's overall runtime complexity is $O(N^5)$. Recall that we have $N = \max\{|V^G|, |V^H|\}$, if the edit cost functions are quasimetric, and $N = |V^G| + |V^H|$, otherwise. Moreover, note that BRANCH-TIGHT's runtime complexity does not depend on the input graphs' maximum degrees, because missing edges are replaced by dummy edges. BRANCH-TIGHT hence does not benefit from sparseness, which explains why, on some datasets, BRANCH-TIGHT is slower than methods with a higher runtime complexity (cf. Section 6.8.4).

### 6.3.2 Correctness of BRANCH-TIGHT

For proving the correctness of BRANCH-TIGHT, we start with a lemma which states that the preprocessing carried our by BRANCH-TIGHT in lines 1 to 6 leaves GED invariant and ensures that there are optimal node maps without node deletions and insertions.

**Lemma 6.1.** *Let $G$ and $H$ be two graphs and $G'$ and $H'$ be the graphs obtained from $G$ and $H$ by executing the lines 1 to 6 of Figure 6.4. Then the following statements hold:*

*(i) There is an optimal node map $\pi \in \Pi(G', H')$ that contains no node insertions or deletions, i.e., we have $\text{GED}(G', H') = \min_{\pi \in \Pi_{N,N}} c(P_\pi)$.*

*(ii) It holds that $\text{GED}(G, H) = \text{GED}(G', H')$.*

*Proof.* We start by proving the first part of the lemma. First assume that the edit costs are quasimetric and let $\pi \in \Pi(G', H')$ be an optimal node map with a minimum number of node deletions. We want to show that $\pi$ does not contain any node deletions. Assume that this is not the case. Since $G'$ and $H'$ have the same number of nodes, we then know that there are nodes $u_i \in V^{G'}$ and $v_k \in V^{H'}$ with $(u_i, \epsilon), (\epsilon, v_k) \in \pi$. Let $\pi' \in \Pi(G', H')$ be defined as $\pi' := (\pi \setminus \{(u_i, \epsilon), (\epsilon, v_k)\}) \cup \{(u_i, v_k)\}$. Note that $\pi'$ contains fewer node deletions than $\pi$. Because the edit cost functions are quasimetric, we have $c(P_{\pi'}) \leq c(P_\pi)$. This implies that $\pi'$ is optimal, which contradicts the choice of $\pi$.

Next, we show the first part of the lemma for edit cost functions that are not quasimetric. Let $\pi \in \Pi(G', H')$ be an optimal node map. We will modify $\pi$ in a way that leaves its induced edit cost invariant and eliminates all node deletions and insertions. To this purpose, we partition the nodes of $G'$ and $H'$ as $V^{G'} = V_{\text{sub}}^G \cup V_{\text{del}}^G \cup \mathcal{E}_{\text{sub} \to V^H}^G \cup \mathcal{E}_{\text{sub} \to \mathcal{E}^H}^G \cup \mathcal{E}_{\text{del}}^G$ and $V^{H'} = V_{\text{sub}}^H \cup V_{\text{ins}}^H \cup \mathcal{E}_{V^G \to \text{sub}}^H \cup \mathcal{E}_{\mathcal{E}^G \to \text{sub}}^H \cup \mathcal{E}_{\text{ins}}^H$. $V_{\text{sub}}^G$ and $V_{\text{sub}}^H$ contain the original nodes of $G$ and $H$ that are substituted by $\pi$ (either with an original node or with a dummy node), $V_{\text{del}}^G$ and $V_{\text{ins}}^H$ contain the original nodes of $G$ and $H$ that are deleted and inserted by $\pi$, $\mathcal{E}_{\text{sub} \to V^H}^G$ and $\mathcal{E}_{V^G \to \text{sub}}^H$ contain the dummy nodes contained in $G'$ and $H'$ that $\pi$ substitutes with original nodes, $\mathcal{E}_{\text{sub} \to \mathcal{E}^H}^G$ and $\mathcal{E}_{\mathcal{E}^G \to \text{sub}}^H$ contain the dummy nodes contained in $G'$ and $H'$ that $\pi$ substitutes with dummy nodes, and $\mathcal{E}_{\text{del}}^G$ and $\mathcal{E}_{\text{ins}}^H$ contain the dummy nodes contained in $G'$ and $H'$ that are deleted and inserted by $\pi$.

As deleting a dummy node, inserting a dummy node, and substituting a dummy node by another dummy node does not incur any edit cost, we can assume w.l.o.g. that $\mathcal{E}_{\text{sub} \to \mathcal{E}^H}^G = \mathcal{E}_{\mathcal{E}^G \to \text{sub}}^H = \emptyset$ (if this is not the case, just delete and insert all dummy nodes contained in $\mathcal{E}_{\text{sub} \to \mathcal{E}^H}^G$ and $\mathcal{E}_{\mathcal{E}^G \to \text{sub}}^H$, respectively). This implies $|V_{\text{del}}^G| + |\mathcal{E}_{V^G \to \text{sub}}^H| \leq |V^G| = |\mathcal{E}_{\text{ins}}^H| + |\mathcal{E}_{V^G \to \text{sub}}^H|$ and hence $|V_{\text{del}}^G| \leq |\mathcal{E}_{\text{ins}}^H|$. For each deleted original node $u_i \in V_{\text{del}}^G$, we can hence pick an inserted dummy node $v_k \in \mathcal{E}_{\text{ins}}^H$ and update $\pi$ as $\pi := (\pi \setminus \{(u_i, \epsilon), (\epsilon, v_k)\}) \cup \{(u_i, v_k)\}$. Since inserting a dummy node does not incur any cost and substituting a real node by a dummy node incurs the same cost as deleting it, this update leaves $\pi$'s induced edit cost invariant. After carrying out all of these updates, we have $V_{\text{del}}^G = \emptyset$. Analogously, we can ensure $V_{\text{ins}}^H = \emptyset$.

Once $\pi$ has been modified such that $V_{\text{del}}^G = V_{\text{ins}}^H = \emptyset$, we have $|V_{\text{sub}}^G| + |\mathcal{E}_{\text{sub} \to V^H}^G| + |\mathcal{E}_{\text{del}}^G| = |V_{\text{sub}}^H| + |\mathcal{E}_{V^G \text{sub}}^H| + |\mathcal{E}_{\text{ins}}^H|$ and hence $|\mathcal{E}_{\text{del}}^G| = |\mathcal{E}_{\text{ins}}^H|$, since node substitution come in pairs. Therefore, we can enforce $\mathcal{E}_{\text{del}}^G = \mathcal{E}_{\text{ins}}^H = \emptyset$ by grouping the dummy nodes contained in $\mathcal{E}_{\text{del}}^G$ and $\mathcal{E}_{\text{ins}}^H$ into pairs and replacing the insertions and deletions with substitutions. Again, this modification leaves $\pi$'s induced edit cost invariant.

The second part of the lemma immediately follows from the fact that dummy nodes and edges can be deleted and inserted for free.  □

Using Lemma 6.1, we are now in the position to prove Theorem 6.3. This theorem states that `BRANCH-TIGHT` computes a series of monotonously in-creasing lower bounds that are at least as good as the lower bound computed by the algorithm `BRANCH` presented in Section 6.2.1.

**Theorem 6.3 (Properties and Correctness of `BRANCH-TIGHT`).** *Let G and H be two graphs and $(LB_r)_{r=1}^{I'}$ be the sequence of lower bounds computed by Figure 6.4. Then the following statements hold:*

(i) *$LB_1$ equals the lower bound $\mathrm{BED}(G, H)$ for $\mathrm{GED}(G, H)$ computed by the algorithm `BRANCH` presented in Section 6.2.1.*
(ii) *The computed values are indeed lower bounds, i.e., $LB_r \leq \mathrm{GED}(G, H)$ holds for all $r \in [I']$.*
(iii) *The lower bounds increase monotonously, i.e., $LB_{r-1} \leq LB_r$ holds for all $r \in [I'] \setminus \{1\}$.*

*Proof.* The first part of the theorem immediately follows from Proposition 4.2, the equations (6.7) to (6.8), and the way `BRANCH` constructs its LSAPE instance **C** (cf. Section 6.2.1 above).

For proving the second part of the theorem, let $G'$ and $H'$ be the graphs obtained from $G$ and $H$ by executing the lines 1 to 6 of Figure 6.4 and $\pi \in \Pi(G', H')$ be an optimal node map without node insertions or deletions. By Lemma 6.1, we know that such a node map exists. Note that, since $\pi$ contains neither node insertions nor node deletions, $\pi$ can equivalently be viewed as a perfect matching $\pi^r \in \Pi_{N,N}$. Moreover, $\pi$ does not encode any edge insertions or deletions, because $G'$ and $H'$ are complete graphs.

For all $r \in [I']$, we introduce the operator

$$Q_\pi(r) := \sum_{u_i \in V^{G'}} c_V(u_i, \pi(u_i)) + \sum_{(u_i, u_j) \in E^{G'}} \left[ c_{j,\pi(j)}^{i,\pi(i),r} + c_{i,\pi(i)}^{j,\pi(j),r} \right],$$

and prove

$$Q_\pi(r) = \mathrm{GED}(G', H') \tag{6.9}$$

by induction on $r$. For proving the base case $r = 1$, observe that we have $Q_\pi(1) := \sum_{u_i \in V^{G'}} c_V(u_i, \pi(u_i)) + \sum_{(u_i, u_j) \in E^{G'}} c_E((u_i, u_j), \pi(u_i, u_j)) = c(P_\pi)$, where the second equality follows from the fact that $\pi$ only contains node substitutions and edge substitutions. Since $\pi$ is an optimal node map, this proves equation (6.9) for the case $r = 1$. For the inductive step $r \to r + 1$, consider the following chain of equalities (the last equality is the inductive assumption):

$$Q_\pi(r+1) = \sum_{u_i \in V^{G'}} c_V(u_i, \pi(u_i)) + \sum_{(u_i,u_j) \in E^{G'}} \left[ c_{j,\pi(j)}^{i,\pi(i),r+1} + c_{i,\pi(i)}^{j,\pi(j),r+1} \right]$$

$$= \sum_{u_i \in V^{G'}} c_V(u_i, \pi(u_i)) + \sum_{(u_i,u_j) \in E^{G'}} \left[ c_{j,\pi(j)}^{i,\pi(i),r} + c_{i,\pi(i)}^{j,\pi(j),r} + s_{j,\pi(j)}^{i,\pi(i),r} \right.$$

$$\left. - s_{j,\pi(j)}^{i,\pi(i),r} + s_{i,\pi(i)}^{j,\pi(j),r} - s_{i,\pi(i)}^{j,\pi(j),r} + \frac{s_{i,\pi(i)}^r - s_{i,\pi(i)}^r + s_{j,\pi(j)}^r - s_{j,\pi(j)}^r}{N-1} \right]$$

$$= \sum_{u_i \in V^{G'}} c_V(u_i, \pi(u_i)) + \sum_{(u_i,u_j) \in E^{G'}} \left[ c_{j,\pi(j)}^{i,\pi(i),r} + c_{i,\pi(i)}^{j,\pi(j),r} \right]$$

$$= Q_\pi(r) = \mathrm{GED}(G', H')$$

We can now finish the proof of the second part of the theorem, i.e., show that `BRANCH-TIGHT` indeed computes a series of lower bounds for GED:

$$LB_r = \mathrm{LSAP}(\mathbf{C}^r) = \min_{\pi' \in \Pi_{N,N}} \mathbf{C}^r(\pi') \leq \mathbf{C}^r(\pi)$$

$$= \sum_{u_i \in V^{G'}} c_V(u_i, \pi(u_i)) + \min_{\pi' \in \Pi_{N-1,N-1}} \mathbf{C}^{i,\pi(i),r}(\pi')$$

$$\leq \sum_{u_i \in V^{G'}} c_V(u_i, \pi(u_i)) + \mathbf{C}^{i,\pi(i),r}(\pi)$$

$$= \sum_{u_i \in V^{G'}} c_V(u_i, \pi(u_i)) + \sum_{u_j \in N^{G'}(u_i)} c_{j,\pi(j)}^{i,\pi(i),r}$$

$$= \sum_{u_i \in V^{G'}} c_V(u_i, \pi(u_i)) + \sum_{(u_i,u_j) \in E^{G'}} \left[ c_{j,\pi(j)}^{i,\pi(i),r} + c_{i,\pi(i)}^{j,\pi(j),r} \right]$$

$$= Q_\pi(r) = \mathrm{GED}(G', H') = \mathrm{GED}(G, H)$$

It remains to be shown that the lower bounds computed by `BRANCH-TIGHT` increase monotonously. For this, we make use of the following Observation 6.1, which immediately follows from standard duality theory [36].

**Observation 6.1.** *Let $\mathbf{C} \in \mathbb{R}^{n \times n}$ be an LSAP instance, $\pi \in \Pi_{n,n}$ be an optimal perfect matching for $\mathbf{C}$, and $\mathbf{S} \in \mathbb{R}^{n \times n}$ be the slack matrix associated with $\pi$. Then $\pi$ is optimal for $\mathbf{C} - \mathbf{S}$, too, and it holds that $\mathrm{LSAP}(\mathbf{C}) = \mathrm{LSAP}(\mathbf{C} - \mathbf{S})$.*

Let $\pi^{r+1} \in \Pi_{N,N}$ and $\pi^r \in \Pi_{N,N}$ be perfect matchings which are optimal for $\mathbf{C}^{r+1}$ and $\mathbf{C}^r$, respectively, i.e., satisfy $\mathbf{C}^{r+1}(\pi^{r+1}) = \mathrm{LSAP}(\mathbf{C}^{r+1})$ and $\mathbf{C}^r(\pi^r) = \mathrm{LSAP}(\mathbf{C}^r)$. Consider the following chain of (in-)equalities:

$$LB_{r+1} = \mathrm{LSAP}(\mathbf{C}^{r+1}) = \mathbf{C}^{r+1}(\pi^{r+1}) \sum_{u_i \in V^{G'}} c_{i,\pi^{r+1}(i)}^{r+1}$$

$$\begin{aligned}
&= \sum_{u_i \in V^{G'}} c_V(u_i, \pi^{r+1}(u_i)) + \mathrm{LSAP}(\mathbf{C}^{i,\pi^{r+1}(i),r+1}) \\
&= \sum_{u_i \in V^{G'}} c_V(u_i, \pi^{r+1}(u_i)) + \min_{\pi' \in \Pi_{N-1,N-1}} \sum_{u_j \in N^{G'}(u_i)} c_{j,\pi'(j)}^{i,\pi^{r+1}(i),r+1} \\
&= \sum_{u_i \in V^{G'}} c_V(u_i, \pi^{r+1}(u_i)) + \min_{\pi' \in \Pi_{N-1,N-1}} \sum_{u_j \in N^{G'}(u_i)} \Big[ c_{j,\pi'(j)}^{i,\pi^{r+1}(i),r} \\
&\quad - s_{j,j,\pi'(j)}^{i,\pi^{r+1}(i),r} - \frac{s_{i,\pi^{r+1}(i)}^r}{N-1} + s_{i,\pi^{r+1}(i)}^{j,\pi'(j),r} + \frac{s_{j,\pi'(j)}^r}{N-1} \Big] \\
&\geq \sum_{u_i \in V^{G'}} c_V(u_i, \pi^{r+1}(u_i)) + \min_{\pi' \in \Pi_{N-1,N-1}} \sum_{u_j \in N^{G'}(u_i)} \Big[ c_{j,\pi'(j)}^{i,\pi^{r+1}(i),r} \\
&\quad - s_{j,j,\pi'(j)}^{i,\pi^{r+1}(i),r} - \frac{s_{i,\pi^{r+1}(i)}^r}{N-1} \Big] \\
&\stackrel{(a)}{=} \sum_{u_i \in V^{G'}} c_V(u_i, \pi^{r+1}(u_i)) + \min_{\pi' \in \Pi_{N-1,N-1}} \sum_{u_j \in N^{G'}(u_i)} \Big[ c_{j,\pi'(j)}^{i,\pi^{r+1}(i),r} - s_{j,j,\pi'(j)}^{i,\pi^{r+1}(i),r} \Big] \\
&\quad - s_{i,\pi^{r+1}(i)}^r \\
&= \sum_{u_i \in V^{G'}} c_V(u_i, \pi^{r+1}(u_i)) + \mathrm{LSAP}(\mathbf{C}^{i,\pi^{r+1}(i),r}) - s_{i,\pi^{r+1}(i)}^r \\
&= \sum_{u_i \in V^{G'}} c_{i,\pi^{r+1}(i)}^r - s_{i,\pi^{r+1}(i)}^r \stackrel{(b)}{\geq} \sum_{u_i \in V^{G'}} c_{i,\pi^r(i)}^r - s_{i,\pi^r(i)}^r \stackrel{(c)}{=} \sum_{u_i \in V^{G'}} c_{i,\pi^r(i)}^r \\
&= \mathbf{C}^r(\pi^r) = \mathrm{LSAP}(\mathbf{C}^r) = LB_r
\end{aligned}$$

Here, equation (a) follows from Observation 6.1, and inequality (b) and equation (c) follow from Observation 6.1 and the optimality of $\pi^r$ for $\mathbf{C}^r$. All other equations and inequalities are directly implied by the involved definitions. □

## 6.4 LSAPE, Rings, and Machine Learning

In this section, we present the algorithms `RING` and `RING-ML`. Both of them are implementations of the paradigm `LSAPE-GED` that are designed to yield tight upper bounds. Moreover, both `RING` and `RING-ML` make use of a new kind of local structures — namely, rings of fixed size rooted at the nodes of the input graphs. Intuitively, rings are sequences of disjoint sets of nodes and edges that are at fixed distances from a root node. Like the subgraph and walks structures used by the algorithms `SUBGRAPH` and `WALKS`, rings are designed to capture more topological information than the local structures used by algorithms `BP`, `STAR`, `BRANCH-CONST`, `BRANCH`, and `BRANCH-FAST`. The advantage w. r. t. subgraphs is that rings can be processed in polynomial time. The advantage w. r. t. walks is that rings model general edit costs and avoid redundancies due to multiple inclusions of nodes and edges.

The difference between the algorithms `RING` and `RING-ML` consists in the way they use the rings to construct their LSAPE instances **C**. `RING` adopts the classical approach also employed by `BP`, `STAR`, `BRANCH-CONST`, `BRANCH`, and `BRANCH-FAST`, `SUBGRAPH`, and `WALKS`, i. e., constructs **C** via a suitably defined ring distance measure. In contrast to that, `RING-ML` uses machine learning techniques to construct its LSAPE instance **C**: During training, the rings are used to construct feature vectors for all possible node assignments and a machine learning framework — e. g., a support vector classifier (SVC), a one class support vector machine (1-SVM), or a deep neural network (DNN) — is trained to output a value close to 0 if a node assignment is predicted to be contained in an optimal edit path and a value close to 1, otherwise. At runtime, the output of the machine learning framework is fed into the LSAPE instance **C**.

The remainder of this section is organized as follows: In Section 6.4.1, we formally describe the classical and the machine learning based strategy for populating the LSAPE instance **C** of an instantiation of `LSAPE-GED`. In Section 6.4.2, we introduce rings and show how to efficiently construct them via breadth-first search. In Section 6.4.4, we present the algorithms `RING` and `RING-ML`.

Throughout this section, we let $\mathfrak{I} := \{(G, u) \mid G \in \mathbb{G} \land u \in (V^G \cup \epsilon)\}$ denote the set of all graph-node incidences and $\mathfrak{A} := \{(G, H, u, v) \mid (G, u) \in \mathfrak{I} \land (H, v) \in \mathfrak{I} \land (u \neq \epsilon \lor v \neq \epsilon)\}$ denote the set of all node assignments.

### 6.4.1 Two Approaches for Populating the LSAPE Instance C

#### 6.4.1.1 Classical Approach

Classical instantiations of the paradigm `LSAPE-GED` construct their LSAPE instance **C** by using local structures rooted at the nodes and the distance between them. Formally, they define local structure functions $\mathcal{S} : \mathfrak{I} \to \mathfrak{S}$ that map graph-nodes incidences to elements of a suitably defined space of local structures $\mathfrak{S}$, and distance measures $d_{\mathfrak{S}} : \mathfrak{S} \times \mathfrak{S} \to \mathbb{R}_{\geq 0}$ for the local structures. Given input graphs $G$ and $H$, the LSAPE instance $\mathbf{C} \in \mathbb{R}^{(|V^G|+1) \times (|V^H|+1)}$ is defined by setting

$$c_{i,k} := d_\mathfrak{S}(\mathcal{S}(G, u_i), \mathcal{S}(H, v_k))$$
$$c_{i,|V^H|+1} := d_\mathfrak{S}(\mathcal{S}(G, u_i), \mathcal{S}(H, \epsilon))$$
$$c_{|V^G|+1,k} := d_\mathfrak{S}(\mathcal{S}(G, \epsilon), \mathcal{S}(H, v_k))$$

for all $(i, k) \in [|V^G|] \times [|V^H|]$. This classical strategy for populating $\mathbf{C}$ is adopted by the existing heuristics NODE [60], BP [83], STAR [111], BRANCH-CONST [113], BRANCH [15], BRANCH-FAST [15], WALKS [50], and SUBGRAPH [32], as well as by the algorithm RING proposed in Section 6.4.4.1 below. Also the heuristic CENTRALITIES [88, 102] can be subsumed under this model; here, the "local structures" are simply the nodes' centralities.

#### 6.4.1.2 Machine Learning Based Approach

Instead of using local structures, LSAPE instances can be constructed with the help of feature vectors associated to good and bad node assignments. This strategy is inspired by the existing algorithms PREDICT [90] and NGM [39]. However, as detailed below, both PREDICT and NGM fall short of completely instantiating it.

**Definition 6.1 (Good and Bad Node Assignments).** A node assignment $(G, H, u, v) \in \mathfrak{A}$ is called good if and only if it is contained in an optimal node map, i.e., if there is a node map $\pi \in \Pi(G, H)$ with $c(P_\pi) = \text{GED}(G, H)$ and $(u, v) \in \pi$. The set of all good node assignments is denoted by $\mathfrak{A}^\star$. Node assignments contained in $\mathfrak{A} \setminus \mathfrak{A}^\star$ are called bad.

If machine learning techniques are used for populating the LSAPE instance $\mathbf{C} \in \mathbb{R}^{(|V^G|+1) \times (|V^H|+1)}$, in a first step, feature vectors $\mathcal{F} : \mathfrak{A} \to \mathbb{R}^d$ for the node assignments have to be defined. Subsequently, a function $p^\star : \mathbb{R}^d \to [0, 1]$ has to be learned, which maps feature vectors $\mathbf{x} \in \mathcal{F}[\mathfrak{A}^\star]$ to large values and feature vectors $\mathbf{x} \in \mathcal{F}[\mathfrak{A} \setminus \mathfrak{A}^\star]$ to small values. Informally, $p^\star(\mathbf{x})$ can be viewed as an estimate of the probability that a feature vector $\mathbf{x} \in \mathcal{F}[\mathfrak{A}]$ is associated to a good node map. Once $p^\star$ has been learned, $\mathbf{C}$ is defined by setting

$$c_{i,k} := 1 - p^\star(\mathcal{F}(G, H, u_i, v_k))$$
$$c_{i,|V^H|+1} := 1 - p^\star(\mathcal{F}(G, H, u_i, \epsilon))$$

$$c_{|V^G|+1,k} := 1 - p^\star(\mathcal{F}(G,H,\epsilon,v_k))$$

for all $(i,k) \in [|V^G|] \times [|V^H|]$.

For learning the probability estimate $p^\star$, several strategies can be adopted. Given a set $\mathcal{G}$ of training graphs, one can mimic PREDICT and compute optimal node maps $\pi_{G,H}$ for the training graphs. These node maps can be used to generate training data

$$\mathcal{T} := \{(\mathcal{F}(G,H,u,v), \delta_{(u,v) \in \pi_{G,H}}) \mid (G,H,u,v) \in \mathfrak{A}[\mathcal{G}]\},$$

where $\mathfrak{A}[\mathcal{G}]$ is the restriction of $\mathfrak{A}$ to the graphs contained in $\mathcal{G}$. Finally, a kernelized SVC with probability estimates [71] can be trained on $\mathcal{T}$. Alternatively, one can proceed like NGM, i.e., use $\mathcal{T}$ to train a fully connected feed-forward DNN with output from $[0,1]$, and define the probability estimate $p^\star$ as the output of the DNN.

The drawback of these approaches is that some feature vectors contained in $\mathcal{T}$ are incorrectly labeled as bad if there is more than one optimal node map. To see why, assume that, for training graphs $G$ and $H$, there are two optimal node maps $\pi_{G,H}$ and $\pi'_{G,H}$, and that the exact algorithm used for generating $\mathcal{T}$ computes $\pi_{G,H}$. Let $(G,H,u,v)$ be a node assignment such that $(u,v)$ is contained in $\pi'_{G,H} \setminus \pi_{G,H}$. According to Definition 6.1, $(G,H,u,v)$ is a good node assignment, but in $\mathcal{T}$, its feature vector $\mathbf{x} := \mathcal{F}(G,H,u,v)$ is labeled as bad.

A straightforward but computationally infeasible way for tackling this problem is to compute all optimal node maps between the training graphs. Instead, we suggest to train a one class support vector machine (1-SVM) [97] with RBF kernel to estimate the support of the feature vectors associated to good node maps, and then to manipulate the trained decision function to obtain the probability estimate $p^\star$. This has the advantage that, given a set $\mathcal{G}$ of training graphs and initially computed optimal node maps $\pi_{G,H}$ for all $G, H \in \mathcal{G}$, we can use training data

$$\mathcal{T}^\star := \{\mathcal{F}(G,H,u,v) \mid (G,H,u,v) \in \mathfrak{A}[\mathcal{G}] \wedge (u,v) \in \pi_{G,H}\},$$

which contains only feature vectors associated to good node assignments and is hence correct even if there are multiple optimal node maps.

For the definition of $p^\star$, recall that the decision function of a trained 1-SVM with RBF kernel is $\text{sgn}(h(\mathbf{x}))$, where $h(\mathbf{x}) = [\sum_{i=1}^{|\mathcal{T}^\star|} \alpha_i \exp(-\gamma \|\mathbf{x}^i - \mathbf{x}\|_2^2)] - \rho$, $\alpha_i$ is the dual variable associated to the training vector $\mathbf{x}^i \in \mathcal{T}^\star$, $\rho$ defines the separating hyperplane in the feature space induced by the RBF kernel, and $\gamma > 0$ is a tuning parameter of the RBF kernel. The following remark immediately follows from the definition of $h$:

**Remark 6.1 (Properties of 1-SVM).** Let $(\boldsymbol{\alpha}, \rho)$ be a 1-SVM with RBF kernel and tuning parameter $\gamma$ that has been trained on data $\mathcal{T}^\star$. Then $h(\mathbf{x}) \in (-\rho, \mathbf{1}^\mathsf{T}\boldsymbol{\alpha} - \rho)$ holds for all $\mathbf{x} \in \mathbb{R}^d$, and $\mathbf{x} \mapsto (\gamma/\pi)^{d/2}(\mathbf{1}^\mathsf{T}\boldsymbol{\alpha})^{-1}(h(\mathbf{x}) + \rho)$ is the density function of the multivariate Gaussian mixture model $\mathcal{M}(\boldsymbol{\alpha}, \gamma) := \sum_{i=1}^{|\mathcal{T}^\star|}(\mathbf{1}^\mathsf{T}\boldsymbol{\alpha})^{-1}\alpha_i \mathcal{N}(\mathbf{0}, (2\gamma)^{-1}\mathbf{I})$ for the feature vectors $\mathcal{F}[\mathfrak{A}^\star]$ associated to good node assignments.

Remark 6.1 tells us how to transform the output of a trained 1-SVM into a probability estimate $p^\star$. We simply define $p^\star(\mathbf{x})$ as the likelihood of the feature vector $\mathbf{x}$ under the model $\mathcal{M}(\boldsymbol{\alpha}, \gamma)$ learned by the 1-SVM, i.e., set $p^\star(\mathbf{x}) := (\gamma/\pi)^{d/2}(\mathbf{1}^\mathsf{T}\boldsymbol{\alpha})^{-1}(h(\mathbf{x}) + \rho)$.

We conclude this section by briefly summarizing why the existing heuristics `NGM` and `PREDICT` fail to fully instantiate the machine learning based transformation strategy described above. There are two problems with `NGM`: Firstly, its feature vectors are defined only for graphs whose node labels are real-valued vectors. Secondly, no feature vectors for node deletions and insertions can be constructed. This implies that `NGM` cannot populate the last row and the last column of its LSAPE instance and can hence be used only for graphs whose optimal node maps are known upfront not to contain node insertions or deletions. Unlike `NGM`, `PREDICT` defines feature vectors that cover node deletions and insertions and are defined for general node and edge labels. However, `PREDICT` is designed to predict if a node assignment is good rather than to use the decision value for populating an LSAPE instance. Therefore, instead of learning a probability estimate $p^\star : \mathbb{R}^d \to [0, 1]$, `PREDICT` uses a kernelized SVC without probability estimates to learn a decision function $f^\star : \mathbb{R}^d \to \{0, 1\}$ which maps feature vectors $\mathbf{x} \in \mathcal{F}[\mathfrak{A}^\star]$ to 1 and feature vectors $\mathbf{x} \in \mathcal{F}[\mathfrak{A} \setminus \mathfrak{A}^\star]$ to 0.

### 6.4.2 Rings as Local Structures

In this section, we introduce rings of size $L$ as a new kind of local structures (Section 6.4.2.1). Subsequently, we show how to to construct them (Section 6.4.3).

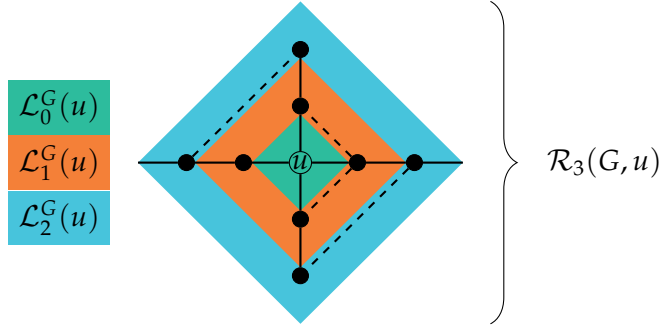

**Figure 6.5.** Visualization of Definition 6.2. Inner edges are dashed, outer edges are solid. Layers are displayed in different colors.

#### 6.4.2.1 Rings: Definition

We define the rings rooted at the nodes of a graph $G$ as $L$-sized sequences of layers $\mathcal{L}^G = (N^G, OE^G, IE^G)$, where $N^G \subseteq V^G$ is a subset of the nodes, and $OE^G, IE^G \subseteq E^G$ are subsets of the edges of $G$. Formally, the space of all $L$-sized rings for graphs from a domain $\mathbb{G}$ is defined as $\mathfrak{R}_L := \{(\mathcal{L}_l)_{l=0}^{L-1} \mid \mathcal{L}_l \in \bigcup_{G \in \mathbb{G}} \mathfrak{L}(G)\}$, where $\mathfrak{L}(G) := \mathcal{P}(V^G) \times \mathcal{P}(E^G) \times \mathcal{P}(E^G)$. Next, we specify a function $\mathcal{R}_L : \mathfrak{I} \to \mathfrak{R}_L$ which maps a graph-node incidence $(G, u)$ to a ring of size $L$.

**Definition 6.2 (Rings, Layers, Outer and Inner Edges).** Given a constant $L \in \mathbb{N}_{>0}$, the function $\mathcal{R}_L : \mathfrak{I} \to \mathfrak{R}_L$ maps a graph-node incidence $(G, u)$ to the ring $\mathcal{R}_L(G, u) := (\mathcal{L}_l^G(u))_{l=0}^{L-1}$ rooted at $u$ in $G$ (Figure 6.5). $\mathcal{L}_l^G(u) := (N_l^G(u), OE_l^G(u), IE_l^G(u))$ denotes the $l^\text{th}$ layer rooted at $u$ in $G$, where:

- $N_l^G(u) := \{u' \in V^G \mid d_V^G(u, u') = l\}$ is the set of nodes at distance $l$ of $u$,
- $IE_l^G(u) := E^G \cap (N_l^G(u) \times N_l^G(u))$ is the set of inner edges connecting two nodes in the $l^\text{th}$ layer, and
- $OE_l^G(u) := E^G \cap (N_l^G(u) \times N_{l+1}^G(u))$ is the set of outer edges connecting a node in the $l^\text{th}$ layer to a node in the $(l+1)^\text{th}$ layer.

For the dummy node $\epsilon$, we define $\mathcal{R}_L(G, \epsilon) := ((\emptyset, \emptyset, \emptyset)_l)_{l=0}^{L-1}$.

It is easy to see that the ring $\mathcal{R}_1(G, u)$ of a node $u \in V^G$ corresponds to the branch structure used by the `LSAPE-GED` instantiations `BP`, `BRANCH`, `BRANCH-FAST`, and `BRANCH-CONST`. Further properties of rings and layers are summarized in Proposition 6.3.

**Proposition 6.3 (Properties of Rings and Layers).** *Let $u \in V^G$ be a node and $\mathcal{R}_L(G, u) = ((N_l^G(u), OE_l^G(u), IE_l^G(u))_l)_{l=0}^L$ be the ring of size L rooted at u. Then the following statements follow from the involved definitions:*

(i) *The node set $N_l^G(u)$ is empty if and only if $l > e^G(u)$, the edge set $IE_l^G(u)$ is empty if $l > e^G(u)$, and the edge set $OE_l^G(u)$ is empty if and only if $l > e^G(u) - 1$.*
(ii) *All node sets $N_l^G(u)$ and all edge sets $OE_l^G(u)$ and $IE_l^G(u)$ are disjoint.*
(iii) *The equalities $\bigcup_{l=0}^{L-1} N_l^G(u) = V^G$ and $\bigcup_{l=0}^{L-1} (OE_l^G(u) \cup IE_l^G(u)) = E^G$ hold for all $u \in V^G$ if and only if $L > \mathrm{diam}(G)$.*

*Proof.* All three statements of the proposition immediately follow from the involved definitions. □

### 6.4.3 Rings: Construction

Figure 6.6 shows how to construct a ring $\mathcal{R}_L(G, u)$ via breadth-first search. The algorithm maintains the level $l$ of the currently processed layer along with the layer's node and edge sets $N$, $OE$, and $IE$, a vector $\mathtt{d}$ that stores for each node $u' \in V^G$ the distance to the root $u$, flags $\mathtt{discovered}[e]$ that indicate if the edge $e \in E^G$ has already been discovered by the algorithm, and a FIFO queue $\mathtt{open}$ which is initialized with the root $u$. Throughout the algorithm, $\mathtt{d}[u'] = d_V^G(u, u')$ holds for all nodes $u'$ which have already been added to $\mathtt{open}$, while newly discovered nodes $u''$ have $\mathtt{d}[u''] = \infty$.

If a node $u'$ is popped from $\mathtt{open}$, we check if its distance is larger than the level $l$ of the current layer. If this is case, we store the current layer, increment $l$, and clear the node and edge sets $N$, $OE$, and $IE$. Next, we add the node $u'$ to the node set $N$ and iterate through its undiscovered incident edges $(u', u'')$. We mark $(u', u'')$ as discovered and push the node $u''$ to $\mathtt{open}$ if it has not been discovered yet and its distance to the root $u$ is less than $L$. If this distance equals the level of the current layer, the edge $(u', u'')$ is added to the inner edges $IE$; otherwise, it is added to the outer edges $OE$. Once $\mathtt{open}$ is empty, the last layer is stored and the complete ring is returned. Since nodes and edges are processed at most once, the algorithm runs in $O(|V^G| + |E^G|)$ time.

**Input**: Graph $G$, node $u \in V^G$, constant $L \in \mathbb{N}_{>0}$.
**Output**: Ring $\mathcal{R}_L(G, u)$ rooted at $u$.

1  $l := 0; N := \emptyset; OE := \emptyset; IE := \emptyset; \mathcal{R}_L(G, u) := ((\emptyset, \emptyset, \emptyset)_l)_{l=0}^{L-1}$;  // initialize ring
2  $\text{d}[u] := 0;$ **for** $u' \in V^G \setminus \{u\}$ **do** $\text{d}[u'] := \infty;$   // initialize distances to root
3  **for** $e \in E^G$ **do** $\texttt{discovered}[e] := \texttt{false};$   // mark all edges as undiscovered
4  $\texttt{open} := \{u\};$   // initialize FIFO queue
5  **while** $\texttt{open} \neq \emptyset$ **do**   // main loop
6     $u' := \texttt{open.pop}();$   // pop node from queue
7     **if** $\text{d}[u'] > l$ **then**   // the $l^{\text{th}}$ layer is complete
8        $\mathcal{R}_L(G, u)_l := (N, OE, IE); l := l + 1;$   // store $l^{\text{th}}$ layer and increment $l$
9        $N := \emptyset; OE := \emptyset; IE := \emptyset;$   // reset nodes, inner, and outer edges
10    $N := N \cup \{u'\};$   // $u'$ is node at $l^{\text{th}}$ layer
11    **for** $(u', u'') \in E^G$ **do**   // iterate through neighbors of $u'$
12       **if** $\texttt{discovered}[(u', u'')]$ **then continue**;   // skip discovered edges
13       $\texttt{discovered}[(u', u'')] := \texttt{true};$   // mark $(u', u'')$ as discovered
14       **if** $\text{d}[u''] = \infty$ **then**   // found new node
15          $\text{d}[u''] := l + 1;$   // set distance of new node
16          **if** $\text{d}[u''] < L$ **then** $\texttt{open.push}(u'');$   // add close new node to queue
17       **if** $\text{d}[u''] = l$ **then** $IE := IE \cup \{(u', u'')\};$// $(u', u'')$ is inner edge at $l^{\text{th}}$ layer
18       **else** $OE := OE \cup \{(u', u'')\};$   // $(u', u'')$ is outer edge at $l^{\text{th}}$ layer
19 $\mathcal{R}_L(G, u)_l := (N, OE, IE);$   // store last layer
20 **return** $\mathcal{R}_L(G, u);$   // return ring

**Figure 6.6.** Construction of rings via breadth-first search.

### 6.4.4 Two Ring Based Heuristics

In this section, we present two new heuristics for ring based transformations to LSAPE. The heuristic `RING` (Section 6.4.4.1) uses the classical transformation strategy presented in Section 6.4.1.1, the heuristic `RING-ML` (Section 6.4.4.2) employs the machine learning based approach presented in Section 6.4.1.2.

#### 6.4.4.1 `RING`: A Classical Instantiation of `LSAPE-GED`

`RING` is a classical instantiation of the paradigm `LSAPE-GED` which uses rings of size $L$ as local structures. Therefore, what remains to be done is to define a distance measure $d_{\mathfrak{R}_L} : \mathfrak{R}_L \times \mathfrak{R}_L \to \mathbb{R}_{\geq 0}$ for the rings. We will define such a distance measure in a bottom-up fashion: Ring distances are defined in terms of layer distances, which, in turn, are defined in terms of node and edge set distances.

Assume that, for all pairs of graphs $(G, H) \in \mathfrak{G} \times \mathfrak{G}$, we have access to measures $d_{\mathcal{P}(V)}^{G,H} : \mathcal{P}(V^G) \times \mathcal{P}(V^H) \to \mathbb{R}_{\geq 0}$ and $d_{\mathcal{P}(E)}^{G,H} : \mathcal{P}(E^G) \times \mathcal{P}(E^H) \to \mathbb{R}_{\geq 0}$ that compute distances between subsets of the nodes and edges of $G$ and $H$. Then we can define layer distance measures $d_{\mathfrak{L}}^{G,H} : \mathfrak{L}(G) \times \mathfrak{L}(H) \to \mathbb{R}_{\geq 0}$ as

$$d_{\mathfrak{L}}^{G,H}(\mathcal{L}^G, \mathcal{L}^H) := \frac{\alpha_0 d_{\mathcal{P}(V)}^{G,H}(N^G, N^H)}{\max\{|N^G|, |N^H|, 1\}} + \frac{\alpha_1 d_{\mathcal{P}(E)}^{G,H}(IE^G, IE^H)}{\max\{|IE^G|, |IE^H|, 1\}} + \frac{\alpha_2 d_{\mathcal{P}(E)}^{G,H}(OE^G, OE^H)}{\max\{|OE^G|, |OE^H|, 1\}},$$

where $\alpha \in \Delta^2$ is a simplex vector of weights associated to the distances between nodes, inner edges, and outer edges. We normalize by the sizes of the involved node and edge sets in order not to over-represent large layers. Using the layer distances and a simplex weight vector $\lambda \in \Delta^{L-1}$ for the layer distances at the different levels, we define the ring distance measure $d_{\mathfrak{R}_L}$ as follows:

$$d_{\mathfrak{R}_L}((\mathcal{L}_l^G)_{l=0}^{L-1}, (\mathcal{L}_l^H)_{l=0}^{L-1}) := \sum_{l=0}^{L-1} \lambda_l d_{\mathfrak{L}}^{G,H}(\mathcal{L}_l^G, \mathcal{L}_l^H)$$

The next step is to define the node and edge set distances $d_{\mathcal{P}(V)}^{G,H}$ and $d_{\mathcal{P}(E)}^{G,H}$. In order to arrive at a tight upper bound for GED, it is desirable to define these distances in such a way that $d_{\mathfrak{R}_L}(\mathcal{R}_L(G, u), \mathcal{R}_L(H, v))$ is small just in case the node assignment $(G, H, u, v)$ induces a small edit cost. We suggest two strategies that meet this desideratum. Both of them make crucial use of the edit cost functions $c_V$ and $c_E$.

**LSAPE Based Definitions of $d_{\mathcal{P}(V)}^{G,H}$ and $d_{\mathcal{P}(E)}^{G,H}$.** A straightforward way for defining distances between sets of nodes and edges is to use the edit cost functions to populate LSAPE instances and then define the distances in terms of the costs of optimal or greedy LSAPE solutions. For spelling out this approach in detail, we consider node sets $N^G = \{u_1, \ldots, u_{|N^G|}\} \subseteq V^G$ and $N^H = \{v_1, \ldots, v_{|N^H|}\} \subseteq V^H$ and define an LSAPE instance $\mathbf{C} \in \mathbb{R}^{(|N^G|+1) \times (|N^H|+1)}$ by setting $c_{i,k} := c_V(\ell_V^G(u_i), \ell_V^H(v_k))$, $c_{i,|N^H|+1} := c_V(\ell_V^G(u_i), \epsilon)$, and $c_{|N^G|+1,k} := c_V(\epsilon, \ell_V^H(v_k))$ for all $(i, k) \in [|N^G|] \times [|N^H|]$. Then we compute a solution $\pi \in \Pi(\mathbf{C})$—either optimally in $O(\min\{|N^G|, |N^H|\}^2 \max\{|N^G|, |N^H|\})$ time or greedily in $O(|N^G||N^H|)$ time—and define the distance between $N^G$ and $N^H$ as $d_{\mathcal{P}(V)}^{G,H}(N^G, N^H) := \mathbf{C}(\pi)$. The edge set distance $d_{\mathcal{P}(E)}^{G,H}$ can be defined analogously.

**Multiset Intersection Based Definitions of $d_{\mathcal{P}(V)}^{G,H}$ and $d_{\mathcal{P}(E)}^{G,H}$.** Using LSAPE to define $d_{\mathcal{P}(V)}^{G,H}$ and $d_{\mathcal{P}(E)}^{G,H}$ yields fine-grained distance measures but incurs a relatively high computation time — especially, if optimal LSAPE solutions are computed. As an alternative, we suggest a faster, multiset intersection based approach which computes a proxy for the LSAPE based distances. Again consider node sets $N^G \subseteq V^G$ and $N^H \subseteq V^H$. The distance between $N^G$ and $N^H$ can be defined as $d_{\mathcal{P}(V)}^{G,H} := \Gamma(\ell_V^G[\![N^G]\!], \ell_V^H[\![N^H]\!], \bar{c}_V^{\text{sub}}, \bar{c}_V^{\text{del}}, \bar{c}_V^{\text{ins}})$, where $\bar{c}_V^{\text{del}}$, $\bar{c}_V^{\text{ins}}$, and $\bar{c}_V^{\text{sub}}$ are the average costs of deleting a node in $N^G$, inserting a node in $N^H$, and substituting a node in $N^G$ by a differently labeled node in $N^H$. Since multiset intersections can be computed in quasilinear time [111], the dominant operation is the computation of $\bar{c}_V^{\text{sub}}$ which requires $O(|N^G||N^H|)$ time. Again, the edge set distance $d_{\mathcal{P}(E)}^{G,H}$ can be defined analogously. The following Lemma 6.2 relates the LSAPE based definitions of $d_{\mathcal{P}(V)}^{G,H}$ and $d_{\mathcal{P}(E)}^{G,H}$ to the ones based on multiset intersection and justifies our claim that the latter can be viewed as proxies for the former.

**Lemma 6.2.** *Let $N^G \subseteq V^G$ and $N^H \subseteq V^H$ be subsets of the nodes of graphs $G, H \in \mathfrak{G}$ and assume that $c_V$ is quasimetric between $N^G$ and $N^H$. Then the multiset intersection based definition of $d_{\mathcal{P}(V)}^{G,H}(N^G, N^H)$ and the one based on optimally solving LSAPE incur the same number of node insertions, deletions, and substitutions. If, additionally, $c_V$ is constant between $N^G$ and $N^H$, the two definitions coincide. For the edge set distances $d_{\mathcal{P}(E)}^{G,H}$, analogous statements hold.*

*Proof.* We assume w.l.o.g. that $|N^G| \leq |N^H|$. Let $\mathbf{C} \in \mathbb{R}^{(|N^G|+1) \times (|N^G|+1)}$ be the LSAPE instance constructed as shown above and $\pi^\star \in \Pi_{|N^G|,|N^H|,\epsilon}$ be an optimal solution for $\mathbf{C}$. From Proposition 4.2 and the assumption that $c_V$ is quasimetric between $N^G$ and $N^H$, we know that $\pi^\star$ does not contain deletions and contains exactly $|N^H| - |N^G|$ insertions. This proves the first part of the lemma. If we additionally have constant node edit costs between $N^G$ and $N^H$, the optimal cost $\mathbf{C}(\pi^\star)$ is reduced to the cost of $|N^H| - |N^G|$ insertions plus $c_V^{\text{sub}} = \bar{c}_V^{\text{sub}}$ times the number of non-identical substitutions. This last quantity is provided by $|N^G| - |\ell_V^G[\![N^G]\!] \cap \ell_V^H[\![N^H]\!]|$. We thus have $\mathbf{C}(\pi^\star) = \bar{c}_V^{\text{ins}}(|N^H| - |N^G|) + \bar{c}_V^{\text{sub}}(|N^G| - |\ell_V^G[\![N^G]\!] \cap \ell_V^H[\![N^H]\!]|) = \Gamma(\ell_V^G[\![N^G]\!], \ell_V^H[\![N^H]\!], \bar{c}_V^{\text{sub}}, \bar{c}_V^{\text{del}}, \bar{c}_V^{\text{ins}})$, as required. The proof for $d_{\mathcal{P}(E)}^{G,H}$ is analogous. □

We conclude the presentation of RING by describing an algorithm in Figure 6.7 that, given a set of training graphs $\mathcal{G}$ and fixed node and edge set

**Input**: Set of graphs $\mathcal{G}$, distance measures $d_{\mathcal{P}(V)}^{G,H}$ and $d_{\mathcal{P}(E)}^{G,H}$, tuning parameter $\mu$.
**Output**: Optimized parameters $L$, $\boldsymbol{\alpha}$, $\boldsymbol{\lambda}$.
1 $L := 1 + \max_{G \in \mathcal{G}} |V^G|$;                              // set L to upper bound for ring sizes
2 build rings for all $G \in \mathcal{G}$ and all $u \in V^G$;           // cf. Figure 6.6
3 $L := 1 + \max_{G \in \mathcal{G}} \text{diam}(G)$;                    // discard layers that are empty in all rings
4 $(\boldsymbol{\alpha}, \boldsymbol{\lambda}) := \arg\min\{obj_{L,\mu}(\boldsymbol{\alpha}, \boldsymbol{\lambda}) \mid \boldsymbol{\alpha} \in \Delta^2 \wedge \boldsymbol{\lambda} \in \Delta^{L-1}\}$;   // run blackbox optimizer
5 $L := 1 + \max \text{supp}(\boldsymbol{\lambda})$;   // discard layers that are not needed for computing $d_{\mathfrak{R}_L}$

**Figure 6.7.** Optimization of RING's meta-parameters $L$, $\boldsymbol{\alpha}$, and $\boldsymbol{\lambda}$.

distances $d_{\mathcal{P}(V)}^{G,H}$ and $d_{\mathcal{P}(E)}^{G,H}$, learns good values for $L$, $\boldsymbol{\alpha}$, and $\boldsymbol{\lambda}$. In a first step, $L$ is set to an upper bound for the ring sizes, and the algorithm in Figure 6.6 is used to compute all rings of size $L$ rooted at the nodes of the graphs $G \in \mathcal{G}$. Subsequently, $L$ is lowered to 1 plus the largest $l < L$ such that there is a graph $G \in \mathcal{G}$ and a node $u \in V^G$ with $\mathcal{R}_L(G, u)_l \neq (\emptyset, \emptyset, \emptyset)$. Note that, by Proposition 6.3, this $l$ equals the maximal diameter of the graphs contained in $\mathcal{G}$. Next, a blackbox optimizer [95] is called to minimize the objective

$$obj_{L,\mu}(\boldsymbol{\alpha}, \boldsymbol{\lambda}) := \left[\mu + (1-\mu)\left(\frac{|\text{supp}(\boldsymbol{\lambda})|-1}{\max\{1, L-1\}}\right)\right] \sum_{(G,H)\in\mathcal{G}^2} \text{RING}_{\boldsymbol{\alpha},\boldsymbol{\lambda},L}(G,H)$$

over all simplex vectors $\boldsymbol{\alpha} \in \Delta^2$ and $\boldsymbol{\lambda} \in \Delta^{L-1}$. $\text{RING}_{L,\boldsymbol{\alpha},\boldsymbol{\lambda}}(G, H)$ is the upper bound for $\text{GED}(G, H)$ returned by RING if called with parameters $L$, $\boldsymbol{\alpha}$ and $\boldsymbol{\lambda}$; and $\mu \in [0, 1]$ is a tuning parameter that should be small if one wants to optimize for tightness and large if one wants to optimize for runtime. We include $|\text{supp}(\boldsymbol{\lambda})| - 1$ in the objective, because only levels which are contained in the support of $\boldsymbol{\lambda}$ contribute to $d_{\mathfrak{R}_L}$. Therefore, only few layer distances have to be computed if $\boldsymbol{\lambda}$'s support is small. Once optimized parameters $\boldsymbol{\alpha}$ and $\boldsymbol{\lambda}$ have been computed, $L$ can be further lowered to $L = 1 + \max \text{supp}(\boldsymbol{\lambda})$.

### 6.4.4.2 RING-ML: A Machine Learning Based Instantiation of LSAPE-GED

If LSAPE-GED is instantiated with the help of machine learning techniques, feature vectors associated to the node assignments have to be defined. The heuristic RING-ML uses rings of size $L$ to accomplish this task. Formally, RING-ML defines $\mathcal{F} : \mathfrak{A} \to \mathbb{R}^{6L+10}$ that maps node assignments to feature vectors with six features per layer and ten global features. Let $(G, H, u, v) \in \mathfrak{A}$ be a node assignment and $\mathcal{R}_L(G,u)$ and $\mathcal{R}_L(H,v)$ be the rings rooted at $u$ in $G$ and at $v$ in $H$, respectively. For each level $l \in \{0,\ldots,L-1\}$, a feature vector $\mathbf{x}^l \in \mathbb{R}^6$ is constructed by comparing the layers $\mathcal{R}_L(G,u)_l = (N_l^G(u), OE_l^G(u), IE_l^G(u))$ and $\mathcal{R}_L(H,v)_l = (N_l^H(v), OE_l^H(v), IE_l^H(v))$ at level $l$. The feature vector $\mathbf{x}^l$ is defined as follows:

$$\begin{aligned}
\mathbf{x}_0^l &:= |N_l^G(u)| - |N_l^H(v)| \\
\mathbf{x}_1^l &:= |OE_l^G(u)| - |OE_l^H(v)| \\
\mathbf{x}_2^l &:= |IE_l^G(u)| - |IE_l^H(v)| \\
\mathbf{x}_3^l &:= d_{\mathcal{P}(V)}^{G,H}(N_l^G(u), N_l^H(v)) \\
\mathbf{x}_4^l &:= d_{\mathcal{P}(E)}^{G,H}(OE_l^G(u), OE_l^H(v)) \\
\mathbf{x}_5^l &:= d_{\mathcal{P}(E)}^{G,H}(IE_l^G(u), IE_l^H(v))
\end{aligned}$$

The first three features compare the layers' topologies. The last three features use node and edge set distances $d_{\mathcal{P}(V)}^{G,H}$ and $d_{\mathcal{P}(E)}^{G,H}$ defined in terms of the edit costs to express the similarity of the involved node and edge labels.

Furthermore, `RING-ML` constructs a vector $\mathbf{x}^{G,H} \in \mathbb{R}^{10}$ of ten global features that only depend on $G$ and $H$: the number of nodes and edges of $G$ and $H$, the average costs for deleting nodes and edges from $G$, the average costs for inserting nodes and edges into $H$, and the average costs for substituting nodes and edges in $G$ by nodes and edges in $H$. The complete feature vector $\mathcal{F}(G,H,u,v)$ is then defined as the concatenation of the global features $\mathbf{x}^{G,H}$ and the features $\mathbf{x}^l$ associated to the levels $l \in \{0,\ldots,L-1\}$.

## 6.5 Enumeration of Optimal LSAPE Solutions

In this section, we present `MULTI-SOL`. `MULTI-SOL` is an extension of the paradigm `LSAPE-GED`, which tightens the upper bounds produced by any of its instantiations.

Recall that, for computing their upper bound $UB$, instantiations of `LSAPE-GED` first construct a suitably defined LSAPE instance $\mathbf{C} \in \mathbb{R}^{(|V^G|+1) \times (|V^H|+1)}$, then use an optimal or a greedy LSAPE solver to compute a cheap LSAPE solution $\pi \in \Pi_{|V^G|,|V^H|,\epsilon}$, and finally set $UB$ to the cost $c(P_\pi)$ of the edit path induced by $\pi$ (cf. lines 1 to 3 of Figure 6.1 presented in Section 6.1.1 above).

`MULTI-SOL` improves the upper bound *UB* returned by instantiations of `LSAPE-GED` by considering not only one, but rather up to *K* optimal LSAPE solutions, where $K \geq 2$ is a constant.[3] Let *S* be the number of optimal LSAPE solutions for **C**. Once the first optimal LSAPE solution $\pi_0 \in \Pi_{|V^G|,|V^H|,\epsilon}$ has been computed, `MULTI-SOL` uses a variant of the algorithm suggested in [106] for enumerating $K' := \min\{K, S\} - 1$ optimal LSAPE solutions $\{\pi_l\}_{l=1}^{K'}$, all of which are pairwise different and different from $\pi_0$. Since *K* is a constant, this enumeration requires only $O(|V^G| + |V^H|)$ additional time. Subsequently, the upper bound for GED is improved from $UB := c(P_{\pi_0})$ to $UB := \min_{l=0}^{K'} c(P_{\pi_l})$.

## 6.6 A Local Search Based Upper Bound

Reconsider the local search based algorithm `REFINE` [111] presented in Section 6.1.3.1 above, which improves an initial node map $\pi \in \Pi(G, H)$ by systematically varying $\pi$ via 2-swaps. In this section, we extend and improve `REFINE` in three ways. Firstly, we suggest a generalization `K-REFINE` of `REFINE`. Instead of considering only 2-swaps, `K-REFINE` considers all $K'$-swaps for all $K' \in [K] \setminus \{1\}$, where $K \in \mathbb{N}_{\geq 2}$ is a constant. Secondly, we show that for computing the induced cost $c(P_{\pi'})$ of a node map $\pi'$ obtained from $\pi$ via a $K'$-swap *C*, it suffices to consider the nodes and edges that are incident with *C*. This observation yields an improved implementation of `K-REFINE`, which is two to four times faster than the naïve implementation suggested in [111]. Thirdly, we suggest to include the dummy assignment $(\epsilon, \epsilon)$ into the initial node map $\pi$ before enumerating the swaps. This modification has the advantage that it allows the number of node substitutions to decrease and hence improves the quality of the obtained upper bound.

The remainder of this section is organized as follows: In Section 6.6.1, we present the generalized algorithm `K-REFINE`. In Section 6.6.2, we show how to efficiently compute the swap costs. In Section 6.6.3, we demonstrate that including a dummy assignment into the initial node map improves the quality of the obtained upper bound.

---

[3]`MULTI-SOL` cannot be used in combination with greedy LSAPE solvers, as it requires that the LSAPE instance **C** is solved optimally in line 2 of Figure 6.1.

**Input**: Graphs $G$ and $H$, edit cost functions $c_V$ and $c_E$, an initial node map
$\pi \in \Pi(G, H)$, a constant $K \in \mathbb{N}_{\geq 2}$.
**Output**: An improved node map $\pi' \in \overline{\Pi}(G, H)$ with $c(P_{\pi'}) \leq c(P_\pi)$.

1    $K' := 2$;      // initialize current swap size
2    $C^\star := \emptyset; \Delta^\star := 0$;      // initialize best swap
3    **while** $\Delta^\star < 0 \lor K' \leq K$ **do**      // main loop
4      **for** $C \in \mathcal{C}_{\pi,K'}$ **do**      // enumerate all swaps of current swap size
5        $\Delta := \texttt{SWAP-COST}(\pi, C)$;      // compute swap cost
6        **if** $\Delta < \Delta^\star$ **then**      // found better swap
7          $C^\star := C; \Delta^\star := \Delta$;      // update best swap
8      **if** $\Delta^\star < 0$ **then** // found better node map reachable via swap of current swap size
9        $\pi' := \texttt{SWAP}(\pi, C^\star,)$;      // compute swapped node map
10        $c(P_{\pi'}) := c(P_\pi) - \Delta^\star$;      // set induced cost of swapped node map
11        $\pi := \pi'$;      // update current node map
12        $K' := 2$;      // reset current swap size
13      **else**
14        $K' := K' + 1$;      // increment current swap size
15      $C^\star := \emptyset; \Delta^\star := 0$;      // reset best swap
16    **return** $\pi' := \pi$;      // return improved node map

**Figure 6.8.** The algorithm K-REFINE.

### 6.6.1 The Algorithm K-REFINE

Figure 6.8 gives an overview of the algorithm K-REFINE, which generalizes the local search method REFINE [111] presented in Section 6.1.3.1. Given graphs $G$ and $H$, an initial node map $\pi \in \Pi(G, H)$, and a maximal swap size $K \in \mathbb{N}_{\geq 2}$, K-REFINE starts by initializing the current swap size $K'$, the best swap $C^\star$, and the best swap cost $\Delta^\star$ as $K' := 2$, $C^\star := \emptyset$, and $\Delta^\star := 0$ (lines 1 to 2). Subsequently, K-REFINE enters its main while-loop and iterates until no improved node map has been found and the current swap size exceeds the maximal swap size (line 3).

Inside the main while-loop, the algorithm K-REFINE first enumerates the set

$$\mathcal{C}_{\pi,K'} := \{C \subseteq A_\pi \mid C \text{ is cycle of length } 2K' \text{ in } G_\pi\}$$

of all $K'$-swaps of $\pi$ (line 4, cf. Figure 6.9 for an illustration of $K$-swaps). The auxiliary directed bipartite graph $G_\pi := (V_\pi^G \cup V_\pi^H, A_\pi)$ is defined as in Section 6.1.3.1 above, i.e., we have $V_\pi^G := \{u_s \mid s \in [|\pi|]\}$, $V_\pi^H := \{v_s \mid s \in [|\pi|]\}$, and $A_\pi := \pi \cup \{(v_s, u_{s'}) \in V_\pi^H \times V_\pi^G \mid (s, s') \in [|\pi|] \times [|\pi|] \land s \neq s'\}$,

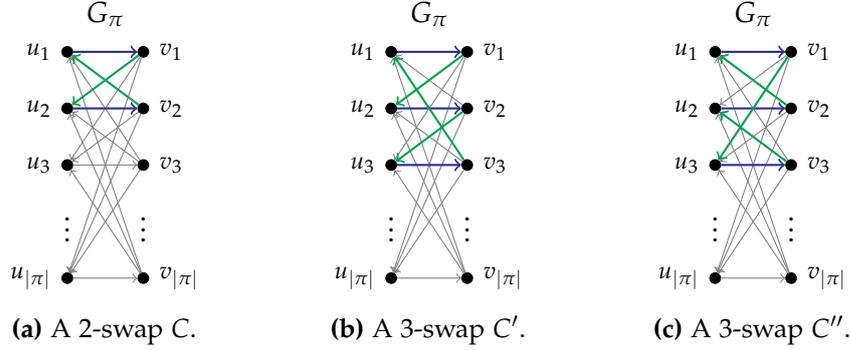

**(a)** A 2-swap $C$.  **(b)** A 3-swap $C'$.  **(c)** A 3-swap $C''$.

**Figure 6.9.** Illustration of *K*-swaps. By applying the swaps to the node map $\pi$, the blue node assignments are replaced by the green node assignments. As exemplified in Figure 6.9b and Figure 6.9c, if $K \geq 3$, two non-identical *K*-swaps can share the same forward arcs.

where $((u_s, v_s))_{s=1}^{|\pi|}$ is an arbitrary ordering of the node assignments contained in $\pi$. For each $K'$-swap $C \in \mathcal{C}_{\pi,K'}$, let

$$F(C) := C \cap \pi \tag{6.10}$$
$$B(C) := \{(u,v) \in (V^G \cup \{\epsilon\}) \times (V^H \cup \{\epsilon\}) \mid (v,u) \in C \setminus \pi\} \tag{6.11}$$

be the sets of node assignments corresponding to forward and backward arcs contained in $C$, respectively. K-REFINE computes the swap cost

$$\texttt{SWAP-COST}(\pi, C) := c(P_\pi) - c(P_{\texttt{SWAP}(\pi,C)}), \tag{6.12}$$

where

$$\texttt{SWAP}(\pi, C) := (\pi \setminus F(C)) \cup \{(u,v) \in B(C) \mid (u,v) \neq (\epsilon, \epsilon)\} \tag{6.13}$$

is the node map obtained from $\pi$ by carrying out the swap encoded by $C$ (line 5). If $C$ yields an improvement, K-REFINE updates the best swap $C^\star$ and the best swap cost $\Delta^\star$ (lines 6 to 7).

Once all $K'$-swaps $C \in \mathcal{C}_{\pi,K'}$ have been visited, K-REFINE checks whether one of them yields an improvement w.r.t. the current node map $\pi$ (line 8). If this is the case, K-REFINE updates $\pi$ (lines 9 to 11) and resets the current swap size to $K' := 2$ (line 12). Otherwise, $K'$ is incremented (line 14). Subsequently, K-REFINE resets the best swap and the best swap cost to $C^\star := \emptyset$ and $\Delta^\star := 0$, respectively (line 15). Upon termination of the main while-loop, K-REFINE returns the current node map $\pi$ (line 16).

Assume that SWAP-COST($\pi, C$) can be computed in $O(\omega)$ time (cf. Section 6.6.2 for details). Furthermore, let $I \in \mathbb{N}$ be the number of times K-REFINE finds an improved node map in line 8. Note that, if the edit costs are integral, it holds that $I \leq c(P_\pi)$, where $\pi$ is K-REFINE's initial node map. Proposition 6.4 below implies that, for all $K' \in [K] \setminus \{1\}$ and each node map $\pi \in \Pi(G, H)$, we have $|\mathcal{C}_{\pi,K'}| = O((|V^G| + |V^H|)^{K'})$. Therefore, K-REFINE's overall runtime complexity is $O(I \cdot (|V^G| + |V^H|)^K \omega)$.

**Proposition 6.4 (Number of $K'$-Swaps).** *For each node map $\pi \in \Pi(G, H)$ and each $K' \in \mathbb{N}_{\geq 2}$, it holds that $|\mathcal{C}_{\pi,K'}| = \binom{|\pi|}{K'}(K'-1)!$.*

*Proof.* By construction, the forward arcs contained in $A_\pi$ are just the node assignments contained in $\pi$. As each $C \in \mathcal{C}_{\pi,K'}$ contains exactly $K'$ forward arcs, this implies $\mathcal{C}_{\pi,K'} = \bigcup_{S \in \binom{\pi}{K'}} \mathcal{C}_{\pi,K',S}$, where $\mathcal{C}_{\pi,K',S} := \{C \in \mathcal{C}_{\pi,K'} \mid C \cap \pi = S\}$ is the set of $K'$-swaps whose forward arcs are the ones contained in a fixed $K'$-element subset $S$ of $\pi$. Since $\mathcal{C}_{\pi,K',S} \cap \mathcal{C}_{\pi,K',S'} = \emptyset$ holds for all $S, S' \subseteq \pi$ with $|S| = |S'| = K'$ and $S \neq S'$, this observation implies the statement of the proposition, if we can show that

$$|\mathcal{C}_{\pi,K',S}| = (K-1)! \tag{6.14}$$

holds for all $S \subseteq \pi$ with $|S| = K'$.

Let $((u_s, v_s))_{s=1}^{K'}$ be an enumeration of a fixed $K'$-element subset $S \subseteq \pi$. By construction, $A_\pi$ contains backward arcs $(v_s, u_{s'})$ for all $s, s' \in [K']$ with $s \neq s'$. In order to construct a cycle of length $2K'$ in $G_\pi$ whose forward arcs are the ones contained in $S$, we start with an arbitrary forward arc $(u_s, v_s) \in S$. We have $|[K'] \setminus \{s\}| = K' - 1$ options for picking the first backward arc $(v_s, u_{s'})$, $|[K'] \setminus \{s, s'\}| = K' - 2$ options for picking the second backward arc $(v_{s'}, u_{s''})$, and so on. This consideration implies equation (6.14) and hence completes the proof. □

### 6.6.2 Efficient Computation of Swap Costs

Given a node map $\pi \in \Pi(G, H)$ and a $K'$-swap $C \in \mathcal{C}_{\pi,K'}$, let $\pi' := \text{SWAP}(\pi, C)$ be the node map obtained from $\pi$ by swapping the forward and backward arcs contained in $C$. Assume that $c(P_\pi)$ has already been computed. By equation (6.12), the swap cost SWAP-COST($\pi, C$) can be computed naïvely by computing the induced costs $c(P_{\pi'})$ of the swapped node map and then considering the difference between $c(P_\pi)$ and $c(P_{\pi'})$. By equation (2.1), this requires $O(\max\{|E^G|,|E^H|\})$ time. Since SWAP-COST$(\pi,C)$ has to be computed in every iteration of K-REFINE's inner for-loop, it is highly desirable to implement SWAP-COST$(\cdot,\cdot)$ more efficiently. The following Proposition 6.5 provides the key ingredient of a more efficient implementation.

**Proposition 6.5 (Efficient Computation of Swap Costs).** *Let $\pi \in \Pi(G,H)$ be a node map, $K' \in \mathbb{N}_{\geq 2}$ be a constant, and $C \in \mathcal{C}_{\pi,K'}$ be a $K'$-swap. Furthermore, let $V_C^G := \{u \in V^G \mid \exists v \in V^H \cup \{\epsilon\} : (u,v) \in F(C)\}$, $V_C^H := \{v \in V^H \mid \exists u \in V^G \cup \{\epsilon\} : (u,v) \in F(C)\}$, $E_C^G := \{e \in E^G \mid e \cap V_C^G \neq \emptyset\}$, and $E_C^H := \{f \in E^H \mid f \cap V_C^H \neq \emptyset\}$ be the sets of nodes and edges of G and H that are affected by the swap C. Then it holds that*

$$\begin{aligned}
\text{SWAP-COST}(\pi,C) = &\sum_{\substack{u \in V_C^G \\ \pi'(u) \neq \epsilon}} c_V(u, \pi'(u)) - \sum_{\substack{u \in V_C^G \\ \pi(u) \neq \epsilon}} c_V(u, \pi(u)) \\
&+ \sum_{\substack{u \in V_C^G \\ \pi'(u) = \epsilon}} c_V(u, \epsilon) - \sum_{\substack{u \in V_C^G \\ \pi(u) = \epsilon}} c_V(u, \epsilon) \\
&+ \sum_{\substack{v \in V_C^H \\ \pi'^{-1}(v) = \epsilon}} c_V(\epsilon, v) - \sum_{\substack{v \in V_C^H \\ \pi^{-1}(v) = \epsilon}} c_E(\epsilon, v) \\
&+ \sum_{\substack{e \in E_C^G \\ \pi'(e) \neq \epsilon}} c_E(e, \pi'(e)) - \sum_{\substack{e \in E_C^G \\ \pi(e) \neq \epsilon}} c_E(e, \pi(e)) \\
&+ \sum_{\substack{e \in E_C^G \\ \pi'(e) = \epsilon}} c_E(e, \epsilon) - \sum_{\substack{e \in E_C^G \\ \pi(e) = \epsilon}} c_E(e, \epsilon) \\
&+ \sum_{\substack{f \in E_C^H \\ \pi'^{-1}(f) = \epsilon}} c_E(\epsilon, f) - \sum_{\substack{f \in E_C^H \\ \pi^{-1}(f) = \epsilon}} c_E(\epsilon, f),
\end{aligned}$$

*where $\pi' :=$ SWAP$(\pi,C)$ is the node map obtained from $\pi$ via C.*

*Proof.* By construction of $V_C^G$ and $V_C^H$, we have $\pi(u) = \pi'(u)$, for all $u \in V^G \setminus V_C^G$, and $\pi^{-1}(v) = \pi'^{-1}(v)$, for all $v \in V^H \setminus V_C^H$. Similarly, $\pi(e) = \pi'(e)$ holds for all $e \in E^G \setminus E_C^G$, and $\pi^{-1}(f) = \pi'^{-1}(f)$ holds for all $f \in E^H \setminus E_C^H$. This implies the statement of the proposition. □

Proposition 6.5 implies that, for computing SWAP-COST$(\pi, C)$, it suffices to consider the nodes and edges contained in $V_C^G$, $V_C^H$, $E_C^G$, and $E_C^H$. By construction, we have $|V_C^G|, |V_C^H| \leq K'$, $|E_C^G| \leq K' \cdot \max \deg(G)$, and $|E_C^H| \leq$ $K' \cdot \max \deg(H)$. Since $K'$ is a constant, SWAP-COST$(\pi, C)$ can hence be computed in $O(\Delta_{\max}^{G,H})$ time. This is a significant improvement w. r. t. the naïve computation, which, as mentioned above, requires $O(\max\{|E^G|, |E^H|\})$ time.

### 6.6.3 Improvement of Upper Bound via Inclusion of Dummy Assignment

For each node map $\pi \in \Pi(G, H)$, let $S(\pi) := |\{(u,v) \in \pi \mid u \neq \epsilon \wedge v \neq \epsilon\}|$ denote the number of node substitutions contained in $\pi$. Now assume that K-REFINE as specified in Figure 6.8 is run from an initial node map $\pi \in \Pi(G, H)$ that does not contain the dummy assignment $(\epsilon, \epsilon)$. Since $\pi$ and $\pi \cup \{(\epsilon, \epsilon)\}$ induce the same edit path, this assumption is likely to hold in most implementations of K-REFINE. The following Proposition 6.6 shows that, under this assumption, the search space of K-REFINE is restricted in the sense that it includes only node maps $\pi' \in \Pi(G, H)$ with $S(\pi') \geq S(\pi)$. This has a negative effect on the quality of the upper bound produced by K-REFINE, as some potentially promising node maps are excluded a priori.

**Proposition 6.6 (Search Space of K-REFINE Without Dummy Assignment).** *Let $\pi \in \Pi(G, H)$ be a node map that satisfies $(\epsilon, \epsilon) \notin \pi$ and $\pi' \in \Pi(G, H)$ be the improved node map obtained from $\pi$ by running K-REFINE as specified in Figure 6.8. Then it holds that $S(\pi') \geq S(\pi)$.*

*Proof.* Let $\pi \in \Pi(G, H)$ be a node map that satisfies $(\epsilon, \epsilon) \notin \pi$, $K' \in \mathbb{N}_{\geq 2}$ be a constant, and $C \in \mathcal{C}_{\pi, K'}$ be a $K'$-swap. By definition of $B(C)$, we have $(\epsilon, \epsilon) \notin \text{SWAP}(\pi, C)$. Therefore, the proposition follows by induction on the number of times K-REFINE finds an improved node map in line 8, if we can show that $S(\text{SWAP}(\pi, C)) \geq S(\pi)$.

To show this inequality, we define $S_F(C) := |\{(u,v) \in F(C) \mid u \neq \epsilon \wedge v \neq \epsilon\}|$, $D_F(C) := |\{(u,v) \in F(C) \mid u \neq \epsilon \wedge v = \epsilon\}|$, $I_F(C) := |\{(u,v) \in F(C) \mid u = \epsilon \wedge v \neq \epsilon\}|$, and $S_F^\epsilon(C) := |\{(u,v) \in F(C) \mid u = \epsilon \wedge v = \epsilon\}|$. $S_B(C)$, $D_B(C)$, $I_B(C)$, and $S_B^\epsilon(C)$ are defined analogously. Furthermore, we introduce $E(C)$ as the number of dummy nodes contained in $C$. By construction, the following equations hold:

$$K' = |F(C)| = S_F(C) + D_F(C) + I_F(C) + S_F^\epsilon(C) \qquad (6.15)$$
$$K' = |B(C)| = S_B(C) + D_B(C) + I_B(C) + S_B^\epsilon(C) \qquad (6.16)$$
$$E(C) = D_F(C) + I_F(C) + 2 \cdot S_F^\epsilon(C) \qquad (6.17)$$

$$E(C) = D_B(C) + I_B(C) + 2 \cdot S_B^\epsilon(C) \tag{6.18}$$

By definition of SWAP$(\pi, C)$, we have $S(\text{SWAP}(\pi, C)) = S(\pi) + S_B(C) - S_F(C)$. It hence remains to be shown that $S_B(C) \geq S_F(C)$. The equations (6.15) to (6.15) imply $S_F(C) = K' - E(C) + S_F^\epsilon(C)$ and $S_B(C) = K' - E(C) + S_B^\epsilon(C)$. Since $(\epsilon, \epsilon) \notin \pi$, we additionally know that $S_F^\epsilon(C) = 0$. We hence obtain $S_B(C) = S_F(C) + S_B^\epsilon(C) \geq S_F(C)$, as required. □

The proof of Proposition 6.6 tells us how we have to modify K-REFINE in order to ensure that node maps with fewer node substitutions than the initial node map are contained its search space: We simply have to update the current node map $\pi$ as $\pi := \pi \cup \{(\epsilon, \epsilon)\}$ before enumerating all $K'$-swaps $C \in \mathcal{C}_{\pi,K'}$ in line 4 of Figure 6.8. With this modification, the dummy assignment $(\epsilon, \epsilon)$ can appear as a forward arc in a $K'$-swap $C$, which allows the number of node substitutions to decrease.

## 6.7 Generation of Initial Solutions for Local Search

In this section, we present RANDPOST, a framework that extends the paradigm LS-GED presented in Section 6.1.3 and can be used to improve any local search based algorithm for upper bounding GED. Intuitively, RANDPOST iteratively runs a given local search algorithm. In each iteration, previously computed locally optimal node maps are combined stochastically to obtain new promising initial node maps to be used in the next iteration.

Figure 6.10 provides an overview of the framework. Given a set of initial node maps $\mathcal{S}_0 \subseteq \Pi(G, H)$ with $|\mathcal{S}_0| = K$, a constant $\rho \in (0, 1]$, and a local search algorithm ALG, RANDPOST computes a set $\mathcal{S}_0' \subseteq \Pi(G, H)$ of improved node maps with $|\mathcal{S}_0'| = \lceil \rho \cdot K \rceil$ by (parallelly) running ALG on all initial node maps and terminating once $\lceil \rho \cdot K \rceil$ runs have converged (line 1). Subsequently, the upper bound $UB$ is initialized as the cost of the cheapest induced edit path encountered so far (line 2). Note that, up to this point, RANDPOST is equivalent to the LS-GED extension MULTI-START described in Section 6.1.3.5 above.

RANDPOST now initializes a matrix $\mathbf{M} \in \mathbb{R}^{(|V^G|+1) \times (|V^H|+1)} := \mathbf{0}_{|V^G|+1, |V^H|+1}$ that contains scores $m_{i,k}$ for each possible node assignment $(u_i, v_k) \in (V^G \cup \{\epsilon\}) \times (V^H \cup \{\epsilon\})$ (line 3). The score for each substitution $(u_i, v_k) \in V^G \times V^H$ is represented by the value $m_{i,k}$, while the scores for the deletion $(u_i, \epsilon)$ and

**Input**: Graphs $G$ and $H$, constants $K \in \mathbb{N}_{\geq 1}$, $L \in \mathbb{N}$, $\rho \in (0,1]$, and $\eta \in [0,1]$,
    local search algorithm `ALG`, initial node map set $\mathcal{S}_0 \subseteq \Pi(G,H)$ with
    $|\mathcal{S}_0| = K$, lower bound $LB$ for $\mathrm{GED}(G,H)$.
**Output**: Upper bound $UB$ for $\mathrm{GED}(G,H)$.

1. $\mathcal{S}'_0 := \mathtt{ALG}(\mathcal{S}_0, \rho)$;  // run local search on initial node maps
2. $UB := \min_{\pi' \in \mathcal{S}'_0} c(P_{\pi'})$;  // set first upper bound
3. $\mathbf{M} := \mathbf{0}_{|V^G|+1, |V^H|+1}$;  // initialize scores matrix
4. **for** $r \in [L]$ **do**  // main loop
5.     $\mathbf{M} := \mathtt{UPDATE\text{-}SCORES}(\mathbf{M}, \mathcal{S}'_{r-1}, \eta, LB, UB)$;  // update scores matrix
6.     $\mathcal{S}_r := \mathtt{GENERATE\text{-}NODE\text{-}MAPS}(\mathbf{M}, K)$;  // generate new initial node maps
7.     $\mathcal{S}'_r := \mathtt{ALG}(\mathcal{S}_r, \rho)$;  // run local search on new initial node maps
8.     $UB := \min\{UB, \min_{\pi' \in \mathcal{S}'_r} c(P_{\pi'})\}$;  // update upper bound
9. **return** $UB$;  // return upper bound

**Figure 6.10.** The framework `RANDPOST`.

the insertion $(\epsilon, v_k)$ are represented by the values $m_{i,|V^H|+1}$ and $m_{|V^G|+1,k}$, respectively. Throughout the algorithm, $\mathbf{M}$ is maintained in such a way that $m_{i,k}$ is large just in case the corresponding node assignment appears in many cheap locally optimal node maps.

After initializing $\mathbf{M}$, `RANDPOST` carries out $L$ iterations of its main for-loop, where $L \in \mathbb{N}$ is a meta-parameter (lines 4 to 8). Inside the $r^{\mathrm{th}}$ iteration, `RANDPOST` starts by updating the scores matrix $\mathbf{M}$ by calling $\mathtt{UPDATE\text{-}SCORES}(\mathbf{M}, \mathcal{S}'_{r-1}, \eta, LB, UB)$, where $\mathbf{M}$ is the current scores matrix, $\mathcal{S}'_{r-1} \subseteq \Pi(G,H)$ is the set of improved node maps obtained from the previous iteration, $\eta \in [0,1]$ is a meta-parameter that can be used to give greater weight to cheap node maps, $LB$ is a previously computed lower bound for $\mathrm{GED}(G,H)$ (possibly 0), and $UB$ is the current upper bound (line 5). Let the matrix $\mathbf{M}' \in \mathbb{R}^{(|V^G|+1) \times (|V^H|+1)}$ be defined as $\mathbf{M}' := \mathtt{UPDATE\text{-}SCORES}(\mathbf{M}, \mathcal{S}'_{r-1}, \eta, LB, UB) \in \mathbb{R}^{(|V^G|+1) \times (|V^H|+1)}$. Then $\mathbf{M}'$ is given as

$$\mathbf{M}' := \mathbf{M} + \sum_{\pi' \in \mathcal{S}'_{r-1}} \left[ (1-\eta) + \eta \frac{UB - LB}{c(P_{\pi'}) - LB} \right] \mathbf{X}',$$

where $\mathbf{X}' \in \Pi_{|V^G|, |V^H|, \epsilon}$ is the matrix representation of the improved node map $\pi' \in \mathcal{S}'_{r-1}$, i.e., for all $u_i \in V^G$ and all $v_k \in V^H$, we have $x'_{i,k} = 1$ just in case $(u_i, v_k) \in \pi'$, $x'_{i,|V^H|+1} = 1$ just in case $(u_i, \epsilon) \in \pi'$, and $x'_{|V^G|+1,k} = 1$ just in case $(\epsilon, v_k) \in \pi'$. Note that, if $\eta = 0$, $m_{i,k}$ represents the number of converged local optima that contain the corresponding assignment. If $\eta > 0$, assignments that appear in node maps with lower costs receive higher scores.

Once **M** has been updated, RANDPOST creates a new $K$-sized set $\mathcal{S}_r \subseteq \Pi(G, H)$ of initial node maps by calling GENERATE-NODE-MAPS(**M**, $K$) (line 6). GENERATE-NODE-MAPS(**M**, $K$) works as follows: For each of the first $|V^G|$ rows $\mathbf{M}_i$ of **M**, RANDPOST draws a column $k \in [|V^H| + 1]$ from the distribution encoded my $\mathbf{M}_i$. If $k = |V^H| + 1$, the node deletion $(u_i, \epsilon)$ is added to the node map $\pi$ that is being constructed. Otherwise, the substitution $(u_i, v_k)$ is added to $\pi$, the score $m_{j,k}$ is temporarily set to 0 for all $j \in [|V^G|] \setminus [i]$, and the column $k$ is marked as covered. Once all nodes of $G$ have been processed, node insertions $(\epsilon, v_k)$ are added to $\pi$ for all uncovered columns $k \in [|V^H|]$. This process is repeated until $K$ different node maps have been created.

After creating the set $\mathcal{S}_r$ of new initial node maps, RANDPOST re-runs MULTI-START from $\mathcal{S}_r$. That is, RANDPOST computes a new $\lceil \rho \cdot K \rceil$-sized set $\mathcal{S}'_r \subseteq \Pi(G, H)$ of improved node maps by (parallelly) running the local search algorithm ALG on the initial node maps contained in $\mathcal{S}_r$ and terminating once $\lceil \rho \cdot K \rceil$ runs have converged (line 7). Subsequently, the upper bound is updated as the minimum of the current upper bound and the cost of the cheapest edit path induced by one of the newly computed improved node maps (line 8). Finally, RANDPOST returns the best encountered upper bound (line 9).

## 6.8 Empirical Evaluation

In this section, we report the results of four series of experiments which we carried out to empirically evaluate the heuristic algorithms presented in this chapter. In Section 6.8.1, we evaluate the quality of the lower bounds produced by the algorithms BRANCH, BRANCH-FAST, and BRANCH-TIGHT presented in Sections 6.2 to 6.3. In Section 6.8.2, we compare the LSAPE based algorithms RING and RING-ML presented in Section 6.4 to other instantiations of LSAPE-GED that are designed to yield tight upper bounds, and evaluate the extension MULTI-SOL of the paradigm LSAPE-GED proposed in Section 6.5. In Section 6.8.3, we evaluate how the local search algorithm K-REFINE presented in Section 6.6 performs w. r. t. other instantiations of LS-GED, and test to what extent the framework RANDPOST presented in Section 6.7 improves the upper bounds produced by heuristics based on local search. In Section 6.8.4 we jointly evaluate the newly proposed techniques, compare them to the existing

GED heuristics presented in Section 6.1, and address the meta-questions Q-1 and Q-2.

### 6.8.1 Evaluation of BRANCH, BRANCH-FAST, and BRANCH-TIGHT

In a first series of experiments, we evaluated the quality of the lower bounds produced by the algorithms BRANCH, BRANCH-FAST, and BRANCH-TIGHT. In Section 6.8.1.1 we describe the setup of the experiments; in Section 6.8.1.2, we report the results; and in Section 6.8.1.3, we concisely summarize the most important experimental findings.

#### 6.8.1.1 Setup and Datasets

**Datasets and Edit Costs.** We tested on the datasets AIDS and PROTEIN. Tests were carried out on both uniform and non-uniform edit costs. We used the numeric attributes of the graphs contained in AIDS and PROTEIN to define the non-uniform, Euclidean node and edge substitution costs, and parameterized the deletion and insertion costs as well as the importance of edge edit operations w.r.t. node edit operations. More precisely, we set $c_V(\alpha, \epsilon) := c_V(\epsilon, \alpha) := \rho \max\{c_V(\beta, \gamma) \mid \beta, \gamma \in \Sigma_V\}$, for all $\alpha \in \Sigma_V$, and $c_E(\alpha, \epsilon) := c_E(\epsilon, \alpha) := \rho \max\{c_E(\beta, \gamma) \mid \beta, \gamma \in \Sigma_E\}$, for all $\alpha \in \Sigma_E$. Increasing $\rho$ increases the importance of deletions and insertions w.r.t. substitutions. We also introduced a multiplicative parameter $\mu$ for the edge edit costs $c_E$, which allows to vary the importance of edit operations on edges w.r.t. edit operations on nodes.

**Compared Methods.** We evaluated BRANCH, BRANCH-FAST, and BRANCH-TIGHT by comparing them to state of the art competitors. For non-uniform metric edit costs, we compared BRANCH, BRANCH-FAST, and BRANCH-TIGHT to NODE and ADJ-IP [60]. For uniform edit costs, we compared BRANCH-TIGHT to NODE, ADJ-IP, BRANCH-CONST, and HYBRID. We did not include BRANCH and BRANCH-FAST in the tests on uniform edit costs, because, for uniform edit costs, they are equivalent to BRANCH-CONST.

**Protocol and Test Metrics.** Before comparing BRANCH-TIGHT to its competitors, we tested how its performance is affected by the termination criterion $(LB_{r+1} - LB_r)/LB_r < \varepsilon$ and the maximum number of iterations $I$. To this purpose, we varied $\varepsilon$ and $I$ over the sets $\{2^k \mid k \in \mathbb{Z} \land -13 \leq k \leq -3\}$ and $\{5k \mid k \in \mathbb{N} \land k \leq 20\}$, respectively. For both datasets, we randomly selected a sample $\mathcal{G}$ of size 100, ran `BRANCH-TIGHT` on each pair $\{G, H\} \in \binom{\mathcal{G}}{2}$, and averaged the resulting lower bound over all test runs. When comparing `BRANCH`, `BRANCH-FAST`, and `BRANCH-TIGHT` to their competitors, we measured the accuracy of an algorithm `ALG` as the relative deviation

$$\mathrm{dev}(\mathtt{ALG}) := LB^\star / LB(\mathtt{ALG})$$

of its lower bound from the best lower bound produced by one of the tested algorithms. If $\mathrm{dev}(\mathtt{ALG})$ is close to 1, $LB(\mathtt{ALG})$ is tight, while $\mathrm{dev}(\mathtt{ALG}) \gg 1$ means that $LB(\mathtt{ALG})$ is loose. For each dataset $\mathcal{D}$, we randomly selected disjoint groups $\mathcal{G}_i \subset \mathcal{D}$ of size at most 10 containing graphs $G$ with $10(i-1) < |V^G| \leq 10i$, for all $i \in [N(\mathcal{D})]$. $N(\mathcal{D})$ is defined as $N(\mathcal{D}) := \max_{G \in \mathcal{D}} \lceil |V^G|/10 \rceil$, and we hence have $N(\mathrm{AIDS}) = 10$ and $N(\mathrm{PROTEIN}) = 13$ (cf. Table 2.3). For each group $\mathcal{G}_i$, we ran each tested algorithm `ALG` on each pair $\{G, H\} \subset \mathcal{G}_i$ and averaged the runtime and relative deviation $\mathrm{dev}(\mathtt{ALG})$ over all test runs associated to $\mathcal{G}_i$. For non-uniform edit costs, we also evaluated the effect of the cost parameters $\rho$ and $\mu$ by varying them over the sets $\{2^k \mid k \in \mathbb{Z} \land -1 \leq k \leq 9\}$ and $\{2^k \mid k \in \mathbb{Z} \land -5 \leq k \leq 5\}$, respectively. We randomly selected a sample $\mathcal{G}$ containing 100 graphs from each dataset, ran each tested algorithm `ALG` on each pair $\{G, H\} \subset \mathcal{G}$, and averaged its relative deviation and runtime over all test runs. Recall that increasing $\rho$ increases the importance of node and edge insertions and deletions w. r. t. substitutions, while increasing $\mu$ increases the importance of edit operations on edges w. r. t. the importance of operations on nodes.

**Implementation and Hardware Specifications.** For implementing `ADJ-IP`, we used the industrial LP-solver Gurobi Optimization [54]. `HYBRID` was implemented with the parameter $T$ set to 4. We deviated from the suggested value $T = 8$, because implementing `HYBRID` with $T > 4$ is not well documented in [113] and the authors did not reply to our request to share their implementation. All algorithms were set up to return the currently maintained lower bound if a time limit of one minute was exceeded. All algorithms were implemented in C++ and tests were run on a MacBook Pro with a 2.7 GHz Intel Core i5 processor and 8 GB of main memory running OS X El Capitan.

#### 6.8.1.2 Results of the Experiments

**Effect of Termination Criteria on `BRANCH-TIGHT`.** Figure 6.11 shows the effect of $I$ and $\varepsilon$ for non-uniform edit costs. As expected, `BRANCH-TIGHT`'s runtime was linear in $I$. On both datasets, `BRANCH-TIGHT` converged very fast. For $\varepsilon < 2^{-10}$, the lower bound improved only marginally while the increase in the number of iterations became more and more significant. We therefore set $\varepsilon := 2^{-10} \approx 1\,\text{\textperthousand}$ for all other experiments, which amounts to running `BRANCH-TIGHT` with $I \approx 20$. The results for uniform edit costs are similar.

**Effect of Graph Sizes for Uniform Edit Costs.** Figure 6.12 shows the effect of the graph sizes for uniform edit costs. In terms of runtime, our algorithm `BRANCH-TIGHT` performed similarly to `ADJ-IP` and, unsurprisingly, worse than `NODE` and `BRANCH-CONST`, which, unlike `BRANCH-TIGHT`, are optimized for runtime. `BRANCH-TIGHT` was faster than the exponential algorithm `HYBRID`, which performed erratically. While `HYBRID` was very fast for some input graphs, it exceeded the time limit of one minute for around 2 % of all test runs on PROTEIN and for around 22 % of all test runs in AIDS. At the same time, `BRANCH-TIGHT`'s maximal runtime was 16.43 seconds on PROTEIN and 5.66 seconds on AIDS. In terms of accuracy, `BRANCH-TIGHT` performed excellently and clearly produced the tightest lower bound. On both datasets and across all test groups, dev(`BRANCH-TIGHT`) is very close or even equal to 1. The relatively bad performance of `HYBRID` on AIDS is due to the fact that, on around 24 % of all test runs, it did not terminate within the given time limit of one minute and that the partition-based lower bound it maintains before termination is typically loose.

**Effect of Graph Sizes for Non-Uniform Edit Costs.** Figure 6.13 shows the effect of the graph sizes for non-uniform edit costs. Just as for uniform edit costs, `BRANCH-TIGHT` and `ADJ-IP` performed similarly in terms of runtime, while `BRANCH`, `BRANCH-FAST`, and `NODE` were significantly faster. What is surprising at a first glance is that `BRANCH-FAST` was slightly faster than `NODE`, although `NODE` has a smaller runtime complexity than `BRANCH-FAST` (cf. Section 6.1.1.1 and Section 6.2.2). The explanation is that `BRANCH-FAST` invests more effort in the computation of discriminative auxiliary edge costs and that the LSAPE solver called by both algorithms is faster if the variance in the costs is high. `BRANCH-TIGHT` again produced the tightest lower bound of all tested algorithms. On AIDS, we have dev(`BRANCH`) = dev(`BRANCH-FAST`) > dev(`ADJ-IP`), while, on PROTEIN, dev(`BRANCH-FAST`) > dev(`ADJ-IP`) > dev(`BRANCH-FAST`) holds. The explanation for the same deviations of BRANCH and BRANCH-FAST on AIDS is that the dataset contains only two different edge labels, which implies that both algorithms compute the same lower bound (cf. Section 6.2.2). The explanation for dev(ADJ-IP) < dev(BRANCH) on AIDS, while the opposite is observed on PROTEIN, is that the accuracy of ADJ-IP significantly decreases with increasing importance of edge edit operations (cf. Figure 6.14). As the average degree of the PROTEIN graphs is greater than the one of the AIDS graphs (cf. Table 2.3), edit operations on edges are more important on PROTEIN than on AIDS.

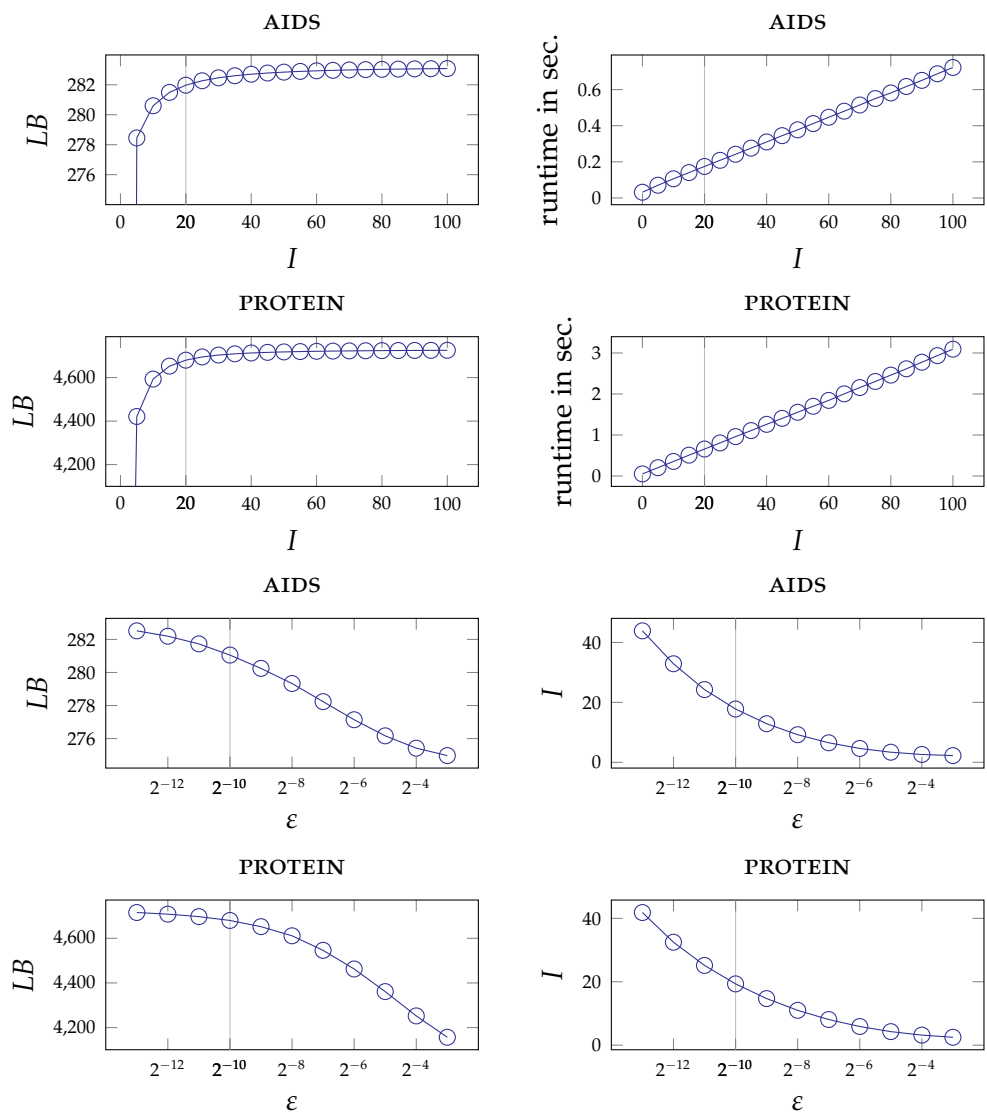

**Figure 6.11.** Effect of $I$ and $\varepsilon$ on `BRANCH-TIGHT` for non-uniform edit costs and cost parameters $\rho = \mu = 1$.

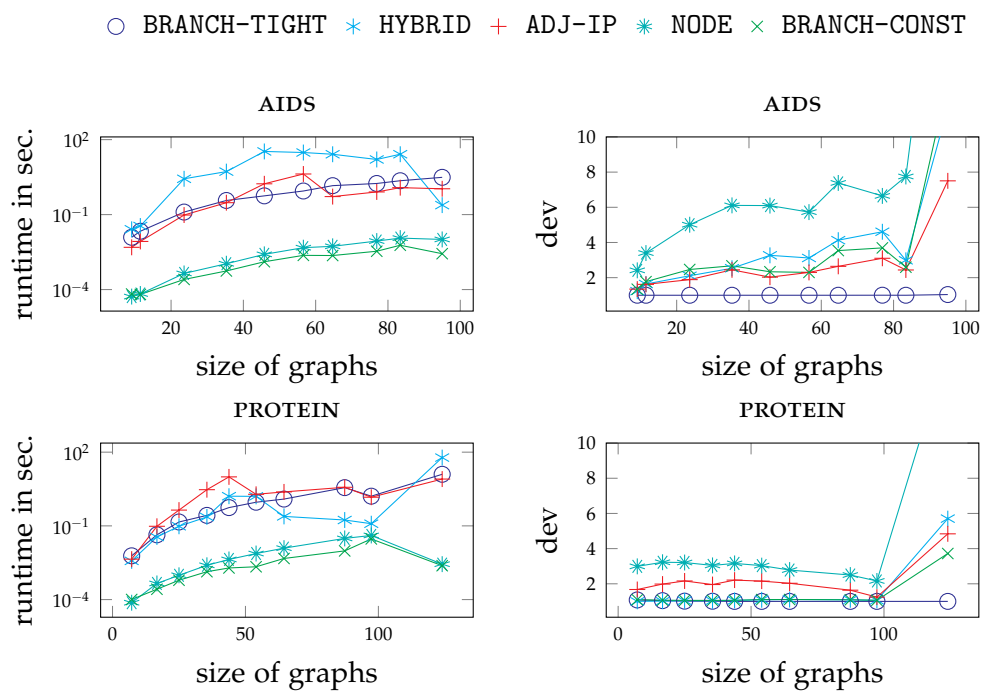

**Figure 6.12.** Effect of graph sizes for uniform edit costs.

**Effect of Cost Parameters.** Figure 6.14 shows the effect of the cost parameters. For better readability, no curve for the deviation of NODE is shown. NODE was always much looser than all other algorithms. We see that, independently of $\rho$ and $\mu$, BRANCH-TIGHT produced the tightest lower bound. Moreover, the performances of BRANCH-FAST and ADJ-IP significantly deteriorated with decreasing $\rho$ and increasing $\mu$. The reason for this is that ADJ-IP does not model edge substitutions at all and that BRANCH-FAST models them only in a rough way. At the same time, they both fully model edge insertions and deletions as well as edit operations on nodes. While, on AIDS, BRANCH computes the same lower bound as BRANCH-FAST (cf. Figure 6.13 and Figure 6.12), on PROTEIN, it reacts much less sensibly to decreasing $\rho$ and increasing $\mu$ than BRANCH-FAST and ADJ-IP. This is because, unlike ADJ-IP and BRANCH-FAST, BRANCH fully models edge substitutions. The runtimes of all algorithms turned out to be independent of $\rho$ and $\mu$.

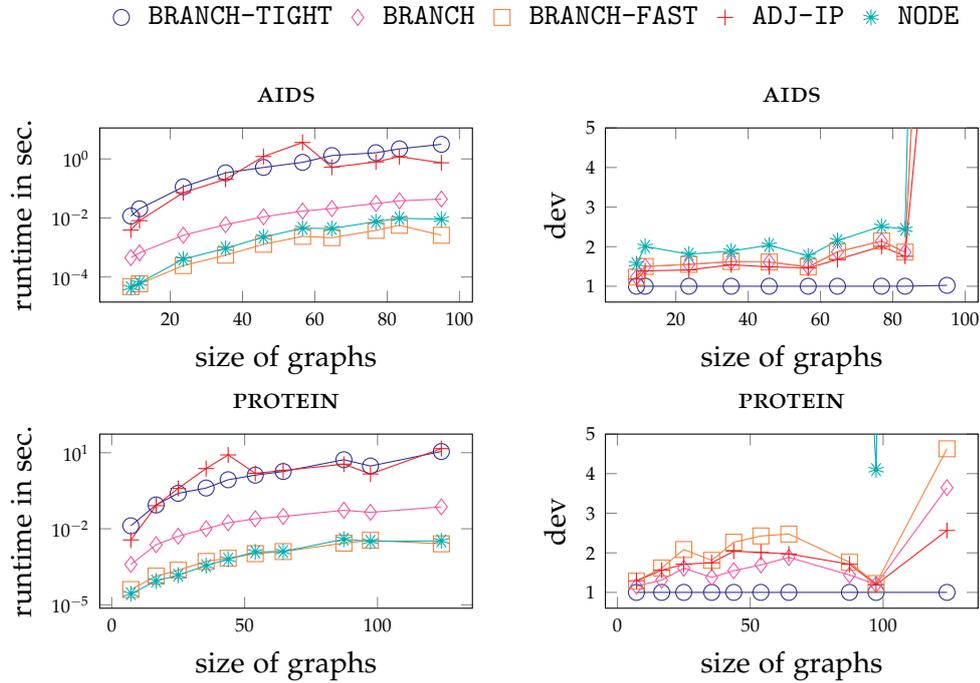

**Figure 6.13.** Effect of graph size for non-uniform edit costs and cost parameters $\rho = \mu = 1$.

**Overall Performance of Compared Methods.** Figure 6.15 shows the excellent runtime/accuracy-tradeoffs of our algorithms. Averages are taken over all test runs and dotted lines demarcate the regions that are dominated by our algorithms. For both uniform and non-uniform edit costs, all proposed algorithms are Pareto optimal: There are no competitors that performed better both in terms of runtime and in terms of accuracy. Furthermore, BRANCH-TIGHT always produced the tightest lower bound. In particular, on both AIDS and PROTEIN with uniform edit costs, BRANCH-TIGHT dominates HYBRID. ADJ-IP, the second competitor for tight lower bounds, is dominated on PROTEIN and almost dominated on AIDS, where it was 1.06 times faster but 2.72 times looser than `BRANCH-TIGHT`. On PROTEIN with non-uniform edit costs, `BRANCH-FAST` dominates `NODE` and `BRANCH-TIGHT` dominates `ADJ-IP`. On AIDS with non-uniform edit costs, `NODE` is dominated, too, while `ADJ-IP` is almost dominated as it was 1.20 times faster and 2.69 times less accurate than `BRANCH-TIGHT`.

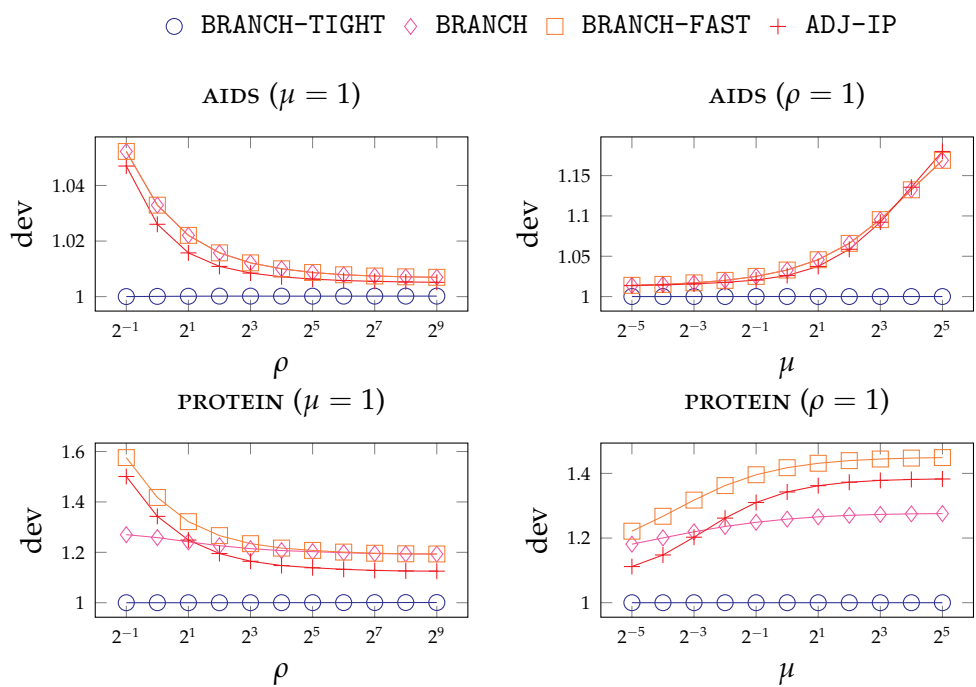

**Figure 6.14.** Effect of cost parameters $\rho$ and $\mu$.

#### 6.8.1.3 Upshot of the Experiments

The empirical evaluation of `BRANCH`, `BRANCH-FAST`, and `BRANCH-TIGHT` shows that all three algorithms yield excellent tradeoffs between runtime and tightness of the produced lower bounds. In particular, `BRANCH-TIGHT` always produced the tightest lower bound of all tested algorithm, and performed much better than all competitors when edit costs were used that emphasize the importance of edge edit operations.

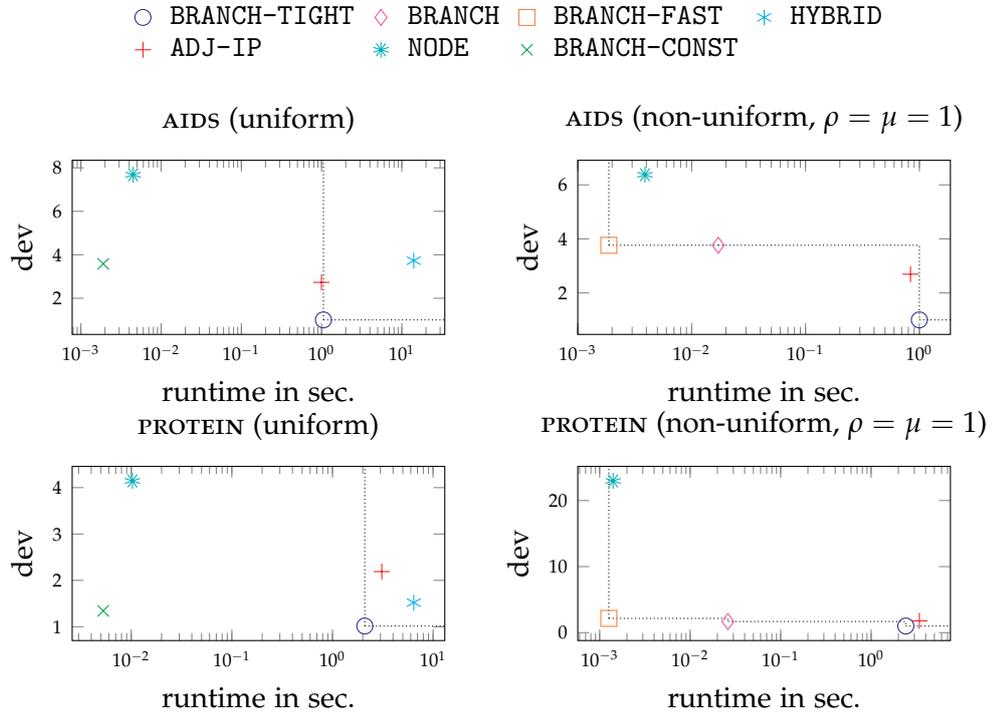

**Figure 6.15.** Relative deviation vs. runtime.

### 6.8.2 Evaluation of `RING`, `RING-ML`, and `MULTI-SOL`

In a second series of experiments, we evaluated the quality of the upper bounds produced by the algorithms `RING` and `RING-ML`. Moreover, we test how the `LSAPE-GED` extension `MULTI-SOL` affects the performance of LSAPE based heuristics. In Section 6.8.2.1 we describe the setup of the experiments; in Section 6.8.2.2, we report the results; and in Section 6.8.2.3, we concisely summarize the most important experimental findings.

#### 6.8.2.1 Setup and Datasets

**Datasets and Edit Costs.** We tested on the datasets LETTER (H), PAH, and AIDS. For LETTER (H), we used the edit costs suggested in [84]. For PAH and AIDS, we employed the edit costs defined in [1] (cf. Section 2.4).

**Compared Methods.** We tested three variants of our classical `LSAPE-GED` instantiation `RING`: `RING-OPT` uses optimal LSAPE for defining the set distances $d_{\mathcal{P}(V)}^{G,H}$ and $d_{\mathcal{P}(E)}^{G,H}$, `RING-GD` uses greedy LSAPE, and `RING-MS` uses the multiset intersection based approach. Our method `RING-ML` was tested with three different machine learning techniques: SVCs with RBF kernel and probability estimates [90], fully connected feed-forward DNNs [39], and 1-SVMs with RBF kernel.

We compared our methods to the `LSAPE-GED` instantiations `BP` (cf. Section 6.1.1.2), `SUBGRAPH` (cf. Section 6.1.1.5), and `WALKS` (cf. Section 6.1.1.6). Like `RING` and `RING-MS`, `SUBGRAPH` and `WALKS` are designed to yield tight upper bounds. `BP` was included as a baseline. As `WALKS` assumes that the costs of all edit operation types are constant, we slightly extended it by averaging the costs before each run. To handle the exponential complexity of `SUBGRAPH`, we enforced a time limit of 1 ms for computing a cell $c_{i,k}$ of its LSAPE instance. Furthermore, we compared to the machine learning based method `PREDICT` (cf. Section 6.4.1.2), which we tested with the same probability estimates as `RING-ML`. Since some of our test graphs have symbolic labels and not all of them are of the same size, we did non include the method `NGM` suggested in [39]. For all methods, we varied the number of threads and the maximal number of LSAPE solutions over the set $\{1, 4, 7, 10\}$, and we parallelized the construction of the LSAPE instance **C** (cf. Section 6.4.1.1 and Section 6.4.1.2).

**Choice of Meta-Parameters and Training of Machine Learning Based Methods.** For learning the meta-parameters of `RING-OPT`, `RING-GD`, `RING-MS`, `SUBGRAPH`, and `WALKS`, and training the DNNs, the SVCs, and the 1-SVMs, we picked a training set $\mathcal{S}_1 \subset \mathcal{D}$ with $|\mathcal{S}_1| = 50$ for each dataset $\mathcal{D}$. Following [32, 50], we learned the parameter $L$ of the methods `SUBGRAPH` and `WALKS` by minimizing the average upper bound on $\mathcal{S}_1$ over $L \in \{1, 2, 3, 4, 5\}$. For choosing the meta-parameters of the variants of `RING`, we set the tuning parameter $\mu$ to 1 and initialized our blackbox optimizer with 100 randomly constructed simplex vectors $\boldsymbol{\alpha}$ and $\boldsymbol{\lambda}$ (cf. Section 6.4.4.1).

For determining the network structure of the fully connected feed-forward DNNs, we carried out 5-fold cross validation, varying the number of hidden layers, the number of neurons per hidden layers, and the activation function at hidden layers over the grid $[10] \times [20] \times \{\text{ReLU}, \text{Sigmoid}\}$. Similarly, we determined the meta-parameters $C$ and $\gamma$ of the SVC via 5-fold cross-validation on the grid $\{10^k \mid k \in \mathbb{Z} \wedge -3 \leq k \leq 3\} \times \{10^k \mid k \in \mathbb{Z} \wedge -3 \leq k \leq 3\}$. For the 1-SVM, we set the meta-parameter $\gamma$ to $1/\dim(\mathcal{F})$, where $\dim(\mathcal{F})$ is the dimensionality of the feature vectors. For the ground truth, we used `IPFP` with `MULTI-START` (cf. Section 6.1.3.4 and Section 6.1.3.5) to compute a node map $\pi \in \Pi(G, H)$ which is close to optimal for each $(G, H) \in \mathcal{S}_1 \times \mathcal{S}_1$. In order to ensure that the training data $\mathcal{T}$ used by DNN and SVC is balanced, we randomly picked only $|\pi|$ node assignments $(u, v) \notin \pi$ for each ground truth node map $\pi \in \Pi(G, H)$.

**Protocol and Test Metrics.** For each dataset $\mathcal{D}$, we randomly selected a test set $\mathcal{S}_2 \subseteq \mathcal{D} \setminus \mathcal{S}_1$ with $|\mathcal{S}_2| = \min\{100, |\mathcal{D} \setminus \mathcal{S}_1|\}$, and ran each method on each pair $(G, H) \in \mathcal{S}_2 \times \mathcal{S}_2$ with $G \neq H$. We recorded the average runtime in seconds ($t$), the average value of the returned upper bound for GED ($d_{UB}$), and the ratio of graphs which are correctly classified if the returned upper bound is employed in combination with a 1-NN classifier ($r_{UB}$).

**Implementation and Hardware Specifications.** All methods are implemented in C++ and use the same implementation of the optimal LSAPE solver proposed in [23]. We used NOMAD [64] as our blackbox optimizer, LIBSVM [33] for implementing SVCs and 1-SVMs, and FANN [80] for implementing DNNs.

The test sources and datasets are distributed with GEDLIB (cf. Appendix A).[4] Tests were run on a machine with two Intel Xeon E5-2667 v3 processors with 8 cores each and 98 GB of main memory running GNU/Linux.

#### 6.8.2.2 Results of the Experiments

**Effect of Machine Learning Techniques.** Table 6.2 shows the performances of different machine learning techniques when used in combination with the feature vectors defined by `RING-ML` and `PREDICT`. We see that, in terms of classification ratio and tightness of the produced upper bounds, the best results were achieved by 1-SVMs with RBF kernels. Using DNNs had a beneficial effect on the runtime but results in dramatically lower classification ratios and looser upper bounds. Using SVCs with RGB kernels instead of 1-SVMs negatively affected all three test metrics. In the following, we will therefore only report the results for 1-SVMs and DNNs.

---

[4]Sources and datasets are available at `https://github.com/dbblumenthal/gedlib`.

**Table 6.2.** Effect of machine learning techniques on `RING-ML` and `PREDICT`. Number of threads and maximal number of LSAPE solutions are fixed to 10. Results for AIDS are similar.

| machine learning technique | $d_{UB}$ | $t$ | $r_{UB}$ | $d_{UB}$ | $t$ | $r_{UB}$ |
|---|---|---|---|---|---|---|
| | | LETTER (H) | | | | |
| *feature vectors* | RING-ML [this thesis] | | | PREDICT [90] | | |
| DNN [39] | 8.24 | **2.99 · 10⁻⁴** | 0.20 | 8.19 | **1.48 · 10⁻⁴** | 0.22 |
| SVC [90] | **5.68** | 6.47 · 10⁻³ | 0.73 | **5.22** | 2.82 · 10⁻³ | 0.76 |
| 1-SVM [this thesis] | 6.07 | 2.58 · 10⁻³ | **0.81** | 6.07 | 2.07 · 10⁻³ | **0.81** |
| | | PAH | | | | |
| *feature vectors* | RING-ML [this thesis] | | | PREDICT [90] | | |
| DNN [39] | 25.29 | **5.69 · 10⁻³** | 0.65 | 44.03 | **1.23 · 10⁻³** | 0.56 |
| SVC [90] | 31.91 | 7.19 · 10⁻¹ | 0.62 | 36.68 | 3.40 · 10⁻¹ | 0.65 |
| 1-SVM [this thesis] | **24.55** | 2.12 · 10⁻¹ | **0.71** | **24.56** | 1.14 · 10⁻¹ | **0.71** |

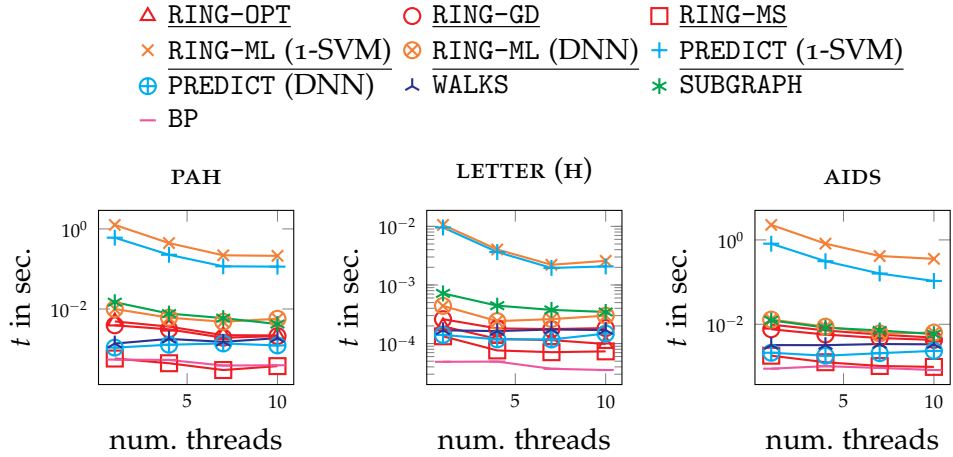

**Figure 6.16.** Number of threads vs. runtimes. Maximal number of LSAPE solutions is fixed to 10. Underlined methods use techniques proposed in this thesis.

**Effect of Number of Threads and Maximal Number of LSAPE Solutions.**
Figure 6.16 and Figure 6.17 show the effect of varying the number of threads and the maximal number of LSAPE solutions. We see that, unsurprisingly, the slower the method, the more it benefited from parallelization, with the only exception of `WALKS`. This can be explained by the fact that computing the local structure distances employed by `WALKS` requires a lot of pre-computing that cannot be parallelized (cf. [50] and Section 6.1.1.6 for details). Increasing the maximal number of LSAPE solutions tightened the upper bounds of mainly those methods that yielded loose upper bounds if run with only one solution. The outlier of SUBGRAPH on the LETTER (H) dataset is due to the fact that SUBGRAPH was run with a time limit on the computation of the subgraph distances; and that the algorithm used for solving these subproblems is guaranteed to always return the same value only if it is run to optimality. Also note that increasing the maximal number of LSAPE solutions significantly increased the runtimes of only the fastest algorithms.

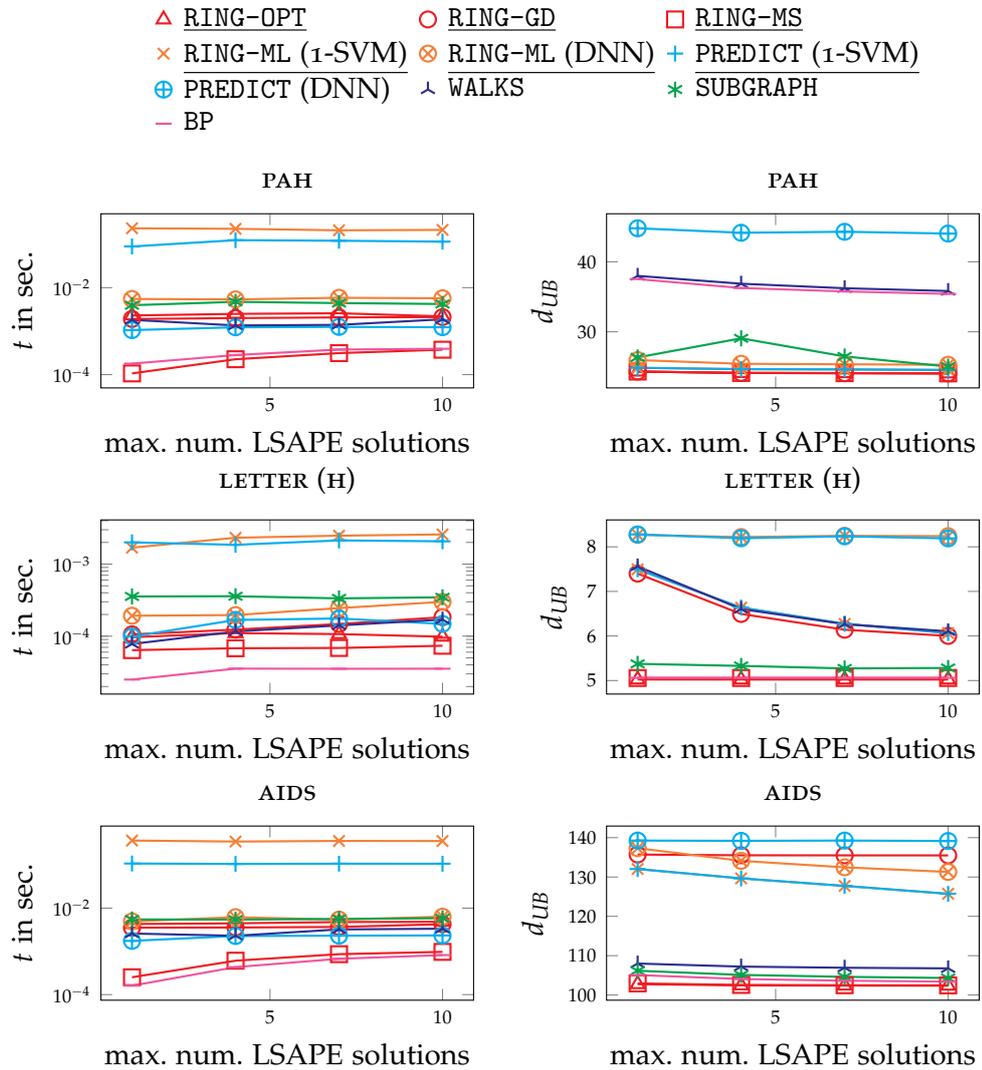

**Figure 6.17.** Maximal number of LSAPE solutions vs. runtimes and upper bounds. Number of threads is fixed to 10. Underlined methods use techniques proposed in this thesis.

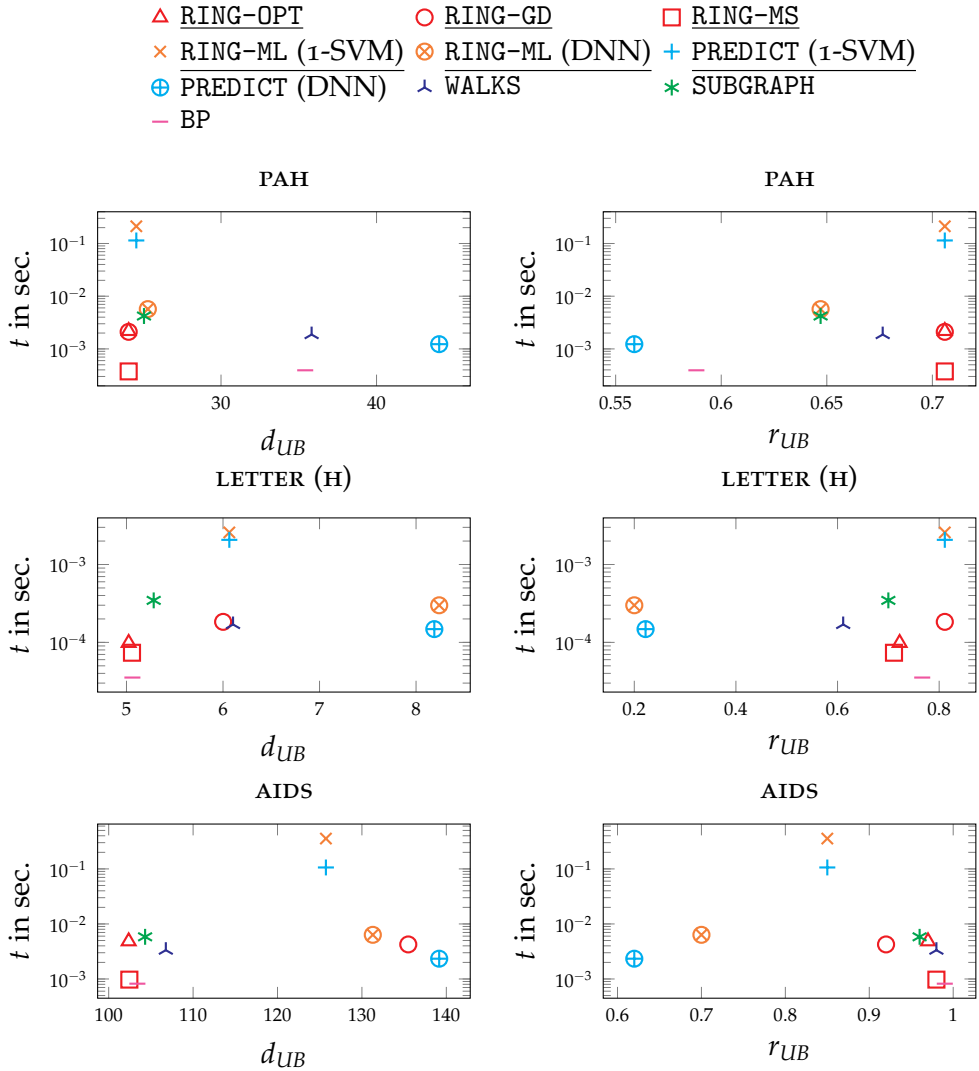

**Figure 6.18.** Runtimes vs. upper bounds and classification ratios. Number of threads and maximal number of LSAPE solutions are fixed to 10. Underlined methods use techniques proposed in this thesis.

**Overall Performance of Compared Methods.** Figure 6.18 summarizes the overall performances of the compared methods on the three datasets with the number of threads and maximal number of LSAPE solutions fixed to 10. We see that, across all datasets, `RING-OPT` yielded the tightest upper bound of all tested methods. `RING-MS`, i. e., the variant of `RING` which uses the multiset intersection based approach for computing the layer distances, performed excellently, too, as it was significantly faster than `RING-OPT` and yielded only slightly looser upper bounds. On the contrary, using a greedy, suboptimal LSAPE solver for computing the layer distances as done by `RING-GD` turned out not to be a good idea, as doing so did not significantly reduce the runtimes and led to much looser upper bounds on the datasets LETTER (H) and AIDS. If run with 1-SVMs with RBF kernel, the machine learning based methods `PREDICT` and `RING-ML` performed very similarly in terms of classification ratio and tightness of the produced upper bounds. Both methods yielded very promising classification ratios on the datasets PAH and LETTER (H). Running `RING-ML` and `PREDICT` with DNNs instead of 1-SVMs dramatically improved the runtimes but led to looser upper bounds and worse classification ratios. If run with DNNs, `RING-ML` produced tighter upper bounds than `PREDICT`. However, globally, all machine learning based methods were outperformed by classical instantiations of `LSAPE-GED`.

### 6.8.2.3 Upshot of the Experiments

The experimental results presented in the previous section lead us to four takeaway messages: Firstly, if one wants to use instantiations of `LSAPE-GED` for computing tight upper bounds for GED, the algorithms `RING-OPT` and `RING-MS` proposed in Section 6.4 are the best choices. Secondly, it is always a good idea to increase the number of LSAPE solutions as suggested in Section 6.5. Doing so only slightly increases the runtime and at the same time significantly improves the upper bounds of methods that yield loose upper bounds if run with only one LSAPE solution. Thirdly, machine learning based `LSAPE-GED` instantiations such as `RING-ML` and `PREDICT` should be run with 1-SVMs as suggested in Section 6.4.1.2 if one wants to optimize for classification ratio and tightness of the produced upper bound, and with DNNs as suggested in [39] if one wants to optimize for runtime behavior. Fourthly, `RING-ML` and `PREDICT` show promising potential but cannot yet compete with classical instantiations of `LSAPE-GED`. If run with 1-SVMs, they are competitive in terms of quality (classification ratio and tightness of the produced upper bound) but not in terms of runtime; if run with DNNs, the opposite is the case. The open challenge for future work is therefore to develop new machine learning frameworks that exploit the information encoded in `RING-ML`'s and `PREDICT`'s feature vectors such that the resulting GED heuristics are competitive both w. r. t. quality and w. r. t. runtime behavior.

### 6.8.3 Evaluation of `K-REFINE` and `RANDPOST`

In a third series of experiments, we evaluated our local search algorithm `K-REFINE` and tested how our `LS-GED` extension `RANDPOST` affects the performance of local search based heuristics. In Section 6.8.3.1 we describe the setup of the experiments; in Section 6.8.3.2, we report the results; and in Section 6.8.3.3, we concisely summarize the most important experimental findings.

#### 6.8.3.1 Setup and Datasets

**Datasets and Edit Costs.** We tested on the datasets MUTA, PROTEIN, GREC, and FINGERPRINT. For all datasets, we used the metric edit costs suggested in [84]. For MUTA, we additionally defined non-metric edit costs by setting the costs of node and edge deletions and insertions to 1, and setting the costs of node and edge substitutions to 3 (cf. Section 2.4).

**Compared Methods.** We tested two versions `2-REFINE` and `3-REFINE` of our local search algorithm `K-REFINE`, which use swaps of maximum size two and three, respectively. We compared them to the existing local search algorithms `REFINE`, `IPFP`, `BP-BEAM`, and `IBP-BEAM`. As suggested in [92] and [44], we set the beam size employed by `BP-BEAM` and `IBP-BEAM` to 5 and the number of iterations employed by `IBP-BEAM` to 20. `IPFP` was run with convergence threshold set to $10^{-3}$ and maximum number of iterations set to 100, as proposed in [12]. In order to evaluate `RANDPOST`, we ran each algorithm with $K := 40$ initial solutions, and varied the pair of meta-parameters $(L, \rho)$ on the set $\{(0,1), (1,0.5), (3,0.25), (7,0.125)\}$. Recall that $L$ is the number of `RANDPOST` loops and $\rho$ is defined such that each iteration produces exactly $\lceil \rho \cdot K \rceil$ locally optimal node maps. Therefore, our setup ensures that each configuration produces exactly 40 local optima. For each algorithm and each dataset, we conducted pre-tests where we varied the penalty parameter $\eta$ on the set $\{n/10 \mid n \in \mathbb{N}_{\leq 10}\}$, and then picked the value of $\eta$ for the main

Table 6.3. K-REFINE vs. REFINE without RANDPOST.

| REFINE | | 2-REFINE | | 3-REFINE | |
|---|---|---|---|---|---|
| $d$ | $t$ | $d$ | $t$ | $d$ | $t$ |
| GREC | | | | | |
| 859.92 | $2.46 \cdot 10^{-2}$ | 857.89 | $1.15 \cdot 10^{-2}$ | 857.12 | $3.85 \cdot 10^{-2}$ |
| FINGERPRINT | | | | | |
| 2.82 | $5.34 \cdot 10^{-4}$ | 2.82 | $4.49 \cdot 10^{-4}$ | 2.82 | $1.58 \cdot 10^{-3}$ |
| PROTEIN | | | | | |
| 295.61 | $3.43 \cdot 10^{-1}$ | 295.55 | $9.07 \cdot 10^{-2}$ | 295.29 | $4.89 \cdot 10^{-1}$ |
| MUTA (metric) | | | | | |
| 74.12 | $1.22 \cdot 10^{-1}$ | 74.12 | $3.92 \cdot 10^{-2}$ | 73.61 | $1.87 \cdot 10^{-1}$ |
| MUTA (non-metric) | | | | | |
| 49.49 | $1.14 \cdot 10^{-1}$ | 49.11 | $3.58 \cdot 10^{-2}$ | 48.44 | $1.81 \cdot 10^{-1}$ |

experiments that yielded the best average upper bound across all RANDPOST configurations.

**Protocol and Test Metrics.** For each dataset, subsets of 50 graphs were chosen randomly. An additional subset of 10 graphs having exactly 70 nodes was extracted from MUTA and is denoted by MUTA-70. Upper bounds for GED were computed for each pair of graphs in the subsets, as well as for each graph and a shuffled copy of itself. In the following, $d$, $\hat{d}$, and $t$ denote the average upper bound, the average upper bound between graphs and their shuffled copies, and the average runtime in seconds, respectively. Note that the test metric $\hat{d}$ gives us a hint to how close to optimality each algorithm is, as the optimal value, namely 0, is known.

**Implementation and Hardware Specifications.** All methods were implemented in GEDLIB (cf. Appendix A) and run in 20 parallel threads.[5] The tests to compare K-REFINE to the baseline REFINE were carried out on a computer using an Intel Xeon E5-2620 v4 2.10GHz CPU. The remaining tests were run on a machine using an Intel(R) Xeon E5-2640 v4 2.4GHz CPU.

#### 6.8.3.2 Results of the Experiments

---

[5]Sources and datasets are available at `https://github.com/dbblumenthal/gedlib/`.

`K-REFINE` **vs. `REFINE`.** In a first series of experiments, we compared the versions `2-REFINE` and `3-REFINE` of our improved and generalized local search algorithm `K-REFINE` to the baseline algorithm `REFINE`. All algorithms were run without `RANDPOST`, i.e., with $(L, \rho) = (0, 1)$. Table 6.3 shows the results. By comparing $t(\mathtt{2\text{-}REFINE})$ and $t(\mathtt{REFINE})$, we see that efficiently computing the swap costs as suggested in Section 6.6.2 indeed significantly improves the runtime performance. Unsurprisingly, the speed-up is especially large on the datasets PROTEIN and MUTA containing the larger graphs. Comparing $d(\mathtt{2\text{-}REFINE})$ and $d(\mathtt{REFINE})$ shows that the inclusion of the dummy assignment proposed in Section 6.6.3 slightly improves the quality of the produced upper bound. As expected, the percentual improvement is largest on the dataset MUTA with non-metric edit costs. Finally, we observe that running `K-REFINE` with swaps of size three slightly improves the upper bounds on all datasets, but significantly increases the runtime of the algorithm.

**Behavior of `RANDPOST` Framework.** In a second series of experiments, we evaluated the behavior of `RANDPOST` by running each algorithm with four different pairs of meta-parameters $(L, \rho) \in \{(0, 1), (1, 0.5), (3, 0.25), (7, 0.125)\}$. We remind that the case $(L, \rho) = (0, 1)$ amounts to a basic multi-start framework with no `RANDPOST` loop. Figure 6.19 visualizes the results. Since `2-REFINE` was always faster and more accurate than the baseline `REFINE`, we do not show plots for `REFINE`.

Figure 6.19 indicates that, on the datasets FINGERPRINT, GREC, and PROTEIN containing small graphs, near-optimality is reached by most algorithms when run with $L \geq 1$ number of `RANDPOST` loops. In these contexts, our algorithm `2-REFINE` with `RANDPOST` configuration $(L, \rho) = (1, 0.5)$ provides the best tradeoff between runtime and accuracy, as it reaches the same accuracy as best algorithms, and, in terms of runtime, outperforms all algorithms except for `BP-BEAM` by approximately one order of magnitude. The only faster algorithm `BP-BEAM` computes much more expensive node maps, even in the `RANDPOST` settings with higher number of loops. We also note that our algorithms `2-REFINE` and `3-REFINE` are already among the best local search algorithms when run in a simple multi-start setting without `RANDPOST` (i.e., when $L = 0$), both in terms of distance and computing time.

On the datasets MUTA and MUTA-70 containing larger graphs, the behavior of the `RANDPOST` framework appears clearly and independently of the local search algorithm it is applied to. In all cases, a higher number of `RANDPOST` loops—and lower number of computed solutions per loop—leads to a higher computation time (the computation time is approximately doubled whenever the number of loops is doubled), and to a lower average distance. In other words, the framework `RANDPOST` provides a very useful algorithmic tool in situations where some time can be dedicated to compute tight upper bounds on big graphs.

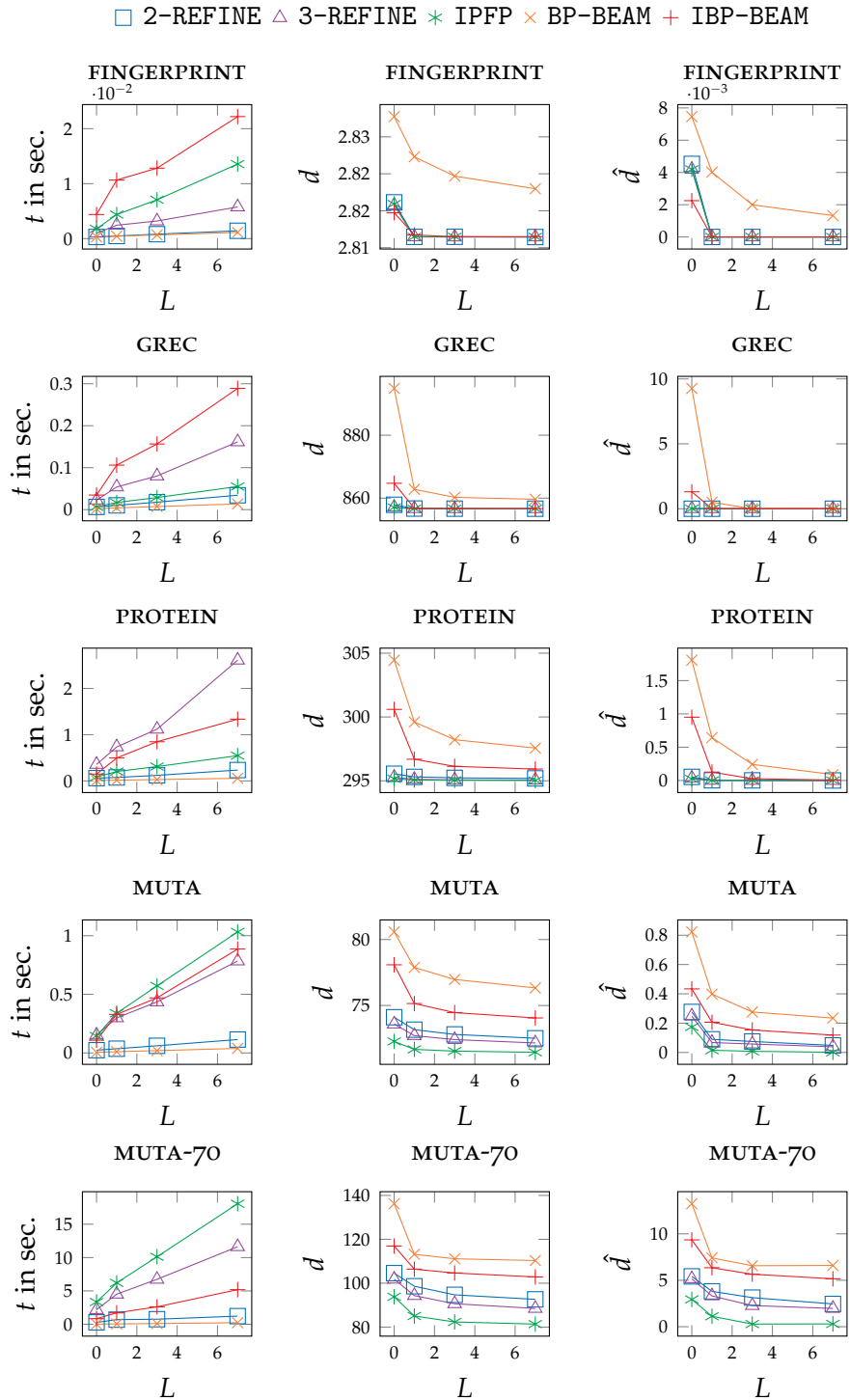

**Figure 6.19.** Effect of RANDPOST on local search algorithms.

#### 6.8.3.3 Upshot of the Experiments

Our experiments showed that both `K-REFINE` and `RANDPOST` perform excellently in practice: On small graphs, `K-REFINE` is among the algorithms computing the tightest upper bounds and, in terms of runtime, clearly outperforms all existing algorithms that yield similar accuracy. On larger graphs, `K-REFINE` provides a very good tradeoff between runtime and accuracy, as it is only slightly less accurate but much faster than the most accurate algorithms. The framework `RANDPOST` is particularly effective on larger graphs, where it significantly improves the upper bounds of all local search algorithms.

### 6.8.4 Joint Evaluation of All Heuristics

In a fourth series of experiments, we jointly evaluated all heuristic algorithms presented in this chapter and addressed the meta-questions Q-1 and Q-2 introduced above:

Q-1  Is it indeed beneficial to use GED as a guidance for the design of graph distance measures, if these distance measures are to be used within pattern recognition frameworks?

Q-2  Do graph distance measures defined by upper bounds for GED or graph distance measures defined by lower bounds for GED perform better when used within pattern recognition frameworks?

In Section 6.8.4.1, we describe the setup of the experiments; in Section 6.8.4.2, we report the results; and in Section 6.8.4.3, we concisely summarize the most important experimental findings.

#### 6.8.4.1 Setup and Datasets

**Datasets and Edit Costs.** We tested on the datasets AIDS, MUTA, PROTEIN, LETTER (H), GREC, and FINGERPRINT. For all datasets, we used the edit cost functions suggested in [2, 84] (cf. Section 2.4). In order to be able to compare all heuristics on all datasets, we used the technique described in Section 6.1 for extending the heuristics with cost constraints to general edit costs. Note that, on all datasets except for FINGERPRINT, the employed edit costs are defined in such a way that node edit operations are more expensive than edge edit operations. Moreover, the edge edit costs for LETTER (H), AIDS, and MUTA suggested in [2, 84] are constant, which implies that, on these datasets BRANCH, BRANCH-FAST, and BRANCH-CONST yield the same lower bounds.

**Compared Methods.** In our experiments, we compared all heuristic algorithm presented in this chapter. Below, we give a detailed description of how we chose the algorithms' options and meta-parameters.

- *SUBGRAPH and WALKS:* As suggested in [32] and [50], for each dataset, we determined the parameters $K$ of SUBGRAPH and WALKS as the $K \in [5]$ that yielded the tightest average upper bounds on a set of training graphs. In order to cope with SUBGRAPH's exponential runtime complexity, we enforced a time limit of 1 ms for the computation of each cell of its LSAPE instance **C**.
- *RING:* As highlighted in Section 6.8.2, RING performs best if the node and edge set distances are computed via optimal LSAPE solvers or multiset intersection based proxies. We included both options in our experiments; the resulting heuristics are denoted as RING-OPT and RING-MS, respectively. For both variants and each dataset, the meta-parameters $\lambda_l$, $\alpha_s$, and $K$ were determined by running a blackbox optimizer on a set of training graphs, as suggested in Section 6.4.
- *RING-ML and PREDICT:* As highlighted in Section 6.8.2, the machine learning based heuristics RING-ML and PREDICT perform best if one-class support vector machines with RBF kernel or fully connected feed-forward deep neural networks are used for training. We included both variants in our experiments; the resulting heuristics are denoted as RING-ML-SVM, RING-ML-DNN, PREDICT-SVM, and PREDICT-DNN, respectively.

- *REFINE and K-REFINE:* As pointed out in Section 6.8.3, `K-REFINE` with maximum swap size $K := 2$ outperforms `REFINE` both in terms of runtime and in terms of accuracy of the produced upper bound. We therefore did not include `REFINE`. Instead, we tested the two versions `2-REFINE` and `3-REFINE` of `K-REFINE` that use swaps of maximum size two and three, respectively.
- *BP-BEAM and IBP-BEAM:* As suggested in [92] and [44], we set the beam size employed by `BP-BEAM` and `IBP-BEAM` to $K := 5$ and the number of iterations employed by `IBP-BEAM` to $I := 20$.
- *IPFP:* As highlighted in Section 3.5, the best performing variant of `IPFP` that can cope with general edit costs is the one suggested in [25]. In our experiments, we therefore only included this variant. As in the implementations used for the experiments of the original publications, we set the maximal number of iterations to $I := 100$ and the convergence threshold to $\varepsilon := 10^{-3}$.
- *BRANCH-TIGHT:* As suggested in Section 6.8.1, we set the number of iterations carried out by `BRANCH-TIGHT` to $I := 20$.
- *SA:* As suggested in [94], we set the number of iterations carried out by `SA` to $I := 100$ and used start and end probabilities $p_1 := 0.8$ and $p_I := 10^{-2}$. We used `BRANCH` for computing `SA`'s initial LSAPE instance **C**.
- *PARTITION and HYBRID:* In [113], it is suggested to set the maximal size of the substructures employed by `PARTITION` and `HYBRID` to $K := 8$. However, how to implement these heuristics with $K > 3$ is not well documented in [113] and the authors did not reply to our request to share their implementation. We therefore used $K := 3$ for our experiments. In order to cope with `HYBRID`'s exponential runtime complexity, we enforced a time limit of 1 s and set up `HYBRID` to return the maximum of the lower bounds computed by `PARTITION` and `BRANCH-CONST` whenever it did not terminate within the time limit.
- *MULTI-SOL and CENTRALITIES:* In order to test `MULTI-SOL` and `CENTRALITIES`, we ran all instantiations of `LSAPE-GED` with all configurations $(K, \gamma) \in \{1, 3, 7, 10\} \times \{0, 0.7\}$, where $K$ is the maximal number of solutions computed by `MULTI-SOL` and $\gamma$ is the weight of the pagerank centralities used by `CENTRALITIES`. We used pagerank centralities with $\gamma = 0.7$, because in [88] this setup is reported to yield the best results among all variants of `CENTRALITIES`. Note that `MULTI-SOL` is used just in case $K \neq 1$ and `CENTRALITIES` is used just in case $\gamma \neq 0$.

- *MULTI-START and RANDPOST*: We tested `MULTI-START` and `RANDPOST` by running each instantiation of `LS-GED` with all configurations $(K, \rho, L, \eta) \in (\{1, 10, 20, 30, 40\} \times \{(1, 0, 0)\}) \cup (\{(40, 0.5, 1), (40, 0.25, 3), (40, 0.125, 7)\} \times \{0, 1\})$, where $K$ is the number of initial node maps constructed by `MULTI-START`, $\lceil \rho \cdot K \rceil$ is the number of completed runs from initial node maps, $L$ is the number of `RANDPOST` loops, and $\eta$ is the penalty for expensive converged node maps employed my `RANDPOST`. The initial node maps were constructed randomly under the constraint that they contain exactly $\min\{|V^G|, |V^H|\}$ node substitutions. Note that `MULTI-START` is used just in case $K \neq 1$, `RANDPOST` is used just in case $L \neq 0$, and each configuration that uses `RANDPOST` in total carries out exactly 40 runs from different initial node maps.

**Protocol and Test Metrics.** For each test dataset $\mathcal{D}$, we randomly selected a training set $\mathcal{D}_{\text{train}} \subseteq \mathcal{D}$ and a testing set $\mathcal{D}_{\text{test}} \subseteq \mathcal{D} \setminus \mathcal{D}_{\text{train}}$. We ensured that both sets are balanced w.r.t. the classes of the contained graphs and set their sizes to the largest integers not greater than, respectively, 50 (for training) and 100 (for testing) that allowed balancing. For instance, for the dataset AIDS which contains 1500 graphs that model molecules with or without activity against HIV, $\mathcal{D}_{\text{test}}$ contains 50 randomly selected active and 50 randomly selected inactive molecules. All algorithms that require training were trained on $\mathcal{D}_{\text{train}}$. Subsequently, we ran all compared algorithms on all pairs of graphs $(G, H) \in \mathcal{D}_{\text{test}} \times \mathcal{D}_{\text{test}}$. Recall that we compared various configurations of the extensions `MULTI-SOL` and `CENTRALITIES` for the instantiation of `LSAPE-GED`; and that we tested various configurations of the extensions `MULTI-START` and `RANDPOST` for the instantiation of `LS-GED`. In the following, the expression "algorithm" denotes a heuristic together with its configuration.

For all compared algorithms `ALG`, we recorded the average runtime $t(\texttt{ALG})$. Moreover, we recorded the average lower bound $d_{LB}(\texttt{ALG})$ and the classification coefficient $c_{LB}(\texttt{ALG})$ for all algorithms that yield lower bounds, and the average upper bound $d_{UB}(\texttt{ALG})$ and the classification coefficient $c_{UB}(\texttt{ALG})$ for all algorithms that yield upper bounds. The classification coefficients $c_{LB}$ and $c_{UB}$ were computed as

$$c_{LB}(\texttt{ALG}) := (d_{LB}^{\text{inter}}(\texttt{ALG}) - d_{LB}^{\text{intra}}(\texttt{ALG})) / \max LB(\texttt{ALG})$$
$$c_{UB}(\texttt{ALG}) := (d_{UB}^{\text{inter}}(\texttt{ALG}) - d_{UB}^{\text{intra}}(\texttt{ALG})) / \max UB(\texttt{ALG}),$$

where $d_{LB}^{\text{inter}}(\texttt{ALG})$ and $d_{UB}^{\text{inter}}(\texttt{ALG})$ are the average lower and upper bounds between graphs with different classes, $d_{LB}^{\text{intra}}(\texttt{ALG})$ and $d_{UB}^{\text{intra}}(\texttt{ALG})$ are the average lower and upper bounds between graphs with the same class, and $\max LB(\texttt{ALG})$ and $\max UB(\texttt{ALG})$ denote the maximal lower and upper bounds computed by $\texttt{ALG}$. The reason for defining the classification coefficients in this way is that pattern recognition frameworks based on distance measures perform well just in case the intra-class distances are significantly smaller than the inter-class distances. Hence, large classification coefficients $c_{LB}(\texttt{ALG})$ and $c_{UB}(\texttt{ALG})$ imply that the respective lower or upper bounds are fit for use within distance based pattern recognition frameworks. We normalized by the maximal lower and upper bounds in order to ensure $c_{LB}(\texttt{ALG}), c_{UB}(\texttt{ALG}) \in [-1, 1]$ and hence render the classification coefficients comparable across different datasets. We rounded $t(\texttt{ALG})$ to microseconds and $d_{LB|UB}(\texttt{ALG})$ as well as $c_{LB|UB}(\texttt{ALG})$ to two decimal places.

After running all algorithms, we computed a joint score $s_{LB}(\texttt{ALG}) \in [0,1]$ for all algorithms that yield lower bounds and a joint score $s_{UB}(\texttt{ALG}) \in [0,1]$ for all algorithms that yield upper bounds. The joint scores are defined as

$$s_{LB}(\texttt{ALG}) := \frac{d_{LB}(\texttt{ALG})}{3 \cdot d_{LB}^\star} + \frac{t_{LB}^\star}{3 \cdot t(\texttt{ALG})} + \frac{c_{LB}(\texttt{ALG})}{3 \cdot c_{LB}^\star}$$

$$s_{UB}(\texttt{ALG}) := \frac{d_{UB}^\star}{3 \cdot d_{UB}(\texttt{ALG})} + \frac{t_{UB}^\star}{3 \cdot t(\texttt{ALG})} + \frac{c_{UB}(\texttt{ALG})}{3 \cdot c_{UB}^\star},$$

where $d_{LB}^\star$, $t_{LB}^\star$, and $c_{LB}^\star$ denote the best (i.e., largest) average lower bound, the best average runtime, and the best classification coefficient yielded by any algorithm that computes a lower bound. Analogously, $d_{UB}^\star$, $t_{UB}^\star$, and $c_{UB}^\star$ denote the best (i.e., smallest) average upper bound, the best average runtime, and the best classification coefficient yielded by any algorithm that computes an upper bound. Note that, with this definition, each evaluation criterion contributes a quantity between 0 and 1/3 to the joint score, and an algorithm has joint score 1 if and only if it performs best w. r. t. all three criteria.

We partially ordered the compared algorithms w. r. t. the Pareto dominance relations $\succ_{LB}$ and $\succ_{UB}$. For two algorithms $\texttt{ALG}_1$ and $\texttt{ALG}_2$ that compute lower bounds, we say that the lower bound computed by $\texttt{ALG}_1$ dominates the one produced by $\texttt{ALG}_2$ on a given dataset (in symbols: $\texttt{ALG}_1 \succ_{LB} \texttt{ALG}_2$) just in case $\texttt{ALG}_1$ performs at least as good as $\texttt{ALG}_2$ w. r. t. to all three evaluation criteria $d_{LB}$, $t$, and $c_{LB}$, and strictly better than $\texttt{ALG}_2$ w. r. t. at least one of them. The dominance relation $\succ_{UB}$ for the upper bounds is defined analogously. Note that, with these definitions, $\texttt{ALG}_1 \succ_{LB} \texttt{ALG}_2$ implies $s_{LB}(\texttt{ALG}_1) > s_{LB}(\texttt{ALG}_2)$ and $\texttt{ALG}_1 \succ_{UB} \texttt{ALG}_2$ implies $s_{UB}(\texttt{ALG}_1) > s_{UB}(\texttt{ALG}_2)$, but the inverse implications do not hold. The joint scores $s_{LB}$ and $s_{UB}$ hence allow to compare algorithms that are Pareto optimal.

Using the partial orders $\succ_{LB}$ and $\succ_{UB}$, we computed aggregated joint lower bound scores $\widehat{s_{LB}}(\texttt{HR})$ for all heuristics HR that compute lower bounds, as well as aggregated joint upper bounds score $\widehat{s_{UB}}(\texttt{HR})$ and $\widehat{s_{UB}}(\texttt{EXT})$ for all heuristics HR that compute lower bounds and all extensions EXT of the paradigms LSAPE-GED and LS-GED. The aggregated joint scores were computed as

$$\widehat{s_{LB}}(\texttt{HR}) := \delta_{\mathcal{C}(\texttt{HR}) \cap \text{MAX}_{\succ_{LB}} \neq \emptyset} \max_{\texttt{ALG} \in \mathcal{C}(\texttt{HR}) \cap \text{MAX}_{\succ_{LB}}} s_{LB}(\texttt{ALG})$$

$$\widehat{s_{UB}}(\texttt{HR}) := \delta_{\mathcal{C}(\texttt{HR}) \cap \text{MAX}_{\succ_{UB}} \neq \emptyset} \max_{\texttt{ALG} \in \mathcal{C}(\texttt{HR}) \cap \text{MAX}_{\succ_{UB}}} s_{UB}(\texttt{ALG})$$

$$\widehat{s_{UB}}(\texttt{EXT}) := \frac{\delta_{\mathcal{C}(\texttt{PAR}(\texttt{EXT})) \cap \text{MAX}_{\succ_{UB}} \neq \emptyset} \sum_{\texttt{ALG} \in \mathcal{C}(\texttt{EXT}) \cap \text{MAX}_{\succ_{UB}}} s_{UB}(\texttt{ALG})}{\sum_{\texttt{ALG} \in \mathcal{C}(\texttt{PAR}(\texttt{EXT})) \cap \text{MAX}_{\succ_{UB}}} s_{UB}(\texttt{ALG})},$$

where $\mathcal{C}(\texttt{HR})$ is the set of compared algorithms that are configurations of the heuristic HR, $\mathcal{C}(\texttt{EXT})$ is the set of compared algorithms that use the extension EXT, $\mathcal{C}(\texttt{PAR}(\texttt{EXT}))$ is the set of compared algorithms that instantiate the paradigm extended by EXT, and $\text{MAX}_{\succ_{LB}}$ and $\text{MAX}_{\succ_{UB}}$ are the set of maxima w.r.t. the partial orders $\succ_{LB}$ and $\succ_{UB}$, respectively. In other words, we set the aggregated joint scores $\widehat{s_{LB}}(\texttt{HR})$ and $\widehat{s_{UB}}(\texttt{HR})$ of a heuristic HR to the maximal scores of Pareto optimal configurations of HR, and to 0 if no configurations of HR were Pareto optimal. The aggregated joint upper bound score $\widehat{s_{UB}}(\texttt{EXT})$ of an extension EXT of the paradigms LSAPE-GED and LS-GED was set to the sum of the joint upper bound scores of Pareto optimal algorithms that use EXT divided by the sum of the joint upper bound scores of Pareto optimal algorithms that instantiate the paradigm extended by EXT, and to 0 if no algorithms that instantiate the paradigm extended by EXT were Pareto optimal. We also computed vectors $\chi_{LB}(\texttt{HR}) \in \{0,1\}^3$ for all heuristics that yield lower bounds and vectors $\chi_{UB}(\texttt{HR}), \chi_{UB}(\texttt{EXT}) \in \{0,1\}^3$ for all heuristics that yield upper bounds and all extensions of the paradigms LSAPE-GED and LS-GED. These vectors indicate whether a heuristic or an extension has a configuration that performed best w.r.t. one or several of the observed metrics $f_{1_{LB|UB}} := d_{LB|UB}$, $f_{2_{LB|UB}} := t_{LB|UB}$, and $f_{3_{LB|UB}} := c_{LB|UB}$. That is, the indicator vectors were computed as follows:

$$\chi_{LB}(\mathtt{HR}) := (\delta_{\exists \mathtt{ALG} \in \mathcal{C}(\mathtt{HR}): f_{r_{LB}}(\mathtt{ALG}) = f^\star_{r_{LB}}})^3_{r=1}$$
$$\chi_{UB}(\mathtt{HR}) := (\delta_{\exists \mathtt{ALG} \in \mathcal{C}(\mathtt{HR}): f_{r_{UB}}(\mathtt{ALG}) = f^\star_{r_{UB}}})^3_{r=1}$$
$$\chi_{UB}(\mathtt{EXT}) := (\delta_{\exists \mathtt{ALG} \in \mathcal{C}(\mathtt{EXT}): f_{r_{UB}}(\mathtt{ALG}) = f^\star_{r_{UB}}})^3_{r=1}$$

Finally, for each dataset, we trained linear regression models $c_{LB} \sim d_{LB} := (a_{LB}, m_{LB})$ and $c_{UB} \sim d_{UB} := (a_{UB}, m_{UB})$ defined as

$$(a_{LB}, m_{LB}) := \underset{(a,m) \in \mathbb{R} \times \mathbb{R}}{\arg\min} \sum_{\mathtt{ALG}} [c_{LB}(\mathtt{ALG}) - (a + m \cdot d_{LB}(\mathtt{ALG}))]^2$$
$$(a_{UB}, m_{UB}) := \underset{(a,m) \in \mathbb{R} \times \mathbb{R}}{\arg\min} \sum_{\mathtt{ALG}} [c_{UB}(\mathtt{ALG}) - (a + m \cdot d_{UB}(\mathtt{ALG}))]^2$$

that relate the tightnesses of the computed upper and lower bounds to the obtained classification coefficients: Tightness of lower bounds is positively correlated to high classification coefficients just in case the slope $m_{LB}$ is positive; tightness of upper bounds is positively correlated to high classification coefficients just in case the slope $m_{UB}$ is negative. We also computed the $p$-values $p_{LB}$ and $p_{UB}$ of the models $c_{LB} \sim d_{LB}$ and $c_{UB} \sim d_{UB}$, respectively, which tell us whether the correlations between bounds and classification coefficients are statistically significant. Table 6.4 provides an overview of all test metrics.

**Implementation and Hardware Specifications.** To ensure comparability, we re-implemented all compared heuristics in C++. Our implementation builds upon the Boost Graph Library [66] and Eigen [53] for efficiently managing graphs and matrices. For solving LSAPE, we used the solver suggested in [24], which is efficiently implemented in the LSAPE toolbox available at https://bougleux.users.greyc.fr/lsape/. We used the blackbox optimizer library NOMAD [64] for training RING-OPT and RING-MS, the support vector machine library LIBSVM [33] for training RING-ML-SVM and PREDICT-SVM, the artificial neural network library FANN [80] for training RING-ML-DNN and PREDICT-DNN, and the mathematical programming library Gurobi [54] for implementing the instantiations of LP-GED.

**Table 6.4.** Test metrics used for global evaluation of heuristic algorithms.

| syntax | semantic |
|---|---|
| *observed metrics for compared algorithms* | |
| $d_{LB\|UB}$ | average lower and upper bounds |
| $t$ | average runtime |
| $c_{LB\|UB}$ | classification coefficients of lower and upper bounds |
| *inferred metrics for compared algorithms* | |
| $s_{LB\|UB}$ | joint lower and upper bound scores |
| *inferred metrics for compared heuristics and extensions* | |
| $\widehat{s_{LB\|UB}}$ | aggregated joint lower and upper bound scores |
| $\chi_{LB\|UB}$ | indicate whether heuristics and extensions have configuration that are optimal w. r. t. observed metrics |
| *inferred metrics for test datasets* | |
| $d^{\star}_{LB\|UB}$ | tightest average lower and upper bounds |
| $t^{\star}_{LB\|UB}$ | average runtimes of fastest algorithms producing lower and upper bounds |
| $c^{\star}_{LB\|UB}$ | best classification coefficients of lower and upper bounds |
| $m_{LB\|UB}$ | slopes of linear regression models $c_{LB\|UB} \sim d_{LB\|UB}$ |
| $p_{LB\|UB}$ | $p$-values of linear regression models $c_{LB\|UB} \sim d_{LB\|UB}$ |

At runtime, all compared heuristics were given access to six threads. More specifically, instantiations of `LSAPE-GED` were set up to parallelly construct their LSAPE instance **C**, instantiations of `LS-GED` were implemented to parallelly carry out runs from several initial solutions, and instantiations of `LP-GED` were allowed to use multi-threading when solving their LP via calls to Gurobi. For the miscellaneous heuristics, we used the following parallelization techniques: Like the instantiations of `LSAPE-GED`, `HED` was set up to construct its LSAPE instance **C** in parallel. Similarly, `BRANCH-TIGHT` was implemented to parallelize the construction phases of all of its LSAP instances $\mathbf{C}^r$. `SA` and `HYBRID` were set up to use the parallelized versions of, respectively, `BRANCH` and `BRANCH-CONST` as subroutines. `BRANCH-COMPACT` and `PARTITION` do not allow straightforward parallelizations and where hence run in only one thread.

The test sources and datasets are distributed with GEDLIB (cf. Appendix A).[6] Tests were run on a machine with two Intel Xeon E5-2667 v3 processors with 8 cores each and 98 GB of main memory running GNU/Linux.

---

[6]Sources and datasets are available at `https://github.com/dbblumenthal/gedlib`.

**Table 6.5.** Results for lower bounds on LETTER (H), MUTA, and AIDS.

| heuristic | LETTER (H) | | MUTA | | AIDS | |
|---|---|---|---|---|---|---|
| | $\chi_{LB}$ | $\widehat{s_{LB}}$ | $\chi_{LB}$ | $\widehat{s_{LB}}$ | $\chi_{LB}$ | $\widehat{s_{LB}}$ |
| *instantiations of the paradigm LSAPE-GED* | | | | | | |
| NODE | $(0,0,0)$ | 0.00 | $(0,\mathbf{1},0)$ | **0.59** | $(0,\mathbf{1},\mathbf{1})$ | **0.92** |
| BRANCH | $(0,0,0)$ | 0.00 | $(0,0,0)$ | 0.00 | $(0,0,\mathbf{1})$ | 0.00 |
| BRANCH-FAST | $(0,0,0)$ | 0.00 | $(0,0,0)$ | 0.00 | $(0,0,\mathbf{1})$ | 0.00 |
| BRANCH-CONST | $(0,\mathbf{1},0)$ | **0.92** | $(0,0,0)$ | **0.41** | $(0,0,\mathbf{1})$ | **0.76** |
| STAR | $(0,0,0)$ | 0.00 | $(0,0,\mathbf{1})$ | 0.00 | $(0,0,0)$ | 0.00 |
| *instantiations of the paradigm LP-GED* | | | | | | |
| F-1 | $(0,0,0)$ | 0.00 | $(0,0,\mathbf{1})$ | 0.00 | $(0,0,\mathbf{1})$ | 0.00 |
| F-2 | $(0,0,0)$ | 0.00 | $(0,0,0)$ | **0.32** | $(0,0,\mathbf{1})$ | **0.66** |
| COMPACT-MIP | $(0,0,0)$ | 0.00 | $(0,0,0)$ | 0.00 | $(0,0,0)$ | 0.00 |
| ADJ-IP | $(\mathbf{1},0,\mathbf{1})$ | **0.67** | $(\mathbf{1},0,\mathbf{1})$ | **0.67** | $(\mathbf{1},0,\mathbf{1})$ | **0.67** |
| *miscellaneous heuristics* | | | | | | |
| HED | $(0,0,0)$ | 0.00 | $(0,0,0)$ | 0.00 | $(0,0,0)$ | 0.00 |
| BRANCH-TIGHT | $(0,0,\mathbf{1})$ | **0.68** | $(0,0,\mathbf{1})$ | 0.00 | $(0,0,\mathbf{1})$ | 0.00 |
| BRANCH-COMPACT | $(0,0,0)$ | 0.00 | $(0,0,\mathbf{1})$ | **0.63** | $(0,0,0)$ | 0.00 |
| PARTITION | $(0,0,0)$ | 0.00 | $(0,0,0)$ | 0.00 | $(0,0,0)$ | 0.00 |
| HYBRID | $(0,0,0)$ | 0.00 | $(0,0,0)$ | 0.00 | $(0,0,\mathbf{1})$ | 0.00 |

#### 6.8.4.2 Results of the Experiments

**Lower Bounds.** Table 6.5 and Table 6.6 show the aggregated joint lower bound scores $\widehat{s_{LB}}(\text{HR})$, as well as the indicator vectors $\chi_{LB}(\text{HR})$ for all heuristics HR that compute lower bounds on the datasets LETTER (H), MUTA, and AIDS (Table 6.5), and PROTEIN, FINGERPRINT, and GREC (Table 6.6). We see that the fast but relatively imprecise LSAPE-GED instantiations BRANCH-CONST and NODE were Pareto optimal on six (BRANCH-CONST) respectively five (NODE) out of six datasets. The heuristics BRANCH and BRANCH-FAST, which are designed to exhibit a good tradeoff between precision and runtime performance, were Pareto optimal on the datasets PROTEIN, FINGERPRINT, and GREC, i.e., on all datasets where they are not equivalent to BRANCH-CONST. Among the heuristics that are optimized for precision, our anytime algorithm BRANCH-TIGHT and the LP-GED instantiations ADJ-IP and F-2 performed best. In particular, we see that BRANCH-TIGHT computed the tightest lower bounds on FINGERPRINT — the only dataset where edge edit operations are more important than node edit operations.

In Figure 6.20, the results are further aggregated by averaging the scores and summing the indicator vectors over all datasets. Instantiations of LSAPE-GED are displayed blue, instantiations of LP-GED are displayed red, and miscellaneous heuristics are displayed green. Unsurprisingly, NODE and BRANCH-CONST performed best in terms of runtime (cf. Figure 6.20c). Moreover, they achieved the best aggregated joint lower bound scores, i.e., exhibited the best tradeoffs between tightness of the obtained lower bound, runtime, and classification coefficient (cf. Figure 6.20a). In terms of tightness of the obtained lower bound, the LP based heuristics ADJ-IP performed best, followed by BRANCH-TIGHT and F-2 (cf. Figure 6.20b). BRANCH-TIGHT and ADJ-IP also were the best performing heuristics w.r.t. the classification coefficient (cf. Figure 6.20d).

**Table 6.6.** Results for lower bounds on PROTEIN, FINGERPRINT, and GREC.

| heuristic | PROTEIN | | FINGERPRINT | | GREC | |
|---|---|---|---|---|---|---|
| | $\chi_{LB}$ | $\widehat{s_{LB}}$ | $\chi_{LB}$ | $\widehat{s_{LB}}$ | $\chi_{LB}$ | $\widehat{s_{LB}}$ |
| *instantiations of the paradigm* LSAPE-GED | | | | | | |
| NODE | (0,**1**,**1**) | **0.97** | (0,**1**,**1**) | **0.93** | (0,**1**,0) | **0.94** |
| BRANCH | (0,0,**1**) | 0.68 | (0,0,0) | 0.68 | (0,0,**1**) | 0.73 |
| BRANCH-FAST | (0,0,**1**) | **0.71** | (0,0,**1**) | **0.74** | (0,0,**1**) | **0.79** |
| BRANCH-CONST | (0,0,**1**) | **0.74** | (0,0,**1**) | **0.75** | (0,0,**1**) | **0.86** |
| STAR | (0,0,0) | 0.00 | (0,0,0) | 0.00 | (0,0,0) | 0.00 |
| *instantiations of the paradigm* LP-GED | | | | | | |
| F-1 | (0,0,**1**) | 0.00 | (0,0,0) | 0.00 | (0,0,0) | 0.00 |
| F-2 | (**1**,0,**1**) | **0.67** | (0,0,0) | 0.00 | (0,0,0) | **0.65** |
| COMPACT-MIP | (0,0,0) | 0.00 | (0,0,0) | 0.00 | (0,0,0) | 0.00 |
| ADJ-IP | (0,0,**1**) | 0.00 | (0,0,**1**) | 0.00 | (**1**,0,0) | **0.66** |
| *miscellaneous heuristics* | | | | | | |
| HED | (0,0,0) | 0.00 | (0,0,0) | 0.00 | (0,0,0) | 0.00 |
| BRANCH-TIGHT | (0,0,**1**) | 0.00 | (**1**,0,0) | **0.64** | (0,0,**1**) | **0.66** |
| BRANCH-COMPACT | (0,0,0) | 0.00 | (0,0,0) | 0.00 | (0,0,0) | 0.00 |
| PARTITION | (0,0,0) | 0.00 | (0,0,0) | 0.00 | (0,0,0) | 0.00 |
| HYBRID | (0,0,**1**) | 0.00 | (0,0,**1**) | 0.00 | (0,0,**1**) | 0.00 |

**Upper Bounds.** Table 6.7 and Table 6.8 show the aggregated joint lower bound scores $\widehat{s_{UB}}(\text{HR})$ and $\widehat{s_{UB}}(\text{EXT})$, as well as the indicator vectors $\chi_{UB}(\text{HR})$ and $\chi_{UB}(\text{EXT})$ for all heuristics HR that compute upper bounds, and all extensions EXT of the paradigms LSAPE-GED and LS-GED on the datasets LETTER (H), MUTA, and AIDS (Table 6.7), and PROTEIN, FINGERPRINT, and GREC (Table 6.8). The LSAPE-GED instantiation NODE and the LS-GED instantiation

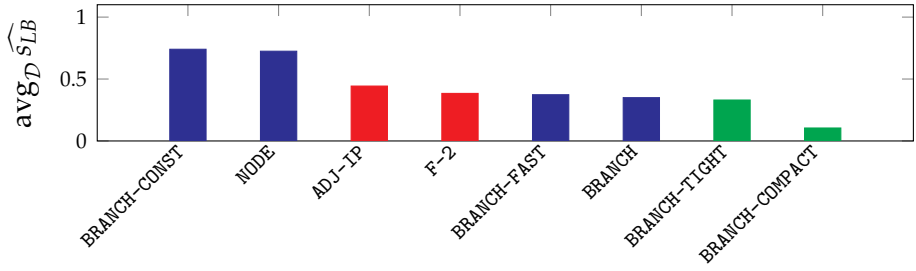

**(a)** Average aggregated joint lower bound scores.

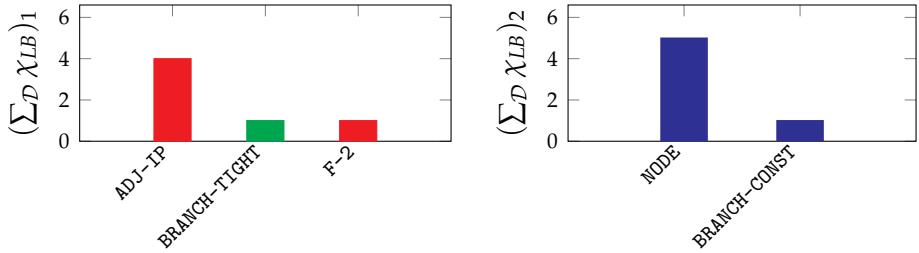

**(b)** Optimality w. r. t. tightness of lower bound.  **(c)** Optimality w. r. t. runtime among heuristics producing lower bounds.

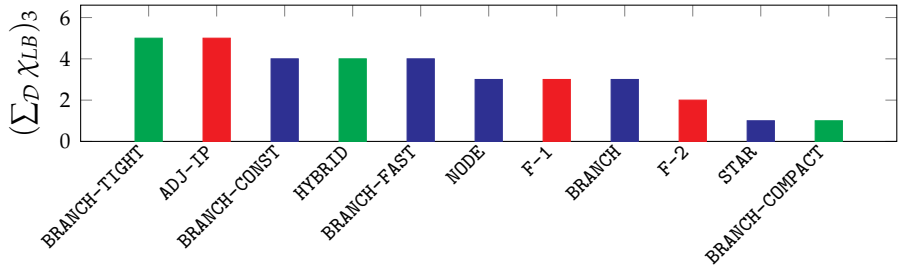

**(d)** Optimality w. r. t. lower bound classification coefficient.

**Figure 6.20.** Overview of results for lower bounds. Only non-zero statistics are displayed.

IPFP performed particularly well: IPFP was Pareto optimal on all datasets, as it always computed the tightest upper bound. NODE was the fastest heuristic on five out of six datasets. It also achieved very high aggregated joint upper bound scores, which can be explained by the fact that the employed edit costs strongly emphasize node edit operations, as mentioned above. Our instantiation 2-REFINE of LS-GED performed well, too, as it was Pareto optimal on all datasets except for GREC and achieved high aggregated joint upper bound scores. On the negative side, we observe that the instantiations of LP-GED and the miscellaneous heuristics performed very poorly, as they were almost always dominated by other heuristics.

**Table 6.7.** Results for upper bounds on LETTER (H), MUTA, and AIDS.

| heuristic | LETTER (H) | | MUTA | | AIDS | |
|---|---|---|---|---|---|---|
| | $\chi_{UB}$ | $\widehat{s_{UB}}$ | $\chi_{UB}$ | $\widehat{s_{UB}}$ | $\chi_{UB}$ | $\widehat{s_{UB}}$ |
| *instantiations of the paradigm* LSAPE-GED | | | | | | |
| NODE | $(0,0,0)$ | 0.00 | $(0,\mathbf{1},\mathbf{1})$ | **0.94** | $(0,\mathbf{1},0)$ | **0.89** |
| BP | $(0,0,0)$ | 0.00 | $(0,0,\mathbf{1})$ | 0.00 | $(0,0,0)$ | 0.00 |
| BRANCH | $(0,0,0)$ | 0.00 | $(0,0,\mathbf{1})$ | 0.00 | $(0,0,0)$ | 0.00 |
| BRANCH-FAST | $(0,0,0)$ | **0.78** | $(0,0,\mathbf{1})$ | 0.00 | $(0,0,0)$ | 0.00 |
| BRANCH-CONST | $(0,\mathbf{1},0)$ | **0.93** | $(0,0,\mathbf{1})$ | **0.68** | $(0,0,0)$ | 0.00 |
| STAR | $(0,0,0)$ | 0.00 | $(0,0,\mathbf{1})$ | 0.00 | $(0,0,0)$ | 0.00 |
| SUBGRAPH | $(0,0,0)$ | 0.00 | $(0,0,\mathbf{1})$ | 0.00 | $(0,0,0)$ | 0.00 |
| WALKS | $(0,0,0)$ | 0.00 | $(0,0,\mathbf{1})$ | 0.00 | $(0,0,0)$ | 0.00 |
| RING-OPT | $(0,0,0)$ | **0.63** | $(0,0,\mathbf{1})$ | 0.00 | $(0,0,0)$ | 0.00 |
| RING-MS | $(0,0,0)$ | 0.00 | $(0,0,\mathbf{1})$ | **0.63** | $(0,0,0)$ | **0.58** |
| RING-ML-SVM | $(0,0,\mathbf{1})$ | 0.00 | $(0,0,\mathbf{1})$ | 0.00 | $(0,0,0)$ | 0.00 |
| RING-ML-DNN | $(0,0,0)$ | 0.00 | $(0,0,0)$ | 0.00 | $(0,0,0)$ | 0.00 |
| PREDICT-SVM | $(0,0,\mathbf{1})$ | **0.55** | $(0,0,\mathbf{1})$ | 0.00 | $(0,0,0)$ | 0.00 |
| PREDICT-DNN | $(0,0,0)$ | 0.00 | $(0,0,0)$ | 0.00 | $(0,0,0)$ | 0.00 |
| *extensions of the paradigm* LSAPE-GED | | | | | | |
| MULTI-SOL | $(0,0,0)$ | 0.00 | $(0,0,\mathbf{1})$ | **0.46** | $(0,0,0)$ | **0.59** |
| CENTRALITIES | $(0,0,0)$ | **0.48** | $(0,0,\mathbf{1})$ | **0.58** | $(0,0,0)$ | **0.44** |
| *instantiations of the paradigm* LP-GED | | | | | | |
| F-1 | $(0,0,0)$ | 0.00 | $(0,0,\mathbf{1})$ | 0.00 | $(0,0,0)$ | 0.00 |
| F-2 | $(0,0,0)$ | 0.00 | $(0,0,\mathbf{1})$ | 0.00 | $(0,0,0)$ | 0.00 |
| COMPACT-MIP | $(0,0,0)$ | 0.00 | $(0,0,0)$ | 0.00 | $(0,0,0)$ | 0.00 |
| ADJ-IP | $(0,0,0)$ | 0.00 | $(0,0,\mathbf{1})$ | 0.00 | $(0,0,0)$ | 0.00 |
| *instantiations of the paradigm* LS-GED | | | | | | |
| 2-REFINE | $(\mathbf{1},0,0)$ | **0.64** | $(0,0,\mathbf{1})$ | **0.66** | $(0,0,0)$ | **0.64** |
| 3-REFINE | $(\mathbf{1},0,0)$ | **0.63** | $(0,0,\mathbf{1})$ | 0.00 | $(0,0,0)$ | 0.00 |
| BP-BEAM | $(\mathbf{1},0,0)$ | **0.62** | $(0,0,0)$ | **0.30** | $(0,0,0)$ | **0.60** |
| IBP-BEAM | $(\mathbf{1},0,0)$ | 0.00 | $(0,0,\mathbf{1})$ | 0.00 | $(0,0,0)$ | 0.00 |
| IPFP | $(\mathbf{1},0,0)$ | **0.63** | $(\mathbf{1},0,\mathbf{1})$ | **0.67** | $(\mathbf{1},0,\mathbf{1})$ | **0.67** |
| *extensions of the paradigm* LS-GED | | | | | | |
| MULTI-START | $(\mathbf{1},0,0)$ | **0.40** | $(\mathbf{1},0,\mathbf{1})$ | **0.91** | $(\mathbf{1},0,\mathbf{1})$ | **0.85** |
| RANDPOST | $(\mathbf{1},0,0)$ | 0.00 | $(\mathbf{1},0,\mathbf{1})$ | **0.29** | $(\mathbf{1},0,\mathbf{1})$ | **0.32** |
| *miscellaneous heuristics* | | | | | | |
| BRANCH-TIGHT | $(0,0,0)$ | **0.62** | $(0,0,\mathbf{1})$ | 0.00 | $(0,0,0)$ | 0.00 |
| SA | $(0,0,0)$ | 0.00 | $(0,0,\mathbf{1})$ | 0.00 | $(0,0,0)$ | 0.00 |

In Figure 6.21, the results are further aggregated by averaging the scores and summing the indicator vectors over all datasets. Instantiations and extensions of LSAPE-GED are displayed blue, instantiations of LP-GED are displayed red, instantiations and extensions of LS-GED are displayed orange,

**Table 6.8.** Results for upper bounds on PROTEIN, FINGERPRINT, and GREC.

| heuristic | PROTEIN | | FINGERPRINT | | GREC | |
|---|---|---|---|---|---|---|
| | $\chi_{UB}$ | $\widehat{s_{UB}}$ | $\chi_{UB}$ | $\widehat{s_{UB}}$ | $\chi_{UB}$ | $\widehat{s_{UB}}$ |
| *instantiations of the paradigm LSAPE-GED* | | | | | | |
| NODE | $(0,\mathbf{1},\mathbf{1})$ | **0.99** | $(0,\mathbf{1},0)$ | **0.92** | $(0,\mathbf{1},0)$ | **0.98** |
| BP | $(0,0,\mathbf{1})$ | 0.00 | $(0,0,0)$ | 0.00 | $(0,0,0)$ | 0.00 |
| BRANCH | $(0,0,\mathbf{1})$ | **0.67** | $(0,0,0)$ | 0.00 | $(0,0,0)$ | **0.68** |
| BRANCH-FAST | $(0,0,\mathbf{1})$ | 0.00 | $(0,0,0)$ | **0.65** | $(0,0,0)$ | **0.76** |
| BRANCH-CONST | $(0,0,\mathbf{1})$ | **0.75** | $(0,0,0)$ | 0.00 | $(0,0,0)$ | **0.81** |
| STAR | $(0,0,0)$ | 0.00 | $(0,0,0)$ | **0.70** | $(0,0,0)$ | 0.00 |
| SUBGRAPH | $(0,0,0)$ | 0.00 | $(0,0,0)$ | 0.00 | $(0,0,0)$ | 0.00 |
| WALKS | $(0,0,0)$ | 0.00 | $(0,0,0)$ | 0.00 | $(0,0,0)$ | 0.00 |
| RING-OPT | $(0,0,\mathbf{1})$ | 0.00 | $(0,0,0)$ | 0.00 | $(0,0,0)$ | 0.00 |
| RING-MS | $(0,0,\mathbf{1})$ | 0.00 | $(0,0,0)$ | **0.64** | $(0,0,0)$ | 0.00 |
| RING-ML-SVM | $(0,0,0)$ | 0.00 | $(0,0,0)$ | 0.00 | $(0,0,0)$ | 0.00 |
| RING-ML-DNN | $(0,0,0)$ | 0.00 | $(0,0,0)$ | 0.00 | $(0,0,0)$ | 0.00 |
| PREDICT-SVM | $(0,0,0)$ | 0.00 | $(0,0,0)$ | 0.00 | $(0,0,0)$ | 0.00 |
| PREDICT-DNN | $(0,0,0)$ | 0.00 | $(0,0,0)$ | 0.00 | $(0,0,0)$ | 0.00 |
| *extensions of the paradigm LSAPE-GED* | | | | | | |
| MULTI-SOL | $(0,0,\mathbf{1})$ | **0.62** | $(0,0,0)$ | **0.38** | $(0,0,0)$ | **0.62** |
| CENTRALITIES | $(0,0,\mathbf{1})$ | **0.49** | $(0,0,0)$ | **0.41** | $(0,0,0)$ | **0.58** |
| *instantiations of the paradigm LP-GED* | | | | | | |
| F-1 | $(0,0,\mathbf{1})$ | 0.00 | $(0,0,0)$ | 0.00 | $(0,0,0)$ | 0.00 |
| F-2 | $(0,0,\mathbf{1})$ | 0.00 | $(0,0,0)$ | 0.00 | $(0,0,0)$ | 0.00 |
| COMPACT-MIP | $(0,0,0)$ | 0.00 | $(0,0,0)$ | 0.00 | $(0,0,0)$ | 0.00 |
| ADJ-IP | $(0,0,\mathbf{1})$ | 0.00 | $(0,0,0)$ | 0.00 | $(0,0,0)$ | 0.00 |
| *instantiations of the paradigm LS-GED* | | | | | | |
| 2-REFINE | $(0,0,\mathbf{1})$ | **0.66** | $(0,0,\mathbf{1})$ | **0.67** | $(0,0,\mathbf{1})$ | 0.00 |
| 3-REFINE | $(0,0,\mathbf{1})$ | 0.00 | $(0,0,\mathbf{1})$ | 0.00 | $(0,0,\mathbf{1})$ | 0.00 |
| BP-BEAM | $(0,0,\mathbf{1})$ | 0.00 | $(0,0,0)$ | **0.63** | $(0,0,\mathbf{1})$ | 0.00 |
| IBP-BEAM | $(0,0,\mathbf{1})$ | 0.00 | $(0,0,\mathbf{1})$ | 0.00 | $(0,0,\mathbf{1})$ | 0.00 |
| IPFP | $(\mathbf{1},0,\mathbf{1})$ | **0.67** | $(\mathbf{1},0,\mathbf{1})$ | **0.67** | $(\mathbf{1},0,\mathbf{1})$ | **0.67** |
| *extensions of the paradigm LS-GED* | | | | | | |
| MULTI-START | $(\mathbf{1},0,\mathbf{1})$ | **0.80** | $(\mathbf{1},0,\mathbf{1})$ | **0.68** | $(\mathbf{1},0,\mathbf{1})$ | **0.83** |
| RANDPOST | $(\mathbf{1},0,\mathbf{1})$ | **0.40** | $(\mathbf{1},0,\mathbf{1})$ | 0.00 | $(\mathbf{1},0,\mathbf{1})$ | **0.17** |
| *miscellaneous heuristics* | | | | | | |
| BRANCH-TIGHT | $(0,0,\mathbf{1})$ | 0.00 | $(0,0,0)$ | 0.00 | $(0,0,0)$ | 0.00 |
| SA | $(0,0,\mathbf{1})$ | 0.00 | $(0,0,0)$ | 0.00 | $(0,0,0)$ | 0.00 |

and miscellaneous heuristics are displayed green. We observe that, globally, NODE, IPFP, and 2-REFINE achieved the best aggregated joint lower bound scores, i.e., exhibited the best tradeoffs between tightness of the obtained upper bound, runtime, and classification coefficient (cf. Figure 6.21a). In

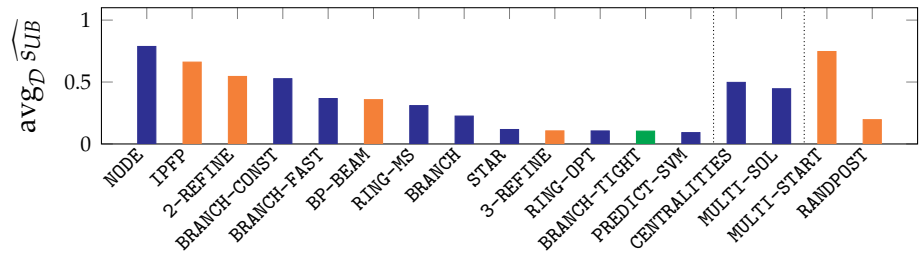

**(a)** Average aggregated joint upper bound scores.

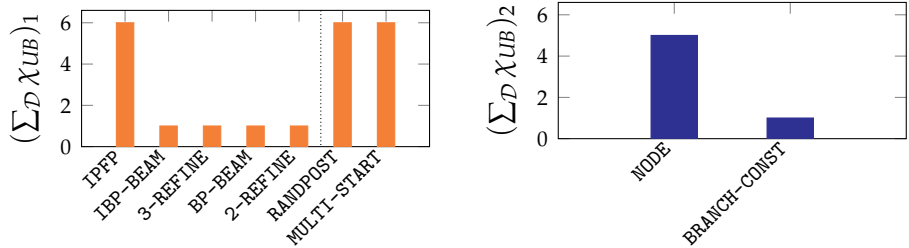

**(b)** Optimality w. r. t. tightness of upper bound.  **(c)** Optimality w. r. t. runtime among heuristics producing upper bounds.

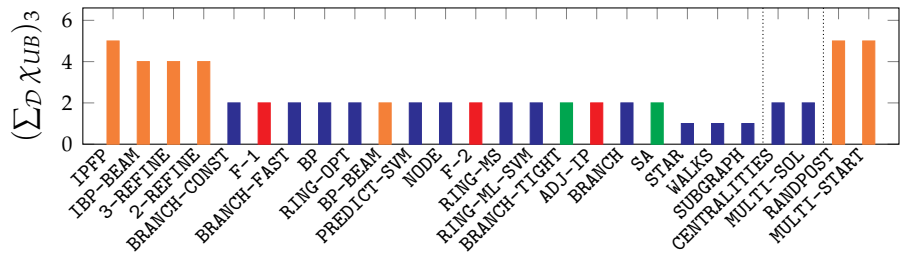

**(d)** Optimality w. r. t. upper bound classification coefficient.

**Figure 6.21.** Overview of results for upper bounds. Only non-zero statistics are displayed.

terms of runtime, NODE and BRANCH-CONST performed best (cf. Figure 6.21c). In terms of classification coefficient and tightness of the obtained upper bound, the instantiations of LS-GED performed best, with IPFP as the best performing heuristic among them (cf. Figure 6.21b and Figure 6.21d).

The average aggregated joint upper bound scores of both extensions CENTRALITIES and MULTI-SOL of the paradigm LSAPE-GED turned out to be smaller than 0.5 (cf. Figure 6.21a). That is, on average, instantiations of LSAPE-GED did not benefit from the extensions CENTRALITIES and MULTI-SOL. However, on each dataset, some instantiations of LSAPE-GED did benefit from the extensions, as some algorithms using CENTRALITIES and MULTI-SOL were

**Table 6.9.** Tightest average lower and upper bounds for all datasets and the gaps between them.

| dataset | $d^\star_{LB}$ | $d^\star_{UB}$ | gap in % |
|---|---:|---:|---:|
| AIDS | 73.45 | 76.18 | 3.58 |
| MUTA | 93.76 | 97.90 | 4.23 |
| PROTEIN | 302.80 | 307.65 | 1.58 |
| LETTER (H) | 4.72 | 4.75 | 0.63 |
| GREC | 898.83 | 904.70 | 0.65 |
| FINGERPRINT | 3.04 | 3.08 | 1.30 |
| average | — | — | 1.99 |

Pareto optimal on almost all datasets (cf. Tables 6.7 to 6.8).

We also observe that the average aggregated joint upper bound scores of the extensions `MULTI-START` and `RANDPOST` of the paradigm `LS-GED` are, respectively, clearly larger and clearly smaller than 0.5 (cf. Figure 6.21a). That is, on average, instantiations of `LS-GED` benefited from `MULTI-START` but not from `RANDPOST`. However, `RANDPOST` still turned out to be used by Pareto optimal algorithms on all datasets except for the datasets LETTER (H) and FINGERPRINT, which contain very small graphs. `MULTI-START` was used by Pareto optimal algorithms on all datasets (cf. Tables 6.7 to 6.8). Moreover, we see that, on all six datasets, algorithms using `MULTI-START` and `RANDPOST` yielded the tightest upper bounds and the best classification coefficients (cf. Figure 6.21b and Figure 6.21d).

**Gaps Between Lower and Upper Bounds.** Table 6.9 shows the tightest average lower and upper bounds $d^\star_{LB}$ and $d^\star_{UB}$ for all datasets and the percentual gaps between them. We see that the best upper bounds overestimate the best lower bounds (and hence, a fortiori, the exact GED) by at most 4.23 % and only 1.99 % on average. Given the hardness of the problem of exactly computing GED (cf. Section 3.1.1), this is a remarkable result. As mentioned above, the tightest upper bound $d^\star_{UB}$ was computed by `IPFP` on all datasets. On the dataset FINGERPRINT, the tightest lower bound $d^\star_{LB}$ was provided by `BRANCH-TIGHT`; on PROTEIN, it was provided by `F-2`; and on all other datasets, it was provided by `ADJ-IP`.

**Classification Coefficients vs. Tightness of Lower and Upper Bounds.** Figure 6.22 and Table 6.10 relate the lower bounds of the algorithms producing lower bounds to the obtained lower bound classification coefficients. Figure 6.22 contains plots for all datasets. In each of them, each black dot represents an algorithm that yields a lower bound and the red line visualizes the obtained linear regression model $c_{LB} \sim d_{LB}$. For each dataset, Table 6.10 summarizes the slopes and *p*-values of the models, as well as the maximum and average lower bound classification coefficients.

**Table 6.10.** Maximum and average lower bound classification coefficients for all datasets, and slopes and *p*-values of the linear regression models $c_{LB} \sim d_{LB}$.

| dataset | $c_{LB}^{\star}$ | avg $c_{LB}(\texttt{ALG})$ | $m_{LB}$ | $p_{LB}$ |
|---|---|---|---|---|
| AIDS | 0.15 | 0.13 | $1.11 \cdot 10^{-3}$ | $2.51 \cdot 10^{-5}$ |
| MUTA | 0.01 | 0.00 | $-2.33 \cdot 10^{-5}$ | $4.76 \cdot 10^{-1}$ |
| PROTEIN | 0.04 | 0.03 | $6.27 \cdot 10^{-5}$ | $1.07 \cdot 10^{-4}$ |
| LETTER (H) | 0.29 | 0.23 | $4.16 \cdot 10^{-2}$ | $2.87 \cdot 10^{-9}$ |
| GREC | 0.37 | 0.32 | $1.97 \cdot 10^{-4}$ | $9.29 \cdot 10^{-5}$ |
| FINGERPRINT | 0.12 | 0.09 | $3.38 \cdot 10^{-2}$ | $1.11 \cdot 10^{-9}$ |
| average | 0.16 | 0.14 | $1.28 \cdot 10^{-2}$ | — |

The first thing we note is that, on MUTA, all obtained classification coefficients either equal 0.00 or 0.01. This can be explained by the fact that, for both of its classes, MUTA contains graphs with very different numbers of nodes. This leads to large average distances within the classes and hence to a small difference between intra- and inter-class distances. As we have $c_{LB}(\texttt{ALG}) \in \{0.00, 0.01\}$ for all algorithms ALG that produce lower bounds, the obtained linear regression model has a very high *p*-value and hence is not statistically significant.

For all other datasets, the obtained linear regression models have *p*-values smaller than $10^{-3}$ and are hence highly significant. Furthermore, all linear regression models except the statistically insignificant model for MUTA have a positive slope. That is, tight lower bounds tend to go hand in hand with good classification coefficients.

Figure 6.23 and Table 6.11 relate the upper bounds of the algorithms producing upper bounds to the obtained upper bound classification coefficients. Figure 6.23 contains plots for all datasets. In each of them, each black dot represents an algorithm that yields an upper bound and the red line visualizes the obtained linear regression model $c_{UB} \sim d_{UB}$. For each dataset, Table 6.11 summarizes the slopes and *p*-values of the models, as well as the maximum and average upper bound classification coefficients.

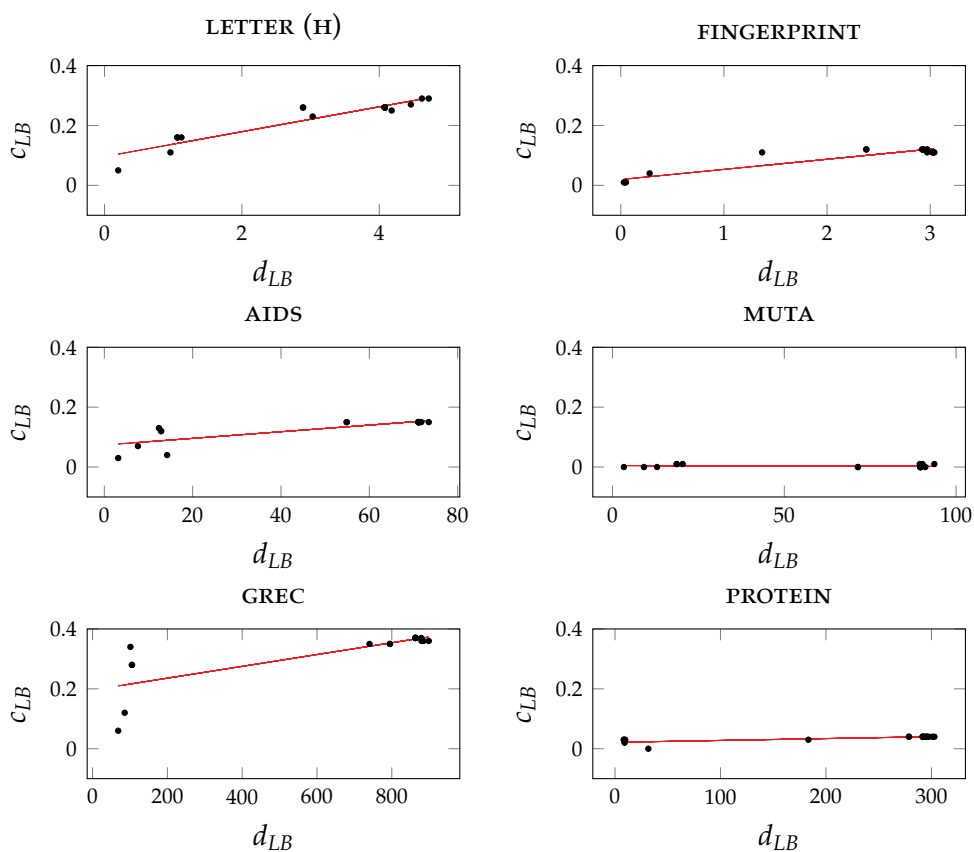

**Figure 6.22.** Average lower bounds vs. lower bound classification coefficients. Each black dot represents one algorithm that computes a lower bound. The linear regression model $c_{LB} \sim d_{LB}$ is displayed in red.

We again note that, on MUTA, all obtained classification coefficients either equal 0.00 or 0.01. Since we tested many more algorithms that compute upper bounds than algorithms that yield lower bounds,[7] the linear regression model $c_{UB} \sim d_{UB}$ for MUTA nonetheless has a $p$-value smaller than $10^{-3}$ and is hence still highly statistically significant. However, its $p$-value is much larger than the $p$-values of the linear regression models $c_{UB} \sim d_{UB}$ we obtained for the other datasets.

We observe that, for all datasets, the slopes of the linear regression models

---

[7]To be precise, we tested 19 algorithms that compute lower bounds and 173 algorithms that compute upper bounds. The reason for this is that the extensions of the paradigms LSAPE-GED and LS-GED only affect the upper bounds.

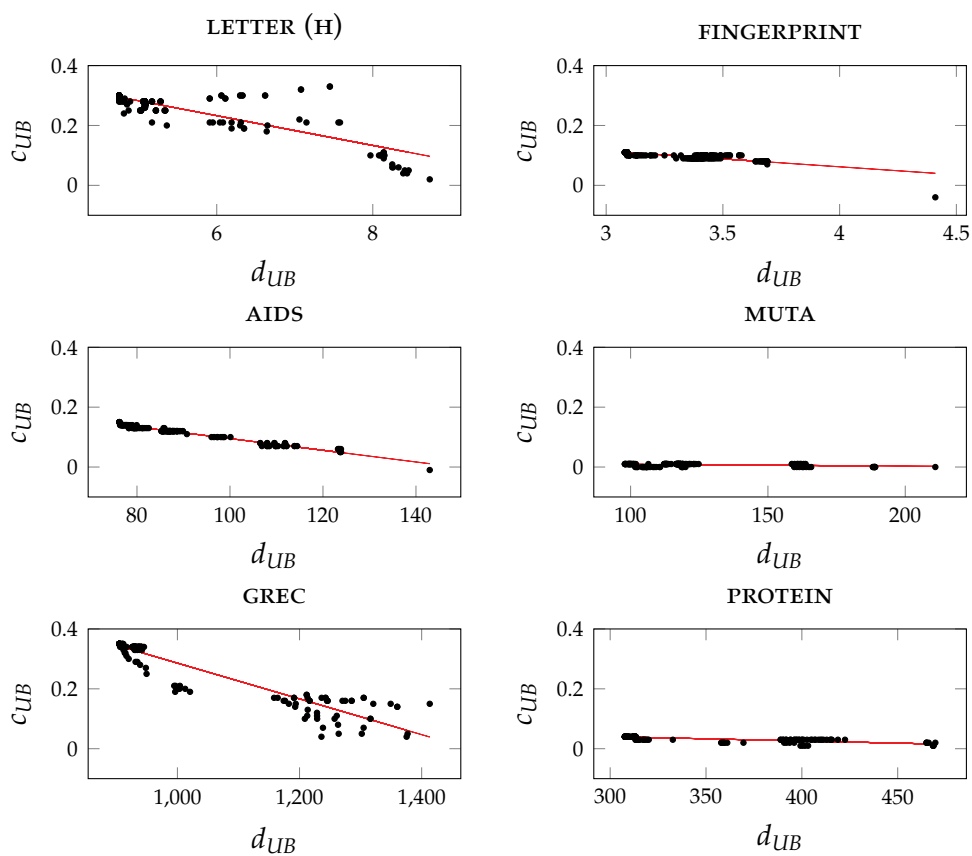

**Figure 6.23.** Average upper bounds vs. upper bound classification coefficients. Each black dot represents one algorithm that computes an upper bound. The linear regression model $c_{UB} \sim d_{UB}$ is displayed in red.

$c_{UB} \sim d_{UB}$ are positive. We can hence draw the same conclusion as for the lower bounds, namely, that tight upper bounds tend to go hand in hand with good classification coefficients. These findings allow us to positively answer the meta-question Q-1 raised in the introduction: It is indeed beneficial to use GED as a guidance for the design of graph distance measures that are to be used within pattern recognition frameworks.

The second meta-question Q-2 asked whether lower or upper bounds for GED are better suited for use as graph distance measures within classification frameworks. Since the classification coefficients induced by the lower and upper bounds turned out to be very similar, this question cannot be answered as straightforwardly as Q-1. However, there is a tendency: While the average lower bound classification coefficients were slightly better than the average upper bound classification coefficients, the opposite can be observed for the maximum lower and upper bound classification coefficients. Moreover, on average, the slopes of the linear regression models $c_{UB} \sim d_{UB}$ are slightly steeper than the slopes of the linear regression models $c_{LB} \sim d_{LB}$. Together, these observations suggest that the upper bound classification coefficients benefit more from tight upper bounds than the lower bound classification coefficients benefit from tight lower bounds. As a rule of thumb, we can hence conclude that tight upper bounds for GED (e. g., the upper bound computed by `IPFP`) should be used for classification purposes, if one is willing to invest a lot of time in the computation of the graph distance measure. Otherwise, a quickly computable lower bound such as the one produced by `BRANCH-CONST` or `BRANCH-FAST` should be employed.

**Table 6.11.** Maximum and average upper bound classification coefficients for all datasets, and slopes and $p$-values of the linear regression models $c_{UB} \sim d_{UB}$.

| dataset | $c^\star_{UB}$ | avg $c_{UB}(\texttt{ALG})$ | $m_{UB}$ | $p_{UB}$ |
|---|---|---|---|---|
| AIDS | 0.15 | 0.11 | $-1.95 \cdot 10^{-3}$ | $5.47 \cdot 10^{-136}$ |
| MUTA | 0.01 | 0.01 | $-5.36 \cdot 10^{-5}$ | $3.76 \cdot 10^{-5}$ |
| PROTEIN | 0.04 | 0.03 | $-1.45 \cdot 10^{-4}$ | $3.91 \cdot 10^{-43}$ |
| LETTER (H) | 0.33 | 0.25 | $-4.97 \cdot 10^{-2}$ | $3.67 \cdot 10^{-43}$ |
| GREC | 0.35 | 0.27 | $-5.97 \cdot 10^{-4}$ | $3.12 \cdot 10^{-82}$ |
| FINGERPRINT | 0.11 | 0.10 | $-5.29 \cdot 10^{-2}$ | $1.69 \cdot 10^{-35}$ |
| average | 0.17 | 0.13 | $-1.76 \cdot 10^{-2}$ | — |

### 6.8.4.3 Upshot of the Experiments

Our experiments show that, on the selected datasets and edit costs, the instantiation `ADJ-IP` of `LP-GED` computed the tightest lower bounds, followed by the `LP-GED` instantiation `F-2` and the anytime algorithm `BRANCH-TIGHT` presented in Section 6.3 above. Our heuristics `BRANCH` and `BRANCH-FAST` produced Pareto optimal lower bounds on all datasets except for the ones with constant edge edit costs, where their lower bounds are equivalent to the one produced by the faster heuristic `BRANCH-CONST`. The tightest upper bounds were computed by the instantiation `IPFP` of `LS-GED`, run with `RANDPOST` and `MULTI-START`. Our local search algorithm `2-REFINE` performed excellently, too, since it achieved a very good tradeoff between runtime, tightness, and classification performance, and, on the datasets containing small graphs, produced just or almost as tight upper bounds as `IPFP`.

Furthermore, our experiments provide thorough evidence to support the assumption that the tighter a lower or upper bound for GED, the better its performance when used as a graph distance measure within pattern recognition frameworks. They hence justify the ongoing competition for tight upper and lower bounds for GED. The experiments also indicate that if bounds for GED are to be used for classification purposed, one should resort to tight upper bounds, if one wants to optimize for classification performance, and to quickly computable lower bounds, if runtime performance is critical. Finally, the experiments show that, on sparse small to medium sized graphs such as the ones used in the experiments, the gaps between the tightest currently available lower and upper bounds for GED is very small. Given the hardness of the problem of computing GED, this is a surprising and remarkable result.

## 6.9 Conclusions and Future Work

In this chapter, we provided an systematic overview of state of the art methods for heuristically computing GED. Moreover, we presented the new LSAPE based heuristics `BRANCH`, and `BRANCH-FAST`, `RING`, and `RING-ML`, the anytime algorithm `BRANCH-TIGHT`, the extension `MULTI-SOL` of LSAPE based heuristics, the local search algorithm `K-REFINE`, and the framework `RANDPOST` for improving the upper bounds produced by local search algorithms. Extensive experiments were carried out to demonstrate the practical usefulness of the proposed methods. In particular, the experiments show that `BRANCH-TIGHT` produces the tightest available lower bound for GED in settings where editing edges is more expensive than editing nodes; that `BRANCH` and `BRANCH-FAST` yield excellent tradeoffs between runtime performance and tightness of the produced lower bound; that `RING` computes the tightest upper bound of all existing LSAPE based heuristics; that, on small graphs, `K-REFINE` is as accurate as but much faster than the most accurate existing local search algorithms; and that, on larger graphs, `RANDPOST` significantly tightens the upper bounds of all local search algorithms.

Moreover, the experiments confirm that, when used within classification frameworks, tight lower and upper bounds for GED perform better than loose bounds, and show that, in practice, the gap between the tightest available lower and upper bounds is very small. This last result implies that there is little room for further tightening lower or upper bounds for GED. Instead, we suggest that future work on the heuristic computation of GED should focus on the task of speeding up those existing heuristics that yield the tightest currently available bounds.

# — 7 —
# Conclusions and Future Work

In this thesis, we proposed various new techniques for exactly and heuristically computing GED and tested them against various competitors, which were re-implemented within an integrated environment. Moreover, we harmonized the GED definitions which are used, respectively, in the database and in the pattern recognition community, showed that computing GED is hard even on very simple graphs, presented a new reduction from GED to QAP for quasimetric edit costs, and provided a new efficient, generic, and easily implementable solver for LSAPE.

Let us briefly recapitulate some of the most interesting conclusions that can be drawn from the experiments reported in this thesis. Firstly, we have seen that exactly computing GED is feasible only for very small graphs and that the methods that scale best use generic IP solvers. This reflects the high theoretical complexity of computing GED but also indicates that there might be room for improvement, as faster, specialized algorithms might be in reach.

Secondly, it turned out that, on the test graphs considered in this thesis, GED can in practice be bounded within very tight margins via heuristics that compute lower and upper bounds. The tightest lower bounds are produced by our anytime algorithm `BRANCH-TIGHT`, if editing edges is more expensive than editing nodes, and by the LP based approaches `ADJ-IP` and `F-2`, otherwise. Our LSAPE based method `BRANCH-FAST` yields an excellent tradeoff between runtime and tightness of the obtained lower bound. The tightest upper bounds are produced by the local search algorithm `IPFP`. Our local search algorithm `K-REFINE` yields an excellent tradeoff between runtime and accuracy, as it is always much faster than `IPFP` and yields upper bounds which are as tight as the ones produced by `IPFP` on small graphs and only slightly looser on larger graphs. Moreover, on larger graphs, our framework RANDPOST significantly tightens the upper bounds of all local search based heuristics.

Thirdly, the evaluation of our new LSAPE solver FLWC showed that FLWC is the method of choice whenever an exact or heuristic GED algorithm has to solve LSAPE a subproblem. Not only is FLWC the fastest and most stable generic algorithm, but it can also be implemented much more easily than HNG-E, which constitutes the only generic competitor that is not outperformed by orders of magnitude.

We conclude this thesis by pointing out to two blind spots that are prevalent in virtually all works on GED—also in this thesis. Firstly, almost all exact or heuristic GED algorithms that have been proposed in the literature have been evaluated on the graphs contained in the IAM Graph Database Repository or in GREYC's Chemistry Dataset. However, these graphs are far from general, as all of them are very sparse and similar w. r. t. important topological properties such as cyclicity, planarity, and the number of connected components (cf. Table 2.3 and Table 2.4). It would therefore be very beneficial for the research community to generate new benchmark datasets which also contain denser graphs, and to use these datasets for systematically testing how the input graphs' topological properties affect the performances of algorithms for exactly or heuristically computing GED.

Secondly, although computing GED is hard even on very sparse graphs, there are specialized polynomially computable distance measures for restricted classes of graphs (cf. Section 3.2). These distance measures are mostly ignored in works on GED. To further evaluate the relevance of GED as a distance measure, it would hence be very interesting to compare the classification performance and runtime of (lower or upper bounds for) GED to the classification performance and runtime of these specialized, efficiently computable distance measures.

# — A —

# GEDLIB: A C++ Library for Graph Edit Distance Computation

In this appendix, we present GEDLIB, an easily extensible C++ Library for exact and heuristic GED computation. GEDLIB facilitates the implementation of competing GED algorithm within the same environment and hence fosters fair empirical comparisons. To the best of our knowledge, no currently available software is designed for this purpose. GEDLIB is freely available on GitHub:

> `https://github.com/dbblumenthal/gedlib`

In its current version, GEDLIB contains implementations of 9 different edit cost functions and all heuristic GED algorithms presented in this thesis. Further algorithms and edit costs can be implemented easily by implementing abstract classes contained in GEDLIB. For this, the user has access to standard libraries for blackbox optimization, mixed integer linear programming, the linear sum assignment problem with and without error-correction, deep neural networks, and support vector machines. GEDLIB provides a parser to load graphs given in the GXL file format. Alternatively, graphs with user-specified node ID, node, and edge label types can be constructed from within GEDLIB. Internally, GEDLIB uses the Boost Graph Library [66] for building the graphs and Eigen [53] for matrix operations.

GEDLIB has previously been presented in the following article:

– D. B. Blumenthal, S. Bougleux, J. Gamper, and L. Brun, "GEDLIB: A C++ library for graph edit distance computation", in *GbRPR 2019*, D. Conte,

J.-Y. Ramel, and P. Foggia, Eds., ser. LNCS, vol. 11510, Cham: Springer, 2019, pp. 14–24. DOI: 10.1007/978-3-030-20081-7_2

The remainder of this appendix is organized as follows: In Appendix A.1, the overall architecture of GEDLIB is sketched. In Appendix A.2, the user interface is presented. In Appendix A.3 and Appendix A.4, the abstract classes for implementing GED algorithms and edit cost functions are described. Appendix A.5 concludes the chapter. Note that the purpose of this chapter is to provide an overview of GEDLIB. Details, examples, and installation instructions can be found in the documentation.

## A.1 Overall Architecture

Figure A.1 shows the overall architecture of GEDLIB in a UML diagram. The entire library is contained in the namespace ged. The template parameters `UserNodeID`, `UserNodeLabel`, and `UserEdgeLabel` correspond to the types of the node IDs, the node labels, and the edge labels of the graphs provided by the user.

- The class template `ged::GEDEnv` provides the user interface. Via the public member functions of `ged::GEDEnv`, graphs can be constructed manually or loaded from GXL files, edit costs can be set, the algorithms implemented in GEDLIB can be run, and the results of the runs can be obtained. For users who do not want to provide extensions for GEDLIB, it suffices to get familiar with this class template.
- The abstract class template *ged::GEDMethod* provides a generic interface for implementing algorithms that exactly or approximately compute GED.
- The abstract class templates *ged::LSBasedMethod*, *ged::MIPBasedMethod*, and *ged::LSAPEBasedMethod* are derived from the generic interface provided by *ged::GEDMethod*. They provide more specialized interfaces for implementing algorithms using local search, mixed integer linear programming, and transformations to the linear sum assignment problem with error-correction.
- The abstract class template *ged::MLBasedMethod* is derived from the interface *ged::LSAPEBasedMethod*. It provides an interface for implementing algorithms that use deep neural networks or support vector machines for carrying out transformations from GED to the linear sum assignment problem with error-correction.
– The class template `ged::GEDData` contains the normalized input data on which all GED algorithms contained in GEDLIB operate. Via the public member functions of `ged::GEDData`, derived classes of *ged::GEDMethod* have access to the graphs that have been added to the environment and to the edit cost functions selected by the user.
– The abstract class template *ged::EditCosts* provides a generic interface for implementing edit cost functions.

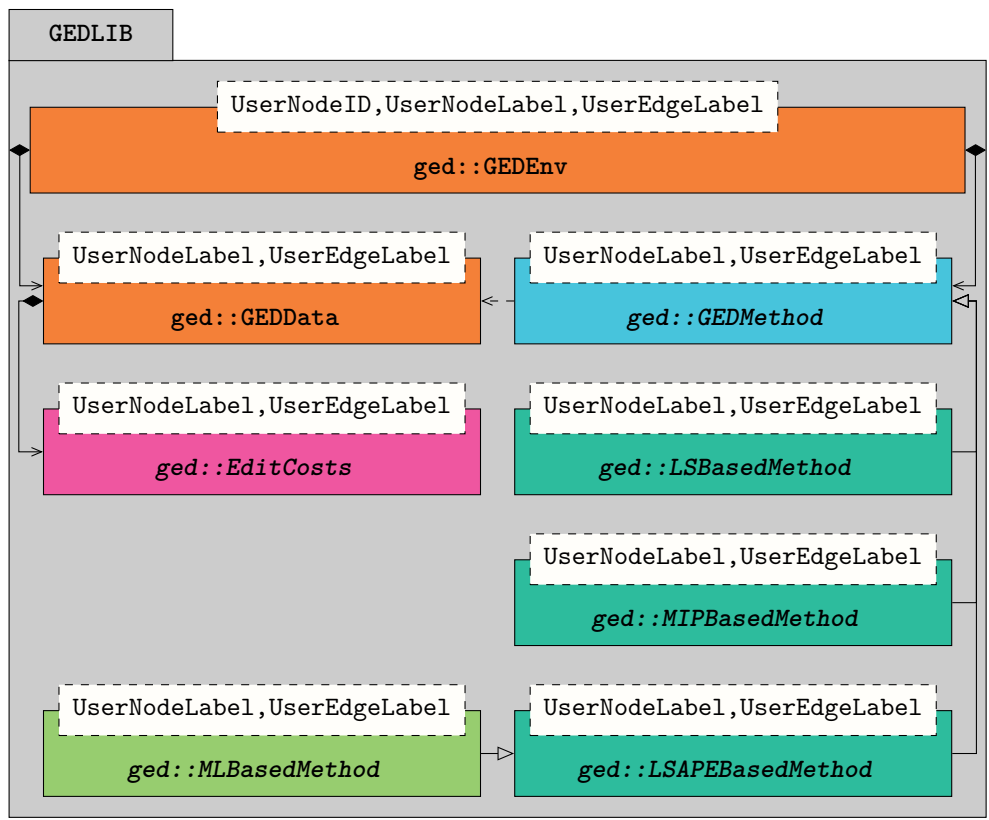

**Figure A.1.** Overall architecture of GEDLIB as a UML diagram.

## A.2 User Interface

In Figure A.2, the class template `ged::GEDEnv`, which constitutes the user interface of GEDLIB, is displayed in detail. By calling `add_graph()`,

`add_node()`, and `add_edge()`, the user can add labeled graphs to the environment. Alternatively, `load_gxl_graphs()` can be used to load graphs given in the GXL file format. For this, the template parameter `UserNodeID` must be set to `ged::GXLNodeID` a.k.a. `std::string`, and the template parameters `UserNodeLabel` and `UserEdgeLabel` must be set to `ged::GXLLabel` a.k.a. `std::map<std::string,std::string>`.

Calls to `set_edit_costs()` add edit cost functions to the environment. The user can either select one of the predefined edit cost functions or use her own implementation of *ged::EditCosts*. Calls to `init()` initialize the environment eagerly or lazily. If eager initialization is chosen, all edit costs between graphs contained in the environment are precomputed. Otherwise, the edit cost functions are evaluated on the fly. The member function `set_method()` selects one of the GED algorithms available in GEDLIB. Some algorithms accept options, which can be passed to `set_method()` as a string of the form `"[-<option> <arg>] [...]"`. Calls to `init_method()` initialize the selected method for runs between graphs contained in the environment, and calls to `run_method()` run the method between two specified graphs. The results of the runs (lower and upper bounds, runtimes, etc.) can be accessed via various getter member function.

| UserNodeID, UserNodeLabel, UserEdgeLabel | |
|---|---|
| **ged::GEDEnv** | |
| ... | *// misc. variables* |
| + add_graph() | *// adds a graph to the environment* |
| + add_node() | *// adds a node to a previously added graph* |
| + add_edge() | *// adds an edge to a previously added graph* |
| + load_gxl_graphs() | *// loads graphs given as GXL files* |
| + set_edit_costs() | *// selects the edit costs* |
| + init() | *// initializes the environment* |
| + set_method() | *// selects the GED method* |
| + init_method() | *// initializes the selected GED method* |
| + run_method() | *// runs the selected GED method* |
| ... | *// misc. member functions* |

**Figure A.2.** User interface.

## A.3 Abstract Classes for Implementing GED Algorithms

### A.3.1 Generic Interface

Figure A.3 details the abstract class template *ged::GEDMethod*, which provides the generic interface for implementing GED. The interface is defined by the virtual member functions starting with the prefix *ged_*. We here describe only the most important virtual member functions; the remaining ones are detailed in the documentation: *ged_run_()* runs the method between two input graphs, *ged_init_()* initializes the methods for the graphs that have been added to the environment, and *ged_parse_option_()* parses the options of the method. The following existing algorithms already implemented in GEDLIB are directly derived classes of *ged::GEDMethod*: ged::BranchTight (Section 6.3), ged::BranchCompact [113], ged::AnchorAwareGED [34], ged::HED [46], ged::Partition [113], ged::Hybrid [113], ged::SimulatedAnnealing [94].

| UserNodeLabel,UserEdgeLabel | |
|---|---|
| *ged::GEDMethod* | |
| ... | // misc. variables |
| − *ged_run_()* | // runs the method between two graphs |
| − *ged_init_()* | // initializes the method for the graphs in `ged_data_` |
| − *ged_parse_option_()* | // parses the options |
| ... | // misc. member functions |

**Figure A.3.** Generic interface for implementing GED algorithms.

### A.3.2 Interface for Algorithms Based on LSAPE

A popular approach for approximating GED is to use transformations to the *linear sum assignment problem with error-correction* (LSAPE). An instance of LSAPE consists of a cost matrix $\mathbf{C} = (c_{i,k}) \in \mathbb{R}_{\geq 0}^{(n+1) \times (m+1)}$. The task is to compute a mapping $\pi$ from rows to columns, such that each row except for $n + 1$ and each column expect for $m + 1$ is covered exactly once and $\mathbf{C}(\pi) := \sum_{(i,k) \in \pi} c_{i,k}$ is minimized. LSAPE can be solved optimally in cubic time [23]; in GEDLIB, we use the LSAPE toolbox [20] for solving LSAPE.[1]

If LSAPE is used for approximating GED$(G, H)$, $n$ and $m$ are set to $|V^G|$ and $|V^H|$, the first $|V^G|$ rows of **C** are associated with the nodes of $G$, the first $|V^H|$ columns of **C** are associated with the nodes of $H$, and the last rows and columns are associated with dummy nodes used for codifying node insertions and deletions. With this setup, each LSAPE solution $\pi$ corresponds to a *node map between G and H*, which, in turn, induces an edit path and hence an upper bound for GED$(G, H)$ [16]. LSAPE based heuristics for GED try to achieve tight upper bounds by encoding structural information of the input graphs into **C**. Moreover, some of them construct **C** such that $\min_\pi \mathbf{C}(\pi)$ lower bounds GED.

Figure A.4 shows the abstract class template *ged::LSAPEBasedMethod*, which provides the interface for implementing heuristics of this kind. The interface is defined by the virtual member functions starting with the prefix *lsape_*. The most important one is *lsape_populate_instance_()*, which populates the LSAPE instance **C**. The following algorithms implemented in GEDLIB are directly derived classes of *ged::LSAPEBasedMethod*: ged::Node [60], ged::Bipartite [83], ged::BranchUniform [113], ged::Subgraph [32], ged::Branch (Section 6.2), ged::BranchFast (Section 6.2), ged::Ring (Section 6.4), ged::Walks [50]. Additionally, all derived classes of *ged::LSAPEBasedMethod* can be run with the node centralities suggested in [88].

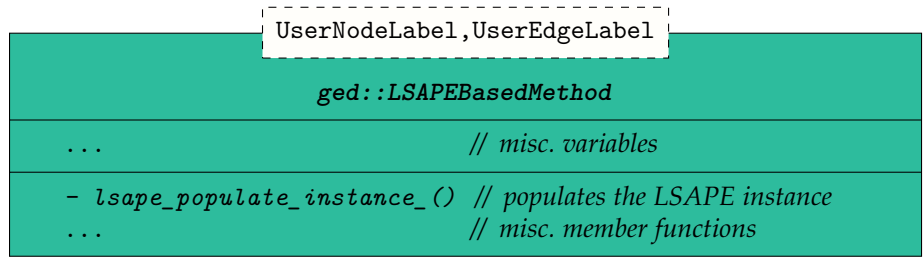

**Figure A.4.** Interface for implementing algorithms based on LSAPE.

---

[1]The LSAPE toolbox is available at https://bougleux.users.greyc.fr/lsape/.

### A.3.3 Interface for Algorithms Based on LSAPE and Machine Learning

Recently, it has been suggested to use deep neural networks or support vector machines for carrying out the transformation from GED to LSAPE. Given two graphs $G$ and $H$, feature vectors are constructed for all node substitutions, deletions, and insertions, and the matrix **C** is defined as $c_{i,k} := 1 - p^\star(i,k)$. Here, $p^\star(i,k)$ is the confidence of a machine learning framework (either a deep neural network or a support vector machine) that the feature vector associated to the node edit operation corresponding to row $i$ and column $k$ is contained in an optimal node map.

Figure A.5 details the abstract class template *ged::MLBasedMethod*, which provides the interface for algorithm adopting this paradigm. For implementing the interface, it suffices to override the virtual member functions starting with the prefix *ml_*. The most important ones are the three virtual member functions of the form *ml_populate_*_feature_vector_()*, which construct the feature vectors associated to the node edit operations. Derived classes of *ged::MLBasedMethod* do not have to implement the machine learning frameworks, as *ged::MLBasedMethod* offers support for artificial deep neural networks (using FANN [80]) and support vector machines (using LIBSVM [33]).[2] The following algorithms implemented in GEDLIB are directly derived classes of *ged::MLBasedMethod*: ged::BipartiteML [90], ged::RingML (Section 6.4).

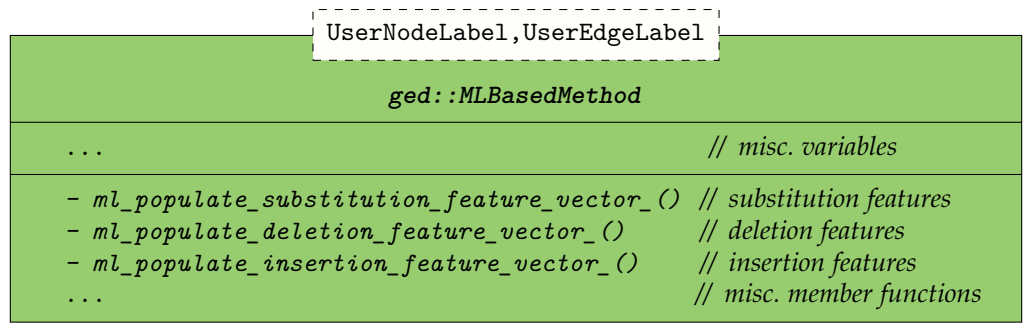

**Figure A.5.** Interface for implementing algorithms based on LSAPE and machine learning.

---

[2]FANN is available at http://leenissen.dk/fann/wp/; LIBSVM is available at https://www.csie.ntu.edu.tw/~cjlin/libsvm/.

### A.3.4 Interface for Algorithms Based on MIP

Another approach for exactly or approximately computing GED is to rephrase the problem of computing GED$(G, H)$ as a mixed integer programming (MIP) problem. GED$(G, H)$ can then be computed exactly by calling an MIP solver. Alternatively, lower bounds for GED$(G, H)$ can be obtained by solving the linear programming (LP) relaxations of the MIP formulations.

Figure A.6 shows the abstract class template *ged::MIPBasedMethod*, which provides the interface for GED algorithms that use MIP formulations. The virtual member functions that define the interface start with the prefix *mip_* . The most important one is *mip_populate_model_()*, which constructs the employed MIP formulation and must be overridden by all derived classes. In GEDLIB, we use Gurobi [54] as our MIP and LP solver. Gurobi is commercial software but offers a free academic license. For users who cannot obtain a license for Gurobi, the installation script distributed with GEDLIB offers the option to install GEDLIB without *ged::MIPBasedMethod* and its derived classes. The following algorithms implemented in GEDLIB are directly derived classes of *ged::MIPBasedMethod*: ged::CompactMIP (Section 5.4), ged::F1 [69], ged::F2 [69], ged::BLPNoEdgeLabels [60].

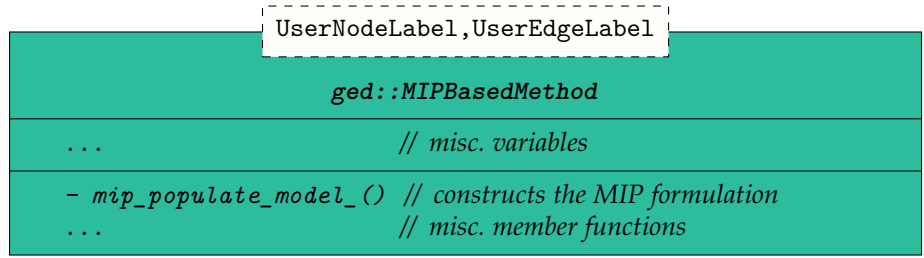

**Figure A.6.** Interface for implementing algorithms based on MIP.

### A.3.5 Interface for Algorithms Based on Local Search

Another popular approach for upper bounding GED is to use variants of local search to systematically vary a previously computed or randomly generated node map, such that the cost of the induced edit path decreases. Figure A.7 shows the abstract class template *ged::LSBasedMethod*, which provides the interface for algorithms using local search. The prefix *ls_* marks the virtual member functions defining the interface. The most important one is *ls_run_from_initial_solution_()*, which runs the local search from an initial node map. The following algorithms implemented in GEDLIB are directly derived classes of *ged::LSBasedMethod*: ged::IPFP [12, 22, 25], ged::BPBeam [44, 92], ged::Refine [18, 111]. Note that ged::Refine not only implements the algorithm REFINE, but also the improved and generalized version K-REFINE proposed in Section 6.6. Moreover, *ged::LSBasedMethod* provides support for running all derived classes with parallel MULTI-START as suggested in [41], and with the stochastic generator RANDPOST presented in Section 6.7.

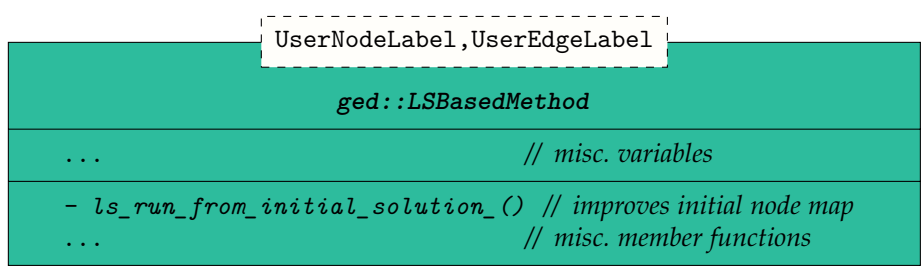

**Figure A.7.** Interface for implementing algorithms based on local search.

## A.4 Abstract Class for Implementing Edit Costs

Figure A.8 shows the abstract class template *ged::EditCosts,* which provided the interface for implementing edit cost functions. The virtual member functions *\*_del_cost_fun()* compute the cost of deleting a node or an edge with a given label, the functions *\*_ins_cost_fun()* compute the insertions costs, and the functions *\*_rel_cost_fun()* compute the costs for relabeling a node or an edge. The functions *vectorize_\*_label()* return vector representations of the node and the edge labels, which are required by some methods. In GEDLIB, edit costs are available for the datasets AIDS, FINGERPRINT, GREC, LETTER, MUTAGENICITY, and PROTEIN from the IAM Graph Database [87], for the datasets ACYCLIC, ALKANE, PAH, and MAO from GREYC's Chemistry Dataset, and for the dataset CMU-GED from the Graph Data Repository for Graph Edit Distance [2].[3] We also provide constant edit cost functions that can be used with any data.

---

[3]The IAM Graph Database is available at http://www.fki.inf.unibe.ch/databases/iam-graph-database; GREYC's Chemistry Dataset is available at https://brunl01.users.greyc.fr/CHEMISTRY/; and the Graph Data Repository for Graph Edit Distance is available at http://www.rfai.li.univ-tours.fr/PublicData/GDR4GED/home.html.

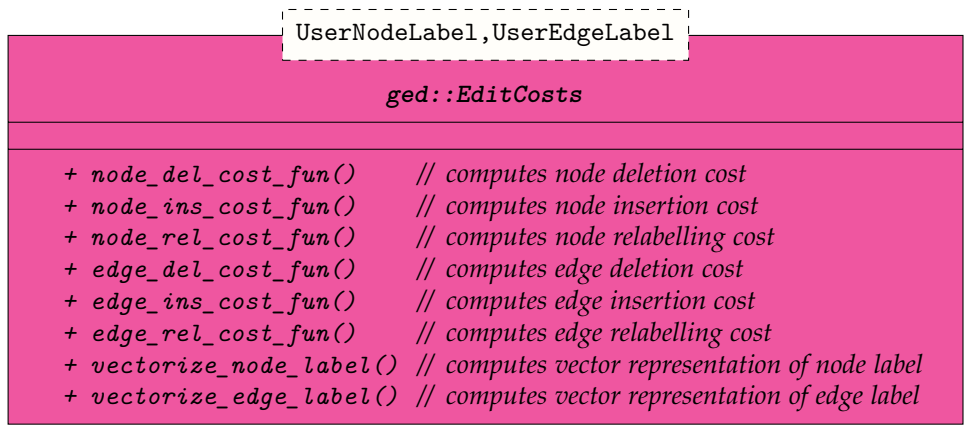

**Figure A.8.** Interface for implementing edit costs.

## A.5 Conclusions and Future Work

In this paper, we have presented GEDLIB, a C++ library for GED computations. GEDLIB currently implements 24 different GED algorithms and 9 different edit cost functions designed for datasets which are widely used in the research community. In the future, we will provide Python and MATLAB bindings for better usability. Moreover, we would like to encourage authors of algorithms and edit costs that are not implemented in GEDLIB to commit their work to GEDLIB.

# List of Definitions

# List of Theorems, Propositions, and Corollaries

# List of Tables

# LIST OF FIGURES

# List of Algorithms

**A**

A\*  Node based best-first algorithm for exactly computing GED presented in [89].

ADJ-IP  IP formulation of GED presented in [60].

**B**

BP  Instantiation of LSAPE-GED presented in [83].

BP-BEAM  Instantiation of LS-GED presented in [92].

BRANCH  Instantiation of LSAPE-GED presented in Section 6.2.1.

BRANCH-COMPACT  Heuristic for lower bounding GED presented in [113].

BRANCH-CONST  Instantiation of LSAPE-GED presented in [113, 114].

BRANCH-TIGHT  Heuristic for lower and upper bounding GED presented in Section 6.3.

BRANCH-FAST  Instantiation of LSAPE-GED presented in Section 6.2.2.

**C**

CENTRALITIES  Extension of LSAPE-GED presented in [40, 88].

COMPACT-MIP  MIP formulation of GED presented in Section 5.4.

CSI-GED  Edge based depth-first algorithm for exactly computing GED presented in [52] and generalized in Section 5.3.

**D**

DFS-GED  Node based depth-first algorithm for exactly computing GED presented in [4] and improved in Section 5.2.

**E**

EBP  Optimal LSAPE solver presented in [82, 83].

**F**

| | |
|---|---|
| `F-3` | IP formulation of GED presented in [42]. |
| `F-2` | IP formulation of GED presented in [69]. |
| `F-1` | IP formulation of GED presented in [68, 69]. |
| `FBP` | Optimal LSAPE solver presented in [100]. |
| `FBP-0` | Optimal LSAPE solver presented in [100]. |
| `FLWC` | Optimal LSAPE solver presented in Section 4.2. |

**H**

| | |
|---|---|
| `HED` | Heuristic for lower bounding GED presented in [46]. |
| `HNG-E` | Optimal LSAPE solver presented in [24]. |
| `HYBRID` | Heuristic for lower bounding GED presented in [113]. |

**I**

| | |
|---|---|
| `IBP-BEAM` | Instantiation of `LS-GED` presented in [44]. |
| `IPFP` | Instantiation of `LS-GED` presented in [22, 25] and Section 3.3. |
| `IPFP-QAPE` | Variant of `IPFP` that uses the reduction to QAPE presented in [25]. |
| `IPFP-B-QAP` | Variant of `IPFP` that uses the reduction to QAP presented in [22]. |
| `IPFP-C-QAP` | Variant of `IPFP` that uses the reduction to QAP presented in Section 3.3. |

**K**

| | |
|---|---|
| `K-REFINE` | Instantiation of `LS-GED` presented in Section 6.6. |

**L**

| | |
|---|---|
| `LP-GED` | Paradigm for lower and upper bounding GED via linear programming presented in Section 6.1.2. |
| `LS-GED` | Paradigm for upper bounding GED via local search presented in Section 6.1.3. |
| `LSAPE-GED` | Paradigm for lower and upper bounding GED via transformations to LSAPE presented in Section 6.1.1. |

**M**

| | |
|---|---|
| `MULTI-START` | Extension of `LS-GED` presented in [41]. |
| `MULTI-SOL` | Extension of `LSAPE-GED` presented in Section 6.5. |

**N**

| | |
|---|---|
| `NGM` | Incomplete instantiation of `LSAPE-GED` presented in [39]. |
| `NODE` | Instantiation of `LSAPE-GED` presented in [60]. |

**P**

| | |
|---|---|
| `PARTITION` | Heuristic for lower bounding GED presented in [113]. |
| `P-DFS-GED` | Parallel version of `DFS-GED` presented in [3]. |
| `PREDICT` | Incomplete instantiation of `LSAPE-GED` presented in [90]. |
| `PREDICT-DNN` | Variant of `PREDICT` that uses DNNs to compute the probability estimates. |
| `PREDICT-SVM` | Variant of `PREDICT` that uses 1-SVMs to compute the probability estimates. |

**R**

| | |
|---|---|
| `RANDPOST` | Extension of `LS-GED` presented in Section 6.7. |
| `REFINE` | Instantiation of `LS-GED` presented in [111]. |
| `RING` | Instantiation of `LSAPE-GED` presented in Section 6.4.4.1. |
| `RING-ML-DNN` | Variant of `RING-ML` that uses DNNs to compute the probability estimates. |
| `RING-ML-SVM` | Variant of `RING-ML` that uses 1-SVMs to compute the probability estimates. |
| `RING-MS` | Variant of `RING` that uses multiset intersection to compute the node and edge set distances. |
| `RING-GD` | Variant of `RING` that uses a greedy LSAPE solver to compute the node and edge set distances. |
| `RING-OPT` | Variant of `RING` that uses an optimal LSAPE solver to compute the node and edge set distances. |
| `RING-ML` | Instantiation of `LSAPE-GED` presented in Section 6.4.4.2. |

**S**

| | |
|---|---|
| `SA` | Heuristic for upper bounding GED presented in [94]. |
| `SFBP` | Optimal LSAPE solver presented in [99, 101]. |
| `STAR` | Instantiation of `LSAPE-GED` presented in [111]. |
| `SUBGRAPH` | Instantiation of `LSAPE-GED` presented in [32]. |

**W**

| | |
|---|---|
| `WALKS` | Instantiation of `LSAPE-GED` presented in [50]. |

# List of Acronyms

**Symbols**
**1-SVM**           One-class support vector machine.
**3-PARTITION**     3-partition problem.

**B**
**BED**             Branch edit distance.

**D**
**DNN**             Deep neural network.

**G**
**GED**             Graph edit distance.
**GI**              Graph isomorphism problem.
**GT**              Graph transformation problem.

**I**
**IP**              Integer linear program.

**L**
**LP**              Linear program.
**LSAP**            Linear sum assignment problem.
**LSAPE**           Linear sum assignment problem with error-correction.

**M**
**MIP**             Mixed integer linear program.

**Q**
**QAP**             Quadratic assignment problem.
**QAPE**            Quadratic assignment problem with error-correction.

**S**
**SGI**        Subgraph isomorphism problem.
**SVC**        Support vector classifier.

**T**
**TED**        Tree edit distance.